\documentclass[12pt]{article} 
\pdfoutput=1
\usepackage[nosort]{cite}

 \usepackage{draft} 
 \usepackage{graphicx,color,subfig,upgreek,mathtools}
\usepackage{amsfonts,amssymb,amsmath,multirow}
\usepackage{mathabx,epsfig}
\usepackage{pifont}
\usepackage{slashed}
\usepackage{mciteplus}
\usepackage{skak}
\DeclareFontFamily{OT1}{pzc}{}
\DeclareFontShape{OT1}{pzc}{m}{it}{<-> s * [1.10] pzcmi7t}{}
\DeclareMathAlphabet{\mathpzc}{OT1}{pzc}{m}{it}

\usepackage{physics}
\newcommand{\qbin}[2]{\begin{bmatrix}{#1}\\ {#2}\end{bmatrix}}

\def\be#1\ee{\begin{align}#1\end{align}}

\def\CA{{\mathcal A}}
\def\CB{{\mathcal B}}
\def\CC{{\mathcal C}}

\def\CH{{\mathcal H}}
\def\CI{{\mathcal I}}

\def\CL{{\mathcal L}}
\def\CM{{\mathcal M}}
\def\CN{{\mathcal N}}

\def\CP{{\mathcal P}}

\def\CS{{\mathcal S}}
\def\CT{{\mathcal T}}

\def\ZVW{Z_{\mathrm{VW}}}

\def\Pol{\mathrm{Pol}}
\def\ker{\mathrm{ker}}
\def\im{\mathrm{im}}

\def\C{\mathbb{C}}
\def\L{\Lambda}
\def\Z{\mathbb{Z}}
\def\R{\mathbb{R}}

\def\acts{\mathrel{\reflectbox{$\righttoleftarrow$}}}

\def\tilde{\widetilde}
\renewcommand{\bar}{\overline}
\renewcommand{\hat}{\widehat}
\def\^{{\wedge}}
\def\*{{\star}}

\definecolor{ao}{rgb}{0.13, 0.55, 0.13}

\newcommand{\cmark}{{\color{ao}\ding{51}}} 
\newcommand{\xmark}{{\color{red}\ding{55}}}

\usepackage{tikz-cd}
\usepackage{datetime}
\begin{document}

\begin{titlepage}


%

\begin{center}

\title{Generalized Global Symmetries of $T[M]$ Theories: \\ Part II}

\vspace{-.2in}
\author{Sergei Gukov$^{1,a}$, Po-Shen Hsin$^{2,b}$, Du Pei$^{3,c}$\\ \vspace{.05in}
with an appendix by Sunghyuk Park$^{4,d}$
}
\vspace{-.3in}
\address{${}^{1}$Walter Burke Institute for Theoretical Physics, California Institute of Technology,
Pasadena, CA 91125, USA}
\vspace{-.1in}
\address{${}^2$ Department of Mathematics, King’s College London, Strand, London WC2R 2LS, UK.}
\vspace{-.1in}
\address{${}^3$ Centre for Quantum Mathematics, University of Southern Denmark, Campusvej 55, 5230 Odense, Denmark}
\vspace{-.1in}
\address{${}^4$ Department of Mathematics and Center of Mathematical Sciences and Applications, Harvard University, Cambridge, MA 02138, USA}
\vspace{-.1in}
 \email{$^a$gukov@theory.caltech.edu, $^b$po-shen.hsin@kcl.ac.uk, $^c$dpei@imada.sdu.dk, $^d$sunghyukpark@math.harvard.edu}

\end{center}

%


\abstract{
We continue the investigation of symmetries and anomalies of $T[M]$ theories obtained by compactifying 6d SCFTs on an internal manifold $M$. We extend the notion of ``polarizations on a manifold $M$'' to cases where $M$ may have boundaries or defects. Through examples with $M$ of dimension two, three, and four, we illustrate recurring themes in compactifications---for instance, the important roles played by Kaluza--Klein modes, and how the generalized symmetries (including higher-group and non-invertible ones) of $T[M]$, together with their anomalies, arise from non-trivial combinations of the parent 6d symmetries and the geometric structures of the internal manifold.  For each dimension, we also focus on several topics that are especially interesting in that setting. These include: for 2-manifolds, the geometry of the ``full moduli space'' of $T[M_2]$ and its interaction with polarizations and symmetries; for 3-manifolds, the effect of torsion in homology on the spectrum of line operators in $T[M_3]$, together with applications to the study of quantum invariants such as $\hat Z_a(M_3, q)$; and for 4-manifolds, predictions for VOA$[M_4]$ following from symmetries of $T[M_4]$, as well as the construction of a new invariant of 4-manifolds that depends on two ``$q$-parameters.'' Along the way, we discuss a range of topics that are of independent interest, such as how non-invertible symmetries in higher dimensions can become invertible under compactification, how to classify defects in quantum field theory via their response to a change of framing, and the interplay between $\hat Z_a$ and volume conjectures.}

\vfill


\vfill

\end{titlepage}

\eject
\tableofcontents

\unitlength = .8mm

\setcounter{tocdepth}{3}

\section{Introduction}

In this paper, as a sequel to our earlier work \cite{Gukov:2020btk}, we continue the investigation of compactifications of six-dimensional superconformal field theories (6d SCFTs) to lower-dimensional systems through the lens of generalized symmetries. Although we explore a broad range of topics and encounter a variety of phenomena, several recurring themes provide organizing principles that unify the different threads of the discussion.

\subsubsection*{The bulk perspective}

We will study various symmetries, both invertible and non-invertible, and their 't Hooft anomalies, using the bulk topological quantum field theory (TQFT) whose boundary supports the interacting system. This method of characterizing symmetries in quantum systems with bulk TQFTs has been used extensively in the recent literature, including \cite{Ji:2019jhk,Gaiotto:2020iye,Apruzzi:2021nmk,Kaidi:2022cpf,Freed:2022qnc,Zhang:2023wlu,Cordova:2023bja}.
There are several advantages of the ``bulk perspective,'' such as:
\begin{itemize}
    \item The bulk topological defects can reveal hidden symmetry (both invertible and non-invertible) on the boundary, and give constraints on the possibly strongly-interacting boundary physics via the correlation functions of the defects. 

    \item We can organize polarizations, which specify versions of the boundary theories with physically sensible spectrum of operators (see Section~\ref{sec:reviewpolarizationclosed} for a review and \cite{Witten:1998wy,Freed:2012bs} for earlier work), to the topological boundary conditions of the bulk theory.\footnote{In general, it can be a topological domain wall between the bulk theory and an invertible phase, which describes the anomaly of the remaining symmetries. The discussion can also be framed as constructing ``absolute'' (or, more generally, ``projective''---but we will not attempt to distinguish them at the level of terminology) theories from a ``relative'' theory. The relative theory is well defined as a boundary condition, but not as a standalone QFT by itself. For example, its anomaly polynomials can have fractional coefficients (see Section~\ref{sec:FractionalA}), similar to the fractional quantum hall response from anomalous one-form symmetry~\cite{Cheng:2022nji}.}

    \item Conversely, higher-dimensional gapped systems above (2+1)d are not very well understood, and the boundary theory can provide information about the gapped theory in the bulk, similar to how the representation theory for the rational chiral conformal field theory in (1+1)d provides insights into the bulk Chern--Simons theory.
\end{itemize}

However, if one directly compactifies the 7d bulk theory $\CT^{\rm bulk}$ to obtain the theory $\CT^{\rm bulk}[M_d]$ in $7-d$ dimensions, the resulting bulk description generally captures only a subset of the symmetries of the boundary theory $T[M_d]$. In fact, there can be additional symmetries that are already present in 6d but not encoded in $\CT^{\rm bulk}$, with one interesting example being the ``universal $\Z_2$'' symmetry of 6d $(1,0)$ theories discussed in Section~\ref{sec:Universal}. Another source, leading to the next point, is the emergence of additional symmetries under compactification.

\subsubsection*{Emergent symmetries}

One scenario for new symmetries to emerge is when the internal manifold $M$ possesses isometries. This often leads to higher-group symmetries resulting from the mixture of the isometry with other global symmetries  \cite{Gukov:2020btk} 
(see {e.g.}~\cite{Kapustin:2013uxa,Cordova:2018cvg,Benini:2018reh} for reviews of higher group symmetries and some applications in physics), which can be understood by decomposing the background gauge field $C$ for an $n$-form symmetry in terms of cocycles on the internal manifolds $M$,
\begin{equation}
    C_{n+1}=\sum_{p_i\leq n+1} B_{n+1-p_i}\wedge \eta_{p_i},\quad \eta_{p_i}\in H^{p_i}(M)~.
\end{equation}
When we incorporate the 0-form symmetry from isometries on $M$, $\eta$ will be modified to be equivariant cocycles, while the condition $dC_{n+1}=0$ implies non-trivial mixing between the isometry and the symmetries with background gauge fields $B_{n+1-p_i}$.\footnote{
This is related to symmetries in the sigma model with target space $M_d$ (see Section~\ref{sec:sigma}), and reminiscent of the loop group symmetry that arises from compactification with extra circle direction, which is present {e.g.}~Fermi liquid or ``Ersatz Fermi liquids" \cite{PhysRevX.11.021005}. 
The above decomposition can be viewed as an $(n+1)$-dimensional sigma model with target space $M$ with symmetries generated by (1) defects of lower dimensions $p_i\leq n+1$ that are decorated with $p_i$-dimensional Berry phase $H^{p_i}(M)$, and (2) isometry on $M$. The condition $dC_{n+1}=0$ is the non-anomalous condition for the symmetries: the coupling to background gauge fields in the sigma model only depend on $n+1$ dimensional manifold where the sigma model lives. If the symmetry is anomalous in the sigma model, this implies that $dC_{n+1}\neq 0$, and thus in higher dimensional theory before compactification the $n$-form symmetry described by $C_{n+1}$ is already a $(n+1)$-group symmetry.}

Another closely related source---discussed in Section~\ref{sec:accidental} which we refer to as “predictable accidental symmetries”---arises when $M$ admits a fibration.

We also frequently encounter non-invertible symmetries (see, e.g.,~\cite{Shao:2023gho,Schafer-Nameki:2023jdn} for reviews), generated by topological defects that do not obey group-like fusion rules. Such non-invertible symmetries have been investigated in superconformal field theories, for example in~\cite{Bashmakov:2022jtl,Antinucci:2022cdi,Bhardwaj:2022yxj}. We will show that, after compactification, non-invertible symmetries can sometimes give rise to emergent {\it invertible} symmetries. This provides another motivation for studying non-invertible symmetry in compactifications: they can be responsible for hidden invertible symmetries in the compactified theory that do not originate from any invertible symmetry in higher dimensions.

\subsubsection*{Boundaries and defects}

In real experiments, all systems have boundaries, and boundary conditions can provide significant insight into the dynamics of the systems themselves. In this work, one of our primary focuses is the compactification on manifolds with boundaries. Such compactifications produce coupled bulk-boundary systems, which are constrained by the bulk TQFT (cf.~Figure~\ref{fig:TMopen}). 

In Section~\ref{sec:boundary}, we generalize our previous definition of a polarization on a closed manifold to one that also encompasses manifolds with either a boundary or a defect, and discuss how these structures affect the symmetries of the coupled system. A large class of examples is explored in Section~\ref{sec:4d}, where $M$ is taken to be a 2-manifold with boundaries.

Defects, which include and generalize boundary conditions, play a variety of interesting roles in compactification. In Section~\ref{sec:symmetry} alone, we discuss condensation defects, twisted compactification with defect insertions, defects associated with the action of the mapping class group, and the classification of defects via their framing anomalies. 

Since our primary interest is in compactifications that preserve supersymmetry, there is a class of supersymmetric defects that becomes particularly important. This leads to the next point.

\subsubsection*{Moduli spaces and BPS objects}

One of the most effective tools for understanding the dynamics of a supersymmetric theory is the study of its moduli space of vacua and its spectrum of BPS operators. Both are sensitive to global features determined by the choice of polarization, making them a particularly fruitful venue for our discussion. As a consequence, much of Section~\ref{sec:4d},~\ref{sec:3d}, and~\ref{sec:2d} is devoted to these two topics. These discussions include:
\begin{enumerate}
    \item A description of the ``full'' moduli space of $T[M_2]$---which becomes a combined moduli space of the bulk-boundary coupled system when $M_2$ is not closed---and its relation to the moduli spaces of class-$S$ SCFTs;
    \item  The interplay between moduli spaces and BPS states in the ``4d symplectic duality,'' which relates the geometry of the Coulomb branch to a sector of BPS operators;
    \item The spectrum of extended operators in the $T[M_d]$ theory when the homology of $M_d$ contains torsion;
    \item   Properties of the $\hat Z$-invariant, which counts BPS states in $T[M_3]$, and its connection to the volume conjecture;
    \item How global structures can be detected via modules of VOA$[M_4]$---a BPS-protected subsector of $T[M_4]$.
\end{enumerate}
Here the $T[M_d]$ theory is implicitly assumed to be one obtained from a 6d $(2,0)$ theory, except in the third point, where the statement is more general and continues to hold even in the non-supersymmetric setting.

\subsubsection*{KK modes}

Closely related to the two topics above are the Kaluza--Klein (KK) modes that arise in compactifications and play important roles in the resulting theories. They influence the global structure of moduli spaces while contributing to BPS spectra. A careful treatment of them leads to a proposal for partition functions of 6d theories on $M_4\times T^2$ ``with two $q$'s.'' 

For instance, the compactification of a 6d $(2,0)$ theory on $T^2$ produces a gauge theory at low energy---given locally by an $\CN=4$ super--Yang--Mills theory but differing in global aspects---which carries an instanton number. In any partition function that sums over instanton number sectors, one can weigh the sum by a fugacity parameter $q_{\rm gauge}$, which is related to the low-energy gauge coupling. In addition, there is also a tower of KK modes, whose masses are controlled by another parameter $q_\text{KK}$. Although the ``natural'' values for these parameters are equal, both given by $q=e^{2\pi i\tau}$ with $\tau$ the complex modulus of the $T^2$, we argue in Section~\ref{sec:2dqq} that they can actually be made independent. This allows one to define new invariants of 4-manifolds that depend on two $q$-parameters using this deformed partition function, $Z[M_4\times T^2;q_\text{gauge},q_\text{KK}]$. 

Separating the two $q$'s is not only useful for producing more refined invariants, but also resolves a paradox concerning the modular weight of the Vafa--Witten partition function. In particular, it explains the difference between the modular anomaly predicted from the anomalies of the 6d theory and that observed in \cite{Vafa:1994tf} for the generating function of the Euler characteristics of instanton moduli spaces, with the discrepancy precisely corresponding to the contribution of KK modes.

\subsubsection*{Organization of the paper}

In Section \ref{sec:boundary}, we study the compactification of 6d theories on manifolds with boundaries and defects. In Section \ref{sec:symmetry}, we discuss a collection of topics related to the symmetries of the compactified theories.
In Section~\ref{sec:4d},~\ref{sec:3d}, and~\ref{sec:2d}, we examine in more detail the compactifications on 2-, 3-, and 4-manifolds, respectively, and analyze aspects of the resulting theories in 4, 3, and 2 spacetime dimensions. 
In Section \ref{sec:2dqq}, we synthesize some of the earlier points to introduce a new invariant of 4-manifolds that depends on two $q$-parameters.

\noindent \textit{A note to the reader:}  although this work is the second part of our series, we have made each section largely self-contained, and much of the material can be read independently of Part I or of the other sections. The reader is therefore encouraged to jump directly to the part that interests them the most.

\section{Compactification on manifolds with boundaries}
\label{sec:boundary}

In this section, we review general aspects of compactification of 6d theories on a $d$-dimensional manifold $M_d$ discuss in \cite{Vafa:1994tf,kapustin2007electric,Tachikawa:2013hya,Eckhard:2019jgg,Gukov:2020btk} (see also \cite{Cvetic:2024dzu} for a more recent work), and extend it to incorporate boundaries and defects.

\subsection{Review of polarizations and symmetries of $T[M_d]$}
\label{sec:reviewpolarizationclosed}

We start by briefly reviewing \cite{Gukov:2020btk} to set up the notation that we will use later.

A relative theory on the boundary of a TQFT has operators whose correlation functions are ambiguous with branch cuts that can be resolved by extending the operators to the bulk TQFT. A choice of polarization projects out some of the operators, such that the correlation functions become unambiguous, and this produces an absolute theory that does not require a non-trivial bulk TQFT.
As discussed in \cite{Gukov:2020btk}, the polarizations are in one-to-one correspondence with the topological boundary conditions of the bulk TQFT,\footnote{
In this work, topological boundary conditions refer to the topological domain walls that separate the theory from an invertible TQFT.
} and the absolute theory corresponding to the polarization can be obtained by putting the bulk TQFT on an interval with the relative theory on one end, a topological boundary condition on the other end, and colliding the topological boundary condition with the relative theory by shrinking the interval. This produces an absolute theory that does not live on the boundary of a non-invertible TQFT.

We start with a 6d/7d coupled system with the 7d TQFT described by a three-form Abelian Chern--Simons theory, which has 3-dimensional volume operators that form an Abelian group $D$ under fusion. $D$ is also called the defect group for the 6d boundary theory \cite{DelZotto:2015isa}. The volume operators have non-trivial braiding, which induces a bi-linear pairing on $D$:
\begin{equation}
    \langle\cdot,\cdot \rangle:\quad D\times D\rightarrow U(1)~.
\end{equation}
The 6d boundary has strings, which are the ending surface of the bulk volume operators.
For the 6d ${\cal N}=(2,0)$ theory labeled by Lie algebra ${\frak g}$, the strings are valued in the weight lattice of ${\frak g}$. The charges of the strings do not obey the Dirac quantization condition.
The correlation functions of the strings on the boundary are ambiguous, the ambiguity is the bulk braiding $\langle\cdot,\cdot\rangle$ that induces the pairing
\begin{equation}
    H_3(M_6,D)\times H_3(M_6,D)\rightarrow U(1)~,
\end{equation}
where $M_6$ is the six-dimensional boundary manifold.
To obtain a well-defined 6d theory, we need to choose a polarization, 
which, up to a choice of a ``quadratic refinement,'' is a maximal isotropic subgroup $\Lambda\subset H_3(M_6)$ with respect to the above pairing.\footnote{When a finite Abelian group $H$ has a non-degenerate pairing $H\times H \rightarrow \mathbb{Q}/\Z$, we will use the term ``maximal isotropic subgroup'' both when the pairing is antisymmetric or when it is symmetric to refer to a subgroup $G$ that 1) trivializes the pairing in the sense that $(g_1,g_2)=0$ for any $g_1,g_2\in G$, and 2) is maximal in the sense that one cannot find a $h\notin G$ that pairs trivially with every element of $G$.}
The set of polarization is denoted by
\begin{equation}
    \text{Pol}(M_6)=\{\Lambda \subset H_3(M_6,D)|\Lambda\text{ is a maximal isotropic subgroup}\}~.
\end{equation}
For any chosen polarization $\Lambda$, $H_3(M_6,D)$ decomposes, though often non-canonically, as
 $H_3(M_6,D)=\Lambda\oplus\bar\Lambda$ for another $\bar\Lambda\in \text{Pol}(M_6)$. The set of polarizations $\text{Pol}(M_6)$ is also the same as the set of absolute theories at a point obtained by reducing on $M_6$. 

When we compactify the 6d/7d system on $M_d$, we obtain a coupled system with the $(6-d)$-dimensional $T[M_d]$ theory living on the boundary of the $(7-d)$-dimensional topological theory $\CT^{\rm bulk}[M_d]$. A (refined) polarization $\CP$ on $M_d$ specifies an absolute theory $T[M_d,\CP]$ (i.e.~a standalone theory that can be defined without a bulk), and one convenient way to view it is as a topological boundary condition for $\CT^{\rm bulk}[M_d]$.

The compactification of the 7d TQFT produces operators from wrapping the volume operators on various cycles in $M_d$, and the braiding between the volume operators induces non-trivial correlation functions for the resulting lower-dimensional operators. Since the braiding of the volume operators is bilinear, the braiding on the lower-dimensional operators is also bilinear, and this induces a non-trivial bilinear pairing on $H_*(M_{6-d},D)$. 

Moreover, there are various symmetries on the boundary from the reduction of the two-form symmetry on the 6d boundary of the 7d TQFT, given by $H^i(M_d,D)$ for $d-3\leq i\leq 3$. The symmetry generators on the boundary have ambiguous correlation functions given by the pairing due to the non-trivial correlation functions in the bulk, and to obtain a well-defined theory we need to gauge a non-anomalous subgroup symmetry $L$ that is maximal isotropic with respect to the pairing. Such gauging procedure (which can include additional topological action for the gauge fields such as the data of the quadratic refinement) is equivalent to choosing a polarization.

To discuss polarizations more generally, let us review some terminology we introduced in \cite{Gukov:2020btk} for compactification on $M_d$.
\begin{itemize}

\item We define the {\it spectrum group} of a polarization ${\cal P}$, denoted as ${\cal S}({\cal P})$, by the union of the images of the maximal isotropic subgroup $\Lambda\subset H^3(M_{6-d}\times M_d,D)$ under the map
\begin{equation}
    H_*(M_{6-d},\mathbb{Z})\times H^*(M_{6-d},H^*(M_d,D))\{3\}\rightarrow H^*( M_d,D)~,
\end{equation}
where we take the union of the images with respect to all cycles in $H_*(M_{6-d},\mathbb{Z})$ for every $M_{6-d}$, and $\{3\}$ denotes taking the degree-3 class.

    \item We define a {\it pure polarization} as a choice of a maximal isotropic subgroup of $H_*(M_{6-d},D)$ with respect to the bilinear pairing. Concretely, the polarization ${\cal P}$ is pure if and only if its spectrum group ${\cal S}({\cal P})$ has a trivial pairing with itself in $H^*(M_d,D)$.
    
\item 
On the other hand, a {\it mixed polarization} is a family of consistent choices of a maximal isotropic subgroup of $H_3(M_d\times M_{6-d},D)$ for each $M_{6-d}$, but it is not a pure polarization. 
 Concretely, the polarization ${\cal P}$ is mixed if and only if its spectrum group ${\cal S}({\cal P})$ has a non-trivial pairing with itself in $H^*(M_d,D)$. It is easy to see that the compactification on either a 6-manifold or a point (e.g.~not compactifying at all) does not involve mixed polarizations. Therefore, it is a phenomenon that only happens in the ``intermediate dimensions.''

\item 
Among pure and mixed polarizations, we will call a polarization ``geometric" if it can be obtained from a seven-dimensional bulk manifold. For pure polarizations, it will take the form $W_7=W_{d+1}\times M_{6-d}$ such that $\partial W_{d+1}=M_d$. Given any $M_{6-d}$, the maximal isotropic subgroup $\Lambda$ of $H^3(M_d\times M_{6-d},D)$ is given by the image of the restriction map from $H^i(W_{d+1}\times M_{6-d},D)$. This is determined by just the ``$M_d$ part'' which sits in a long exact sequence for relative cohomology,
\begin{equation}
    \cdots \rightarrow H^i(W_{d+1},D)\rightarrow H^i(M_d,D)\rightarrow H^{i+1}(W_{d+1},M_d;D)\rightarrow \cdots,\quad d-3\leq i\leq 3.
\end{equation}

\end{itemize}

Once we fix a choice of polarization, the symmetry of the theory is obtained by reducing the two-form symmetry in 6d. This concludes our brief review for closed $M_d$, and we now proceed to discuss the case with $M_d$ itself having a boundary.

\subsection{Compactification on manifolds with boundaries}\label{sec:PolBoundary}

\begin{figure}[tbh]
    \centering
    \includegraphics[width=0.6\textwidth]{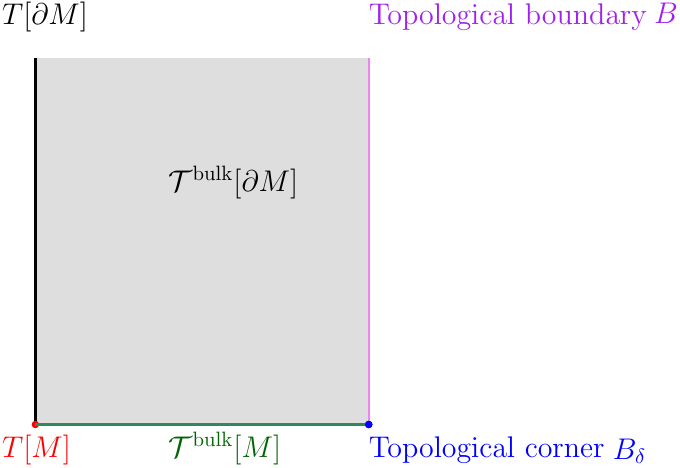}
    \caption{Choosing a polarization in a relative theory with boundary is equivalent to colliding the topological boundary conditions of the corresponding TQFT, that are labelled by (1) a topological boundary condition (purple) of the TQFT (grey) for the relative theory (black), which is Morita equivalent to the TQFT (green) for the boundary relative theory (red), and (2) topological domain wall (blue) between the topological boundary condition (purple) and the TQFT (green) for the boundary relative theory. 
    Here topological boundary conditions refer more generally to the topological domain walls that separate the theory from an invertible TQFT. For the picture to be consistent, the invertible TQFT is the same for the purple and the green boundaries, since otherwise there would be another branch cut extending from the blue corner. 
    }
    \label{fig:TMopen}
\end{figure}

As reviewed above, in Part I of this work, the primary role played by manifolds with a boundary is in the context of geometric polarizations. There, compactifying the 7d theory on $W_{d+1}$ while putting the 6d theory on the boundary $\partial W_{d+1}=M_d$ leads to an absolute theory in $6-d$ dimensions. What we investigate now is another interesting possibility of incorporating manifolds with a boundary, namely we will reduce the 6d theory on them, leading to a coupled system. Namely, compactifying the 6d theory on a manifold $M_d$ with a boundary $\partial M_d =M_{d-1}$ would lead to a theory $T[M_{d-1}]$ with a boundary, where the theory $T[M_d]$ lives. However, reducing the 7d theory on $M_d$ and its boundary leads to another coupled system, which serves as the topological bulk of the previous system. This is illustrated on the left side of Figure~\ref{fig:TMopen}, where $T[M_d]$ can be viewed as an interface between the dynamical theory $T[M_{d-1}]$ and the topological theory $\CT^{\rm bulk}[M_{d}]$, which are two different boundary conditions for $\CT^{\rm bulk}[M_{d-1}]$.

The goal of this section is to understand the choices of polarization and how it determines symmetries of the coupled system. In the next section, we will discuss how to glue along a boundary.

\subsubsection{Polarizations on open manifolds}

The manifold $M_{d-1}$ on the boundary of $M_d$ is itself closed, and the previous discussion of polarization applies. Again, some of the polarizations are geometric, given by a manifold $W_d$ with $\partial W_d=M_{d-1}$. The challenge is then to extend the notion of polarization to the pair $(M_d,M_{d-1})$. 

From the TQFT point of view, the question becomes that of finding a topological boundary condition for the coupled system consisting of $\CT^{\text{bulk}}[M_{d-1}]$ in the bulk and  $\CT^{\text{bulk}}[M_d]$
 on the boundary, as illustrated on the right side of Figure~\ref{fig:TMopen}. Then it is clear that the choice of such a boundary condition would consist of two parts, one is a boundary condition $B$ for $\CT^{\text{bulk}}[M_{d-1}]$, which can be regarded as an element in $\Pol(M_{d-1})$, while the other is a boundary condition $B_\delta$ for $\CT^{\text{bulk}}[M_{d-1}]$ that sits at the corner (hence the subscript ``$\delta$'') in Figure~\ref{fig:TMopen}. In other words, we need to choose
\begin{itemize}
    \item[(1)] A topological boundary condition $B$ of $\CT^{\text{bulk}}[M_{d-1}]$ that is ``Witt/Morita equivalent'' to $\CT^{\text{bulk}}[M_{d}]$, i.e.~they can be connected by a gapped domain wall.
    
    \item[(2)] A topological domain wall $B_\delta$ between $\CT^{\text{bulk}}[M_{d}]$ and the topological boundary condition of $\CT^{\text{bulk}}[M_{d-1}]$.
\end{itemize}
The existence of (2) follows from condition (1). Here, we include in the topological boundary conditions the topological domain walls that separate the theory with an invertible TQFT.  
Then, by colliding the theories $(T[M_{d}],T[M_{d-1}])$  with the topological boundary conditions after ``shrinking the interval,'' we obtain a well-defined absolute theory which is now a coupled system on the open space-time $\R^+\times \R^{6-d}$.\footnote{Note that the boundary theory $T[M_d]$ still lives on the boundary of a dynamical bulk theory obtained from colliding $T[M_{d-1}]$ with $B$, and is itself still a relative theory. However, the full coupled system is now absolute as there is not a topological bulk for which it is a boundary, unlike the pair $(T[M_{d}],T[M_{d-1}])$ that we started with.}
Such a choice $(B,B_\delta)$ of topological boundary conditions is equivalent to choosing a polarization in the relative theory.

As in \cite{Gukov:2020btk}, we can study the theory by compactifying the system all the way to a point.
Denote the open spacetime manifold by $N_{7-d}$, with boundary $\partial N_{7-d}=N_{6-d}$,
reducing the coupled system $(T[M_{d-1}],T[M_{d}])$ on $(N_{7-d},N_{6-d})$ is equivalent to reducing the 6d theory directly on the six-dimensional manifold
 \begin{equation}
     Y_6=(M_{d-1}\times N_{7-d}) \bigcup_{M_{d-1}\times N_{6-d}} (M_{d}\times N_{6-d})~.
 \end{equation}
 Then a choice of a polarization for the open manifold $M_{d}$ is a family of polarizations on the manifold $Y_6$ that are ``functorial'' when $(N_{7-d},N_{6-d})$ is varied. 

Similarly to the case of systems on closed spacetime manifold, let us first discuss pure polarization, while more general mixed polarization is introduced by twisted gauging a global symmetry with additional topological terms. We will also first suppress the choice of the quadratic refinement but will comment on it later.

 \subsubsection{Pure polarizations on manifolds with boundary}\label{sec:3Constraints}

We define a pure polarization on $(M_d,M_{d-1})$ as a pair of subgroups $(L_\delta,L)$:
\begin{itemize}
    \item $L\subset H^{d-4\le*\le 3}(M_{d-1},D)$ gives a pure polarization on $M_{d-1}$, {i.e.}~$L$ trivializes the pairing on $H^*(M_{d-1},D)$ and is maximal isotropic.

    \item $L_{\delta}\subset H^{d-3\le*\le 3}(M_d,D)$.
\end{itemize}

The motivation for this is that given any $(N_{7-d},N_{6-d})$ to compactify on,  $(L_
\delta,L)$ can fix a polarization in $H^3(Y_6,D)$ via the Mayer--Vietoris sequence
     \begin{equation}\label{MVOpen}
     \ldots\rightarrow H^2(M_{d-1}\times N_{6-d},D) \rightarrow H^3(Y_6,D)\rightarrow H^3(M_{d-1}\times N_{7-d},D)\oplus H^3(M_{d}\times N_{6-d},D) \rightarrow\ldots~.
 \end{equation}
by specifying a subgroup on both side. However, since $M_d$ is open, it is not obvious how to impose, or whether one should impose, any isotropic condition on $L_{\delta}$. Another question, which turns out to be closely related, is whether there exists compatibility condition between $L_\delta$ and $L$. 
 
To see how these questions are related, notice that $H^i(M_d,D)$ is dual to the relative homology $H_{d-i}(M_d,M_{d-1};D)$, which is part of the long exact sequence
 \begin{equation}
     \ldots\rightarrow H_{d-i}(M_d,D)\xrightarrow{\iota} H_{d-i}(M_d,M_{d-1};D)\xrightarrow{\partial} H_{d-i-1}(M_{d-1},D)\simeq H^{i}(M_{d-1},D)\rightarrow \ldots
 \end{equation}
 Then $L_\delta$ and $L$ can be regarded respectively as subgroups of the second and the third term. 

 We propose the following consistency conditions on
 the choice $L_\delta$:
 \begin{itemize}
     \item[(1a)] {\bf Compatibility condition}. The image of $L_\delta$ under the connecting morphism is contained in $L$ 
 \begin{equation}\label{LLdelta}
 \partial(L_{\delta})\subset L.
 \end{equation}

 \item[(1b)] {\bf Maximal image condition.} Furthermore, 
  \begin{equation}
 \partial(L_{\delta})= L\cap \im(\partial).
 \end{equation}
 In other words, the image $\partial(L_{\delta})$ is as large as it can be inside $L$.

\item[(2)] {\bf Maximal isotropic condition.} The overlap $L_\delta \cap\mathrm{ker}(\partial)$ will have trivial intersection pairing with $L_\delta$, and is maximal.\footnote{Recall that an element in the kernel of $\partial$ can be represented as a chain in the interior of $M_d$, and such an element will have a well-defined intersection with a relative chain. Via Poincaré duality, this is the same as the intersection between a relative cocycle in $H^*(M_d,M_{d-1};D)$ and one in $H^*(M_d,D)$. Here, maximal means that there is no element in ker$(\partial)$ of degree between $d-3$ and 3 that is not in $L_\delta$ but pairs trivially with all of $L_\delta$. The pairing defined in this way naively depends on the lift of $\alpha\in L_\delta \cap\mathrm{ker}(\partial)$ to $\alpha'\in H_*(M_{d})$, but the difference is required to be in the image of $L$ under $\iota:  H_*(M_{d-1})\rightarrow H_*(M_{d}).$ (In other words, the lift should be in $L_{\text{b.d.}}$---``charges for boundary operators''---which we will introduce later.) But such ambiguity pairs trivially with anything in $H_*(M_d,M_{d-1})$ as one can compute the intersection on $M_{d-1}$, which vanishes as $L$ is isotropic.} This is in addition to the usual maximal isotropic condition imposed on $L$. 
 
  \end{itemize} 

\begin{figure}[t]
  \centering
    \includegraphics[width=0.3\textwidth]{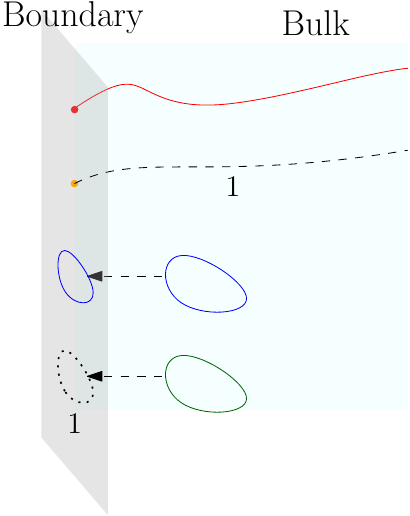}
     \caption{In general, there are two classes of bulk operators in the presence of a boundary---these that can end on the boundary ({\color{red}red}) and these that cannot. One can also classify them by how they behave when moved to the boundary. They can either stay non-trivial ({\color{blue}blue}) or become the identity operator ({\color{teal}green}, which is in fact a subclass of red ones). There are also operators ({\color{orange}orange}) that only live on the boundary, i.e. ending on the trivial bulk operator. From the perspective of polarization on a manifold with boundary, these correspond to the following groups: {\color{red}$\partial(L_\delta)$}, {\color{blue} $L/\partial(L_\delta)$}, {\color{teal} $L\cap\im(\partial)$}, and {\color{orange} $L_\delta\cap\ker (\partial)$}.
     }\label{fig:bulkboundaryopr}
\end{figure}

While (2) constrains the part of $L_\delta$ in the kernel of $\partial$, the conditions (1a) and (1b) collectively constrain the image (hence the choice of numbering), and later we sometimes refer to them as the stronger version of the compatibility condition. Before elaborating on the physical meaning of these conditions, we first comment on the meaning of the various relevant groups. 
 \begin{itemize}
     \item The subgroup $L$ classifies the charges of operators in the theory $T[M_{d-1}]$, as $M_{d-1}$ is itself closed and all the analysis in Part I applies.

     \item      
  $L_{\delta}$ can be viewed as a subgroup of $H_{i}(M_d,M_{d-1};D)$, and its elements can be represented as $i$-chains in $M_d$ with possible boundaries that belong to $M_{d-1}$. Reducing a two-dimensional string in the 6d theory on such an open chain leads to a $(2-i)$-dimensional operator in the theory $T[M_d]$ that is attached to a $(3-i)$-dimensional operator in the bulk theory $T[M_{d-1}]$. The charge of the latter is the image under $\partial$.   
  \item Elements in $\partial(L_\delta)\simeq \im(\partial)\cap L$ are bulk operators that can end on the boundary. Under the next map in the long exact sequence $\iota: H_*(M_{d-1})\rightarrow H_*(M_{d}),$ they become trivial. Hence, these also label bulk operators that become trivial when moved to the boundary. 
     \item The overlap $L_{\delta,\mathrm{ker}}:=L_{\delta}\cap \mathrm{ker}(\partial)$ are elements in $L_\delta$ that become zero under the map $\partial$ and correspond to operators that live solely on the boundary. In fact, this only classify equivalence classes up to next group of operators.
     \item The coset $L/\partial(L_{\delta})$ labels the operators in the bulk theory $T[M_{d-1}]$ that cannot end on the boundary $T[M_d]$, modulo operators that could.  Another way to think about it is as the charges of operators on the boundary that can be obtained from charged operators in the bulk. 
     \item To get all allowed charges for boundary operators, one needs to combine the previous two classes into a subgroup $L_{\text{b.d.}}\subset H_*(M_d)$ via a possibly non-trivial extension 
     \begin{equation}
         L/\partial(L_{\delta})\rightarrow L_{\text{b.d.}}\rightarrow L_{\delta,\mathrm{ker}}.
     \end{equation}

 \end{itemize}
 The relation between bulk and boundary operator is illustrated in Figure~\ref{fig:bulkboundaryopr}.
 
The compatibility condition \eqref{LLdelta} in (1a) is equivalent to the statement that charges of the bulk operators that can end on the boundary are compatible with charges of the boundary operators on the ending loci. The condition (1b) comes from the following consideration. If some of the elements in $L\cap{\rm im}(\partial)$ does not come from $L_\delta$, then they actually cannot end on the boundary. But as they become trivial when moved to the boundary, they should be able to end by moving part of it to the boundary. Therefore, $L\cap{\rm im}(\partial)$ should be a subgroup of $\partial(L_\delta)$, which, when combined with (1a), gives the isomorphism in (1b). In Figure~\ref{fig:bulkboundaryopr}, condition (1b) is the statement that the blue and red bulk operators are actually the same.\footnote{Notice that this is a statement about the operators that comes from compactification of extended operators in higher dimensions. For a more general quantum field theory, this statement is true at the level of charges. Indeed, a ``red operator'' that can end on the boundary must be neutral under boundary symmetries, and is thus also a ``green operator.''}

For an operator living only on the boundary, with charges in $L_{\delta}\cap \mathrm{ker}(\partial)$, it can be viewed as the ending loci of the trivial operator, and conditions (1a) and (1b) will be trivially satisfied, but the maximal isotropic condition (2) will become meaningful. This condition ensures that a pair of charged operators on the boundary, if at least one is not attached to a bulk operator, are mutually local i.e.~the operators obey single-valued correlation functions and do not attach to additional ``branch cut'', which can also be interpreted as a generalized Dirac quantization condition.

We remark that the isotropic condition is not directly imposed on the entire $L_\delta$ due to the fact that the boundary theory $T[M_d]$ itself, if well defined, is in general a relative theory (i.e.~charges of operators can violate the naive Dirac quantization condition), but the bulk theory $T[M_{d-1}]$ is required to be absolute.
 
 We also expect that there is no additional constraint on the pair $(L_\delta,L)$, and any choice for $L_\delta$ satisfying all constraints can be physically realized. One way to see that the constraints are sufficient is by demonstrating that they always lead to a Lagrangian subgroup of $H^3(Y_6,D)$ in \eqref{MVOpen}. This is shown in Appendix \ref{sec:completeness} with a geometric interpretation.

 \subsubsection{Geometric polarizations}
 
 Just as in the case with closed manifolds, one can define a notion of ``geometric polarizations.'' Naively, they are just given by a choice of a pair of manifolds $(W_{d+1},W_{d})$ such that $\partial W_d = M_{d-1}$ and $\partial W_{d+1} = W_d\cup_{M_{d-1}} M_d$, leading to a pure polarization with $L$ and $L_\delta$ given respectively by the image of the map
 \begin{equation}
     H^{d-4\le *\le 3} (W_d)\rightarrow H^{d-4\le *\le 3}(M_{d-1})
 \end{equation}
 and 
  \begin{equation}
     H^{d-3\le *\le 3} (W_{d+1})\rightarrow H^{d-3\le *\le 3}(M_{d}).
 \end{equation}
It is straightforward to check that the conditions (1a) and (2) are both satisfied. The compatibility condition follows from the commutativity of restriction maps, which guarantees that $\partial (L_\delta)\subset L$, and the maximal isotropic condition can be argued, similar to the closed case, by pushing the (relative) cycles into $W_{d+1}$, where they can be made mutually disjoint. 

However, the ``maximal image condition'' (1b) is not automatically satisfied, and would impose an extra condition on $(W_{d+1},W_{d})$ for it to actually define a polarization. Notice that this is in sharp contrast with the closed case $\partial M_d=\emptyset$, where any choice of a bounding $W_{d+1}$ defines a polarization.

The condition (1b) would additionally require that any elements in $L\cap \im(\partial)$ can be lifted to $L_\delta$ and hence in this geometric setup to $H^*(W_{d+1})$. Rephrased using homology groups, it requires that the image of $L_\delta$ under the map $\partial$ not only lies inside $L$ but is also maximal, so that any cycle $\alpha \in L\subset H_{0\le*\le1}(M_{d-1},D)$ that can be written as the boundary of a relative cycle $\beta\in H_{*+1}(M_d,M_{d-1};D)$ must be the boundary of a relative cycle in $L_\delta$. In other words, it can lifted to $H_*(W_{d+1},\partial W_{d+1};D)$.
 
However, notice that this condition can be violated by a general choice of $(W_{d+1},W_{d})$, as such a lift may not exist. The obstruction can be understood in the following way. Let $\gamma\in H_{*+1}(W_d,M_{d-1};D)$ be a bounding relative cycle of $\alpha$ in $W_d$. Then one can ``glue'' $\beta$ and $\gamma$ along $\alpha$ to obtain a class $\beta\cup_\alpha\gamma$ in $H_{*+1}(\partial W_{d+1})$ by lifting along the Mayer--Vietoris sequence,
\begin{equation}
    \cdots\to H_{*+1}(\partial W_{d+1},D)\to H_{*}(M_d,M_{d-1};D) \oplus H_{*+1}(W_d,M_{d-1};D) \to H_{*}(M_{d-1},D)\to\cdots
\end{equation}
The image of that class under $H_{*+1}(\partial W_{d+1})\rightarrow H_{*+1}(W_{d+1})$ is then an obstruction of capping it off in $W_{d+1}$. Therefore, to have $(W_{d+1},W_d)$ giving rise to a geometric polarization, it has to satisfy the property that any 1- and 2-cycles in $\partial W_{d+1}$ obtained by such gluing have to be trivial in $W_{d+1}$. This can be stated more cleanly as the following map
\begin{equation}
    H_*(\partial W_{d+1},D)\rightarrow H_{*}(W_{d+1},D)
\end{equation}
in degree 1 and 2 can always be lifted along 
\begin{equation}
    H_*(M_{d-1},D)\rightarrow H_*(\partial W_{d+1},D).
\end{equation}
This condition ensures that any cycles in $\partial W_{d+1}$ of the relevant degrees that are not capped off in $W_{d+1}$ come from $M_{d-1}$ instead of arising as two relative cycles glued together. 

To get a counter-example, consider $W_{3}=S^1\times D^2$ is a solid torus, while $W_2$ and $M_2$ each looks like a cylinder, glued together along $M_{1}=S^1\sqcup S^1$ to form the boundary torus. Then if the cycle on the boundary $T^2$ that is cut open by $M_{1}$ is not contractible in $W_3$, such a choice of $(W_3,W_2)$ does not lead to a polarization for $(M_2,M_1)$.

What is the problem when this condition is violated? When (1b) is not satisfied, there is an operator with charge $\alpha$ in $T[M_{d-1}]$ that is supposed to be able to end on the boundary $T[M_d]$, as $\alpha$ becomes trivial when pushed to the boundary, but the boundary operator would have the ``wrong charge'' for this to happen. Around the end point, a string in the 6d theory should wrap a relative cycle $\beta$ whose boundary $\partial \beta$ is $\alpha$. However, as $\beta\notin L_\delta$, this is actually not a valid configuration. The reason is that the string lives on the boundary of a 3-dimensional operator, and when such a lift does not exist, the operator attached to the end point does not have a place to go inside $W_{d+1}$.

Another way to state this, in the setting of Figure~\ref{fig:TMopen}, is the following. When condition (1b) is not satisfied, one will not actually have a well-behaved topological corner $B_\delta$ that connects the two topological boundaries $\CT^{\rm bulk}[M_{d}]$ and $B=\CT^{\rm bulk}[W_{d-1}]$ of $\CT^{\rm bulk}[M_{d-1}]$. Instead, there will be a ``branch cut'' that originates from $B_\delta=\CT^{\rm bulk}[W_{d}]$. To detect it, one can consider operators obtained by wrapping the 3-dimensional topological operator on various (relative) cycles. In the bulk, $\alpha$ labels a topological operator that can end on both topological boundaries via $\beta$ and $\gamma$. Then one property that has to be satisfied by $B_\delta$ is that $\beta$ and $\gamma$ can end on it from the two sides. This mutual end point is labeled by the bounding cycle for $\beta\cup_\alpha\gamma$ in $W_{d+1}$. When (1b) is not satisfied, there will be another operator involved, labeled by $\beta\cup_\alpha\gamma$ and living entirely in $B_\delta$. It forms a junction with $\beta$ and $\gamma$. Such operators lead to branch cuts and cause problems with single-valuedness of correlation functions in the dynamical theory after ``shrinking the interval.''\footnote{It is perhaps more convenient to view $B_\delta$ as a boundary of another topological theory $\CT^{\rm bulk}[W'_d]$ where such operator can go. This corresponds to cutting off part of $W_{d+1}$ along a submanifold $W'_d$ such that all such non-trivial cycles are now cut open (e.g.~becoming relative cycles ending on $W'_d$). After shrinking the interval, the $T[M_{d},B_\delta]$ is no longer a boundary of $T[M_{d-1},B]$, but a interface between the dynamical theory and the topological theory $\CT^{\rm bulk}[W'_d]$.}

This concludes our discussion of the physical meaning of the pair $(L_\delta,L)$ from the perspective of charged objects, we will now switch to the ``dual perspective,'' focusing instead on the symmetries that they transform under.\footnote{As a general remark, it is possible that a symmetry does not have charged object, but nevertheless has non-trivial symmetry generator, such as the $\mathbb{Z}_2$ one-form symmetry in $\mathbb{Z}_2\times\mathbb{Z}_2$ gauge theory in 3+1d generated by the gauged SPT phase given by the non-trivial element in $H^2(B\mathbb{Z}_2\times\mathbb{Z}_2,U(1))=\mathbb{Z}_2$
\cite{Hsin:2019fhf}.
}
 
\subsubsection{Symmetries}\label{sec:SymmetryBoundary}

Let us discuss how the global symmetry in the $T[M_{d-1}]$ theory depends on the polarization data $(L,L_\delta)$. We will focus on the invertible symmetries.

The choice of $L$ determine the remaining symmetry $L^\vee$ of the theory $T[M_{d-1}]$, given by the quotient
\begin{equation}
    L\rightarrow H_{d-4\le*\le3}(M_{d-1})\rightarrow L^\vee
\end{equation}
as discussed in Part I. The gauge fields for the symmetries of the theory $T[M_{d-1}]$ can be obtained by decomposing the 3-form field $C$ as a sum of $B_i\wedge \omega_{3-i}$ with various $\omega_{3-i}\in L^\vee$.\footnote{From this point of view, it is more natural to regard $L$ as a subset of the homology of $M_{d-1}$, while $L^\vee$ as a subset of the cohomology. However, we will often not make a sharp distinction between groups related by the Poincar\'e duality.}  However, the story will become richer in the presence of a boundary theory. 

Given a subgroup $L_\delta$ of $H^*(M_{d},D)\simeq H_{d-*}(M_d,M_{d-1};D)$, one can again look at the quotient $H^*(M_{d})/L_\delta$. However, as $M_d$ is not closed, it can no longer be identified as the dual of $L_\delta$. Instead, one could first restrict to the subgroup
\begin{equation}
    L_{\delta,\mathrm{ker}}:=L_\delta\cap \mathrm{ker}(\partial).
\end{equation}
Then $L_{\delta,\mathrm{ker}}^\vee \simeq \ker(\partial)/L_{\delta,\mathrm{ker}}$, and as they are represented as cochains vanishing on  $\partial M_d= M_{d-1}$, they leads to extra symmetries of the boundary theory $T[M_d]$.

The quotient of $L_\delta$ by $L_{\delta,\mathrm{ker}}$ can be identified with $\partial (L_\delta)$. This is a subgroup of $L$, and the existence of the decomposition
\begin{equation}
    \partial (L_\delta)\rightarrow L\rightarrow L/\partial (L_\delta)
\end{equation}
signifies that there is in general a non-trivial interplay between the $L^\vee$ symmetry of the theory $T[M_{d-1}]$ and the symmetry on the boundary theory $T[M_d]$. More precisely, we have
\begin{equation}\label{SESBulk}
    (L/\partial (L_\delta))^\vee\rightarrow L^\vee \rightarrow (\partial (L_\delta))^\vee,
\end{equation}
and $ (L/\partial (L_\delta))^\vee$ and $(\partial (L_\delta))^\vee$  can be respectively identified with the kernel and image of $L^\vee$ under
$H^*(M_{d-1})\rightarrow H^{*+1}(M_{d},M_{d-1})$. Then the group $(\partial (L_\delta))^\vee$ represent a symmetry of one degree less in $T[M_d]$. This is perfectly consistent with the interpretation that $\partial (L_\delta)$ represent charged objects on the boundary theory that are themselves boundaries of bulk operators in $T[M_{d-1}]$. On the other hand, the subgroup $(L/\partial (L_\delta))^\vee$ of $L^\vee$ represents symmetries shared by the boundary and the bulk theory.\footnote{We note that the identity operator is shared between the bulk and the boundary, compatible with the fact that such symmetries form a subgroup.}  This is consistent with the interpretation of $L/\partial (L_\delta)$ as equivalence classes of charges carried by bulk operators in $T[M_{d-1}]$ that cannot be screened by boundary operators. However, it remains a possibility that the action becomes trivial on the boundary, which can happen when a charged bulk operator becomes neutral after moving to the boundary. This is measured by the map $\iota:H_*(M_{d-1},D)\rightarrow H_*(M_d,D)$, and the image of $\alpha\in L\subset H_*(M_{d-1},D)$ may be trivial $\iota(\alpha)=0$. Such non-trivial bulk charges that become trivial on the boundary are classified by $\ker(\iota)\cap L\simeq \im(\partial)\cap L$. Then it is easy to see that the subgroup of $(L/\partial (L_\delta))^\vee$ that acts trivially on the boundary is $(L/\im(\partial))^\vee$.

To summarize, the symmetry group on the boundary consists of three parts:
\begin{itemize}
\item $(L/\partial (L_\delta))^\vee$ consists of symmetries shared by the bulk theory $T[M_{d-1}]$ and the boundary theory $T[M_d]$. Generators for this symmetry in the bulk theory can end on the boundary theory.

\item $(\partial (L_\delta))^\vee$ consists of boundary $n$-form symmetries that ``descend'' from bulk $(n+1)$-form symmetries. Bulk operators only charged under this symmetry (i.e.~neutral under the $(L/\partial (L_\delta))^\vee$ symmetry) can end on the boundary.

    \item $L_{\delta,\mathrm{ker}}^\vee$ consists of extra symmetries on the boundary. Boundary operators charged under this symmetry cannot be moved to the bulk.
\end{itemize}
They can be organized into two short exact sequences, one for the bulk symmetries is \eqref{SESBulk} that we have already seen, and one for ``purely'' boundary symmetries
\begin{equation}
    L_{\delta,\mathrm{ker}}^\vee  \rightarrow L_{\text{b.d.}}^\vee \rightarrow (L/\partial (L_\delta))^\vee,
\end{equation}
where the descendant $(\partial (L_\delta))^\vee$ symmetry is not included but can be incorporated by changing the rightmost term to $L^\vee$ and enlarging the middle term accordingly. 

From the perspective of compactification, the compatibility condition for $(L_\delta,L)$ ensures that the background fields for bulk and boundary symmetries can be ``glued'' together to a 3-form field in 6-dimensions. Conversely, when the boundary and bulk symmetries are not compatible, it is an obstruction for realizing the coupled system via compactification, as the bulk and boundary symmetries cannot have the same origin in higher dimensions. In fact, as the conditions can be argued through consistency of the coupled system in purely lower-dimensional terms after compactification, the ``wrong choices'' for the polarization should lead to problematic physical systems. We will see examples of this in the next part.

\subsubsection{Examples}

We give some examples for polarization on open manifolds. The detailed study of the corresponding physical system can be found in later sections. For simplicity, we take $D=\Z_p$ with $p$ a prime.

\subsubsection*{The case of $M_d=D^2$}

The boundary is $\partial D^2=S^1$, and there are two choices of $L$ given respectively by $H_1(S^1,\Z_p)$ and $H_0(S^1,\Z_p)$. $L_\delta$ on the other hand is a subgroup of
 $H^*(D^2,\Z_p)\simeq H_*(D^2,S^1;\Z_p)$ whose only non-vanishing piece is  $H^0(D^2,\Z_p)\simeq H_2(D^2,S^1;\Z_p)$. Therefore, $L_\delta$ is either 0 or the entire group. However, compatibility conditions only allow the following two choices:
 \begin{itemize}
     \item $L=H_0(S^1,\Z_p)$ and $L_\delta =0$.
     ``(PSU$(p)_\delta,$PSU$(p)$) theory'' with 2-form symmetry $L^\vee$ in the bulk and boundary.
 
     \item $L=H_1(S^1,\Z_p)$ and $L_\delta =H_2(D^2,S^1;\Z_p)$. ``(SU$(p)_\delta$, SU$(p)$) theory'' with 1-form symmetry in the bulk and 0-form symmetry on the boundary that descends from the bulk symmetry.

 \end{itemize}
 There the name reflects the fact that if we compactify a 6d $(2,0)$ theory labeled by $A_{p-1}$ on $D^2$, there is an SU$(p)$ or PSU$(p)$ gauge field in the bulk $T[S^1]$, while an SU$(p)$- or PSU$(p)$-valued compact scalar, which can be thought of as the period of the 2-form field $\int_{D^2} B$ of the 6d theory on $D^2$. Among all physically allowed possibilities, only two are realized by polarizations. What is wrong with the other theories?

 The  ``(SU$(p)_\delta,$PSU$(p)$) theory'' and the ``(PSU$(p)_\delta,$SU$(p)$) theory'' are well defined at the classical level, as coupling either an SU- or PSU-valued scalar to the bulk SU or PSU gauge field via the adjoint action is perfectly fine. However, they are expected to be inconsistent at the quantum level, which we will break down below. 
\begin{itemize}
\item $L=H_1(S^1,\Z_p)$ and $L_\delta =0$. This 
     ``(PSU$(p)_\delta,$SU$(p)$) theory'' would have 1-form symmetry in the bulk and 2-form symmetry on the boundary. However, it does not satisfy the condition (1b). In other words, the Wilson line of the bulk SU$(p)$ theory becomes trivial on the boundary, thus it should be able to end on the boundary, but, on the other hand, there is no boundary operator for it to end on. Another way to think about the inconsistency is by attempting to construct it from the previous two well-defined systems by gauging either the 2-form symmetry in the bulk or 0-form symmetry on the boundary. However, the 2-form symmetry in the (PSU$(p)_\delta,$PSU$(p)$) theory is shared between the bulk and boundary, with the gauge field satisfying a Neumann boundary condition, making it impossible to just gauge the symmetry in the bulk. One encounters a similar problem when trying to gauge the 0-form symmetry of the (SU$(p)_\delta$, SU$(p)$) theory on the boundary. The gauge field for it descends from the 2-form gauge field of the one-form symmetry in the bulk via the Dirichlet boundary condition, and one cannot simply gauge the boundary symmetry without also gauging the bulk symmetry.
    \item $L=H_0(S^1,\Z_p)$ and $L_\delta =H_2(D^2,S^1;\Z_p)$. This 
     ``(SU$(p)_\delta,$PSU$(p)$) theory'' violates the compatibility condition (1a), and as a consequence, there is a boundary operator that violates the charge quantization condition, being mutually non-local with the 2-dimensional string of the PSU theory when the latter is moved to the boundary. And the correlation functions involving them cannot be single-valued. Notice that this is directly related to the fact that $\partial(L_\delta)$ is not contained in $L$ in this example, as, otherwise, the isotropic condition of $L$ will prevent such non-locality from happening. In other words, if one can find a bulk line to attach to the local operator on the boundary, well-definedness of the bulk theory will tell us that such a 2-dimensional string won't exist. One can also detect the inconsistency of the system from the obstruction to constructing it via gauging. The problem is similar to the previous case where one is supposed to only gauge in either the bulk or the boundary, which is not consistent due to the boundary condition of the gauge field.
\end{itemize}
 
     The problem can also be seen from the point of view of compactification. For example, in the (SU$(p)_\delta,$PSU$(p)$) theory, one cannot ``glue'' together the 0-form symmetry on the boundary and the 2-form symmetry in the bulk to a 2-form symmetry in 6d.

\begin{figure}[t]
    \centering
    \includegraphics[width=0.3\textwidth]{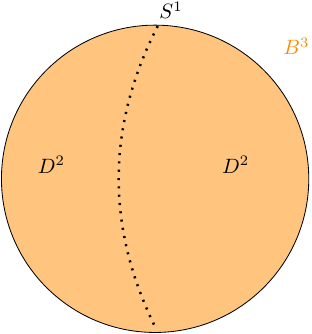}
    \caption{The boundary $S^2$ of $B^3$ obtained by gluing two disks $D^2$ along a great circle.}
    \label{fig:B3D2gluing}
\end{figure}

 Furthermore, the polarization that leads to the (SU$(p)_\delta,$SU$(p)$) theory is geometric and is given by filling in $D^2$ with a $W_3=B^3$ the three-ball, whose boundary is the union of $W_2=D^2$ and $M_2=D^2$ glued along $M_1=S^1$. This is illustrated in Figure~\ref{fig:B3D2gluing}. Then $L_\delta$ is given by the image of $H^*(W_3)\rightarrow H^*(M_2)$, which is $H^0(M_2=D^2)\simeq H_2(D^2,S^1)$ while $L$ is given by the image of $H^*(W_2)\rightarrow H^*(M_1)$, which is $H^0(M_1=S^1)\simeq H_1(S^1)$. On the other hand, it is easy to see that the other polarization that leads to the (PSU$(p)_\delta,$PSU$(p)$) theory is not geometric, as $H^1(S^1)$ cannot be in the image of $H^*(\Sigma)$ for any 2-manifold $\Sigma$ with $\partial \Sigma = S^1$.

 \subsubsection*{The case of $M_{d-1}=S^{d-1}$.} This is the extreme case when $M_d$ can be arbitrarily complicated but little information is captured by the boundary $M_{d-1}=S^{d-1}$. Assuming $d
 \le 4$, $L$ can be either $H_0(S^{d-1})$ or $H_{d-1}(S^{d-1})$. As the map $H_{*+1}(M_d,S^{d-1})\rightarrow H_{*}(S^{d-1})$ only lands in the piece in degree $d-1$, the choice of $L_\delta$ is given by a maximal isotropic subgroup of $H_{d-3\le*\le 3}(M_d\cup_{S^{d-1}} B^d)$ that has to include the top degree piece when $L=H_{d-1}(S^{d-1})$ or does not include the top degree piece when $L=H_0(S^{d-1})$. In the special case of $d=4$ and $L=H_3(S^{3})$, no choice of $L_\delta$ exist for degree reasons. This is actually a general feature for $d=4$, and can be understood as the conflict between the physical fact that ``$(-1)$-form symmetries'' in the bulk theory $T[M_3]$ won't descend to a symmetry on the boundary $T[M_4]$ (as ``$(-2)$-form symmetries'' don't exist in the usual sense) and the geometric fact the fundamental class of $M_3$ is a boundary of the fundamental class of $M_4$.

\subsubsection{Quadratic refinements and mixed polarizations}

As discussed in Part I, choosing a different quadratic refinement corresponds to shifting background fields by a special $2$-torsion element in cohomology (often given by a Stiefel--Whitney class such as $w_2$ or $w_3$ of the 5d spacetime for $T[S^1]$). 
Then for a choice of $(L_\delta,L)$, one can ask again whether one can consider other quadratic refinements labeled by $(\gamma_\delta,\gamma)$ which are 2-torsion elements in $L^\vee_{\delta,\ker}$ and $L^\vee$ respectively. However, there should be again consistency conditions. We conjecture that the only condition to impose is that $\gamma$ is in the subgroup $ (L/\partial(L_\delta))^\vee$. In other words, it has to become trivial under the quotient map
\begin{equation}
    L^\vee\rightarrow (\partial(L_\delta))^\vee.
\end{equation} 
One way to see this is from the boundary condition for the $(\partial(L_\delta))^\vee$-valued background field $B_{n+1}$ associated with an $n$-form symmetry
\begin{equation}
    B_{n+1}|_\partial=dB_n
\end{equation}
with $B_n$ the background field for the $(n-1)$-form symmetry on the boundary that descends from the bulk $n$-form symmetry. However, this boundary condition is not compatible with a shift of $B_{n+1}$ by a non-zero element in cohomology. Another justification for this condition is that when the background field is shifted, the statistics of certain operators can change from bosonic to fermionic, but fermionic operators cannot end on the boundary.

The discussion about other types of polarizations, i.e.~the mixed polarizations, is similar. As discussed in Part I \cite{Gukov:2020btk}, the mixed polarizations can be constructed from pure polarization by coupling to TQFT using
gauging a symmetry and including topological actions for the gauge fields.
Such topological terms
for the boundary theory $T[M_d]$ and the bulk theory $T[M_{d-1}]$ correspond to $L_{\delta,\ker}$ and $L/\partial(L_\delta)$, 
but more generally, there can be bulk topological term that couples to the boundary.

\subsection{Cutting and gluing along boundaries}

One benefit of having the discussion about polarizations on open manifolds is that we can now consider cutting and gluing of polarizations. Namely, for two open manifolds $M_d$ and $M'_d$ with the same boundary $\partial M_d=-\partial M'_d=M_{d-1}$, one can consider the map
\begin{equation}\label{CutGluePol}
    \Pol(M_d)\times_{\Pol(M_{d-1})} \Pol(M'_d)\rightarrow \Pol(\tilde M_d)
\end{equation}
with $\tilde M_d:=M_d\cup_{M_{d-1}} M'_d$. 

In this section, we will construct the map \eqref{CutGluePol} and its  properties. Questions that we are interested in includes 
\begin{itemize}
    \item 
How symmetries behave under gluing,
\item and whether \eqref{CutGluePol} is surjective in the sense that any polarizations on $\tilde M_d$ can be constructed via cutting and gluing.
\end{itemize}

One useful tool of studying such questions for pure polarizations\footnote{Notice that mixed polarizations obtained by adding a topological term to theory given by a pure polarization shares the same symmetries if we restrict to these coming from the 2-form symmetry in 6d. So the discussion below is general from the point of view of symmetries.} is  the following Mayer-Vietoris sequence,
\begin{equation}
    \ldots\rightarrow H^{*-1}(M_{d-1},D) \rightarrow H^*(\tilde M_d,D)\rightarrow H^*(M_d,D)\oplus  H^*(M'_d,D) \rightarrow\ldots
\end{equation}
If the pure polarizations on $M_d$ and $M'_d$ is given by $(L_\delta,L)$ and $(L'_\delta,L)$, then one can construct a unique subset $\tilde L$ of $H^*(\tilde M_d)$ by requiring that it includes all elements with pre-image in $L$ and image in $L_\delta \oplus L'_\delta$. To prove that $\tilde L$ give a pure polarization, we only need to show that $\tilde L$ trivialize the pairing on $H^*(\tilde M_d)$, and is maximal in degree between $d-3$ and $3$. The analysis is similar to that in the previous section, with the pairing being again ``block-diagonalizable'' into three parts. To see this, we will again use $i^*$ for the map
\begin{equation}
    i^*:\quad H^*(M_d)\oplus  H^*(M'_d)\rightarrow H^{*}(M_{d-1})
\end{equation}
which is given by $i^*=\partial^*-\partial'^*$ with $\partial^*$ and $-\partial'^*$ being respectively the restriction of $i^*$ on the two summands. Then we have
\begin{equation}
     0\rightarrow \mathrm{ker}(\partial^*)\oplus  \mathrm{ker}(\partial'^*)\rightarrow\mathrm{ker}(i^*)\rightarrow \mathrm{im}(\partial^*)\cap  \mathrm{im}(\partial'^*)\rightarrow 0.
 \end{equation}
The pairing on $\ker (\partial^*)$ and $\ker (\partial'^*)$ being trivialized by $L_{\delta,\ker}$ and $L'_{\delta,\ker}$. On the other hand, $\mathrm{im}(\partial^*)\cap  \mathrm{im}(\partial'^*)$ is paired with $H^*(M_{d-1})/(\im(\partial^*)\cup \im(\partial'^*))$, and this pairing is trivialized by picking the subgroups $\partial L_\delta\cap \partial'L'_\delta$ and $L/(\partial L_\delta\cup \partial' L'_\delta) $. It is easy to see that $\tilde L$ determined in this way is maximal.

The four groups that appear above as ``building blocks'' of $\tilde L$ account for operators in the theory $T[\tilde M_d]$ with different origins:

\begin{itemize}
    \item $L_{\delta,\ker}$ and $L'_{\delta,\ker}$. They label operators that comes from these living on the two boundaries $T[M_d]$ and $T[M'_d]$.
    \item $\partial L_\delta\cap \partial'L'_\delta$. It classify operators that come from a pair of operators living on the two boundaries that can be connected by the same operator in the bulk. Elements in this group specify elements in $\tilde L$ up to a pair of boundary operators in $L_{\delta,\ker}$ and $L'_{\delta,\ker}$.
    \item $L/(\partial L_\delta\cup \partial' L'_\delta) $. It consists of (equivalence class of) operators that comes from bulk operators in $T[M_{d-1}]$. One has to mod out by those that can end on either boundary as they are screened.
\end{itemize}

Alternatively, one can think of the dual groups and how they are different parts of $\tilde L^\vee$ which is the symmetry group of $T[\tilde M_d]$:

\begin{itemize}
    \item $(L_{\delta,\ker})^\vee$ and $(L'_{\delta,\ker})^\vee$. They are symmetries of the boundary theories $T[M_d]$ and $T[M'_d]$.
    \item $(\partial L_\delta\cap \partial'L'_\delta)^\vee$. It consists of symmetries of the bulk theory that can descend to symmetries on either boundary.
    \item $(L/(\partial L_\delta\cup \partial' L'_\delta))^\vee $. It consists of symmetries of the bulk theory that are unbroken by the boundaries.
\end{itemize}

Among these, $(L_{\delta,\ker})^\vee$ and $(L'_{\delta,\ker})^\vee$ are quotients, $(L/(\partial L_\delta\cup \partial' L'_\delta))^\vee $ is a subgroup, while $(\partial L_\delta\cap \partial'L'_\delta)^\vee$ is a subquotient. In general, $(\tilde L)^\vee$ is formed out of these groups via non-trivial extensions.

See Figure~\ref{fig:bulkboundaryopr} for an illustration of different boundary conditions for various symmetry defects associated with various groups listed above. 

For more general cases with additional topological terms (e.g.~those giving rise to mixed polarizations or a different choice of the quadratic refinement), the discussion is almost completely analogous. The end result will simply be $T[\tilde M_d,\tilde L]$ but with additional topological terms. These additional topological terms won't affect these symmetries coming from the 2-form symmetry of the 6d theory.

\subsection{Defects of higher co-dimensions}

Just as considering a manifold $M_D$ with a boundary $\partial M_D=M_{D-1}$ leads to a coupled system between the $(7-D)$-dimensional theory $T[M_{D-1}]$ and the $(6-D)$-dimensional theory living on its boundary, one can study a ``defect'' along a submanifold $M_d$ of $M_D$ that are in the interior, which in general leads to a coupled system between $T[M_d]$ and $T[M_{D}\backslash M_d]$ as a codimenion-($D-d$) defect. 

\begin{figure}
    \centering
    \includegraphics[width=0.6\linewidth]{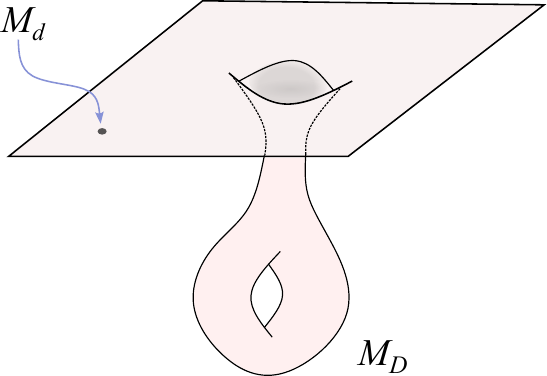}
    \caption{This illustrates how to ``blow up'' along $M_d\subset M_D$ in the case of $M_d$ being a point in $M_D=T^2$. After compactifying on this geometry, one obtains a codimension-$(D-d)$ defect, which we denote as $T[M_D\backslash M_d]$, in the theory $T[M_d]$. }
    \label{fig:BlowupDefect}
\end{figure}

To see this, one can first stretch $M_d$ to infinity. We first assume that the normal bundle of $M_d$ is trivial. Then the geometry looks like $M_d \times \R^{D-d}$ except near the origin. Therefore, after reducing on $M_d$, the system becomes a $T[M_d]$ theory living on $\R^{D-d}\backslash\{0\} \times \R^{6-D}$. Near the origin, the non-trivial geometry of $M_D$ leads to a codimension-$(D-d)$ defect along $\{0\} \times \R^{6-D}$, with the world-volume theory identified with $T[M_D\backslash M_d]$. This is illustrated in Figure~\ref{fig:BlowupDefect} for $M_D=T^2$ and $M_d$ being a point.  When the normal bundle is non-trivial, one can still trivialize it away from a submanifold $M_r$ of $M_{d}$. Therefore, one will have a nested system of quantum field theories where the theory $T[M_d\backslash M_r]$ is itself a defect in a quantum field theory $T[M_r]$ of higher dimensions. For simplicity, we will assume that the normal bundle to $M_d$ is trivial. 

In the case that $T[M_d]$ is a gauge theory, $T[M_D\backslash M_d]$ can have a global symmetry coupling to gauge fields in the bulk. Then as a necessary requirement, this global symmetry has to be either anomaly free, or cancelled by anomaly inflow from the bulk. We will see examples of both kinds in later sections.

\subsubsection{Polarizations with defects}

To have the bulk theory $T[M_d]$ well defined, one needs to choose a polarization in $\Pol(M_d)$, which will also give a well defined theory $T[M_d\times S^{D-d-1}]$. Then $T[M_D\backslash M_d]$ can be viewed as a boundary condition for the theory $T[M_d\times S^{D-d-1}]$, and the previous discussion about polarization of manifold with boundary can be applied to this case. Also, notice that when the normal bundle to $M_d$ is non-trivial, this construction still apply with the bulk theory being $T[M_d\tilde\times S^{D-d-1}]$ where product becomes non-trivial. 

Assuming that the polarization on $M_d$ is given by $L_0\subset H^*(M_d)$, then it determines a subgroup $L\subset H^*(M_d\times S^{D-d-1})$ by taking the  part of $L_0\times H^*(S^{D-d-1})$ in degrees between $D-4$ and $3$. We will write $L=L_0\oplus L_1$ although technically some elements in $L_0$ might not be in $L$ for degree reasons. Assuming that a choice of polarization for the theory $T[M_D\backslash M_d]$ is given by $L_\delta \subset H^*(M_D\backslash \nu(M_d))$ where $\nu(M_d)$ denotes a tubular neighborhood of $M_d$, then it has to be compatible with the following long exact sequence
\begin{equation}
    \ldots \rightarrow H^*(M_D\backslash \nu(M_d))\xrightarrow \partial H^*(M_d\times S^{D-d-1})\rightarrow H^{*+1}(M_D\backslash \nu(M_d),M_d\times S^{D-d-1})\rightarrow \ldots 
\end{equation}
with $\partial L_\delta=L\cap \im(\partial)$. Both $\partial L_\delta$ and $L/\partial L_\delta$ can be decomposed into two parts by intersecting with $L_0$ and $L_1$. We will denote these four groups by $(\partial L_\delta)_0$, $(\partial L_\delta)_1$, $(L/\partial L_\delta)_0$ and $(L/\partial L_\delta)_1$. Then they corresponds to the following types of operators.

\begin{itemize}
    \item $(\partial L_\delta)_0$. These are operators in the bulk theory $T[M_d]$ that can end on the defect theory $T[M_D\backslash M_d]$.
    \item $(L/\partial L_\delta)_0$. This group classifies operators in the bulk theory $T[M_d]$ that can be moved onto the defect theory $T[M_D\backslash M_d]$, up to these operators that can end on the defect.
    \item $(\partial L_\delta)_1$. These are operators in the bulk theory $T[M_d]$ that can ``intersect'' the defect theory $T[M_D\backslash M_d]$. In other words, these operators, after reducing on $S^{D-d-1}$, can end on the defect.
    \item $(L/\partial L_\delta)_1$. This group classifies operators that can ``shrink on $S^{D-d-1}$'' to become operators of $T[M_D\backslash M_d]$.
\end{itemize}

Again, one can consider the duals of these groups, which describe how different symmetries in the bulk theory are related to symmetries on the defect theory. Denoting by $B_n$ and $B_n^\delta$ the background fields for $n-1$ form symmetries in the bulk and defect, and $B_n|_\delta$ the boundary value of bulk field, then the boundary conditions for the background fields are
\begin{equation}\label{eqn:defectbc0}
    B_n|_\delta =dB_{n-1}^\delta
\end{equation}
for $(\partial L_\delta)_0^\vee$,
\begin{equation}
    B_n|_\delta =B_n^\delta
\end{equation}
for $(L/\partial L_\delta)_0^\vee$,\footnote{
Since the volume form is closed, it can be expressed as $B_n|_\delta=d\left(B_{n-D+d}^\delta \text{vol}_{S^{D-d-1}}\right)$, this is a Dirichlet boundary condition for $B_n|_\delta$.
}
\begin{equation}\label{eqn:defectbc1}
    B_n|_\delta =dB_{n-D+d}^\delta \cdot \mathrm{vol}_{S^{D-d-1}}=d\left(B_{n-D+d}^\delta \cdot \mathrm{vol}_{S^{D-d-1}}\right)
\end{equation}
for $(\partial L_\delta)_1^\vee$, and
\begin{equation}\label{eqn:defectbc2}
    B_n|_\delta =B_n^\delta\cdot \mathrm{vol}_{S^{D-d-1}}
\end{equation}
for $(L/\partial L_\delta)_1^\vee$. Here $\mathrm{vol}_{S^{D-d-1}}$ represent the generator in top degree of the group $H^{*}(S^{D-d-1})$, and will lead to a singularity of $B_n$ along the defect. When $D-d-1=1$, then it can be represented as $d\theta$ with $\theta$ being the angular coordinate for the normal directions of the codimension-2 defect.  

In the remainder of this section, we provide some additional remarks about defects in TQFTs.

\subsubsection{Topological boundary condition with defect}

We illustrate the perspective above with an example of 7d TQFT containing a defect supported on an $M_3\subset \R^7$. Near the defect, we consider the decomposition
\begin{equation}
    C=B_3+B_2^i\alpha_1^i~,
\end{equation}
where $\alpha_1^i$ lives on $M_3$. 
Then the action of the 7d TQFT theory decomposes into 
\begin{align}
    &\frac{N}{4\pi}\int CdC=\frac{N}{4\pi}\int B_3dB_3+
    \frac{N}{2\pi}\int \alpha_1^i B_2^idB_3
    +\frac{N}{4\pi}\int \alpha_1^id\alpha_1^j B_2^iB_2^j\cr 
    &=\frac{N}{4\pi}\int B_3dB_3+\frac{N}{2\pi}\int_{\text{PD}(\alpha_1^i)}B_2^idB_3
    +\frac{N}{4\pi}\int_{\text{PD}(\alpha_1^id\alpha_1^j)}B_2^iB_2^j~.
\end{align}
Near the dual of $\alpha_1^i$, the TQFT is effectively a 6d TQFT; similarly, near the dual of $\alpha_1^id\alpha_1^j$ the TQFT is effectively a 4d TQFT.

Consider the topological boundary condition for the fields $B_3,B_2^i$.
The equation of motion for the boundary variation of $B_3$ gives
\begin{equation}
    B_3|+\alpha_1^iB_2^i|=0~.
\end{equation}
This gives an example of the boundary condition discussed above.

We can also change the boundary condition by adding boundary topological terms. For instance, by adding $\frac{N}{2\pi}\alpha_1^iB_2^iB_3$ we find the boundary condition using the equation of motion for the boundary variation of $B_2^i$
\begin{equation}
    \alpha_1^iB_3|=0~.
\end{equation}

We now give some examples.

\paragraph{Wilson line in Chern--Simons theory.}

The Wilson line of charge $q$ inserted at $\gamma$ in $U(1)$ Chern--Simons theory at even level $k$ is equivalent to specifying the background for the one-form symmetry generated by unit Wilson line
\begin{equation}
    B=\frac{2\pi q}{k}\delta(\gamma)^\perp~.
\end{equation}
we can view $\frac{2\pi q}{k}$ with $q\sim q+k$ as a discrete theta angle on the worldline, i.e.~the background for a $\mathbb{Z}_k$ $(-1)$ form symmetry. This is an example of (\ref{eqn:defectbc2}), where the boundary condition is extended to the bulk.

\paragraph{Maxwell theory: surface operators and domain walls.}

Maxwell theory in $d$ spacetime dimension has $U(1)$ magnetic $(d-3)$-form magnetic symmetry generated by the operator ${\theta\over 2\pi}\oint_\Sigma F$. If we insert such surface operator, the background is
\begin{equation}
    B_{d-2}=\theta\delta(\Sigma)^\perp~,
\end{equation}
where $\theta$ is a continuous theta angle i.e.~background of $U(1)$ $(-1)$-form symmetry on the surface defect.
This is an example of (\ref{eqn:defectbc2}), where the boundary condition is extended to the bulk. Consider $(3+1)$-d spacetime, and the domain wall in Maxwell theory defined by continuously varying $\theta\rightarrow\theta+4\pi$.
The domain wall supports a level two $U(1)$ Chern--Simons theory.
The surface operator $F$ remains non-trivial on the domain wall, and thus the background for the magnetic one-form symmetry has the boundary condition
\begin{equation}
    B_2=dB_1~,
\end{equation}
where $B_1$ is the magnetic 0-form symmetry generated by the same operator $F$.
This is an example of the boundary condition \eqref{eqn:defectbc0}.

\subsubsection{Open defects in the bulk}

More generally, we can consider defects that can itself have boundary in the bulk. Such defects can be topological, but it cannot braid non-trivially with another topological operator, since the braiding can be removed by sliding the other operator through the boundary. The topological defects with boundaries generate unbroken symmetry. Such defects can always end on the boundary, with the boundary of defect lies completely on the boundary.
Examples of this arise in finite group two-form gauge theory in 3+1d with non-trivial topological action.
More generally, we can also consider situation where such defects end on the boundary also by an open defect.\footnote{
For instance, consider the 3d-4d theory
\begin{equation}
    \int_{3d}\left(
    \frac{2}{2\pi}Bd\phi + \frac{2}{2\pi} Bu+\frac{2}{4\pi}udu\right)
    +
    \int_{4d}\left(\frac{2}{4\pi}BB+\frac{2}{2\pi}dBa    \right)~.
\end{equation}
Then the operator $\oint a+\int B$ can terminate on the boundary by $\phi+\int u$, where $\phi$ is inserted at the corner where the lines $\int a,\int u$ meet.
}
Example of bulk open defects are the topological defects that live on the boundary of bulk topological domain walls in a TQFT that does not have local operators. In the 7d TQFT there are no point, line and surface operators, and thus the topological defects of codimension one, two and three can be open with topological boundary as there are no defects that can detect these opening, and these topological boundaries give (generally non-invertible) topological defects of dimension 5, 4, 3 in the 6d boundary (although these defects can in general be trivial on the 6d boundary). 
We note that in the case of 5-dimensional defect, the two sides of the domain wall are in general different when we choose a polarization. Thus unless the 6d theory has some duality property, the domain wall is a topological interface once we choose a polarization. On the other hand, for defects of higher codimension there is no such issue and they are defects in a single theory.
More about non-invertible symmetry will be discussed in Section \ref{sec:non-invertiblesymmetry}.

\section{Symmetry in compactifications}
\label{sec:symmetry}

In this section, we study various topics centered around the theme of symmetry. Indeed, they abound in theories obtain from compactifications. For $M_d=S^1$ and $T^2$, see Table~\ref{tab:G4d5d6d} for a summary of some notable symmetries in the compactification of the 6d $(2,0)$ theory. We will discuss higher group symmetry, non-invertible symmetry, symmetry from mapping class group action on the internal manifold $M_d$, ``predictable accidental symmetries'' when $M_d$ has a fibration, and anomalies of various symmetries.

\begin{table}[h]
    \centering
    \begin{tabular}{c|c|c}
    4d     & 5d & 6d  \\ \hline
    electric 1-form $Z(G)$ sym.   &  electric 1-form $Z(G)$ sym.  & 2-form sym.\\
    magnetic $\pi_1(G)$ 1-form sym.   & magnetic 2-form $\pi_1(G)$ sym.   & 2-form sym.\\
    0-form $Z(G)$ sym.  shifting $\varphi$ & electric 1-form $Z(G)$ sym.  & 2-form sym. \\
    2-form $\pi_1(G)$ sym. charge $\oint u_1$ & magnetic 2-form $\pi_1(G)$ sym.  & 2-form sym.\\
    0-form $U(1)$ sym. $j_1=\star \text{Tr}(F d\varphi)$ & 0-form  $U(1)$ sym. $j_1=\star\text{Tr}(FF)$ & isometry\\
    0-form $U(1)$ sym. $j_1=\star\text{Tr}(\star F d\varphi)$ & isometry & isometry\\\hline
    theta angle $\theta \text{Tr}(FF)$ & 0-form $U(1)$ sym. $j_1=\star\text{Tr}(FF)$ & isometry\\
    theta angle $\int \langle w_2^G,\text{Bock}(u_1)\rangle$ 
        & theta angle $\int\langle w_2^G,\text{Bock}(w_2^G)\rangle$ & \\
    theta angle $\int {\cal P}(w_2^G)$ &  $\Gamma(\pi_1(G))$ 0-form sym. charge $\oint {\cal P}(w_2^G)$ & 
    \end{tabular}
    \caption{Symmetries and theta angles (continuous and discrete) in the compcatifiation of the 6d theory on $S^1$ and $T^2$.  $F$ is the field strength for the $U(1)$ gauge field. $\Gamma(\pi_1(G))$ is the universal quadratic group of $\pi_1(G)$ (for a review, see Appendix C of \cite{Benini:2018reh}). The symmetry comes from isometry is related to the kinetic term of the gauge field in the higher dimension that depends on the metric (and thus couples to the graviphoton upon compactification).}
    \label{tab:G4d5d6d}
\end{table}

\subsection{Higher groups in compactification: relation with symmetries in NLSM}\label{sec:sigma}

As discussed in \cite{Gukov:2020btk}, compactification on manifold $M_d$ can give rise to symmetry that combines the isometry 0-form symmetry on $M_d$ with the internal symmetry: the generators of the internal symmetry can wrap cycles on $M_d$, and they are acted on by the isometry of $M_d$. Since the subsequent discussion is general, we will keep the dimension $D$ general in this subsection.

Let us first review the discussion in \cite{Gukov:2020btk} (see also \cite{DeMarco:2025pza,Najjar:2024vmm} for some related discussions). We consider the configuration of the background gauge field $C_n$ for an $(n-1)$-form finite Abelian symmetry ${\cal A}$ in the theory before compactification, which takes the following form of decomposition using the cocycles in $H^*(M_d)$: 
\begin{equation}\label{eqn:bgdecompose}
    C_{n}=\sum \pi^* B_k^\eta\cup  \pi'^*\eta^{(n-k)}~,
\end{equation}
where $\eta^{(n-k)}\in H^{n-k}(M_d,{\cal A})$, and the projections in the tensor product of manifolds
$\pi: M_d\times X\rightarrow X$,
$\pi':M_d\times X\rightarrow M_d$, with $X$ the spacetime after compactification, $B_k^\eta\in H^k(X,{\cal A})$ depends on the generator $\eta^{(n-k)}$ that it pairs with in the decomposition.
We will omit the projections $\pi,\pi'$ in the following to simplify the notation.

The backgrounds $B_k$ in (\ref{eqn:bgdecompose}) correspond to the symmetries whose codimensional-$k$ generator on submanifold $Y_{D-k}$ is equivalent to inserting the generator of the symmetry for $C_n$ at the Poincar\'e dual of $\eta^{(n-k)}$ with respect to $Y_{D-k}$, which has codimension $n$.
If the generator of the symmetry for $C_n$ is described by the $(D-n)$-dimensional operator $\int x_{D-n}$, then the generator for the symmetry of $B_k$ is $\int \eta^{n-k}\cup x_{D-n}$.

In the presence of background for the isometry of $M_d$, $\eta^{(n-k)}$ are modified with respect to equivariant cohomology, such that they are no longer closed; then demanding that $C_{n}$ is closed requires non-trivial relations between $B_k$, and this can be described by higher-group symmetry. 

\subsubsection{Relation to defects in non-linear sigma models}

The discussion of the higher group symmetry is closely related to symmetries in non-linear sigma model with target space $M_d$.
Let us illustrate the relation and give another way to understand the higher group symmetry in compactification.

In non-linear sigma model, there are various ``electric defects" on submanifolds decorated with topological action of the sigma model fields. There are also ``magnetic defects" labeled by boundary condition of the sigma model fields, which can be described by homotopy groups, and they can also be stuck at the junctions of the isometry defects corresponding to isometries of the target space. This is discussed in more detail in \cite{Hsin:2022heo}. As described there, when the ``electric defects" intersect the junction, there are additional electric defects emitted that are lower-dimensional submanifolds decorated with topological action of the sigma model fields. For example, denote the sigma model field by $\phi$ with target space $M_d$, and $\omega\in H^*(M_d,U(1))$, then there are actions given by $\int \phi^*\omega$ integrated on the submanifolds.

In our case, when the junctions of the defects that generate isometries of $M$ intersect $\eta^{(n-k)}$, there is an additional cocycle $\beta$ of lower degree given by cap products of $\eta^{(n-k)}$ (that depends on the codimensions of the junction and $(n-k)$ \cite{Hsin:2022heo}. 
For degree $\ell < n-k$, there are additional defects of higher codimension $(n-\ell)>k$ on submanifold $Y_{D-n+\ell}$ given by inserting the generator of symmetry for $C_n$ at the Poincar\'e dual of $\beta$ with respect to $Y_{D-n+\ell}$. Equivalently, the new defect is the $(D-n+\ell)$-dimensional operator $\int \beta\cup x_{D-n}$.
Thus when the generator of the symmetry for $B_k$ intersect the junction of the defects for isometry symmetry, it produces additional defects of lower dimensions; therefore the defects combine to form higher-group junction.

To be concrete, let us take $n=3$. Then the expansion (\ref{eqn:bgdecompose}) with the closure condition of $C_n$ can be interpreted as finding the symmetry in 2+1d sigma model, where we restrict the background gauge fields $B_k$ such that the anomaly of various symmetries are canceled by the local counterterm $B_3$. 

\subsubsection{Example: compactification on $\mathbb{C}\mathbb{P}^2$}

Let us illustrate the discussion by compactification of two-form symmetry in 6d on $M_4=\mathbb{C}\mathbb{P}^2$.

The cohomology of $\mathbb{C}\mathbb{P}^2$ is generated by K\"ahler form $\Omega_2$. Decompose
\begin{equation}
    C=B_3+B_1\Omega_2~.
\end{equation}
$\mathbb{C}\mathbb{P}^2$ has $PSU(3)$ isometry. To examine the equivariant version of the K\"ahler form, consider an auxiliary model of three complex scalars of charge one coupled to $U(1)$ gauge field, and the scalar condenses due to a $PSU(3)$ singlet potential. The K\"ahler form arises from the magnetic flux of the $U(1)$ gauge field. In the presence of $PSU(3)$ background gauge field, the magnetic flux becomes quantized as $(1/3) w_2^{PSU(3)}$ mod 1, where $w_2^{PSU(3)}$ is the obstruction to lifting the $PSU(3)$ gauge field to $SU(3)$ gauge field. Denote the resulting K\"ahler form coupled to background $PSU(3)$ gauge field $A$ by $\omega_2(A)$. The decomposition of $C$ is
\begin{equation}
    C=B_3+B_1\omega_2(A)~.
\end{equation}
For the gauge transformation $B_1\rightarrow B_1+d\lambda$ to leave $e^{i\oint C}$ invariant, $B_3$ also transforms as $B_3\rightarrow B_3-(1/3) d\lambda w_2^{PSU(3)}(A)$.
The theory has 3-group symmetry with backgrounds satisfying
\begin{equation}
 dB_3=B_1\text{Bock}(w_2^{PSU(3)}(A))   ~.
\end{equation}

The central question here is how the geometric forms such as $\omega\in H^*(M_d,U(1))$ on the internal manifold $M_d$ get modified in the presence of symmetries like $PSU(3)$.  In the context of compactification, this leads to higher group symmetries as above. In the sigma model context, it is about how the symmetries generated by operator $\int \phi^*\omega$ mix with the symmetries such as $PSU(3)$--which is related to whether they have fractional fluxes.

\subsection{Accidental symmetry from fibration}\label{sec:accidental}

Sometimes, the theory $T[M_d]$ can have extra symmetries in addition to isometries of $M_d$ and what is naively obtained from symmetries of the 6d theory via compactification. We will refer to these as ``accidental symmetries.'' In general, it is a hard problem to predict them, but there is a class of ``predictable accidental symmetries'' that we will discuss now.

One example is when the 6d theory is of type $A_{N-1}$ and $M_3=L(k,1)$ with $N$ and $k$ coprime. Then expanding $C_3=B_3+\sum B_2^\eta \eta^{(1)}+\sum B_1^{\eta'}\eta'^{(2)}+\sum B_0^{\eta''}\eta''^{(3)}$, one expect there is no 1-form symmetry (with background $B_2^\eta$) since $H_1(M_3,\Z_N)=0$. However, the theory $T[M_3]$ in this case is conjectured to be the 3d $\CN=2$ $SU(N)_k$ Chern--Simons theory with an adjoint chiral multiplet (see e.g.~\cite{Gukov:2015sna,Pei:2015jsa} for discussions and checks of this proposal), which always has a $\Z_N$ 1-form symmetry. 

This is a quite general phenomenon when $M_d$ is the total space of a non-trivial fibration. Indeed, the lens space $L(k,1)$ is the total space of degree-$k$ $S^1$ bundle over $S^2$, and the symmetry of the theory $T[L(k,1)]$ doesn't depend on $k$. In particular, they all have the same $\Z_N$ 1-form symmetry, which is only expected when $k=0$ and $M_3$ becomes $S^1\times S^2$.\footnote{When $k$ is not coprime with $N$, one also has to choose a polarization to specify the 1-form symmetry of the theory. We will choose one with maximal possible 1-form symmetry. This corresponds to having gauge group $SU(N)$ instead of a quotient of it.} 

In general, when $M_d$ is a fibration with fiber $\mathbf{F}$ and base $\mathbf{B}$, one can obtain the theory $T[M_d]$ by first compactifying on $\mathbf{F}$ and then $\mathbf{B}$. Compactification on $\mathbf{F}$ leads to a theory $T[\mathbf{F}]$. The theory have symmetries given by isometries of $\mathbf{F}$, which will have non-trivial background on $\mathbf{B}$. However, as long as the isometries involved act trivially on $H^*(\mathbf{F},D)$, turning on a non-trivial background will not break the symmetry of the theory that comes from the 2-form symmetry of the 6d theory, as there won't be mixed anomalies, after further reducing on $\mathbf{B}$. Therefore the symmetry of $T[M_d]$ will be the same as that of $T[\mathbf{F} \times \mathbf{B}]$. 

We now give a few more remarks.
\begin{itemize}
    \item The discussion above concerns only the symmetries that originate from the 2-form symmetry of the 6d theory and these coming from isometries of $\mathbf{F}$. There may be other (super-)symmetries in both $T[\mathbf{F} \times \mathbf{B}]$ and $T[M_d]$ that are different from each other. 
    
    \item On the other hand, the argument above can be generalized to any symmetries of $T[\mathbf{F}]$ that is not acted upon by isometries of $\mathbf{F}$ for which we turn on a non-trivial background.
    
    \item The anomaly of the symmetry $T[M_d]$ is in general different from that in $T[\mathbf{F} \times \mathbf{B}]$. This is clear in the example of $T[L(k,1)]$ which will be discussed in more detail later. 
    
    \item The identity component of $\mathrm{Iso}(\mathbf{F})$ always acts trivially on $H^*(\mathbf{F},D)$. The example of $L(k,1)$ belong to this class. One example of isometry that acts non-trivially on homology is the $S$-action on $\mathbf{F}=T^2$. If $M_3$ is the mapping torus of $T^2$ given by $S$, the symmetry of $T[M_3]$ is in general different from that of $T[T^3]$. 
\end{itemize}

One question that remains is how to get $\CT^{\text{bulk}}[M_d]$ in these cases that includes the accidental symmetries. As the action of the topological theory is a function on the homotopy class of maps $[M_d,Y]$, where $Y$ for us is often $BG$ for some (higher) group $G$ or variants thereof that encode additional structures, one can first rewrite it using the fibration structure. The data for the fibration is an element of $[\mathbf{B},B\text{Diff}(\mathbf{F})]$, and $[M_d,Y]$ is then identified with the pre-image of 
\begin{equation}
    [\mathbf{B},[\mathbf{F},Y]_{\text{Diff}(\mathbf{F})}]\rightarrow[\mathbf{B},B\text{Diff}(\mathbf{F})],
\end{equation}
where 
\begin{equation}
[\mathbf{F},Y]_{\text{Diff}(\mathbf{F})}:=[\mathbf{F},Y]\times_{\text{Diff}(\mathbf{F})} E{\text{Diff}(\mathbf{F})}
\end{equation}
is given by the Borel construction. One can check that this indeed leads to the expected result when $Y$ is a point and when $M_d=\mathbf{F}\times \mathbf{B}$. For example, when $M_d$ is a product, we get the pre-image of the trivial map in $[\mathbf{B},B\text{Diff}(\mathbf{F})]$, which is then $[\mathbf{B},[\mathbf{F},Y]]=[\mathbf{B}\times \mathbf{F},Y]$. 
Done in this way, fields on $\mathbf{F}$ are not killed by the fibration, but are instead made Diff$\mathbf{F}$-equivariant, and can survive after the further compactification on $\mathbf{B}$. One can think of $[\mathbf{F},Y]_{\text{Diff}(\mathbf{F})}$ as the classifying space for the higher group that emerge after compactifying on $\mathbf{F}$ analyzed previously.

This may sound fairly abstract, but is in fact just a more formal way to state and generalize what we have done in some simple cases. For example, when $\mathbf{F}$ is $S^1$, it is often good enough to just remember its $U(1)$ isometry group. We described how to make the fields equivariant and part of a higher group in Section~3.5.1 of the first part of this series.

\subsubsection{Example: discrete one-form anomaly from lens space compactification }
\label{sec:oneformsymmetrycompactification}

Let us illustrate how to use TQFT to discuss compactification on lens space $L(k,1)$.

For example, let us start with the $CdC$ TQFT for $A_{N-1}$ theory, which can be dualized to
\begin{equation}
    -\frac{N}{4\pi}C_3dC_3~.
\end{equation}
When we compactify the TQFT on the lens space, we use the decomposition into free part, torsion part and discrete part (see (3.57), (3.59) of \cite{Gukov:2020btk})
\begin{equation}
    C_3=B_3+\alpha_3B_0+\tau_2B_1+\hat \tau_1 B_2~,
\end{equation}
where $k\tau_2=d\hat\tau_1$ for $\tau_2$ dual to the $\mathbb{Z}_k$ torsion 1-cycle on the lens space, and $\alpha_3$ is the volume form. The fields $B_3,B_0$ are the free part. Similarly, the coboundary is
\begin{equation}
    dC_3=dB_3-\alpha_3 dB_0+\tau_2 dB_1+k\tau_2 B_2-\hat \tau_1 dB_2~.
\end{equation}
Using $\int \hat \tau_1 \tau_2=1$, we find that the action of the 7d theory reduced on lens space into
\begin{equation}
-\frac{N}{2\pi}\int B_0dB_3-\frac{N}{2\pi}\int B_1dB_2
    -\frac{Nk}{4\pi}\int B_2B_2~.
\end{equation}
In particular, the one-form symmetry coupled to gauge field $B_2$ in the reduction theory has anomaly described by
\begin{equation}
    -\frac{Nk}{4\pi}\int B_2B_2~.
\end{equation}
If we change the normalization of $B_2$ to be $0,1,\cdots,N-1$, this becomes
\begin{equation}
    -2\pi \frac{k}{2N}\int {\cal P}(B_2)~.
\end{equation}
Such anomaly means that the reduction theory has $\mathbb{Z}_N$ one-form symmetry generated by a line operator with $\mathbb{Z}_N$ fusion rule and spin $-\frac{k}{2N}=\frac{k(N-1)}{2N}$ mod $1/2$. (Since the theory has local fermions, the spin can be modified by $\frac{1}{2}$ by attaching with the transparent fermion line). This is the anomaly for $\mathbb{Z}_N$ one-form symmetry in $SU(N)_k$ Chern--Simons theory \cite{Hsin:2018vcg} (see also \cite{Najjar:2024vmm} for a computation that is similar in spirit to the one above).

\subsection{Non-invertible symmetry: compactification of TQFT fusion coefficients}
\label{sec:non-invertiblesymmetry}

In this section, we will discuss topological operators that do not obey Abelian fusion rules. Such topological operators generate non-invertible symmetry.

The fusion coefficients for non-invertible symmetry generators are in general TQFTs.\footnote{For more examples, see e.g. \cite{Roumpedakis:2022aik,Choi:2022zal,Cordova:2024jlk,Cordova:2024mqg}.}
For example, in $\mathbb{Z}_2$ gauge theory fusing the $k$-dimensional defect $S$ where the electric charge condenses gives
\begin{equation}
    S\times S=(\mathbb{Z}_2\text{ scalar in }k\text{ dimensions}) S~.
\end{equation}
We can describe the fusion coefficients in two ways:
\begin{itemize}
    \item If we describe the fusion using junction of three generators, the fusion coefficient means that the junction lives on the boundary of the fusion coefficient TQFT. 

    \item  If we describe the fusion by bringing the two generators in parallel, the fusion coefficient means that we get a decoupled TQFT in the fusion outcome.

\end{itemize}
The fusion degeneracy comes from the dimension of Hilbert space of the TQFT: when the TQFT has dimension greater than one, there is nontrivial fusion multiplicity. When we compactify theories with non-invertible symmetry, we also need to compactify the TQFT from the fusion coefficient, and this can convert a non-invertible symmetry to an invertible symmetry.
For example, the fusion coefficient TQFT can become invertible. We will give an example of compactification of non-invertible symmetries producing invertible symmetries.

\subsubsection{Condensation defect}

The 6d theory has two-form symmetry $D$, we will study the condition where we can gauge the two-form symmetry on submanifolds of dimension four and five. In other words, we sum over the volume operator insertions on the submanifolds. 
For this to be consistent, the volume operator needs to have trivial F symbol.\footnote{
The anomaly of finite group one-form symmetry in 5d can be described by 6d effective action cubic in the background gauge field.
}
The F symbols for $D=\mathbb{Z}_N$ and anomaly $p$ can be computed from the Abelian 7d three-form Chern--Simons theory following the method in \cite{Kapustin:2010hk}
\begin{equation}
    F^{Q,Q',Q''}=(-1)^{\frac{p}{N}(Q+Q'-[Q+Q'])[Q'']}~,
\end{equation}
where $[Q]=Q\text{ mod }N$.
Thus for even $p$ or odd $N$ there is a well-defined condensation defect associated with two-form symmetry.
More generally, there can be condensation defects for subgroups $D'$ of $D$.

If we denote such defects by ${\cal C}$, 
\begin{equation}
    {\cal C}_\omega(W)=\frac{|H^1(W,D')|}{|H^0(W,D')||H^2(W,D')|}\sum_{{\cal V}\in H_3(W,D')} e^{i\omega({\cal V})} \eta({\cal V})~,
\end{equation}
where $\eta({\cal V})$ is the volume operator that generates two-form symmetry in 6d, and the phase $e^{i\omega({\cal V})}$ is the partition function for an invertible theory. 

\paragraph{4-dimensional condensation defects}
Let's begin with the case that $W$ is 4-dimensional. Denote the Poincar\'e dual of ${\cal V}$ by 1-form gauge field $a\in H^1(W,D')$, then the phase can be classified by $\Omega^4_{SO}(BD')$.
Let us focus on the group cohomology phase $H^4(BD',U(1))$. For $D'=\prod\mathbb{Z}_{N_i}$, denote  $a=(a^1,a^2,\cdots)$ with $\mathbb{Z}_{N_i}$ gauge field $a^i$. The phase can be expressed as follows:
\begin{equation}
 \omega(a)=2\pi\int\left( \sum_{i\neq j} \frac{k_{ij}}{N_i}a^ia^j\text{Bock}(a^j)
+ \sum_{i<j<l}\frac{k_{ijl}}{N_i}
 a^ia^j\text{Bock}(a^l) \right)~,
\end{equation}
where the phases are la belled by $k_{ij}\in\mathbb{Z}_{\gcd(N_i,N_j)}$ and $k_{ijl}\in\mathbb{Z}_{\gcd(N_i,N_j,N_l)}$.

Hereafter we will focus on 5-dimensional condensation defects.

\paragraph{5-dimensional condensation defects}

Now, let's discuss the case when $W$ is 5-dimensional.
Denote the Poincare dual of ${\cal V}$ by $B\in H^2(W,D')$, then we can write the phase as the partition function for the topological action
\begin{equation}
e^{i\omega({\cal V})}=e^{i    \int \langle B,\text{Bock}(B)\rangle}~. 
\end{equation}
The phases are classified by $\Omega^5_{SO}(B^2D')$.
For $D'=\prod \mathbb{Z}_{N_i}$, it is described by $(k_i,k_{ij})\in \prod_i \mathbb{Z}_{\gcd(N_i,2)}\times \prod_{i<j}\mathbb{Z}_{\gcd(N_i,N_j)}$:
\begin{equation}
\omega(B)=2\pi\int \left(\sum_i    \frac{k_i}{N_i} B^i\text{Bock}(B^i)+\sum_{i<j}\frac{ k_{ij}}{N_i}B^i\text{Bock}(B^j)\right),\quad k_{ij}\in\mathbb{Z}_{\gcd(N_i,N_j)}~.
\end{equation}

In 6d ${\cal N}=(2,0)$ theories there are following condensation defects:\footnote{
For $\mathfrak{so}(2n)$, the F symbol is the same as that of the $Spin(2n)_1$ Chern--Simons theory as listed in \cite{Kitaev:2005hzj}.
}
\begin{itemize}
    \item $\mathfrak{su}(N)$ for odd $N$. $D=\mathbb{Z}_N$.
Any subgroup of $\mathbb{Z}_N$, labeled by divisors of $N$, have trivial F symbol, and they give rise to condensation defects. 
    Since the divisors of $N$ are odd, the weight  $\omega$ is trivial for 5-dimensional submanifolds. Thus
    the number of 5-dimensional condensation defects both equal to the number of divisors of $N$ respectively.
    
    \item $\mathfrak{su}(N)$ for even $N$. $D=\mathbb{Z}_N$.
    Only the subgroups of $\mathbb{Z}_{N/2}$, which are generated by even charges, have trivial F symbol, and they give condensation defects.
    For divisor of $N/2$ that are even, the weight $\omega$ can be trivial or non-trivial. Thus the number of 5-dimensional condensation defects equal the sum of number of odd divisors of $\mathbb{Z}_{N/2}$ and twice the number of even divisors of $\mathbb{Z}_{N/2}$. 
    
    \item $\mathfrak{so}(4n+2)$ has $D=\mathbb{Z}_4$.
    Only the $\mathbb{Z}_2$ subgroup has trivial F symbol, and there are two 5-dimensional condensation defects.
    
    \item $\mathfrak{so}(8n+4)$ has $D=\mathbb{Z}_2\times\mathbb{Z}_2$. 
    Only one $\mathbb{Z}_2$ subgroup has trivial F symbol, and there are two 5-dimensional condensation defects .
    
    \item $\mathfrak{so}(8n)$ has $D=\mathbb{Z}_2\times\mathbb{Z}_2$ with trivial F symbol.
    There are $2+2+2+2\times 2\times 2=14$ 5-dimensional condensation defects from the three $\mathbb{Z}_2$ subgroups and $\mathbb{Z}_2\times\mathbb{Z}_2$ subgroup.
    
    \item $\mathfrak{e}_6$ has $D=\mathbb{Z}_3$. There is one 5-dimensional condensation defect.
    
\end{itemize}

\paragraph{Fusion of condensation defect with volume operator.}

Let us fuse ${\cal C}_\omega(W)$ with $\eta({\cal V}')$. 
We have
\begin{align}
    &\eta({\cal V}')\times {\cal C}_\omega(W)= \cr &\!\!\!\!\!\!\!\!\!\!\!\!\!   \frac{|H^1(W,D')|}{|H^0(W,D')||H^2(W,D')|}\sum_{{\cal V}\in H_3(W,D')} e^{i\omega({\cal V}+{\cal V}')-i\left(\omega({\cal V}+{\cal V}')-\omega({\cal V})\right)} \eta({\cal V}+{\cal V}')
    ={\cal C}_{\omega}(W)e^{-i\omega({\cal V}')},
\end{align}
where we used $\int \langle B,\text{Bock}(B')\rangle +\langle B',\text{Bock}(B)\rangle =0$.

\paragraph{Fusion among condensation defect.}
For trivial weight, it obeys the fusion rule
\begin{align}\label{eqn:fusionComega}
    &{\cal C}_\omega(W)\times \bar {\cal C}_\omega(W)\cr
    &=
    \left(\frac{|H^1(W,D')|}{|H^0(W,D')||H^2(W,D')|}\right)^2\sum_{{\cal V},{\cal V}'\in H_3(W,D')}
    e^{i\omega({\cal V})-i\omega({\cal V}')}\eta({\cal V})\eta({\cal V}')\cr 
    &=
    \left(\frac{|H^1(W,D')|}{|H^0(W,D')||H^2(W,D')|}\right)^2\sum_{{\cal V},{\cal V}'\in H_3(W,D')}
    e^{i\omega({\cal V}+{\cal V}')-2i\omega({\cal V}')}\eta({\cal V}+{\cal V}')\cr 
    &=\chi(W,D') Z_\text{TQFT}^{-2\omega}(W){\cal C}_\omega(W)~,
\end{align}
where $Z_\text{TQFT}^{-2\omega}(W)$ is the partition function of two-form $D'$ gauge theory on $W$ with topological action $-2\omega$, and $\chi(W,D')$ is a Euler counterterm.

We note that one can also construct similar condensation defect of dimension four (for dimension three this is the direct sum of volume operator). In four-dimension, the volume operator generates 0-form symmetry, and the symmetry is anomaly free if the analogue of F symbol for fusing five domain wall is trivial. From the parent $CdC$ theory, such associator is trivial (fusing four volume operator has trivial associator by the pentagon identity for $F$ symbol, and thus the associator for fusing more operators is trivial), and thus one can condense the volume operator on four-dimensional locus for every theory. Thus we find that all 6d theories with two-form symmetry has four-dimensional condensation defect, while five-dimensional condensation defect only occurs in certain theories.

\paragraph{Action of condensation defect on other operators.}

We can define action of condensation defect on other operator by surrounding the operator with the condensation defect and shrink the defect. 
Equivalently, we can move the condensation defect passing through other operators.

In general, relativistic $q$-form symmetry can only act on operators supported on submanifolds of dimension greater or equal to $q$. The generator of $q$-form symmetry has trivial correlation function with operators of lower dimensions.
Thus the condensation defect of the $q$-form symmetry generator acts trivially on operators of dimension lower than $q$.
The condensation defect of 2-form symmetry acts trivially on the local operators and line operators, but it can act on surface operators.

Let us consider codimension-one condensation defect for $D'=\mathbb{Z}_N$ two-form symmetry. Take a surface operator $S_Q$ of two-form charge $Q\in\mathbb{Z}_N$ and wrap it with the condensation domain wall $S^3\times S^2$, then shrink the domain wall. Following \cite{Hsin:2019fhf,Roumpedakis:2022aik}, we find that this produces an open volume operator
\begin{equation}
{\cal C}\cdot S_Q=    \frac{1}{N}\sum_{n,m=0}^{N-1} e^{2\pi i n(pm/2+Q)\over N} \eta({\cal V})^m S_Q~,
\end{equation}
where ${\cal V}$ is an open volume with boundary $S_Q$.

Let us consider an example. For ${\cal N}=(2,0)$ $A_{N-1}$ theory, $p=N-1$. Let us consider $N=3$ (note $N$ is odd to define the condensation defect). Then the above formula for $Q=1$ gives
\begin{equation}
{\cal C}\cdot S_1=    \frac{1}{3}\sum_{n,m=0}^{2} e^{2\pi i (m+1)n\over 3} \eta({\cal V})^m S_1= \eta({\cal V})^2S_1=\eta({\cal V})^{-1}S_1~,
\end{equation}
where ${\cal V}$ is an open volume with boundary $S_1$.
Similarly, ${\cal C}\cdot S_2=\eta({\cal V})S_2$.

In compactification, the condensation defect becomes condensation defects in lower dimensions.
We can also consider twisted compactification where the condensation defect wraps a non-trivial cycle on the internal manifolds, with simplest case being a point $p\in M_d$.

\begin{figure}[t]
  \centering
    \includegraphics[width=0.3\textwidth]{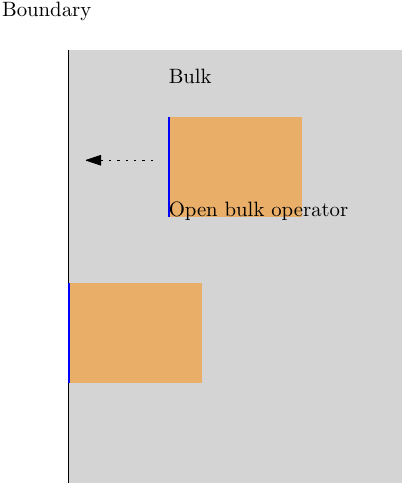}
      \caption{Bulk operator with topological boundary can always end on the boundary to give topological defect on the boundary.}
      \label{fig:openbulkopr}
\end{figure}

\subsubsection{Twist defect from 0-form symmetries in 7d theory}

The 7d $CdC$ theory can have invertible and non-invertible 0-form symmetries that permute the volume operators. 
These symmetries are generated by 6d topological domain walls, which can be classified by topological boundaries of the folded ``double theory'' ${\cal C}\times\overline{{\cal C}}$ of the $CdC$ theory ${\cal C}$, where bar denotes orientation reversal.
These domain walls have gapped boundaries since there are no local operators that can detect the ``holes'' on the topological domain walls.
They are condensation defects in 7d TQFT. 
Let us focus on the invertible symmetries.

Since the 7d TQFT does not have topological local operators, the domain wall defects can have a topological boundary, that describes
5-dimensional topological defect. See Figure \ref{fig:openbulkopr}. If we gauge the symmetry in the bulk, these topological defects become genuine 5-dimensional topological defects in the resulting TQFT.

Alternatively, we can end the 6d domain wall on the boundary using the 5d topological defect, 
then it gives a 5d domain wall on the boundary,
 see Figure \ref{fig:openbulkopr}.
In general, it will be an interface on the boundary that separate different boundary theories, related by polarizations that are acted by the symmetry in the 7d TQFT.
Only when the polarization respects the symmetry that the 5d domain wall becomes a domain wall in the same 6d boundary theory.

As an example, let us consider the 7d TQFT of $A_{N-1}$ type with $N>2$. (See also \cite{Apruzzi:2024cty} for related discussions.)
 
 \paragraph{Charge conjugation symmetry.} 
 The 7d theory has $\mathbb{Z}_2$ charge conjugation symmetry that flips sign of the three-form gauge field, and thus permutes the volume operator $Q\rightarrow -Q$ for $Q\in\mathbb{Z}_N$.

The 6d domain wall in the 7d theory generating this symmetry is given by the condensation of the charge $2$ volume operator. To see this, we note that
 the action on the domain wall on volume operators in the 7d TQFT is
 \begin{itemize}
     \item Even $N$.
 \begin{equation}
     {\cal C}\cdot \eta^Q({\cal V})=\frac{1}{N/2}\sum_{n,m=0}^{N/2-1} e^{2\pi i n((N-1)(m+Q)/(N/2)} \eta^{Q+2m}({\cal V})= \eta^{-Q}({\cal V})~,
 \end{equation}
where the summation over $n$ restricts the sum to $m=-Q$.

     \item Odd $N$.
 \begin{equation}
     {\cal C}\cdot \eta^q({\cal V})=\frac{1}{N}\sum_{n,m=0}^{N-1} e^{2\pi i (2n)((N-1) (m+Q)/N} \eta^{Q-2m}({\cal V})= \eta^{-Q}({\cal V})~,
 \end{equation}
where the summation over $n$ restricts the sum to $m=-Q$.

 \end{itemize}

\paragraph{Twist defect.}
 
Let us denote the open 6d domain wall by ${\cal D}$, and we will focus on even $N$. When it ends on 6d boundary, it gives a 5d domain wall defect. We will call it the twist defect.
The twist which can be expressed as (denote the 6d domain wall by $M_6$ with boundary $\partial M_6=M_5$)
\begin{equation}
    {\cal D}(M_6)=\frac{|H^1(M_6,M_5,\mathbb{Z}_{N/2})|}{|H^0(M_6,M_5,\mathbb{Z}_{N/2})||H^2(M_6,M_5,\mathbb{Z}_{N/2})|}\sum_{{\cal V}\in H_3(M_6,\mathbb{Z}_{N/2})} \eta({\cal V})~,
\end{equation}
where $\eta({\cal V})$ is the generator for $\mathbb{Z}_{N/2}$ subgroup 2-form symmetry, and it satisfies $\eta({\cal V})\eta({\cal V}')= e^{\frac{2\pi i (N-1) 2^2 \#({\cal V},{\cal V}')}{2N}}\eta({\cal V}+{\cal V}')=e^{\frac{2\pi i(N-1) \#({\cal V},{\cal V}')}{N/2}}\eta({\cal V}+{\cal V}')$.

When the polarization respects the charge conjugation symmetry in 7d TQFT $C\mapsto -C$, this twist defect is a topological domain wall in the same 6d theory.

\paragraph{Fusion rule of twist defect.}
Let us compute its fusion with orientation reversal in 7d. When $M_6$ does not have a boundary,\footnote{It can be obtained from the inflow argument in \cite{Hsin:2021mjn,Chen:2021xuc,Barkeshli:2022edm}, or the method in \cite{Choi:2021kmx,Kaidi:2021xfk,Choi:2022zal}.}
using $\eta({\cal V})\eta(-{\cal V}')=e^{-\frac{2\pi i  (N-1) \#({\cal V},{\cal V}')}{N/2}}\eta({\cal V}-{\cal V}')$, we have:
\begin{multline}
    {\cal D}(M_6)\times \overline{{\cal D}(M_6)}
    \\=
    \left(\frac{|H^1(M_6,\mathbb{Z}_{N/2})|}{|H^0(M_6,\mathbb{Z}_{N/2})||H^2(M_6,\mathbb{Z}_{N/2})|}\right)^2\cdot
\sum_{{\cal V},{\cal V}'\in H_3(M_6,\mathbb{Z}_{N/2})} \eta({\cal V}-{\cal V}')e^{-2\pi i(N-1)\#({\cal V},{\cal V}')/(N/2)}\cr 
=    \left(\frac{|H^1(M_6,\mathbb{Z}_{N/2})|}{|H^0(M_6,\mathbb{Z}_{N/2})||H^2(M_6,\mathbb{Z}_{N/2})|}\right)^2\cdot 
\sum_{{\cal V},{\cal V}'\in H_3(M_6,\mathbb{Z}_{N/2})} \eta({\cal V}-{\cal V}')e^{-2\pi i(N-1)\#({\cal V}-{\cal V}',{\cal V}')/(N/2)}\cr
=\frac{|H^1(M_6,\mathbb{Z}_{N/2})|}{|H^0(M_6,\mathbb{Z}_{N/2})||H^2(M_6,\mathbb{Z}_{N/2})|}=\chi(M_6,\mathbb{Z}_{N/2})^{-1}~,
\end{multline}
where we used $\#({\cal V}',{\cal V}')=0$ in the second equality, the property that the summation of ${\cal V}'$ restricts the summation over ${\cal V}$ to be trivial in
 the third equality, and the definition of $\chi$ in the last one (i.e.~when $H^3$ is free, $\chi=(N/2)^{b_0-b_1+b_2}$). Thus for closed $M_6$, the defect is invertible.
Another way to write the last equation is:
\begin{equation}
\left(\frac{|H^1(M_6,\mathbb{Z}_{N/2})|}{|H^0(M_6,\mathbb{Z}_{N/2})||H^2(M_6,\mathbb{Z}_{N/2})|}\right)^2\sum_{b,b'}  \exp\left({2\pi i\over N/2}\int b_3 \cup b_3'\right)=\chi(M_6,\mathbb{Z}_{N/2})^{-1}~,
\end{equation}
where $b_3,b_3'$ are $\mathbb{Z}_{N/2}$ three-form gauge fields.

When $M_6$ has a boundary $\partial M_6=M_5$, the twist defect becomes non-invertible: the fusion with its orientation reversal gives the condensation defect, where on the half-space $x<0$ on the 6d boundary separated by the 5d domain wall at $x=0$ that supports the twist defect we have
\begin{equation}
\frac{2\pi}{N/2}\int_{x<0} b_3\cup b_3'~,
\end{equation}
with Dirichlet boundary condition for the $\mathbb{Z}_{N/2}$ three-form gauge fields $b_3,b_3'$ on the domain wall. This describes the condensation defect of gauging $\mathbb{Z}_{N/2}$ symmetry generated by the volume operator on the 5d domain wall at $x=0$ \cite{Choi:2022zal}.
Let us denote the twist defect by ${\cal D}(M_5)^{6d}$, then it obeys the fusion rule
\begin{equation}\label{eqn:6ddualitydefect}
    {\cal D}(M_5)^{6d}\times \overline{{\cal D}(M_5)^{6d}}={\cal C}(M_5)~,
\end{equation}
where ${\cal C}$ is the condensation defect for the $\mathbb{Z}_{N/2}$ volume operators.

\subsubsection{Invertible symmetries from compactification of non-invertible symmetry}
\label{sec:compactifynoninvertible}

Let us illustrate the phenomenon that the compactification of theory with non-invertible symmetry can give rise to theories where the non-invertible symmetry becomes invertible.

A non-invertible fusion rule means that there are multiple fusion outcomes, 
\begin{equation}
    \sum_i {\cal O}_i(\gamma_i)~,
\end{equation}
where ${\cal O}_i$ is supported on submanifold $\gamma_i$.
If we compactify on an internal manifold such that $\gamma_i=0$ is trivial in cohomology for all $i$ except for one, the fusion becomes invertible after compactification. In particular, if we fuse a symmetry operator with its orientation reversal dual, when the fusion outcome only has contribution from all $\gamma_i=0$, i.e. the identity operator (with suitable normalization factor), the symmetry becomes invertible.

\paragraph{Example: compactification of Maxwell theory.}

For instance, consider compactifying free $U(1)$ gauge theory in 3+1d on $S^3$ to quantum mechanics, at $\tau=iN$. The parent theory has Kramers-Wannier duality defect for gauging the $\mathbb{Z}_N\subset U(1)$ subgroup one-form symmetry: the fusion of the defect with itself gives the charge conjugation symmetry composed with a wall that support the condensation defect of the surface operators that generate the $\mathbb{Z}_N$ one-form symmetry.
On the other hand, there are no non-contractible surfaces, and thus such defects become invertible after compactification. The compactification of Maxwell theory on $S^3$ produces $U(1)$ gauge theory in 0+1d,

\paragraph{Example: compactification of a 6d theory.}

Let us illustrate the discussion in the case of
the compactifications of the 6d theory on internal manifold $M_4$ that does not have any 3-cycle or 2-cycle, such as $M_4=S^4$, the domain wall in 6d is supported on the domain wall in 2d spacetime and wrapping $M_4$. Since there are no non-trivial three-cycles on the domain wall, ${\cal C}(M_5)=1$ in \eqref{eqn:6ddualitydefect}, and
the fusion of the twist defect ${\cal D}$ with its orientation reversal becomes the trivial defect, and thus the non-invertible symmetry in 6d generated by ${\cal D}$ becomes invertible symmetry in the compactified theory $T[M_4]$ for such $M_4$.

In other words, there are symmetries in $T[M]$ theory not from invertible symmetry in 6d or the mapping class group of $M$, but rather from {\it non-invertible} symmetry in 6d.

\subsection{Mapping class group action as invertible defects}\label{sec:MCGDefect}

In Part I of the paper, we have discussed mapping class group action on polarizations,
\begin{equation}
    \mathrm{MCG}(M_d)\acts \mathrm{Pol}(M_d).
\end{equation}
Here we will look again at this action from a different perspective as invertible defects in $\CT^{\text{bulk}}[M_d]$ and then generalize it. 

An action on the internal manifold $M_d$ by an element $g\in \mathrm{MCG}(M_d)$ generates a duality of $\CT^{\text{bulk}}[M_d]$, and one can consider a duality interface $\CI_g$ in this $(7-d)$-dimensional theory. This would be automatically an invertible defect as it would be annihilated by $\CI_{g^{-1}}$. When we have $\CT^{\text{bulk}}[M_d]$ living on an interval with two boundaries being $T[M_d]$ and the topological boundary condition $B_{\CP}$ labeled by a polarization $\CP$, we can insert the pair $\CI_{g^{-1}}$ and $\CI_g$ on two points on the interval without changing what the system would become of after shrinking the interval. We can now move $\CI_g$ to collide with $B_{\CP}$, after which, almost by definition, we get  $B_{g\cdot\CP}$. On the other side, when we collide $\CI_{g^{-1}}$ with the relative theory $T[M_d]$, the theory is unchanged. Therefore we see that $T[M_d,\CP]$ and $T[M_d,g\cdot\CP]$ are indeed dual theories. 

Note that the relative theory can depend on additional data associated with $M_d$, and they can be transformed under the action of the mapping class group. One needs to keep track of this action to arrive at the correct duality. One example is $T[T^2]$, which we will spend quite some time on in the next section. The theory depends on the modulus $\tau$ of $T^2$, and $T[T^2,\tau,\CP]$ is dual to $T[T^2,g^{-1}\cdot\tau,g\cdot\CP]$, which, for $g=S\in {\rm SL}(2,\Z)$, at low energy reproduces the familiar duality of 4d $\CN=4$ super--Yang--Mills theory that  simultaneously changes the global form of the gauge group and the coupling constant.

This picture can be generalized in several interesting ways. First of all, any topological domain wall can act in this manner on polarizations. Second, any invertible topological domain wall in the theory $\CT^{\text{bulk}}[M_d]$ can be used in similar way to get dual theories. To make this precise, one in general need to consider walls that, after colliding with $T[M_d]$, will leave it invariant.

We have actually seen examples of this kind that doesn't comes from mapping class group symmetry. For example, the $\frak{spin}(8)$ theory we talked about have duality already in the 7d/6d system before even compactifying on $M_d$. So for any $M_d$, the triality can act on $\mathrm{Pol}(M_d)$.

Another interesting scenario is when the bulk has a topological domain wall that admits topological boundary condition. 
Then if we let the open domain wall end on the boundary relative theory, this gives a symmetry on the boundary.

\subsection{Framing anomaly and classification of extended operators}\label{sec:framing}

To specify an extended operator insertion in a quantum field theories, one not only need to decide the type of the operator, the location of the insertion, but also its ``framing,'' which is a trivialization of the normal bundle to the operator.\footnote{We have assumed that the normal bundle is trivializable. It would be interesting to remove this assumption and ask whether one can obtain a finer classification. One can also consider several different notions of framing for line and surface operators in higher dimensions. The one considered here describes how the defect in question couples to its normal bundle in the ambient ``bulk'' theory.}  When the framing changes, often a phases factor will arise, which gives a way to classify extended operators.

For line operators (a.k.a.~particles), the classification is well known: in four dimensions and higher, there is a $\Z_2$ classification distinguishing bosons from fermions, while in 3d there is a $U(1)$ (or $\mathbb{Q}/\Z$) classification for the spin of anyons. This is related to the fact that the framing of line operators in 3d is $\mathbb{Z}$-valued, whereas the framing of line operators in 4d or higher is $\Z_2$-valued (this is relevant for the Gluck twist which we will discuss in later sections). The goal of this subsection is to generalize it to operators of arbitrary dimensions and codimensions. Let $S \subset M_d$ be an $n$-dimensional operator in a $d$-dimensional theory on a space-time manifold $M_d$ (here often assumed to be the Euclidean space $\R^d$), we give three classifications that are in a sense gradually more refined:
\begin{enumerate}
    \item Classification by a homotopy group of the special orthogonal group $\pi_n(SO(d-n))$. This is given in Table~\ref{tab:framing}.
    \item Classification by the subgroup of the mapping class group of the tubular neighborhood of $S$ fixing $S$.
    \item Classification by homotopy group of spheres $\pi_d(S^{d-n})$. This is also given in Table~\ref{tab:framing} in blue color, if different from the $\pi_n(SO(d-n))$.
\end{enumerate}

\begin{table}[ht]
	\begin{centering}
		\begin{tabular}{|c||c|c|c|c|c|c|c|c|c|}
			\hline
			$n$ & $d=3$ & $d=4$ & $d=5$ & $d=6$ & $d=7$ & $d=8$ & $d=9$ & $d=10$ & $d=11$  \tabularnewline
			\hline
			\hline
			$1$ & $\mathbb{Z}$ & $\mathbb{Z}_2$ & $\mathbb{Z}_2$ & $\mathbb{Z}_2$ & $\mathbb{Z}_2$ & $\mathbb{Z}_2$ & $\mathbb{Z}_2$ & $\mathbb{Z}_2$ & $\mathbb{Z}_2$
			\tabularnewline
			\hline
			$2$ & --- & $0$({\color{blue}$\Z_2$}) & $0$({\color{blue}$\Z_2$}) & $0$({\color{blue}$\Z_2$}) & $0$({\color{blue}$\Z_2$})& $0$({\color{blue}$\Z_2$})& $0$({\color{blue}$\Z_2$})& $0$({\color{blue}$\Z_2$})& $0$({\color{blue}$\Z_2$})
			\tabularnewline
			\hline
			$3$ & --- & --- & $0$({\color{blue}$\Z_2$}) & $\mathbb{Z}$({\color{blue}$\Z_{12}$}) & $\mathbb{Z}^2$({\color{blue}$\Z\cdot\Z_{12}$}) & $\mathbb{Z}$({\color{blue}$\Z_{24}$}) & $\mathbb{Z}$({\color{blue}$\Z_{24}$}) & $\mathbb{Z}$({\color{blue}$\Z_{24}$}) & $\mathbb{Z}$({\color{blue}$\Z_{24}$}) 
			\tabularnewline
			\hline		
            $4$ & --- & --- & --- & 0({\color{blue}$\Z_{12}$}) & $\mathbb{Z}_2$ & $\mathbb{Z}_2^2$ & $\mathbb{Z}_2$ & 0 & 0
			\tabularnewline
			\hline		
            $ 5 $ & --- & --- & --- & --- & 0({\color{blue}$\Z_{2}$})  & $\mathbb{Z}_2$ & $\mathbb{Z}_2^2$ & $\mathbb{Z}_2$ & $\mathbb{Z}$ 
			\tabularnewline
			\hline		
            $ 6 $ & --- & --- & --- & --- & --- & $0$({\color{blue}$\Z_{2}$})  & $\mathbb{Z}_{12}$({\color{blue}$\Z_{3}$}) & $\mathbb{Z}^2_{12}$({\color{blue}$\Z_{24}\cdot\Z_3$}) & $0$({\color{blue}$\Z_{2}$})
			\tabularnewline
			\hline		
            $7$ & --- & --- & --- & --- & --- & --- & 0({\color{blue}$\Z_{3}$}) & $\mathbb{Z}_2$({\color{blue}$\Z_{15}$}) & $\mathbb{Z}_2^2$({\color{blue}$\Z_{15}$}) 
			\tabularnewline
			\hline		
            $8$ & --- & --- & --- & --- & --- & --- & --- & 0({\color{blue}$\Z_{15}$}) & $\mathbb{Z}_2$ 
			\tabularnewline
			\hline		
		\end{tabular}
		\par\end{centering}
	\caption{\label{tab:framing} Classification of defects via $\pi_n(SO(d-n))$ and $\pi_d(S^{d-n})$ (in blue if different). The defects are classified by the Pontryagin duals of these groups whose elements describe responses of defects to a change of framing. ``---'' means the group vanish for dimensional reasons (i.e.~$n\ge d-1$). The two classification are related by the J-homomorphism, which is neither injective or surjective. As a consequence, although many entries coincide, the two classifications are in general different. }
\end{table}

The first classification comes from the following consideration when $S$ has the topology of an $n$-sphere (which is called an $n$-knot).
The unit normal bundle of $S \subset M_d$ is $S^{d-n-1}$ and its symmetry group is $SO(d-n)$.
Therefore, fixing a particular trivialization of the normal bundle as the reference point, we can define a framing of $S$ to be the homotopy class of a map
\be
S \; \to \; SO(d-n).
\ee
When $S$ is an $n$-knot, this is simply an element of $\pi_n( SO(d-n))$. Then the Pontryagin dual of this group classifies operators in terms of their response to a change of framing. In other words, under a change of framing labeled by $g\in \pi_n( SO(d-n))$, the operator will pick up a phase $f(g)$ where $f$ is in the dual of $\pi_n( SO(d-n))$. 

What happens when $S$ has more interesting topology? Then $\pi_n( SO(d-n))$ characterizes the ``local'' change of framing in the sense that if the map from $S$ to $SO(d-n)$ is trivial outside a disk, one again gets an element in $\pi_n( SO(d-n))$. So the Pontryagin dual of it can still classify operators via how it responds to local change of the framing. 

But how to take into account of global change of framing? The above consideration seems to lead to the set $[S,SO(d-n)]$, which is not naturally an Abelian group and doesn't have a natural notion of the Pontryagin dual. This is where the second classification mentioned above come in. 

Going back for a moment to the familiar notion of framing of line operators in three dimensions, we can identify $\mathbb{Z}$ with the subgroup of $SL(2,\mathbb{Z})$ generated by the $T$ element.
Indeed, if we define a line operator a la 't Hooft, i.e.~remove its tubular neighborhood from the space-time manifold, we then need to impose boundary conditions at $\partial (M_d \setminus S)$.

When $d=3$ and $n=1$, the boundary is a 2-torus, $T^2$, and its mapping class group is $SL(2,\mathbb{Z})$. In principle, one could consider boundary conditions related by arbitrary elements of the mapping class group. However, if we wish to retain the interpretation of $S$ as the location of a line (or, more generally, surface) operator, we should consider only those elements of the mapping class group that leave the meridian of $S$ intact (in particular, non-contractible). These consist of $\mathbb{Z}$ generated by $T$ in the case of 3d line operators, and $\mathbb{Z}_2$ generated by the Gluck twist in the case of 4d line operators.

Another way of getting this information is by considering the mapping class group of the tubular neighborhood of $S$ itself. Again we are only interested in the subgroup of elements that fix $S$ point-wise. When the dimension is larger than 4, one should work with the topological mapping class group as the exotic diffeomorphisms doesn't seem relevant for the framing dependence. This group might be non-Abelian, in which case the classification of operators will be given by its characters.

Is this already the best classification of operators? There are several things that are not satisfactory. First of all, the mapping class groups are hard to compute, and can get arbitrarily complicated when $S$ is complicated. So this is not a simple generalization of the fermion/boson dichotomy. Secondly, we might want a classification that only depends on the type of operators, not where it is inserted. Indeed, we often has the notion of two operators are of the same type although they can be inserted differently, and we hope to have statement analogous to ``electrons are fermionic'' as opposed to ``it actually depends on the world line geometry.'' Lastly, how an operator responds to a change of framing given by different elements in the mapping class group might not be arbitrary or independent. Therefore, this classification might be inefficient as many characters of the mapping class group might not be realizable by any physical operators in any quantum field theory.

The third classification via homotopy groups of spheres exactly tackles these three problems. It is not any more complicated than the first classification as these groups are extremely well-studied in topology. And compared to the first classification, it not only takes into account local change of framing but also global effects.

To motivate this, one first see that the framing of an $n$-knot leads to an element in the homotopy group of spheres $\pi_{d}(S^{d-n})$ via the J-homomorphism,
\begin{equation}
    \pi_n(SO(d-n)) 
    \rightarrow \pi_{d}(S^{d-n}).
\end{equation}
The latter, via the Thom--Pontryagin construction, classifies cobordism classes of $n$-dimensional submanifolds in $\R^d$. One way to see how a framing determines an element in $\pi_{d}(S^{d-n})$ is the following. Consider $S$, an $n$-dimensional submanifold embedded in spacetime $\R^d$, which has tubular neighborhood $N$ of the topology $S\times B^{d-n}$. 
By compactifying with the point at infinity we find the framing of $S^n$ can be described by a map from spacetime $S^d$ to $S^{d-n}$, under which the complement of $N$ is sent to $\infty\in S^{d-n}=\R^{d-n}\cup\{\infty\}$ while every point in $N$ is sent to a point in $\R^{d-n}$ given by the framing. The homotopy class of this map is then described by $\pi_d(S^{d-n})$.\footnote{We thank Pavel Putrov for many discussions regarding $\pi_d(S^{d-n})$ and its relevance to physics.}

As the J-homomorphism is neither injective nor surjective, the two classifications are different. Let's look into some cases where they differ:
\begin{itemize}
    \item Point operators ($n=0$). We have $\pi_d(S^d)=\Z$ while $\pi_0(SO(d))=0$. The dual of this $\Z$ is a $U(1)$ that classifies topological operator given by $e^{i\theta}$. 
    \item Strings / surface operators ($n=2$). As $\pi_d(S^{d-2})=\mathbb{Z}_2$ for $d\geq 4$, there is a $\mathbb{Z}_2$ framing. This framing of the normal bundle in this case (with its dimension $d-n\ge2$) can be identified with a choice of spin structure on the defect, and how the defect responds to such a change is telling us whether the string is bosonic or fermionic. Indeed, such a property is something that requires having non-trivial worldsheet to detect as there is no spin structure on $S^2$, and is therefore invisible to the classification via $\pi_n(SO(d-n))$.
    \item Membranes / volume operators ($n=3$). In the stable range $d\ge 8$, one can shift to any element of the $\Z_{24}\simeq(\pi_3(\mathbb{S}))^\vee$ with a Chern--Simons term for the connection of the normal bundle. As framing of the normal bundle uniquely specifies a choice of stable tangent framing (or equivalently string structure) on the worldvolume, this can also be interpreted as a gravitational Chern--Simons term.\footnote{As usual, one can use the version regularized by the $\eta$-invariant, which is itself framing independent, to cancel the dependence on the metric \cite{Witten:1988hf}. Upon a change of framing, it is shifted by a rational number in $\frac{1}{24}\Z$. This measures the relative $\frac{\hat A}{2}$-number of the spin 4-manifold (or equivalently $-\frac{p_1}{48}$, which makes it easier to see that the term is valued in $\frac{1}{24}\Z$ since $p_1$ is even for the spin case) bounding the world volume. See  \cite{atiyah1990framings,kirby1999canonical,bunke2009secondary} for related discussions.}  The story becomes more interesting in the unstable range. For example, when $d=6$, we have $\pi_3(SO(3))=\Z$ while $\pi_6(S^3)=\Z_{12}$, and one can ask what the different types of objects that they respectively classify are. 
\end{itemize}

We won't attempt to answer this question, as getting into these issues requires a better understanding of the physics of the $\pi_*(S)$ classification, which we will only briefly comment on next. 

As $\pi_{d}(S^{d-n})$ classifies cobordism classes of $n$-dimensional framed submanifolds in $\R^d$,\footnote{The Pontryagin construction applies to a general ambient space $M_d$, with the cobordism class of submanifold classified by $[M_d,S^{d-n}]$. Here we will focus on $M_d=\R^d$, with the infinity added to treat it as $S^d$.} one can attempt to run an argument similar to that in \cite{kapustin2014symmetryprotectedtopologicalphases} relating invertible SPT phases with cobordisms, except that we now have the normal bundle playing the role of $G$-bundle, and a connection of the normal bundle replacing the $G$-connection. One thing that needs to be checked is whether all the relevant topological terms in the action, that are unchanged under cobordism, can be  captured by the Pontryagin--Thom isomorphism. 

In addition, the statistics from framing dependence of extended excitations can also be studied on the lattice \cite{Kobayashi:2024dqj}, where we can move the excitations around in a topologically-nontrivial way using sequences of unitary operators. The Berry phase of such sequence captures the statistics. 
For example, the framing dependence of loop excitations and membrane excitations on the lattice is explored in \cite{Fidkowski:2021unr,Chen:2021xks} and \cite{Feng:2025mdg}, respectively. In particular, it is discovered that membrane excitations can have $\mathbb{Z}_3$-valued statistics in 5+1d and higher spacetime dimensions, which is consistent with $\pi_d(S^{d-3})$ containing $\mathbb{Z}_3$ subgroup for $d\geq 6$.

We hope to explore and better understand this classification and it application in future work. Notice that a similar classification, but for branes in string theory, is discussed in \cite{Sati:2019nli,Sati:2021uhj,Sati:2025ucv}, and it would be interesting to see how these two are related.

\subsection{Anomaly, polarization and condensation on boundary}
\label{sec:anomalypolarization}

As discussed in \cite{Eckhard:2019jgg,Gukov:2020btk}, different polarizations can be related by gauging a non-anomalous invertible symmetry. If the symmetry is anomalous, we cannot gauge it to obtain new polarization. Here, we will provide an alternative perspective via condensation on the topological boundary condition of bulk TQFT.

When an invertible symmetry is non-anomalous, it means that the correlation functions of the symmetry defects are trivial. Thus we can consistently sum over all possible symmetry defect insertion, since reconnecting the defect network --which correspond to correlation functions of symmetry defects--does not change the answer. In other words, we can condense the symmetry defect \cite{Roumpedakis:2022aik}. Since the topological boundary conditions of bulk TQFT are given by condensation on the boundary, there is a new topological boundary condition given by condensing the symmetry defect, i.e.~new polarization. We note that in terms of higher-gauging \cite{Roumpedakis:2022aik}, condensation on the gapped boundary means we ``higher-gauge'' the symmetry not on the entire bulk TQFT but only on the boundary, and the existence of new gapped boundary is due to we can ``higher-gauge'' the symmetry on the boundary.

However, when the invertible symmetry is anomalous, it means that the correlation function of symmetry defects is nontrivial---we cannot consistently sum over all possible defect insertions.
This means that the bulk TQFT does not have the gapped boundary where the symmetry defect condenses, i.e.~there is no new polarization.

\subsection{Fractional Anomaly Polynomial from Bulk TQFT}\label{sec:FractionalA}

The 6d theory can also have perturbative gravitational, R-symmetry, and mixed anomalies. The usual story  of anomaly inflow  (i.e.~when the theory is absolute) tells us that these are described by a characteristic class---often expressed in terms of the anomaly polynomial---in 8d, which determines a 7d Chern--Simons term for the spin connection and the background gauge field for the R-symmetry. We now investigate how this story interacts with the 7d TQFT of the $CdC$ type.

\subsubsection{Review of 3d fractional quantum Hall effect and one-form symmetry}

In integer quantum Hall effect, the 3d bulk is an invertible topological phase, and the boundary is an absolute theory.
The boundary has anomalous $U(1)$ 0-form symmetry described by bulk Chern--Simons term with properly quantized coefficients by the anomaly inflow mechanism. 

When the bulk is a non-trivial TQFT, it can describe fractional quantum Hall effect,  and the boundary is a relative theory. In such relative theory, the $U(1)$ symmetry is anomalous, captured by the bulk fractional quantum Hall coefficient.

The fractional quantum Hall coefficient is related to the anomaly of the one-form symmetry in the 3d bulk TQFT \cite{Cheng:2022nji}. 
Denote the one-form symmetry by $D$, it is generated by topological line operators. (We note that the line operators when restricted to the boundary generate a 0-form symmetry $D$ on the boundary, since the codimension reduces by one with respect to the boundary).

The topological line operators can have non-trivial statistics described by $\pi_1(SO(2))=\mathbb{Z}$, given by
\begin{equation}
    h:\quad D\quad\rightarrow\quad \mathbb{R}/2\pi\mathbb{Z}\cong  U(1)~.
\end{equation}
The statistics $h$ is a quadratic function: the braiding of the topological lines is given by
\begin{equation}
    \langle x,y\rangle= e^{2\pi i\left(  h(x+y)-h(x)-h(y)\right)}\in U(1),\quad \langle x^m,y^n\rangle= \langle x,y\rangle^{m+n}~. 
\end{equation}
The anomaly of one-form symmetry can be described by the 4d topological term
\begin{equation}
    2\pi \int h[B]~,
\end{equation}
where $B$ is the background two-form $D$ gauge field for the one-form symmetry.
The coupling to the $U(1)$ symmetry can be expressed by the following relation with the first Chern class $c_1$ for the $U(1)$ symmetry background gauge field:
\begin{equation}
    B= vc_1~,
\end{equation}
where $v:\mathbb{Z}\rightarrow D$ is a homomorphism, and as it is $D$-valued, $v$ can be represented by a topological line in $D$. 

Physically, the above expression implies charge fractionalization \cite{Barkeshli:2014cna} on the line operators \cite{Benini:2018reh}. For general line operators carry representation $Q\in \text{Hom}(D,U(1))$ under the $D$ one-form symmetry, it is attached to the Wilson surface operator $e^{i\int Q(B)}$, and thus from $B=vc_1$ it
transforms under the $U(1)$ 0-form symmetry with fractional charge
\begin{equation}
    \text{Fractionalization of 0-form symmetry on lines:}\quad Q(v)~.
\end{equation}
We note that since charge is described by 1d Wilson line of the background gauge field for the 0-form symmetry, it can also be viewed as a 1d fractional Chern--Simons term, i.e.~fractional quantum Hall effect in 1d.

Denote the statistics of the topological line $v$ by the  $p/2n$ for integers $p,n$ (one can show this is the most general value of the statistics if $v^n=1$), the bulk term with $B=vc_1$ is
\begin{equation}\label{eqn:4dbulk}
    2\pi \frac{p}{2n}\int c_1^2~,
\end{equation}
which can be cancelled by a fractional Chern--Simons term with level $p/n$. This is the fractional quantum Hall coefficient.
The 4d bulk term (\ref{eqn:4dbulk}) is the anomaly polynomial for the $U(1)$ symmetry on the relative boundary theory.

We remark that the fractional anomaly coefficient means that the bulk being a non-invertible TQFT, and the symmetry acts on the TQFT in a non-trivial way, such that the response current is related to the background gauge field with fractional conductance.

When the bulk is a trivial TQFT (but it can still be nontrivial invertible phase), the statistics becomes trivial $h=0$ mod $\mathbb{Z}$, and the anomaly polynomial is properly quantized.

\subsubsection{7d Fractional quantum Hall effect and three-form symmetry}

The above discussion carries over to 6d/7d setup (see also \cite{Heckman:2017uxe}).
The 7d bulk TQFT has three-form symmetry $D$, generated by topological volume operators. (When the volume operators are restricted to the boundary, they generate two-form symmetry, since the codimension of the operators reduces by one with respect to the boundary).

The topological volume operators in 7d can have statistics described by $\pi_3(SO(4))$ which contains $\mathbb{Z}$ from $SU(2)\cong S^3$ and $\pi_3S^3=\mathbb{Z}$.
The corresponding statistics is again described by a quadratic function
\begin{equation}
    h:\quad D\rightarrow \mathbb{R}/2\pi\mathbb{Z}\cong U(1)~.
\end{equation}
The anomaly for the three form symmetry is described by 8d bulk topological term
\begin{equation}\label{eqn:8dbulk}
    2\pi\int_{8d} h[B]~,
\end{equation}
where $B$ is a background four-form $D$ gauge field for the three-form symmetry.

The continuous ordinary symmetry $G$ couples to the TQFT by
\begin{equation}
    B= \omega_4 (A),\quad \omega_4\in H^4(BG,D)~,
\end{equation}
where we denote the background for the ordinary symmetry by $A$.
Concretely, in our case we can take (denote the generator of $H^4(BG,D)$ by $I_4$)
\begin{equation}
     \omega_4 (A)= vI_4(A)~,
\end{equation}
where $v:\mathbb{Z}\rightarrow D$ is a homomorphism, and as it takes value in $D$, $v$ can be represented by a topological volume operator in $D$.

Physically, the above expression implies charge fractionalization on the volume operators \cite{Benini:2018reh,Hsin:2019fhf}. For general volume operators carry representation $Q\in \text{Hom}(D,U(1))$ under the $D$ three-form symmetry, it is attached to the Wilson 4-dimensional operator $e^{i\int Q(B)}$, and thus from $B=vI_4(A)$ it
transforms under the $U(1)$ 0-form symmetry with 3d fractional quantum Hall coefficient
\begin{equation}
    \text{Fractionalization of 0-form symmetry on volume:}\quad Q(v)~.
\end{equation}

Denote the statistics for the volume operator $v$ by $\frac{p}{2n}$ mod 1 for integers $p,n$ (one can show this is the most general statistics if $v^n=1$), the 8d bulk term (\ref{eqn:8dbulk}) becomes
\begin{equation}
    2\pi \frac{p}{2n}\int_{8d} I_4(A)^2~.
\end{equation}
The is a total derivative and can be cancelled by 7d fractional Chern--Simons term, which gives the fractional quantum Hall coefficient. The above term is also the anomaly polynomial for the 0-form symmetry $G$ on the boundary.

For instance, the anomaly for continuous ordinary symmetry in general 6d SCFTs is discussed in \cite{Gukov:2018iiq}, where the possible anomaly can be parameterized by $\alpha,\beta,\gamma,\delta$ in (2.9) of \cite{Gukov:2018iiq}:
\begin{align}
    I_8&=\alpha c_2(R)^2
 + \beta c_2(R)p_1(T) + \gamma p_1(T)^2
 + \delta p_2(T) + I_\text{flavor}\cr 
 &=\alpha c_2(R)^2
 +3\beta c_2(R)L_1(T)+
 \frac{9}{7}(7\gamma+\delta)L_1(T)^2+\frac{45}{7}\delta L_2+T_\text{flavor}~,
\end{align}
where $R,T$ are the $R$-symmetry bundle and the tangent bundle, and $I_\text{flavor}$ is the anomaly polynomial involves the  flavor symmetry. The anomaly is described by the 7d action
\begin{equation}
    S_\text{7d}=\int_{Y_7} \text{CS}_7=2\pi\int_{Y_8} I_8~,
\end{equation}
where $Y_7=\partial Y_8$.

For absolute 6d theories, the anomaly is described by properly quantized 7d topological action that does not depend on 8d extensions. For absolute theories on oriented spin manifolds with $w_4=0$, the coefficients obey quantization conditions discussed in section 2.11 of \cite{Gukov:2018iiq}:
\begin{equation}
    24\alpha\in\mathbb{Z},\quad 48\beta\in\mathbb{Z},\quad 
    2304(2\gamma+\delta)\in\mathbb{Z},\quad 1440\delta\in\mathbb{Z}~.
\end{equation}

For relative theories, the bulk is a non-trivial TQFT, and it can have fractional quantum Hall effect that gives quadratic fractional anomaly polynomial. 
In the above anomaly polynomial, the coefficients $\alpha,\beta,\gamma$ are quadratic terms, and they can be fractional due to the bulk TQFT, while the coefficient $\delta$ is not a quadratic term and should always be properly quantized. Indeed, the quantization condition for $\delta$, 
\begin{equation}
    1440\delta \in \mathbb{Z}
\end{equation}
is satisfied in all known examples where $\alpha,\beta,\gamma$ can become fractional in relative theories.\footnote{
We thank Pavel Putrov for confirming such a property.
}

\subsubsection{Example: $O(-k)$ theories}

Let us illustrate the discussion with 6d $O(-k)$ theories, which can be engineered by considering F-theory on singular elliptic Calabi--Yau threefolds with the base being the total space of the $O(-k)$-bundle over $\mathbb{P}^1$. Such a theory is relative with the bulk 7d TQFT given by
\begin{equation}\label{eqn:O(-k)TQFT}
    \frac{k}{4\pi}\int CdC~,
\end{equation}
with dynamical $U(1)$ three-form gauge field $C$.

The 7d TQFT can couple to 0-form symmetry bundle by the homomorphism $v$ as above, which can be described by a volume operator $e^{iq\oint C}$ for integer $q$. The statistics of $v$ is $\frac{q^2}{2k}$ mod 1, and thus the fractional anomaly polynomial is\footnote{
Here $I_4$ is related to $C$ by $kdC=I_4$ from the definition of the three-form symmetry in the 7d three-form Chern--Simons theory (\ref{eqn:O(-k)TQFT}).
}
\begin{equation}
    I_8=2\pi \frac{q^2}{2k} I_4 I_4~.
\end{equation}
Having a fractional anomaly polynomial is not contradicting the quantization of the coefficients because what really happens is that, on top of the $CdC$ term in \eqref{eqn:O(-k)TQFT}, one has an additional term coupling $C$ to the background fields,
    \begin{equation}
    \int C\^ I_4~.
\end{equation}
In addition, there will be a coupling of $C$ to Tr$\,F^2$ for the gauge field. As the 7d theory is not invertible, we cannot really integrate out $C$ and get the fractional $I_8$. When we talk about the fractional $I_8$, it is always understood that there is a $CdC$ theory with coupling to backgrounds. 

Concretely, we can substitute
\begin{equation}
I_4=x c_2(R)+yp_1(T)\text{ mod }k~,
\end{equation}
and absorb $q$ by $qx\rightarrow x,qy\rightarrow y$, 
then the quantization condition of the anomaly polynomial will be modified into
\begin{equation}\label{eqn:quantizationrelativeO(-k)}
    24\left(\alpha-\frac{x^2}{2k}\right)\in\mathbb{Z},\quad 
    48\left(\beta-\frac{xy}{k}\right)\in\mathbb{Z},\quad 
    2304\left(2\gamma+\delta-\frac{y^2}{k}\right)\in\mathbb{Z}~.
\end{equation}
This gives non-trivial constraints on the quantization of the coefficient of the anomaly polynomials in the $O(-k)$ theories. Namely, we must have that 
\begin{itemize}
    \item The quantization condition for $\delta$ itself is not modified. This is due to the fact that $C$ cannot couple to $p_2$ for degree reasons.
    \item There is a constraint for the quantization of $\alpha,\beta$ and $\gamma$ as there are only two coupling coefficients, $x$ and $y$.
\end{itemize}

In fact, in this case one can completely determine $x$ and $y$. The process for determining $x$ and $y$ is similar to the discussion in Section 3 of \cite{Ohmori:2014kda}. Namely, one first computes the (mixed) gauge anomaly, which in this case uniquely fix the coupling between $C$ and Tr$\,F^2$ as well as the value for $x$ and $y$. We find $x=3(k-2)$ for $k\neq 7$, and $x=18$ for $k=7$ (since the theory has $E_7$ gauge algebra and half hypermultiplet in $\mathbf{56}$, the latter does not contribute to the R-symmetry anomaly).
In addition, $y=(k-2)/(4k)$.

The anomaly polynomials for $O(-k)$ with $k\geq 3$ are as follows:\footnote{
We thank Kantaro Ohmori and Pavel Putrov for very helpful discussion 
regarding the computation of the anomaly polynomials for $O(-k)$ theories.
}
\begin{itemize}
    \item $k=3$. $\alpha=\frac{29}{24}$, $\beta=\frac{5}{48}$, $\gamma=\frac{3}{640}$, $\delta=-\frac{7}{480}$.
    
    \item $k=4$. $\alpha=\frac{27}{8}$, $\beta=\frac{3}{16}$, $\gamma=\frac{7}{5760}$, $\delta=-\frac{1}{1440}$.

    \item $k=5$. $\alpha=\frac{239}{40}$, $\beta=\frac{23}{80}$, $\gamma=-\frac{17}{5760}$, $\delta=\frac{23}{1440}$.

    \item $k=6$. $\alpha=\frac{211}{24}$, $\beta=\frac{19}{48}$, $\gamma=-\frac{43}{5760}$, $\delta=\frac{49}{1440}$.

    \item $k=7$. $\alpha=\frac{247}{14}$, $\beta=\frac{13}{28}$, $\gamma=-\frac{121}{10080}$, $\delta=\frac{19}{360}$.

    \item $k=8$. $\alpha=\frac{59}{4}$, $\beta=\frac{5}{8}$, $\gamma=-\frac{49}{2880}$, $\delta=\frac{13}{180}$.

    \item $k=12$. $\alpha=\frac{653}{24}$, $\beta=\frac{53}{48}$, $\gamma=-\frac{71}{1920}$, $\delta=\frac{73}{480}$.
\end{itemize}
And it is straightforward to check that they obey the quantization condition (\ref{eqn:quantizationrelativeO(-k)}) with the value of $x$ and $y$ mentioned above. In particular, for $k=5,7$ the quantization condition of the anomaly coefficients uniquely fix $x,y$ up to $(x,y)\rightarrow (-x,-y)$ from charge conjugation on the $C$ field (if we treat $1/4$ as the inverse of $4$, {i.e.} $4,2$, in mod $5$ and mod $7$).

\section{Compactification to 4d}
\label{sec:4d}

In this section, we consider reducing the 6d theory on two-dimensional manifolds $M_2$ to 4d systems. After discussing some general aspects, we analyze in detail the $T[T^2]$ theory. One focus is the moduli space of the full theory and how various symmetries manifest themselves through this moduli space.
 We then consider cases with $M_2$ being of higher genus and having boundaries, which, in general, give rise to 4d boundary systems coupled to 5d bulk systems.  

\subsection{Polarization of $T[M_2]$}

The topology of the internal two-manifold $M_2$ is determined by the genus $g$, the number of crosscaps $n_c$, the number of punctures $n_p$, and the number of boundary components $n_b$.
We will start with the case when the 2-manifold is oriented and closed, $n_c=n_b=n_p=0$. Toward of the end of this section, we will generalize the discussion to the case with boundary.

The classification of polarizations on a two manifold $M_2$ starts with classifying maximal isotropic subgroups of $H^*(M_2,D)=D^{(0)}\oplus (D^{(1)})^{2g}\oplus D^{(2)}$. Decomposing $L\subset H^*(M_2,D)$ as $L^{(0)}\oplus L^{(1)}\oplus L^{(2)}$, it is easy to see that $L^{(0)}\oplus L^{(2)}$ has to be maximal isotropic in $D^{(0)}\oplus  D^{(2)}$ while $L^{(1)}$ should be maximal isotropic in $(D^{(1)})^{2g}$. The classification of maximal isotropic subgroups in $(D^{(1)})^{2g}$ is discussed in \cite{Tachikawa:2013hya} in the context of Class-$S$ theories. On the other hand, it is easy to see that the problem of classifying maximal isotropic subgroup in $D^{(0)}\oplus  D^{(2)}$ is exactly the same as that of classifying maximal isotropic subgroups of $H^*(S^1,D)$, which we encountered in Part I when we discussed polarizations on $S^1$. This is not a coincidence as we will explain later. 

The choice of $L$ leads to $L^\vee \simeq H^*(M_2,D)/L$ symmetries being preserved in 4d. 
To fully specify a pure polarization, one also needs to specify a $\Z_2$-valued function on $L$, or equivalently, a 2-torsion element in $L^\vee$. This can be interpreted as turning on a non-trivial value for certain background field.

\subsubsection{Geometric polarizations}

There is a subclass of pure polarizations for which $L$ is given by the image of
\begin{equation}
    H^*(W_3,D)\rightarrow H^*(M_2,D),
\end{equation}
with $W_3$ being a 3-manifold that bounds $M_2$. Considering such polarizations is very natural if we want the topological boundary condition $B$ of $\CT^\text{bulk}[M_2]$ to arise from compactification as well (in the present case, from $\CT^\text{bulk}$ on $W_3$). Such polarizations are therefore highly constrained. For example, the map above is an isomorphism in degree 0 while zero in degree 2. Therefore, geometric polarizations will have $L^{(0)}=D$ and $L^{(2)}=0$, and consequently the theory will have $D$ 0-form symmetry while no $D$ 2-form symmetry. An interesting question is whether there is any condition on $L^{(1)}$ for the polarization to be geometric. One might guess that it cannot be a torsion subgroup of $H^1(M_2,D)$. (For example, when $D=\Z_4$, $M_2=T^2$, then $L^{(1)}=\Z_2\times \Z_2$ is a torsion subgroup.) However, the example below suggest that geometric polarizations can give rise to torsion subgroups.  

For $D=\Z_N$, take a lens space $L(k,1)$ such that $k|N$, and consider a loop $\gamma$ that represent the generator of $H_1=\Z_k$. Then we remove a tubular neighborhood of the loop to get a three-manifold $W_3$ whose boundary is a $T^2$. $[\gamma]$ still represent a $k$-torsion class in $H_1(W_3,\Z_N)$. $k[\gamma]$ is zero in homology, and is the boundary of a surface $S$, which corresponds to a class in $H_2(W_3,M_2;D)$. Then the image of the map
$H_2(W_3,M_2;D)\rightarrow H_1(M_2,D)$ will be a torsion subgroup $\Z_{n/k}\subset \Z_N$. So we see that torsion subgroups of $H_1(M_2,D)\simeq H^1(M_2,D)$ can come from geometric polarizations.

It is not true either that any pure polarization with $L^{(0)}=D$ is geometric. For example, when $D$ is a product of multiple subgroups, we can combine geometric polarizations for these subgroups. The resulting polarization is in general not geometric unless the polarizations for subgroups are associated with the same 3-manifold $W_3$. Another way of looking at this is that a geometric polarization has to treat all the subgroups of $D$ ``independently and in the same way,'' i.e.,~the induce maps $H^{1}(W_3)\rightarrow H^1(M_2)$ with the coefficient in any sub-factors of $D$ would be determined by the map with $\Z$ coefficient, and different sub-factors would not be able to talk with each other.

To end this subsection on a positive note, a statement mentioned later in subsection~\ref{sec:MCG2} indicates that, when $D=\Z_N$, any polarization with $L^{(0)}=D$ is geometric.

\subsubsection{Quadratic refinements and mixed polarizations}

Classifying $L$ is not exactly the same as classifying polarizations, as one still need to choose a $\Z_2$-valued function on $L$ to completely specify a pure polarization, and there are also mixed polarizations. We will discuss these issues here.

A $\Z_2$-valued function on $L$ can also be interpreted as a 2-torsion element of $L^\vee\simeq H^*(M_2,D)/L$ and can be specified by its projection onto the three different components $(L^{(0)})^\vee$, $(L^{(1)})^\vee$ and $(L^{(2)})^\vee$. These leads to three maps
\begin{equation}
    \gamma_{i+1}: H^{1+i}(M_4,\Z_2) \rightarrow H^{1+i}(M_4,(L^{(i)})^\vee)
\end{equation}
when we consider the theory $T[M_2]$ on a four manifold $M_4$, and they tell us that the background field for the $(L^{(i)})^\vee$ symmetry is shifted by the image of a Stiefel--Whitney class $\gamma_{i+1}(w_{i+1})$. This is completely similar to the 5d case discussed in Part I. When $M_2=T^2$, the two cases are related by dimensional reduction on $S^1$, which we will discuss in the next subsection.

Such polarizations with non-trivial quadratic functions can also be geometric, though with insertions of defects of co-dimension 2 such as \begin{equation}
    \mathrm{exp}\left(\pi i\int C\wedge w_2 \right)
\end{equation}
or a defect of co-dimension 3 given by a similar expression but with $w_1$ in the expression instead. These defects are defined by placing $\int C$ at the intersection of the support of the defect with the Poincar\'e dual of $w_n$, where $n=1,2$ in above.

On the other hand, mixed polarizations are classified by certain discrete angles that turn on topological terms in the theory $T[M_d]$. For $d=2$, all the relevant terms are of the form 
\begin{equation}
    \int B_2\wedge \mathrm{Bock}( B_1 )
\end{equation}
given by a product of a 2-form discrete gauge field and the Bockstein of a 1-form gauge field. The problem of classifying them is almost exactly the same as that encountered in the case of $T[S^1]$ and therefore will not be repeated here. 

\subsubsection{Mapping class group action}\label{sec:MCG2}

The mapping class group of a manifold $M_d$, MCG$(M_d)$, acts non-trivially on polarizations, and therefore on $T[M_d]$. Theories related in this way are ``dual'' in the sense that they are different descriptions of the same quantum field theory, since the physical system remains the same after the mapping class group action. Therefore classifying orbits of the mapping class group action is an interesting problem, as different orbits can potentially be distinct quantum field theories.\footnote{Notice that different orbits can happen to give rise to equivalent theories, as there can be dualities that do not originate from the mapping class group action. As explained in Section~\ref{sec:MCGDefect}, the ``$\mathfrak{spin}(8)$ theory'' considered in Part I is an example of this. This relative theory admits three polarizations already in 6d, with all three being dual to each other. Such 6d dualities will lead to extra dualities in $T[M_d]$ in lower dimensions.}  

The discussion about mapping class group action on polarizations can be separated into two parts. The first part concerns how different choices of $L$ are acted upon by the mapping class group. As the action of MCG$(M_d)$ on different choices of $L$ factors through the action on $H^*(M_d,D)$, for $d=2$ it boils down to the study of Sp$(2g,\Z)$. Then one can discuss how the quadratic function and mixed theta angles transform. Naively, the second part is ``boring,'' as the MCG action on mixed theta angles and the quadratic function is obvious. However, there is subtlety for the quadratic function, which is defined only after we have chosen a splitting of $H^*(M_d)$ into $L\oplus \bar{L}$. When $d=2$, such splitting always exists but is not invariant under the MCG action. As a consequence, the action is not just by pullback, and can have ``anomalies.'' We have discussed this in detail in Part I with $M_6=S^3\times S^3$. One way to determine the MCG action on quadratic refinement for general $M_d$ is by decomposing the partition function in a basis given by the splitting polarization $L\oplus \bar{L}$ and find the action of the mapping class group in this basis. We will demonstrate this in more details in the next subsection.

For the the first part of the problem, namely the MCG action on different choices of $L$, we make the following observations,
\begin{enumerate}
    \item When $D=\Z_p$, there is a single orbit. 
    \item When $D=\Z_N$, there exists one orbit for each divisor of $N$. 
\end{enumerate}

Both statements are universal as they are true for all choices of $M_2$. One can verify them explicitly. For example, for the second statement, one can show that there is only one orbit for each $k|N$  by casting any $L\simeq \Z_{k}\times \Z_{N/k}$ into a standard form with the actions of Dehn twists. When $D=\Z_p$, this is closely related to the statement that the action of Sp$(2g,\Z_p)$ on $\Z^g_p$ is transitive.

Another way to think about the first statement is that, when $D=\Z_p$, every $L$ comes from a geometric polarization given by a handlebody, and all handlebodies with the same boundary are all homeomorphic.

\subsection{Example:  $M_2=T^2$}
Part of the previous discussion might be abstract to some of the readers, and we will now analyze very explicitly and in greater detail the case of $M_2=T^2$ with $D=\Z_N$. This is a familiar example closely related to the 4d $SU(N)$ super--Yang--Mills theory when the 6d theory is chosen to be a $(2,0)$ theory of type $A_{N-1}$. The emphasis of this subsection and the next, where we will discuss $M_2=S^2$, will be placed on illustrating some general phenomena in the compactification of relative theories:
\begin{itemize}
    \item The theory is always non-conformal, and global structures (such as the structure of the KK tower) depend on the polarization.
    \item There are often higher group symmetries.
\end{itemize}
Furthermore, the global structure of the full theory is often much more interesting than what is captured by an IR SCFT. In fact, there is often multiple IR SCFTs that appear at different singularities on the moduli space. The discussion on the structures of the ``full moduli space'' will be another focus of this subsection.
    
\subsubsection{Polarization and global structures}
    
We will start out by clarifying one important subtlety that distinguish $T[T^2]$ and the usual 4d $\CN=4$ super--Yang--Mills theory. The theory $T[T^2]$ would in general have compact scalars given by the holonomy of the $B$ fields in the tensor multiplet along $T^2$. The global form of it will depend both on the 6d theory and on the choice of a polarization. 

When the $(2,0)$ theory is chosen to be of type $A_{N-1}$, and when the polarization is a pure one given by $L\subset H^*(T^2,D)$, this will be an $SU(N)/{L^{(0)}}$-valued scalar. 

This fact differentiates the theory from the more familiar 4d $\CN=4$ super--Yang--Mills theory with $SU(N)$ gauge group, which can be viewed as a limit of $T[T^2]$ by decompactifying this scalar and decoupling all the non-trivial KK modes. 

Notice that this process, although commonly used to ``simplify'' $T[T^2]$, is ``discontinuous'' and often result in discontinuity even in quantities that are supposed to be protected by supersymmetry such as various partition functions. One slightly improved approach is to try to find other values for this group-valued scalar around which one has another massless theory at low energy. This approach has been studied in the literature in the compactification of lower-dimensional theories, and these special values are sometime referred to as ``holonomy saddles'' \cite{Hwang:2018riu}. We remark here that this approach might not be always possible as such values might not be discrete (such as in the Abelian case or the $U(N)$ case), and, even when possible, loses interesting information when removing the massive KK modes. This will be investigated further in Section~\ref{sec:2dqq}, where we illustrate how information about the KK tower can be used to refine the partition function with another variable. Here, following the main theme of the paper, we will emphasize the difference between $T[T^2]$ and the usual 4d $\CN=4$ theory by looking at their symmetries.

Although $T[T^2]$ also have 16 supercharges, the supersymmetry algebra is slightly different from that of the usual SYM theory. In particular $T[T^2]$ would in general have only $SO(5)$ R-symmetry, which can be understood either as ``broken'' from the $SO(6)$ of the 4d $\CN=4$ by a group-valued scalar or as inherited from the 6d $(2,0)$ theory with no further enhancement.  Indeed, the SUSY algebra of $T[T^2]$ is like that of 6d $(2,0)$ theory, but with two translations replaced with two copies of $U(1)_{\text{KK}}$, geometrically identified with the continuous part of the isometry of $T^2$. 

Another point of view is that the algebra, except for the part involving R-symmetry and $U(1)_{\text{KK}}$'s, is exactly the same, and the matter content, instead of respecting the $SO(6)$ symmetry, only transform under $SO(5)$ as they carry non-trivial central charges which are identified with KK momenta.

\subsubsection{The moduli spaces}

One can look at the moduli space of the theory, which, for one choice of polarization, namely $L^{(0)}=0$, is given by $(T_{\C}\times \frak{t}_{\C} \times \frak{t}_{\C})/W$ where $T$ is the Cartan of $G$ and $W$ the Weyl group. The five copies of $\frak{t}\simeq \frak{t}^{\vee}$ is rotated by $SO(5)$, which would enhance to $SO(6)$ if $T$ is decompactified. 

Another symmetry of the moduli space is given by multiplying $T_\C$ with an element of the center. This is exactly the action of the 0-form symmetry given by the center of $G$. For general polarizations, this symmetry will be gauged, and moduli space will be a further quotient by a finite group, with orbifold singularities. These singularities are exactly what correspond to $L^{(0)}$ 2-form symmetries of the theory. The charged objects are ``axion strings'' in the 4d theory---defects around which the vevs of the scalars are required to trace out a non-trivial loop in the moduli space.

One can also consider the moduli space on $S^1\times \R^3$, which adds two $T$ factors associated with the vevs of the electric and magnetic lines. A covering of the moduli space is
$(T_{\C}\times T_{\C} \times T_{\C}\times \frak{t}_{\C})/W$.
Here the 1-form electric and magnetic symmetries act by multiplying the second and third factor by an element of the center.\footnote{Note that this is not in conflict with the fact that only 0-form symmetries natually act on moduli spaces, as there is a $S^1$ in the geometry.} To get the moduli given by a particular polarizaiton, one needs to quotient by the ``extra symmetries'' (i.e.~those killed by a choice of polarization). After this, the two factors in the middle can be naturally written as $T'_\C\times T'^\vee_\C$, where $T'$ and $T'^\vee$ are the Cartan of $G/L^{(0)}$ (which is now not necessary simply-connected) and its dual. Then the charge lattice for the line operators $\pi_1(T'_\C\times T'^\vee_\C)$ will be ``correctly quantized.''

We now describe the moduli space in more detail in low ranks.

\paragraph{The case of $U(1)$.} In this case the moduli space is $S^1\times \R^5$. The rotation of $S^1$ is the action of the $U(1)$ 0-form symmetry in 4d. The dual of $\pi_1=\Z$ is what gives rise to a $U(1)$ 2-form symmetry of the 4d theory $T[T^2,U(1)]$. As the 6d theory is absolute, there is no choice of a polarization involved, and the moduli space will not depend on it.

\begin{figure}
    \centering
    \includegraphics[width=1\linewidth]{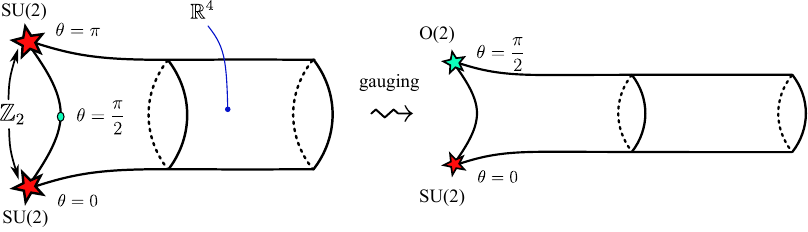}
    \caption{The Coulomb branch of $T[T^2]$ theory with two different polarizations. The red stars are $SU(2)$ singularities with a low energy description given by 4d $\CN=4$ super--Yang--Mills theory with $SU(2)$ gauge groups. The teal star on the right is a 4d $\CN=4$ $O(2)$ theory, which comes from a $U(1)$ theory at the teal dot in the left figure that is fixed by the $\Z_2$ 0-form symmetry.}
    \label{fig:O2Coulomb}
\end{figure}

\paragraph{The case of $SU(2)$, on $\R^4$.} The moduli space is similar to the $U(1)$ case but with a $\Z_2$ quotient:
\begin{equation}
(\theta,x_1,\ldots,x_5)\rightarrow (-\theta,-x_1,\ldots,-x_5).
\end{equation}
The two fixed points are $(0,
\ldots,0)$ and $(\pi,0,\ldots,0)$. Away from the two singularities, the moduli space looks like an $\R^4$ fibered over a semi-infinite cigar. This way of decomposition corresponds to choosing an $\CN=2$ subalgebra, and identifying a ``Coulomb branch,'' which amounts to picking one (combination) of $x_i$ to be combined with $
\theta$ to form a complex scalar to parametrize the Coulomb branch. This geometry is illustrated in the left part of Figure~\ref{fig:O2Coulomb}. The $\Z_2$ symmetry acts by sending $\theta$ to $\theta+\pi$, therefore exchanging the two fixed points. After the quotient, a new orbifold point emerges at $\theta=\pi/2\sim-\pi/2$. This is depicted on the right of Figure~\ref{fig:O2Coulomb}. The two pictures looks similar, but they have different special geoemtries. To see this more clearly, one can look at the moduli space of the theory on $S^1\times\R^3$.
\paragraph{Case of $SU(2)$, on $S^1\times\R^3$.} Now the moduli space will be $(\C^*\times\C^*\times\C^*\times\C)/\Z_2$. The fiber over the previous moduli space is a two-torus, parametrized by $(\phi_1,\phi_2)$, which becomes singular at the two fixed points on the base with $\theta=0$ and $\theta=\pi$ respectively. The singular fiber will be a pillow case with four singularities $(0,0)$, $(0,\pi)$, $(\pi,0)$ and $(\pi,\pi)$. This is a limit of the $I_0^*$ singularity, and is compatible with the fact that the SL$(2,\Z)$ monodromy is given by $-1$ as one can check explicitly. On the other hand, there is no monodromy at infinity.\footnote{The electric and magnetic one-form symmetry will act as $\Z_2\times \Z_2$ shifting the two $\phi$'s by $\pi$. As mentioned before, to get the moduli space of an absolute theory, one will quotient by a $\Z_2$ subgroup. Otherwise charges in the theory will be ``incorrectly quantized,'' signaling that we are having a relative theory. There are three different choices, leading to the usual $SU(2)$, $SO(3)_+$, and $SO(3)_-$ theories. However, the resulting moduli space looks identical, as $I_0^*$ is special with $-1$ monodromy that commute with any $\Z_2$ action. To distinguish them geometrically, one can turn on a mass deformation. Then each $I_0^*$ can become deformed into a $I_4$ and two $I_1$'s. Which fiber become $I_4$ tells one which $\Z_2$ subgroup was used and which theory we are in. This is also explained in detail in Section~4 of \cite{Gukov:2022gei}.} The $\Z_2$ 0-form symmetry acts only on the base. After its quotient, the two singular fibers get identified. In addition, a new sigularity arises from the fixed point at $\theta=\pi/2$. This orbifold singularity is what gives rise to a $\Z_2$ 2-form symmetry after gauging the 0-form symmetry. The orbifold point is in fact another $I_0^*$ singularity, as it is fixed by a $\Z_2$ that is a combination of the $\Z_2$ Weyl group and the symmetry $\Z_2$ and, therefore, will have $-1$ monodromy around it and the fiber over it will be a ``pillow case'' with four $\Z_2$ orbifold points. To be consistent, one again needs to have trivial monodromy at infinity. This is indeed what one finds. Notice that this geometry is very similar with the previous case. However, the geometry is different, since, after the $\Z_2$ quotient, the size of the fiber stays the same while the circumference of the base halved. There appears to be a new $\Z_2$ symmetry that exchange the two singularities, but this is not expected to act on the full theory.

Now an interesting question arises: what is the SCFT at the other new singularity? Appearance of such singularity is quite universal as it happens as long as there is 2-form symmetry, which comes from (real-)codimension-2 singularities on the moduli space. Even without symmetry, we expect that there always have to be some additional fixed points away from the naive SCFT fixed points, simply because the Coulomb branch of $T[T^2]$ is expected to have a cylindrical end, which would not be consistent with having simply a single SCFT point.

What is the nature of these singularities? This is what we will discuss next.

\subsubsection{Other SCFT points: $O(2)$ theory from $A_1$}

We again look at the $A_1$ example with $\Z_2$ 2-form symmetry, and ask what happens if we go to the other singularity. We argue below that it is actually an $\CN=4$ $O(2)$ theory. 

 Given the  supersymemtry, it can only be either another 4d $\CN=4$ super--Yang--Mills theory with $SU(2)$ gauge group, or a $U(1)$ theory with $\Z_2$ quotient. All other possibilities are not viable. For example, the $N_f=4$ theory has the same singularity, but different amount of supersymmetry as well as the dimension of the Higgs branch. The Abelian theory with $\CN=4$ will not correspond to a singular point on the Coulomb branch. However, one can perform a $\Z_2$ quotient without breaking the $\CN=4$ symmetry, and this is exactly what we refer to as the $O(2)$ theory. The matter indeed transform under the adjoint as required by $\CN=4$ supersymmetry,\footnote{Other quotient will instead will not be $\CN=4$, as the matter will no longer be in the adjoint of $U(1)\rtimes \Z_k$ for $k>2$. See \cite{Argyres:2019ngz} for realization of $\CN=3$ theories with $k=3,4$, and $6$.} which means that $O(2)$ acts via $\pi_0(O(2))=\Z_2$ flipping the sign of the matter field. 
 
 So how to decide which theory we get? Intuitively, it is very hard to have a mechanism to enhance the gauge symmetry only in the $L^{(0)}=\Z_2$ but not the $L^{(0)}=0$ case at this point on the moduli space. One naive solution is to have certain KK-modes becoming massless at this special vev that enhanced the gauge symmetry from Abelian to non-Abelian, but the possible candidate of the KK-modes are also present in the $SU(2)$ case, and would also become massless there.

To better learn about the nature of these singularities, one can turn on an $\CN=2^*$ mass deformation. The singularity at the origin will be deformed into an $I_4$ and two $I_2$ singularities. These can be referred to as the ``monopole,'' the ``dyon,'' and the ``quark'' singularities, and which one is $I_4$ depends on the polarization. 

 However, the other singularity is not expected to be deformed. In fact, as turning on the mass commutes with gauging the 0-form $\Z_2$ symmetry, one can first turn on the mass in the theory with $L^{(0)}=0$. The two singularities will deform in a way to maintain $\Z_2$ symmetry, and after the $\Z_2$ quotient, the new singularity would arise. It will have to be the undeformed $I_0^*$. Therefore, it cannot be an $SU(2)$ singularity and has to be the $O(2)$ theory.

 This $O(2)$ can in fact be realized as a subgroup of $SU(2)$, with two components parametrized by
 \begin{equation}
     \begin{pmatrix}
e^{i\theta} & 0 \\
0 & e^{-i\theta} 
\end{pmatrix}, \quad\begin{pmatrix}
0 & e^{i\theta} \\
-e^{-i\theta} & 0 
\end{pmatrix}.
 \end{equation}
 This is in fact the normalizer of the Cartan of $SU(2)$, and the group of components is by definition the Weyl group $\Z_2$. It is also the stabilizer of the image of 
 \begin{equation}
     \begin{pmatrix}
i & 0 \\
0 & -i 
\end{pmatrix}\in SU(2)\rightarrow SO(3)
 \end{equation}
 under the adjoint action of $SU(2)$. 

\begin{figure}[htb!]
    \centering
    \includegraphics[width=\linewidth]{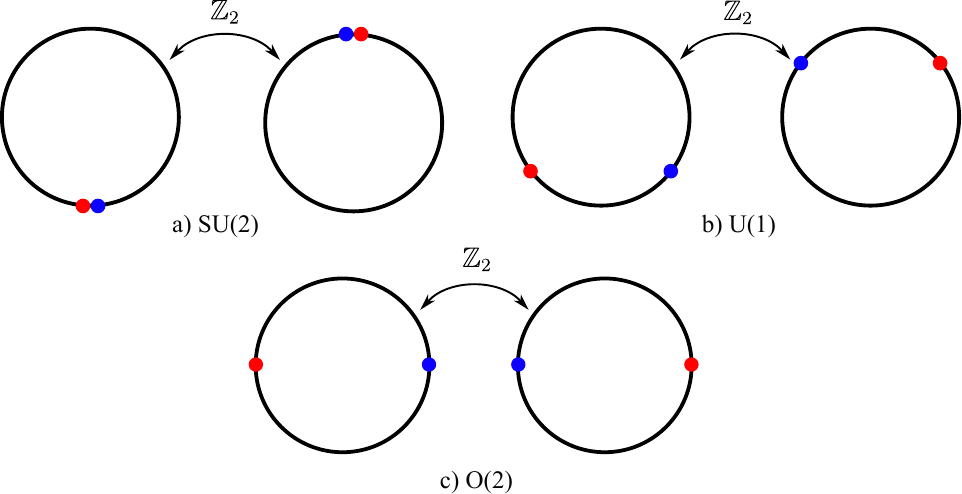}
    \caption{After decoupling the center of mass degree of freedom, the two D3-branes can be placed on the transverse circle in symmetric positions $\theta$ and $-\theta$. a) When $\theta=0$, where the two branes coincide and a non-Abelian gauge group emerges; this is related to $\theta=\pi$ via the $\Z_2$ symmetry. b) For generic values of $\theta$, there is no enhancement. c) When $\theta=\frac{\pi}{2}$, the configuration is fixed by the $\Z_2$; after gauging this discrete symmetry, the gauge group is extended to $O(2)$.}
    \label{fig:O2Brane}
\end{figure}

Another way of confirming the existence of this $O(2)$ theory is via string duality. If we use the equivalence between M-theory on $T^2$ and Type-IIB on $S^1$, we get now from a stack of M5-branes wrapping $T^2$ to a collection of D3 branes distributed over $S^1$. The holonomies of the 2-form gauge fields on M5s become the positions of the D3s on the circle. Consider now the case of two D3s. Decoupling the center-of-mass motion would enable us to put the two D3s in a symmetric configuration as in Figure~\ref{fig:O2Brane}. Then there are indeed two singularities given by $\theta=0$ and $\pi$. Now, consider the $\Z_2$ symmetry of $\theta\mapsto \theta+\pi$. After quotienting it out, we will only need to consider the configurations with the two points in the ``lower half,'' and the two singularities at $\theta=0$ and $\pi$ are identified. However, there is another special configuration that arises when $\theta=\pi/2$. As the two D3-branes are not together, there is no enhancement of gauge symmetry with the W-boson remaining massive. Instead, the $\Z_2$ act non-trivially on the free 4d $\CN=4$ $U(1)$ theory at low energy. For example, $\theta$ itself is a massless scalar, increasing which corresponds to moving both branes upward, and with this $\Z_2$ symmetry, the two branes for $\theta=3\pi/2$ have switched position, and increasing $\theta$ will actually correspond to having them moving downwards. It is then straightforward to check that after gauging the $\Z_2$ action on the theory one gets an $O(2)$ theory.

\subsubsection{Other SCFT points: general case with $L^{(0)}=0$}

 For the $A_{N}$ series with $L^{(0)}=0$, it is also easy to see that the type of ``maximal singularities'' elsewhere on the moduli space is the same as the one at the ``origin,'' but this is not the case in general. For groups of other types, the statement is noticeably more interesting, as the other singularities can be of different types compared with the one at origin. One phenomenon similar to the case of type $A$ is that the singularities appear regardless of choice of polarization. This is indeed the case also for other types of groups. 

For groups of type $D$ and $E$, they can be obtained directly from 6d $(2,0)$ theory on $T^2$, while to get groups of other types, one can wind a duality defect along a cycle of $T^2$ \cite{Vafa:1997mh}. See \cite{Tachikawa:2011ch,Cordova:2015vwa,Duan:2021ges} for discussions on various aspects of this compactification. For a class of polarizations, the construction will factor through a 5d absolute theory, but in general it will not.

For general $G$, the singularities are all on the real slice $T/W$ of the Coulomb branch. It is well known that the singularities are in one-to-one correspondence with proper subsets of the affine Dynkin diagram. The SCFT points with locally maximal gauge group---which we will later simply refer to as ``SCFT points''---are located at the locally most singular points (e.g.~where singularities of lower codimensions intersect) and correspond to subgraphs of the affine Dynkin diagram with only one node removed, which are maximal in the partial order given by inclusion. The gauge group at these singularities are going to have the same rank as $G$ and it will be fully non-Abelian with no Abelian factors.  

Below we list the gauge algebras that can be obtained from $G$:

\begin{itemize}
    \item $A_{N}$: one only gets $N$ other $A_{N}$ theories. This is related to the fact that all weights are minuscule. In general, the affine Dynkin diagram with a node whose fundamental weight is minuscule removed will be isomorphic to the original (non-affine) Dynkin diagram.
    \item $B_N$: Another $B_N$, a $2A_1+B_{N-2}$, $A_3+B_{N-3}$, $D_4+B_{N-4}$,\ldots,  $D_k+B_{N-k}$,\ldots, $D_N$.
    \item $C_N$: Another $C_N$, two $A_1+C_{N-1}$, and $C_{k}+C_{N-k}$ with $2\leq k\leq N-2$ (note $C_2=B_2$).
    \item $D_N$: Three other $D_N$, and $D_k+D_{N-k}$ with $2\leq k\leq N-2$ (note that $D_2=2A_1$ and $D_3=A_3$).
    \item $E_6$: two more $E_6$, three $A_1+A_5$, and one $3A_2$.
    \item $E_7$: one more $E_7$, an $A_7$, two $A_1+D_6$, two $A_2+A_5$, one $A_1+2A_3$.
    \item $E_8$: $A_1+E_7$, $A_2+E_6$, $A_3+D_5$, $2A_4$, $A_1+A_2+A_5$, $A_1+A_7$, $A_8$, and $D_8$.
    \item $F_4$: a $B_4$, an $A_1+A_3$, 
 an $A_2+A_2$, and an $A_1+C_3$.
 \item $G_2$: an $A_1+A_1$ and an $A_2$. 
\end{itemize}

Notice that for group $G$ of rank $N$, there are $N$ additional SCFTs (so the total is $N+1$). The number of SCFTs with gauge group $G$ in total will be the same as the order of the center (which also equals to the number of minuscule weights). Also it is not hard to see that the 0-form symmetry can act non-trivially permuting singularities of the same type. For example, in the $E_6$ case, the three $E_6$ and the three $A_1+A_5$ should form two orbits of the $\Z_3$ symmetry, while the $A_2+A_2+A_2$ singularity will be fixed.  

The exact global form of the gauge groups will be determined by the polarization chosen. When $G$ is simply connected, it might at first appear reasonable to think that one can choose all gauge groups to be simply-connected. However, this would not be consistent with $S$-duality in general. Namely, we should have each factor transforming under SL$(2,\Z)$ in the same way as predicted by how SL$(2,\Z)$ acts on polarizations. This combined with the 0-form symmetry is usually strong enough to completely determine the global form of the gauge group.

For example, the $E_8$ theory is absolute in 6d, and $T[T^2]$ obtained from it will have no dependence on polarizations. Therefore, one expects that the other SCFTs should also be invariant under SL$(2,\Z)$. Therefore, it is natural to conjecture that the gauge group for each singularity is given by:
\begin{align}\nonumber
    &(SU(2)\times E_7)/\Z_2, \quad(SU(3)\times E_6)/\Z_3, \quad(SU(4)\times Spin(10))/\Z_4, \\&\quad(SU(5)\times SU(5))/\Z_5, \quad(SU(2)\times SU(3) \times SU(6))/\Z_6,\\& 
    (SU(2)\times SU(8))/\Z_4, \quad
    SU(9)/\Z_3, \quad Spin(16)/\Z_2 \nonumber
\end{align}
The quotient are always diagonal, and these all in fact comes from absolute theories in 6d.\footnote{For example, in the fifth case, the $\Z_6$ acts on $SU(2)\times SU(3)$ via the isomorphism with the center $\Z_2\times \Z_3$, and for the next one we have picked a $\Z_4$ Lagrangian subgroup of $\Z_2\times \Z_8$ generated by $(1,2)$. One can view the $\Z_4$ as an extension of $\Z_2$ by $\Z_2$, and it is equivalent to write $(SU(2)\times SU(8)/\Z_2)/\Z_2$. For the last one, one gets the ``semi-spin'' group $Ss(16)$, which is different from $SO(16)$ and is another quotient of $Spin(16)$. Notice that each gauge group arising in this way in 4d is expected to be a subgroup of $E_8$. In particular, while the semi-spin group is a subgroup, $SO(16)$ is not and should not appear as the gauge group in lower dimensions.} See a classification in Section 4.2 of Part I, where all of the cases mentioned here feature.

The other absolute 6d theories are of $D$ type: $SO(4n+2)$, $SO(8n-4)$, $SO(8n)$, and $Ss(8n)$ (equivalent to $Sc(8n)$ with duality and further equivalent with $SO(8n)$ when $n=1$ with triality).\footnote{Notice that one should not think of these as gauge groups of the 6d theory. Instead, their character lattices classify charges of strings (as opposed to lines) in the theory. Another point to clarify is that we are not talking about 6d absolute theory on $T^2$ but have gauged the 2-form symmetry. In other words, we still have $L^{(0)}=0$ here, with the compact scalar valued in the Spin group and only the $L^{(1)}$ part looks like what one would get by compactifying an absolute theory. The more general case will be discussed later. } From Table 1 in Section 4 of Part I, it at first glance seems that there are more than one possibilities for the other gauge groups appearing after putting the theory on $T^2$ whenever $D_{4n}$ appears as a subfactor or for $D_{4m-2}\oplus D_{4n-2}$. However, demanding ``naturalness'' (e.g.~they behave nicely when we vary the rank), we conjecture that the global forms are given below:
\begin{itemize}
    \item The $SO(4n+2)$ theory. The  groups appearing are three (four in total) $SO(4n+2)$, two $SO(4)\times SO(4n-2)$, $\ldots$, two $SO(2m)\times SO(4n-2m+2),\ldots,$ two $SO(2n)\times SO(2n+2)$.
    \item The $SO(8n-4)$ theory. Three other $SO(8n-4)$, and $SO(2m)\times SO(8n-2m-4)$ for $m=2,\ldots,4n-4$. 
    \item The $SO(8n)$ theory. Three other $SO(8n)$, $(Spin(4m)\times Spin(8n-4m))/(\Z_2\times\Z_2)$ with $m=1,\ldots,2n-2$, and $(Spin(4l+2)\times Spin(8n-4l-2))/\Z_4$ with $l=1,\ldots,2n-3$ where all actions are diagonal.
\end{itemize}

In the third case, there are more than one ways to get an SL$(2,\Z)$-invariant theory with gauge algebra of the form $D_{2m}\oplus D_{2n}$ (for example, two ways for $D_{4m-2}\oplus D_{4n-2}$ and at least six different ways for $D_{4m}\oplus D_{4n}$ using Table 1 in Part I). However, there is a unique one for $D_{4m-3}\oplus D_{4n-1}$ and requiring that they all fit in a family in a natural way leads to the global forms listed above. 

Similarly one can also try to determine the global form of the gauge group in other cases of type ADE using the SL$(2,\Z)$ action, demanding that the representation one gets for each SCFT point are the same. We will not perform this analysis here.

What is more delicate is the non-simply-laced cases, where one can no longer demand that everything is SL$(2,\Z)$ covariant. This is because that an duality defect has been inserted to get these theories, and it will not be invariant under $S$-duality. It is in fact only preserved by the congruence subgroup $\Gamma_0(2)$ for $BCF$ or $\Gamma_0(3)$ for $G_2$. Then again one can check that there is a choice of global form that makes the mapping class group action consistent. For example, in the $G_2$ case, the $A_2$ for any choice of the global form would be invariant under $\Gamma_0(3)$, while for $A_1\times A_1$, one will have to choose $SO(4)$ as the global form to be invariant.

In these non-simply-laced cases, it might be easier to determine the global form of gauge groups from the purely mathematical perspective. Namely, we are trying to determine the ``locally maximal'' stabilizers of elements $g\in T$ under the adjoint action. Then the statement is that Stab$_G(g)$ transforms covariantly with $G$. In particular, when $G=G^\vee$ is self-dual in the Langlands sense, then Stab$_G(g)=$Stab$_G(g)^\vee$.

A related physics scenario where these subgroups also feature is in the context of surface operators (e.g.~in 4d gauge theory). When we tune the ``ramification parameters'' to special values on the boundary of the parameter space, the defect becomes invariant under a larger group of gauge transformations in the bulk. This is related with the setup being discussed presently as surface defects can be engineered by fivebranes whose world-volume theory is what we are studying.\footnote{The mathematical framework to consider such defects often involves ``parahoric subgroups.'' This can be understood with the following consideration. As one gets these algebras by looking at subsets of the affine Dynkin diagram, it is actually natural to embed all the groups into the loop group of $G$. Then these fixed points are related to ``parahoric subgroups'' of the loop group.}

The next question is to determine the coupling constant. This can be done either via geometry or physics, with both methods relying on consequences of the $\CN=4$ supersymmetry. The geometric way is to simply look at the shape of the fiber over the singularities and its neighborhood, which encode the coupling constant. As the geometry is obtained via a $W$-quotient of a space with trivial fibration, one expect that up to normalization and conventions for different gauge groups, the coupling constant will be the same $\tau$. Notice that because of $S$-duality, one has to fix a precise global form in the SL$(2,\Z)$ orbit to talk about the coupling constant, whereas in the previous part, we only care about the global form \emph{up to} duality. Conjecturally, the right choice is given by asking for what the $G$-stabilizer of $g\in T$ that corresponds to this singularity is.

The physical way is to follow a trajectory on the moduli space. Because of the supersymmetry, the coupling constant is invariant, and the only thing to worry about is the convention for $\tau$ before and after the enhancement of gauge symmetry, which depends on the exact global form of the gauge groups. This physical process has better chance of generalizing to larger class of 6d $(1,0)$ theories on $T^2$.

We will end this subsection with a comment on the consequences of the existence of the other SCFTs.

Very often, when computing the partition function of the 6d $(2,0)$ theory on certain manifolds, one would attempt to turn this into a computation of the 4d $\CN=4$ theory. This has several problems. First of all, it might not be always possible. In particular, when a holonomy of the 0-form symmetry is turned on, then even if there is a $T^2$ in the geometry, one cannot reduce to the 4d SCFT. Furthermore, approximating the 6d theory on $T^2$ by a single 4d SCFT will cause us to loose the information about the global structure of the moduli space, with various other SCFT points. A better approximation is to sum over all the SCFTs. It would be interesting to compare this with the localization computation using the 5d $\CN=2$ theory, which, on a circle, can capture the information about the compact scalar.

\subsubsection{Other SCFT points: general pure polarization}

We will first consider $A_{p-1}$ with $p$ prime and $L^{(0)}=\Z_{p}$, then focus on $A_{N-1}$. Toward the end, we will also remark about generalization to groups of other types.

 For type $A_{p-1}$, if $L^{(0)}$ is non-empty, then it can only be $\Z_{p}$. One would similarly first conclude that there are other points which could be either a 4d $\CN=4$ theory with $SU(N+1)$ gauge group, or an orbifold of a theory with smaller gauge group. Again, a similar argument as before would suggest that there is not really an enhancement of gauge symmetry by massless W-bosons. Instead, the compact scalar with a non-trivial expectation value is now fixed by an additional discrete group. 
 
 The compact scalar $\phi$ is now $PSU(p)$-valued. If we express it in terms of a $p\times p$ special unitary matrix, then we need to remember the equivalence under multiplication with a $p$-th root of unity. Then the special values of $\phi$ are those with
 \begin{equation}\label{Zpaction}
     w\cdot \phi=e^{2\pi i m/p}\phi
 \end{equation}
 for some $w\in S_p$ acting by permuting the eigenvalues and some $m\in\Z_p$. Requiring that $w$ is not in the Weyl group of Stab$_{SU(p)}(\phi)$ means that $m$ can't be 0. Then $w$ has to act on all eigenvalues. It is then easy to see that $w$ is a cyclic permutation. Then we have $\phi=\text{diag}\{1,e^{2\pi i/p},e^{4\pi i/p},e^{6\pi i/p},\ldots,e^{2(p-1)\pi i/p}\}$ up to permutation. Therefore there is a unique $\phi$ with this enhancement. The gauge group is $U(1)^{p-1}\rtimes \Z_p$. The action $\Z_p$ on $U(1)^{p-1}$, parametrized by $a_1,\ldots,a_{p-1}$, is generated by 
 \begin{equation}
     a_1\mapsto a_2, \; a_2\mapsto a_3,\;\ldots,a_{p-2}\mapsto a_{p-1},\; a_{p-1}\mapsto (a_1a_2\ldots a_{p-1})^{-1},
 \end{equation}
 while the action on the adjoint matter fields are given by the infinitesimal version of this action. 
 This being a symmetry of the 4d $\CN=4$ 
 $U(1)^{p-1}$ theory is consistent with the fact that the coupling constants for different $U(1)$'s are not independent, but instead determined by the coupling constant of $SU(p)$.

 In terms of the configuration of D3-branes on a transverse $S^1$, this new singularity corresponds to ``evenly distributing'' the branes along the circle. Another perspective is that the Seiberg-Witten curve covering the $T^2$ have additional automorphism at special values of Coulomb vev, which can be gauged. 

\begin{figure}
    \centering
\includegraphics[width=0.8\linewidth]{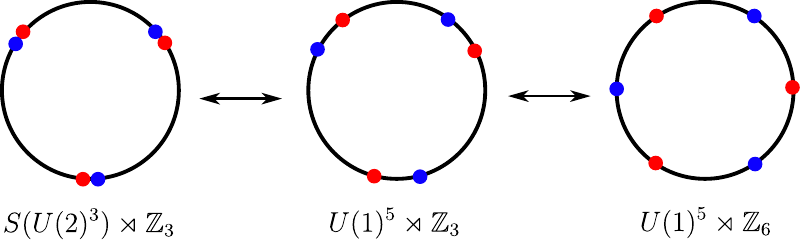}
    \caption{Deformation between two maximal singularities along a path consists of less singular theories. The configuration on the left  is fixed by $\Z_3\subset \Z_6$, which extends the gauge group after gauging the $\Z_6$. The $\Z_2$ quotient will identify it with another configuration given by a $\frac{\pi}{3}$-rotation. One can deform it to the middle one by separating the branes, breaking non-Abelian gauge symmetry. Once they are separated by $\frac{\pi}{3}$, the configuration is fixed by the entire $\bZ_6$.}
    \label{fig:PSUBrane}
\end{figure}

Now assume that $\phi$ is $PSU(N)$-valued. Then there is going to be additional singularities for each divisor of $N$. Assuming $k|N$, then one can have solutions to
\begin{equation}\label{fixedpoints}
    w\cdot \phi=e^{2\pi i m/N}\phi
\end{equation}
where $w$ is a product of $k$ commuting cyclic permutations, each permuting $N/k$ entries, and $\phi$ takes the form of a block-constant diagonal matrix,
\begin{equation}
    \mathrm{diag}\{\mathbf{1}_{k\times k},e^{2\pi i k/N}\cdot\mathbf{1}_{k\times k},\ldots, e^{2\pi i (N-k)/N}\cdot\mathbf{1}_{k\times k}\}.
\end{equation}
Then the gauge group takes the form of $S(U(k)^{N/k})\rtimes \Z_{N/k}$, where $S(U(k)^{N/k})$ is $U(k)^{N/k}$ with the diagonal $U(1)$ removed. If it is not removed, then the $\Z_{N/k}$ action would be obvious, and the $\Z_{N/k}$ action on the central part $U(1)^{N/k-1}$ is the same as in \eqref{Zpaction}. It is not hard to prove that these are the only possible ``maximal'' singularities, meaning that other singularities (e.g.~$S(U(k-m)^{N/k}\times U(m)^{N/k})\rtimes \Z_{N/k}$) can be obtained by deformations from them. This is illustrated in Figure~\ref{fig:PSUBrane}.  Therefore, the statement for $\phi\in PSU(N)$ and $G=SU(N)$ is that there is a singularity for each divisor of $N$, with $1$ and $N$ included.
The case when $\phi\in SU(N)/\Z_k$ and $G=SU(N)/\Z_{k'}$ is similar. One can again use \eqref{fixedpoints} to work out the fixed points, which we will not analyze here. 

The $L^{(0)}\neq0$ case for other types of groups will involve singularities similar to the type-$A$ case, where the theory is given by a discrete gauging of a theory with Abelian factors in the gauge group, as well as new kinds of singularities given by quotients of fully non-Abelian theories. 

For example, in the $E_6$ case, the $3A_2$ SCFT, with gauge group $SU(3)^3/\Z_3$ will itself have a $\Z_3$ symmetry, which in this case permute the three gauge groups. After gauging the $\Z_3$, one gets an $(SU(3)^3/\Z_3)\rtimes \Z_3$ theory. It is a technical though straightforward exercise in Lie theory to work out the structure of all singularities and to determine the gauge group appearing at each of them, which we will not perform here.

\subsubsection{Polarization and S-duality}

In Part I of the paper, we already described the action and the orbit of polarizations under the SL$(2,\Z)$. The goal of this section is to clarify the meaning of the action of SL$(2,\Z)$ on \emph{the theory}.

For a pure polarization, the mapping class group will only act on $L^{(1)}$. So for now we will ignore the effect of $L^{(0)}$ and $L^{(2)}$. Note that the discussion in this section can also be generalized to cases with a quadratic refinement or with mixed polarizations.

We first look at the partition function. We start with a polarization that leads to a $G$ gauge theory in 4d at low energy. The familiar statement of $S$-duality is that the partition function of this theory is the same as that for the $G^\vee$ theory but at a different coupling constant $-1/\tau$,
\begin{equation}\label{SPart}
    Z_G(\tau)=Z_{G^\vee}(-1/\tau).
\end{equation}
This is indeed correct in general for $T[T^2]$ obtained from any relative 6d theories, supersymmetric or not, on any four manifolds, even if we keep all KK modes or when there is not a Lagrangian description, as this equality simply means that the partition function of the 6d theory can be expressed in different ways depending on the choice of basis of $H^1(T^2)$.\footnote{Although this is very general, for simplicity we will use the terminology of $G$ and $G^\vee$ gauge theory, as this is the most familiar context. Also, this is assuming that there are no other hidden SL$(2,\Z)$-breaking choices made in defining and regularizing the partition function.
} In other words, one can choose any cycle to be the ``electric cycle'' and any other compatible cycle (i.e.~with the right pairing) to be the ``magnetic cycle,'' and the partition function will not be affected. What it does affect is what the modulus $\tau$ is, which depend on a choice of A- and B-cycles. Correctly taking into account of what $\tau$ becomes after changing to the $S$-dual basis of $H^1(T^2)$ leads to the previous expression \eqref{SPart}.

However, this expression is not how $S$ acts on the partition function --- it simply identifies the partition function of two apparently different theories. To have something closer to an action, one needs to express the right-hand side as a combination of partition function for the $G$ theory, using the fact that they are related by gauging certain generalized symmetries (assuming $G$ is of type ADE),
\begin{equation}\label{SPart2}
    Z_G(\tau)=\#\sum_B Z^B_{G}(-1/\tau),
\end{equation}
where $B$ denotes the backgrounds (usually referred to as ``'t Hooft fluxes'' for the 1-form electric symmetry) in the $G$ theory to be summed over in order to get to the $G^\vee$ theory. Similarly, with a background flux to start with, one has
\begin{equation}
Z^B_G(\tau)=Z^B_{G^\vee}(-1/\tau)=\#\sum_{B'}e^{(B,B')} Z^{B'}_{G}(-1/\tau)
\end{equation}
where $e^{(B,B')}$ is the $U(1)$-valued pairing on $H^2(M_4,D)$. This is now a genuine action of $S$ on the vector $Z^B_G(\tau)$. Similarly, one can get the action for $T$, as discussed in Part I, leading to a full
SL$(2,\Z)$ action. This action naturally arise from a flat vector bundle of rank $|H^2(M_4,D)|$ over the moduli space of elliptic curves. The connection of the vector bundle is obtained by varying $\tau$ and keeping track of $Z^B_G(\tau)$. This connection is obviously flat, but in general has monodromies, giving rise to exactly the SL$(2,\Z)$ representation described previously. This SL$(2,\Z)$ action is a special case of the MCG$(M_6)$ action on the Hilbert space of the 7d TQFT on $M_6$.

So far we are basically reproducing and rephrasing results well known since \cite{Vafa:1994tf}. But now we will move to ``higher categorical levels.'' The first step is to look at the Hilbert space of the theory on a three-manifold $M_3$. In general, this is infinite dimensional, but, when there is supersymemtry, one can look at certain subspace of BPS states which can be finite dimensional.

If one only cares about the (graded) dimension of $H_{G}(M_3)$, then one can look at the partition function on $M_3\times S^1$. As this is an integer when defined, it cannot depend on $\tau$, and one must have
\begin{equation}\label{DimIso}
\mathrm{dim\;}H^B_G(M_3)=\mathrm{dim\;}H^B_{G^\vee}(M_3).
\end{equation}
Note that this is in line with a conjecture of David Jordan on the dimension of the skein module of $M_3$ \cite{jordan2023langlands}, which is closely related to $H^B_G(M_3)$ with the Marcus (aka GL) twist, although it is expected that the Hilbert space of the physical theory is actually always infinite dimensional \cite{Gukov:2022cxv}. See \cite{Pei:2026lnl} for a recent study on the relation between skein modules and the Hilbert space of the gauge theory.

Even when the Hilbert space is infinite dimensional, one still has a canonical isomorphism,
\begin{equation}\label{HilbertSpaceIso}
H^B_G(M_3;\tau)\simeq H^B_{G^\vee}(M_3;-1/\tau),
\end{equation}
as this is the Hilbert space of the 6d theory on the same five-manifold with the same choice of polarization, but just described in different basis. Of course the interpretation of the same $B$ in the two low-energy gauge-theory descriptions on the two sides are different and are related by electro-magnetic duality. See \cite[Sec.~7]{kapustin2007electric} for more detailed discussion on this aspect.

Again, this is not yet an action on a single Hilbert space, as it just says that two apparently unrelated spaces are isomorphic. To make it an action, one would first transform $H^B_{G^\vee}(M_3;-1/\tau)$ to $H^B_{G^\vee}(M_3;\tau)$, and then, by discrete gauging, map it to $H^B_G(M_3;\tau)$. As the first step depends on a choice of a path, the second step also cannot be canonical. This makes the ``action of SL$(2,\Z)$'' on the Hilbert space ambiguous. Another way to say this is that the connection for Hilbert space bundle over $\mathbb{H}/\mathrm{SL}(2,\Z)$ is in general non-flat. A slightly better situation is when it is flat on certain subspaces or quotients of the full Hilbert space, albeit having non-trivial holonomies globally. This is typically only expected when there is supersymmetry and one looks at supersymmetry-protected subsectors, with the topologically twisted theory being one special example of this. 

As we have assume that the only SL$(2,\Z)$-breaking effect is the choice of a polarization on $T^2$, it is useful to separate it from the ``physical degrees of freedom'' by again representing the system as the 5d TQFT $\CT^{\text{bulk}}[T^2]$ sandwiched between two boundary conditions. One, denoted as $\CB_p$, comes from the boundary theory in 6d, while the other $\CB_{t}$ is a topological boundary condition determining the polarization on $T^2$. The category of boundary conditions of the 5d TQFT is a higher category, but, for our purpose of understanding the Hilbert space on $M_3$, we can effectively work with the category of boundary conditions of $\CT^{\text{bulk}}[M_3\times T^2]$.\footnote{For a TQFT, one often considers a category that only contains objects that are topological boundary conditions. We will instead consider a larger category that at least includes the non-trivial boundary theory $\CB_p$ in addition to all topological boundary conditions.} Denote it as $\CC_{M_3}$, then the Hilbert space is Hom$(\CB_t,\CB_p)$. The mapping class group MCG$(M_3\times T^2)$ acts as auto-equivalences of this category $\CC_{M_3}$. But as $\CB_p$ and $\CB_t$ comes from boundary conditions of the 7d and 5d TQFT respectively, $\CB_p$ is invariant under the entire MCG action, while $\CB_t$ is invariant under MCG$(M_3)$. This gives an isomorphism 
\begin{equation}\label{HomIso}
    \mathrm{Hom}(\CB_t,\CB_p)\simeq \mathrm{Hom}(g\cdot\CB_t,g\cdot\CB_p=\CB_p)
\end{equation}
for any $g\in\mathrm{SL}(2,\Z)$, and \eqref{HilbertSpaceIso} is a special case for this. As for the action on the Hilbert space, it is meaningful when $g$ leaves  $\CB_t$ invariant.

As Pol$(T^2)$ is usually a finite set, given a $\CB_t$, there is always a subgroup with finite index $\Gamma\subset$ SL$(2,\Z)$ that fixes $\CB_t$. For example, when the 6d theory is an $A_1$ $(2,0)$ theory, then $\Gamma$ is congruent to $\Gamma_0(4)$ (and $\Gamma_0(2)$ if we only consider the theory on spin manifolds and therefore do not keep track of the quadratic refinement). One expects that the Hilbert space of the 4d $\CN=4$ SYM as a low-energy description on $M_3$ is a $\Gamma_0(4)$-equivariant subspace of the 6d theory on $T^2\times M_3$, and there is also a $\Gamma_0(4)$-action on it.

The discussion of higher categorical structures is similar. For example, one can consider similar hom spaces in the 2-category $C_\Sigma$ associated with a Riemann surface $\Sigma$ describing boundary conditions of the 3d TQFT $\CT^{\text{bulk}}[\Sigma\times T^2]$, and one now has \eqref{HomIso} as an equivalence of categories for a choice of $\CB_t$ and a $g\in\mathrm{SL}(2,\Z)$. When $\CB_t$ is invariant, this leads to an auto-equivalence of the category Hom$(\CB_t,\CB_p)$. A special case is when the low energy effective theory given by $\CB_t$ and $S\cdot\CB_t$ are sigma models to moduli spaces $\CM_{\text{full}}(G)$ and $\CM_{\text{full}}(G^\vee)$ containing the Hitchin moduli spaces $\CM_{H}(G)$ and $\CM_{H}(G^\vee)$, reducing to the setup for the gauge theory approach to the geometric Langands correspondence \cite{kapustin2007electric}. We will discuss the relation between the moduli spaces and their differences further in later subsections. 

Note that there is a qualitative difference between how partition functions, and ``higher'' structures like Hilbert spaces and categories behave under the mapping class group action. When we talk about partition functions given by polarizations in Pol$(M_6)$, the action of MCG can be represented by an action on a ``partition vector,'' which doesn't have a natural analog for the higher structures. Namely, MCG$(M_d)$ always acts on Pol$(M_d)$ via permutation, but what is special for $d=6$ is that there is an embedding of Pol$(M_d)$ into the Hilbert space (a 0-category) of the 1d TQFT $\CT^{\text{bulk}}[M_6]$. As this embedding is equivariant with respect to the MCG$(M_6)$ action, it is sufficient to understand how MCG$(M_6)$ acts on this Hilbert space. More concretely, one can choose a basis of this Hilbert space, and then find the matrices that represent the MCG action. This is the origin of the partition vector, and, in fact, how such action on Pol$(M_6)$ was studied in the Part I of the paper. Notice that this will not work in any straightforward manner once we move to $d<6$. There, the boundary condition of $\CT^{\text{bulk}}[M_d]$ are objects in a $(6-d)$-category, and, under modest assumptions, auto-equivalence given by MCG$(M_d)$ cannot send a boundary condition to a linear combination of boundary conditions but instead should act by permutation. Therefore, when $d<6$, the statement involving non-invariant polarizations are the isomorphisms in \eqref{HomIso}, while one gets auto-equivalences for the $(5-d)$-category $\mathrm{Hom}(\CB_t,\CB_p)$ when $\CB_t$ is fixed by certain elements in MCG$(M_d)$. When $d=5$, this is again a representation, which gives the action on the Hilbert space of the 6d physical theory on $M_5$ discussed previously. 

As a simple example, consider $M_5=S^3\times T^2$. The action of such a 2d TQFT will be of the form $K_{ij}\int A_id\phi_j$, which, in the present case, involves only a single $A$ and a single $\phi$ obtained by integrating the three-form $C$ over $T^2$ and $S^3$. It is easy to see that this action is invariant under SL$(2,\Z)$, and hence the Hilbert space of the $T[T^2]$ theory with any polarization on $S^3$ always has an SL$(2,\Z)$ action.

This is a special case of a more general phenomenon that Pol$(M)\rightarrow$Pol$(M\times M')$ may not be injective. Indeed, while Pol$(T^2)$ can contain non-trivial orbits of $SL(2,\Z)$, they will collapse once reduced on $S^3$. Conceptually, it is clear why this is happening: $S^3$ lacks 1-cycle and 2-cycle for $B_2^E$ and $B_2^M$ (obtained by integrating $C$ on the two 1-cycles of $T^2$) to be turned on either in two spatial or one spatial and one temporal direction. Similar statement would be true for $M_3$ with $H^1(M_3,D)=H^2(M_3,D)=0$. For such $M_3$, one can improve the statement for the S-duality, as now we have an canonical isomorphism improving \eqref{DimIso}
\begin{equation}
H^B_G(M_3;\tau)\simeq H^B_{G^\vee}(M_3;\tau),
\end{equation}
which can be obtained by discrete gauging. Combining it with \eqref{HilbertSpaceIso} leads to
\begin{equation}
H^B_G(M_3;\tau)\simeq H^B_{G}(M_3;-1/\tau),
\end{equation}
giving the S part of the SL$(2,\Z)$-action on this Hilbert space.

\subsubsection{A ``universal $\Z_2$ symmetry'' of 6d $(1,0)$ theories}\label{sec:Universal}

Until now, we have used the 6d $(2,0)$ theories and their compactifications as our main examples. They are usually very good as illustrating the general phenomena concerning symmetries of polarizations that occur under compactification, but there are some phenomena that are unique to $(1,0)$ theories. We will now discuss one such example here.

When compactifying on $T^2$, we usually assume that the spin structure is the non-bounding one (e.g.~periodic along both circles) so that supersymmetry can be preserved for $T[T^2]$. For other spin structures, naively supersymmetry is broken via the Scherk--Schwarz mechanism. However, one can preserve supersymmetry with a non-trivial background for R-symmetry. This can be done by turning on a holonomy of $-1\in SU(2)_R$ along one (or both) of the cycles of $T^2$ so that the supercharges will still satisfy the periodic boundary condition. Then, instead of having a single theory with $SL(2,\Z)$ duality, one will have three different theories transforming under $SL(2,\Z)$ via the quotient $PSL(2,\Z_2)$. When there is a polarization chosen on $T^2$, the story can be more interesting and there can be new $SL(2,\Z)$ orbit arising. 

In fact, we don't need to go down to $T[T^2]$ to observe this phenomenon. It arises already for $T[S^1]$ and even $T[\mathrm{point}]$. Indeed, what we used here is a ``universal $\Z_2^U$ global symmetry'' of the 6d theory given by the product of $(-1)^F$ and the $-1$ in $SU(2)_R$, and there are in general $|H_1(M_d,\Z_2)|$ different versions of $T[M_d]$ from choosing the holonomy for this symmetry.

What would the theory $T[T^2]$ with a bounding spin structure on $T^2$ looks like?

One can first look at the boundary condition for different multiplets in 6d, and it is easy to see that the only difference is that the hypermultiplet will satisfy an anti-periodic boundary condition. Therefore the Higgs branch is expected to be the fixed point of this $\Z_2^U$ action, and is therefore in general smaller. From the point of view of the Coulomb branch, one expects that the singularities will become less singular. But to ensure that the limit of the large Coulomb parameter to remain the same, it might be required that the structure of other singularities away from the origin are also modified. This is very similar to a mass deformation, except that it is a ``large deformation.'' However, in some cases one indeed expects that it can be embedded in a family of continuous mass deformations.   

In general, there can be three possibilities for how $\Z_2^U$ acts on the theory:
\begin{itemize}
    \item This symmetry acts trivially. This would be the case for a theory without hypermultiplet, or when there are hypermultiplets but this symmetry is gauged.  
    \item This symmetry acts non-trivially, but is part of another continuous global symmetry group $G$. This means that the true symmetry that acts faithfully is $(G\times SU(2)_R \times \Z_2^F)/\Z_2$. One example is $N$ free hypermultiplets which has $Sp(N)$ flavor symmetry but the $\Z_2$ center of $Sp(N)$ acts in the same ways as composition of the center of $SU(2)_R$ and $\Z_2^F$. 
    \item This symmetry acts non-trivially, and is also not part of another continuous global symmetry group. When this happens, this symmetry, although harder to analyze, is probably the most interesting and useful. For example, one might be able to use it to eliminate the zero modes of the theory and make the partition function well defined even when there is not a continuous global symmetry present.  
\end{itemize}

The first class of theories will have the property that they can be made independent of the spin structure of the underlying manifold. Which might be an desirable property for topological applications.

Conjecturally, a non-trivial example of the first case is the (rank-1) E-string theory (see \cite{EString} for a more detailed discussion of this phenomenon). Indeed, the $SU(2)_R$ action on the Higgs branch, which is the minimal nilpotent orbit of $E_8$, factors through $SO(3)$. This is in fact true for any nilpotent orbits  \cite{kronheimer1990instantons} (and hence for the reduced one-instanton moduli space of other gauge groups).\footnote{One check for this is that the Hilbert series has only even $t$ powers. In fact, the Hilbert series of the Higgs branch provides a quick way to check whether this $\Z_2$ action can be trivial. If there are odd powers of $t$ (or half-integer powers in another often-used normalization), then one expects that the center of $SU(2)_R$ has to act non-trivially on the Higgs branch, and therefore the $\Z_2$ action on the theory will be non-trivial. Hilbert series for nilpotent orbits of various groups were computed in \cite{hanany2017quiver} and indeed they are compatible with this criterion.} As a corollary, $\Z_2^U$ would act trivially on all the 5d Seiberg theories and their IR fixed points studied in \cite{seiberg1996five} as they all appear in the RG flow of the rank-1 E-string theory \cite{ganor1997branes}. Indeed, the Higgs branches of the 5d rank-1 theories labeled by ADE are moduli spaces of one instanton \cite{morrison1997extremal} and the $\Z^U_2$ again acts trivially. 

However, if we consider an E-string theory for general rank $Q>1$, it will belong to the second class. This is because the transverse directions of the small $E_8$ instantons form an $\R^4$ and $SO(4)=
(SU(2)_F\times SU(2)_R)/\Z_2$ acts on it. It is obvious that the center of the $SU(2)_R$ still acts non-trivially on the moduli space even after decoupling the center of mass degree of freedom,\footnote{For example, one can just look near the boundary of the moduli space, where the instantons are almost point-like. Then there are non-$\Z_2^U$-invariant configurations for $Q>2$. For $Q=2$, taking into account of internal degrees of freedom of instantons still show that the action is non-trivial. Another check for this is that the Hilbert series for the multi-instanton moduli space now has odd $t$ powers \cite[Sec.~8]{cremonesi2014coulomb}.} and it can be identified with the action of the center of $SU(2)_F$.

For a theory in the second class, $T[T^2]$ and $T[S^1]$ with non-trivial spin structures can be embed into a continuous family where we turn on a holonomy for the bigger symmetry group on $S^1$ or $T^2$. 

The theories demonstrating the third scenario, which are arguably more interesting, can be constructed from theories of the second class via orbifolding. Namely, we can gauge an anomaly-free subgroup of of the global symmetry group which $\Z_2^U$ embeds such that its commutant is discrete. As $\Z_2^U$ is in the center, it will remain unbroken. For the general-rank E-string theory, gauging the a discrete subgroup $\Gamma\subset SU(2)_F$ has the interpretation of replacing the transverse $\C^2$ with the ALE space $\C^2/\Gamma$. When $\Gamma$ is either the binary dihedral, tetrahedral, octahedral or icosahedral group, $SU(2)_F$ will be broken down to $\Z_2^U$.

Beside E-string theories, another interesting class of examples for the second scenario is actually $(2,0)$ theories, whose compactification we will analyze next.

\subsubsection{6d $(2,0)$ theories on $S^1$ and $T^2$ with a bounding spin structure}

6d $(2,0)$ theories can be regarded as $(1,0)$ theories with a global $SU(2)_F$ symmetry.\footnote{One might ask whether there is a similar construction that preserves $(2,0)$ supersymmetry using the center of $Spin(5)_R=\Z_2$. However, this $\Z_2$ appears to be exactly $\Z_2^F$, and would not lead to interesting new theories in lower-dimensions.} This $SU(2)_F$ is the commutator of $SU(2)_R\subset Spin(5)_R$. The $\Z_2$ center of $SU(2)_R$ multiplies four components of the vector representation by $-1$, which is also how the center of $SU(2)_F$ acts. It is straightforward to verify that indeed the true global symmetry preserving $(1,0)$ is at most $(SU(2)_F\times SU(2)_R \times \Z_2^F)/\Z_2$, which becomes $SO(4)$ on bosonic degrees of freedom. The universal $\Z_2^U$ then acts as the center of this $SO(4)$ on bosons.

What happens if we consider $T[S^1]$ and $T[T^2]$ with a bounding spin structure? We expect a genuine 4d $\CN=2$ theory as only half of supercharges are made periodic. For the free $(2,0)$ theory, indeed at low energy one gets a free 4d $\CN=2$ vector multiplet with one of the two scalars being compact. We will now focus on the interacting case with a 6d $(2,0)$ theory labeled by $\frak{g}$ of type ADE.

The analysis above tells us that this can be studied with instead an $SU(2)_F$ holonomy. Near one of the superconformal points for $T[T^2]$, the deformation with a small holonomy for $SU(2)_F$ looks like the 4d $\CN=2^*$ deformation. For example, in the $A_1$ case, the $I_0^*$ singularity of $T[T^2]$ will split into $I_1+I_1+I_4$ with the exact behavior depending on the choice of polarization. Then the question is whether for this ``large deformation,'' some of the singularities will collide to form new singularities. We will argue that this will always be the case as at least part of the Higgs branch will not be lifted.  

Consider now the $\Z_2$ action on the Higgs branch, which is $(\frak{t}\otimes_\R \mathbb{H})/W$. Then there are two cases: either $-1\in W$ and the $\Z_2$ action is trivial (for $A_1$, $B_n$, $C_n$, $D_{\text{even}}$, $F_4$, $G_2$, $E_7$ and $E_8$), or $-1$ fix a proper subset of the Higgs branch ($A_{n>1}$, $D_{\text{odd}}$, and $E_6$).\footnote{One ways to distinguish the two types is to ask whether a representation of $\frak{g}$ and its complex conjugate (i.e.~its dual) is always equivalent. For example, all representations of $A_1$ are either real or pseudo-real, while for $A_{n>1}$, the fundamental is a genuine complex representation.} Naively, one would conclude that they are just like the rank-1 and higher-rank E-string theories, with trivial $\Z_2^U$ action on the first class and non-trivial $\Z_2^U$ action on the second. While there is no doubt that we will have a non-trivial $\Z_2^U$ when it acts non-trivially on the Higgs branch, the other case is more subtle, as we now have mixed branches. In other words, having a non-trivial Coulomb vev of the scalar in the vector multiplet will not kill the Higgs branch. However, this can kill at least part of the Weyl group, and once the $-1\in W$ become broken, the action of $\Z_2^U$ will be non-trivial. Indeed, at a generic point on the Coulomb branch, the mixed branch disappears, and the low-energy effective theory is an 4d $\CN=2$ theory. This is compatible with the fact that there are 8 supercharges that are indeed broken by the Scherk--Schwarz mechanism. 

We now describe the geometry of the Coulomb branch of $T[S^1]$ and $T[T^2]$ for $A_1$ with all possible $\Z_2^U$ holonomy in detail, while generalization for other $G$ are similar in spirit but require more careful analysis. See also \cite{Closset:2023pmc} for related discussions.

\paragraph{Interplay with polarization.} Without any holonomy of $\Z_2^U$, all the versions of $T[T^2]$ for the $A_1$ case obtained by choosing different geometric polarizations are related by duality, as the mapping class group acts transitively on such polarizations (three on spin manifolds and six in total on general 4-manifolds).  However, with non-trivial $\Z_2^U$ holonomy, there are now several physically distinct theories, each with several descriptions related by dualities. One way to talk about these theories in a duality-invariant way is by choosing always the ``electric cycle'' of $T^2$ to be skrinkable in the 7d bulk,\footnote{We will take this opportunity to clarify one potential confusion. For $A_1$, the bulk 7d theory is bosonic. Therefore even when we have chosen a non-trivial spin structure on the 6d boundary, the bulk can still ``cap it off'' (i.e.~giving a null-cobordism of it) as the spin structure on the boundary doesn't have to extend to the bulk for the system to make sense. For example, when we consider $S^1$ with the odd (Ramond) spin structure, it is non-trivial in the spin cobordism group but trivial in the oriented cobordism group, and can be filled in with a disk. This is the geometric polarization that we discussed previously. } and ask about the $\Z_2^U$ holonomy of the electric and magnetic cycles. The latter is well-defined when the $\Z_2^U$ holonomy along the electric cycle is trivial, and not well-defined when that is non-trivial, thus leading to dualities that we will see later. Then there are four theories in 4d $T_{++}$, $T_{+-}$, $T_{-+}$ and $T_{--}$ coming from two different theories in 5d $T_{+}$ and $T_{-}$. (There will be more theories with either quadratic refinement or general non-geometric polarizations, which we will not discuss here.) The theories $T_{+-}$, $T_{-+}$ and $T_{--}$ are related by discrete gauging. Unlike the case without holonomy, where changing polarizations will lead to dual theories, one now has to in general change both the polarization and the $\Z_2^U$ holonomy to get duality, except between $T_{-+}$ and $T_{--}$ which are physically equivalent up to $\tau\mapsto \tau+1$. As $T_{+}$ and $T_{++}$ are the 5d $\CN=2$ theory and its KK-reduction that we discussed earlier, and our task now is to describe the other theories, for which our conjecture is the following.

\paragraph{The 5d theory $T_-$.} This is still a 5d rank-1 $\CN=1$ theory, and its Coulomb branch is $\R_+$. There are two special points, one being the origin $\sigma=0$, where one has an $E_1$ SCFT, and another singular point $\sigma=\sigma_0$ where the low-energy effective theory is $U(1)$ with a charge-2 hypermultiplets. The $E_1$ theory is the UV completion of the strong coupling limit of the $SU(2)$ gauge theory. Its Higgs branch is $\C^2/\Z_2$, agreeing with the expectation that it won't be lifted by the $\Z_2^U$ holonomy. The charge-2 hyper comes from a KK-mode of the off-diagonal components of the adjoint matter, whose KK-mass is compensated by the Coulomb vev to become massless at $\sigma_0=\frac{1}{8\sqrt{2}R_6}$. The profile of $g_{\text{eff}}$ is similar to that of the $S^1$-compactified E-string theory computed in \cite{ganor1997branes} except for factors of $8=2^3$ interpreted as the cube of the charge of the $U(1)$ hyper. In particular, this geometry has the right property that the effective coupling is constant 
\begin{equation}
    \frac{16\pi^2}{g_{\text{eff}}^2}=\frac{1}{R_6}
\end{equation} 
for $\sigma>\sigma_0$, as expected for the KK reduction of a 6d theory. One can also check that the theory has the right continuous and discrete symmetries. For example, the $\Z_2$ 1-form acts as the ``electric symmetry'' on both the $E_1$ theory (identified with the center of the $SU(2)$ gauge group in IR) and the $U(1)$ theory. One can get the theory for the other polarization by gauging this $\Z_2$, and we get a different version of the $E_1$ theory at the origin with a 2-form $\Z_2$ ``magnetic symmetry'' which can flow to an $SO(3)$ theory, and at $\sigma=\sigma_0$ a $U(1)$ theory with a charge-1 hyper, which has all the electric symmetry being screened and the dual $\Z_2$ being now a subgroup of the magnetic $U(1)$ 2-form symmetry. By changing the holonomy of $SU(2)_F$, one can interpolate between this theory and the 5d $\CN=2$ theory. What we expect to happen is that the value of $\sigma_0$ will decrease, and the theory at origin will be an $SU(2)$ theory at finite coupling, and the Higgs branch will disappear. When we finally make the holonomy vanish, $\sigma_0$ will also vanish, and the two singularities will collide to form the 5d $\CN=2$ theory with coupling constant $\frac{16\pi^2}{g^2}=\frac{1}{R_6}$.\footnote{Notice that this is very similar to the behavior of the E-string theory when a flavor holonomy breaking $E_8$ to $D_7$ is turned on. One difference is that, for the E-string theory, there are two singularities with trivial holonomy which combine into a single $D_8$ singularity when the holonomy is tuned to be ``$-1$'' (i.e.~the one preserving $D_8$), the opposite of what happens for the $(2,0)$ theory. There seems to be another---and more meaningful---difference. At the $D_8$ point, the theory is described by a 5d $SU(2)$ $N_f=8$ theory. But unlike the $(2,0)$ counterpart (given by 5d $\CN=2$ super--Yang--Mills), it doesn't seem to capture all the KK-modes of the E-string theory on a circle with holonomy \cite{EString}. For example, if one goes onto the Higgs branch, unlifted by the holonomy, of the E-string theory, the low energy effective theory is given by massless hypermultiplets. Whereas the massless degrees of freedom match with that of the 5d $SU(2)$ $N_f=8$ on its Higgs branch, the KK modes of these hypermultiplets are absent in the 5d theory. This is in contrast with the 6d $(2,0)$ case, whose low-energy effective theory on the moduli space is still an Abelian $(2,0)$ tensor multiplet, and its KK modes can be captured by instanton bound states (see e.g.~\cite{Kim:2011mv}, and notice that as the effective theory has a UV completion, non-commutative deformation is not needed to have instantons in the Abelian theory).}

\paragraph{The theory $T_{-+}$.} This can be obtained from the KK-reduction of the theory mentioned above. The $E_1$ theory at $\sigma=0$ will split into two $I_1$ dyon singularities, plus an $I_2$ singularity where the low-energy effective theory is $U(1)$ coupled to two hypermultiplets. They are monopoles from the point of view of the UV theory, and can be viewed as the result of colliding the two $I_1$ monopole points when the holonomy in $SU(2)_F$ is tuned from a generic value to $-1$. The Higgs branch at this point is $\C^2/\Z_2$, matching that of the 6d and 5d theory. On the other hand, the $U(1)$ theory at $\sigma=\sigma_0$ will lead to two $I_4$ singularities. The fact that it splits into two is related to the fact that the hyper has charge-2, and can be massless with either $\int_{S^1} A_5=0$ or $\pi$.  The effective theory at one of these $I_4$'s is a 4d $\CN=2$ $U(1)$ theory with a charge-2 hypermultiplet. One can check that the geometry can have a cylindrical end, as the deficit angles of all singularities add up to $2\pi$. The theory now have a $\Z_2$ 0-form and $\Z_2$ 1-form symmetry. The former is a geometric symmetry swapping the  two $I_1$'s and the two $I_4$'s, while leaving the $I_2$ fixed, and the latter is the unscreened electric symmetry at the $I_4$ while being the $\Z_2$ subgroup of the $U(1)$ magnetic symmetry at $I_2$ and $I_1$'s. One can again obtain the theories for other polarizations by gauging either or both $\Z_2$'s with possibly topological terms (discrete theta angles) added. Gauging the $\Z_2$ 0-form symmetry will lead to identifying the two $I_1$'s and two $I_4$'s. What happens to the $I_2$ point is more interesting. The deficit angle can be computed to be $\frac{5\pi}{6}$, and one might wonder why this is not on the list of possible rank-1 SCFTs.\footnote{For IR free theories, they are cusps with the ``local deficit angle'' being $2\pi$. The notion of deficit angle that we use is the asymptotic one.} This is because the new theory at this point is again an $O(2)$ theory with the matter being in the natural two-dimensional representation. Stated in the language of $U(1)\rtimes \Z_2$, the two hypermultiplets are exchanged by the $\Z_2$ as $Q_1\mapsto \tilde{Q}_2$ and $Q_2\mapsto \tilde{Q}_1$. This originates from the fact that the $\Z_2$ can be viewed as a composition of an action that simply swap the two hypers with the Weyl $\Z_2$ that acts as charge conjugation.  On the other hand, gauging the $\Z_2$ 1-form symmetry will change the two $I_4$ into two $I_1$ where the hypermultiplets become charge-one, and the $I_2$ to $I_8$ where the two hypermultiplets now have charge two. The dyonic points will stay as $I_1$ due to the presence of a topological term assigning a $\Z_2$-valued phase given by the mod-2 reduction of the Pontryagin square of $B_2$ in $H^4(M_4,\Z_4)$. 

\paragraph{The theory $T_{+-}$.} There are two dual ways to obtain this theory, first as the KK-reduction of the 5d $\CN=2$ SYM on a circle with a non-trivial $\Z_2^U$ holonomy, and the second from $T_{-+}$ via a change in polarization by gauging both the 0-form and 1-form $\Z_2$ symmetry. The agreement of the two can be viewed as a non-perturbative check of the consistency of the 6d theory. Using the embedding of the $\Z_2^U\subset SU(2)_F$, one can see that now the two quark singularities $I_4$ should collide to form $I_8$. To see this, one can make $R_6$ much smaller than $R_5$. Then the 4d effective theory will be weakly coupled $g^2_{\text{4d}}\sim \frac{R_6}{R_5}$ before turning on the holonomy in $SU(2)_F$. With a holonomy turned on, the two groups of two $I_1$'s will be far way from each other, while the $I_4$ will ``move faster.'' This is because the distance of the quark singularity $I_4$ from the origin is proportional to $\frac{m^2}{g^2}$. Then with mass in the presence of a holonomy $m\sim \frac{1}{R_5}$, we have $\frac{m^2}{g^2}\sim R_6^{-2}$ independent of $R_5$ to the leading order. The configuration with four $I_1$ and an $I_8$ is exactly also what one gets from gauging the $\Z_2$ 0- and 1-form symmetry in the $T_{-+}$ theory. Now, one would refer to the $I_8$ singularity as the result of colliding the two $I_4$ monopole point in two 4d $\CN=2^*$ $SO(3)$ theories, demonstrating the fact that we are now in a different duality frame of the same theory. 

\paragraph{The theory $T_{--}$.} One expect that this theory is dual to the $T_{-+}$ theory, with a $I_2$, two $I_1$ and two $I_4$, except that what have collided are two dyon singularities. This difference is of course just the result of working in a different duality frame. One can end up with yet another duality frame by gauging the $\Z_2$ 0- and 1-form symmetry of $T_{+-}$ with a topological term. Then the $I_2$ singularity is interpreted as colliding two $I_1$ quark singularities of two copies of $SO(3)_-$ theory. Then the duality between $T_{-+}$ and $T_{--}$ is analogous to that in the pure 4d case where $SO(3)_+$ and $SO(3)_-$ are related by $\tau\mapsto \tau+1$. 

We hope by this point it is clear that this web of theories related by compactification with $\Z_2^U$ holonomy, dualities, and discrete gauging is highly constrained, and could be useful to better understand and constrain non-perturbative dynamics of more general 6d $(1,0)$ theories. As an application, we give an argument that the discrete theta angle of 5d $\CN=2$ $SU(2)$ theory cannot be lifted to the 6d $(2,0)$ theory, and 6d $(1,0)$ non-Abelian tensor multiplet doesn't exist as an SCFT.

\subsubsection{On the non-existence of certain 6d theories}

One can ponder on the existence of two closely related cousins of the 6d $A_1$ $(2,0)$ theory. One is the version with a discrete theta angle, another is the 6d $(1,0)$ non-Abelian tensor multiplet.

In 5d, an $SU(2)$ gauge theory with matter only in the adjoint (or other even representations) will have a $\Z_2$ discrete theta angle given by the dual of $\pi_4(SU(2))=H^5(BSU(2))=\Z_2$. In 6d, the dynamics in the presence of the self-dual tensor is mysterious, and if one can think of it as a map to $B^2SU(2)$, there is a natural candidate for a theta angle as $H^6(B^2SU(2))=\Z_2$. However, there is no obvious way of turning on such a discrete theta angle in string theory. It is also not known to us whether there is any argument for the non-existence of it (e.g.~why it conflicts with either supersymmetry or self-duality of the field strength). Here, we show that turning on such a discrete theta angle doesn't lead to a consistent web of theories using the $\Z_2^U$ symmetry.

\paragraph{Discrete theta angles in the 6d $A_1$ $(2,0)$ theory.}   In 5d, turning on the theta angle doesn't alter the Higgs branch $\C^2/\Z_2$, and $\Z_2^U$ still act trivially on the Higgs branch. Its 6d lift $T'$ will also have the same Higgs branch with trivial $\Z_2^U$ action. Then a compactification on an $S^1$ with a non-trivial $\Z_2^U$ holonomy will lead to a theory $T'_-$
with Higgs branch $\C^2/\Z_2$. Near the origin of the 5d Coulomb branch $\R^+$, the theory should look like a 5d $\CN=1$ $SU(2)_{\pi}$ theory. However, the strong coupling limit will then be a $\tilde{E}_1$ theory still without a Higgs branch. As the theta angle would not affect the behavior away from the origin, we expect again to have a $U(1)$ theory with a charge-two hyper. In particular, the Higgs branch cannot be at any place away from the origin, as it would then require a $U(1)$ theory with two hypers. But any hypermultiplet in the theory must have even charges, and having two is not compatible with the UV completeness of the 6d theory. 

This argument shows that if there exists a version of the 6d theory with a non-trivial theta angle, it must be more exotic and should not straightforwardly reduce to the $\CN=2$ $SU(2)_\pi$ theory in 5d. One can also turn the argument around to state that the 5d $SU(2)_\pi$ theory cannot be a direct $S^1$ compactification of any 6d SCFTs.

Another class of postulated theories closely related to the previous one is the 6d non-Abelian $(1,0)$ tensor multiplets. To the best of our knowledge, there is currently no embedding of these in string theory, while on the other hand, also no argument why it cannot be consistent quantum mechanically.\footnote{At the level of supersymmetric transformation of fields, the theory appears to be consistent \cite{Chen:2013wya}.} We will argue here that it cannot be an SCFT in the $A_1$ case. So it is either inconsistent or dependent on a scale. 

\paragraph{6d $A_1$ $(1,0)$ tensor multiplet.} In 6d, one cannot obtain this theory from a mass deformation of the $(2,0)$ theory, but it can be done once we put it on a small circle. The mass is just the holonomy of the $SU(2)_F$ used previously. However, as this parameter space is compact, one cannot really separate the hypermultiplets with the KK modes of the tensor multiplet, whose mass are all proportional to the inverse of $R_6$. Therefore the two theories are only the same in the limit $R_6\rightarrow 0$. Using this, we know that when the Coulomb vev $\sigma\ll 1/R_6$, the geometry looks like that of the 5d $SU(2)$ theory. In the region $\sigma \gg 1/R_6$, the effective coupling will remain constant, given by the only scale $R_6$. Then there must be one additional singularity on the Coulomb branch, and the only option is again $U(1)$ with a charge-2 hypermultiplet. There is already a problem here as this massless hyper cannot find a 6d origin, from which one can conclude that the 6d $(1,0)$ $A_1$ tensor multiplet is either inconsistent or at least non-conformal. Putting this issue aside and assuming that there is a mysterious hyper that can become massless, this moduli space would be exactly the same geometry as the 6d $(2,0)$ theory with a non-trivial $SU(2)_F$ holonomy on $S^1$. The holonomy also cannot be $-1$ as that will have a Higgs branch. And due to the non-trivial holonomy, the geometry after reducing on another circle won't be modular under $\Gamma_0(2)$. Then it appears that there is no web of consistent theories in lower dimensions, suggesting the non-existence of the 6d SCFT to start with.

\subsubsection{Higher group symmetry from continuous isometry}

Unsurprisingly, the theory $T[T^2]$ also has higher group symmetries, similar to the case of $T[S^1]$ studied in Part I \cite{Gukov:2020btk}. Indeed, part of the higher group symmetry comes from the 3-group symmetry of the 5d $T[S^1]$ theory.
In fact, if we ignore the other isometry of $T^2$ except for the isometry of one $S^1$, then the higher-group symmetry in $T[T^2]$ is the same as the dimensional reduction of the higher-group symmetry in $T[S^1]$.

Let us expand the background gauge field $C$ in terms of the basis $d\theta^i$ of $H^*(T^2)$. We will focus on the case of a single Abelian $C$ field:
\begin{equation}\label{eqn:T23d}
    C=B_3+{d\theta^i\over 2\pi} B_2^i+{d\theta^1d\theta^2\over (2\pi)^2} B_1~.
\end{equation}

We will discuss the higher group symmetry using the auxiliary $T^2$ sigma model in 2+1d, with the coupling to background fields as given on the right-hand side of (\ref{eqn:T23d}).

\paragraph{$U(1)^2$ isometry}

Let us begin with $U(1)\times U(1)$ isometry of $T^2$, which is connected to the identity.
Let us turn on background gauge field $A^i$. The coupling is modified to be
\begin{equation}
    B_3+\frac{d\theta^i-A^i}{2\pi}B_2^i+\frac{(d\theta^1-A^1)(d\theta^2-A^2)}{(2\pi)^2}B_1~.
\end{equation}

The condition $\int dC\in2\pi\mathbb{Z}$ implies the following conditions on the background fields:
\begin{align}
    &dB_2^1={dA^2 B_1\over 2\pi},\quad dB_2^2=-{dA^2 B_1\over 2\pi}\cr 
    &dB_3={dA^iB_2^i\over 2\pi}-d(A^1A^2B_1/2(\pi)^2)~.
\end{align}
We note that if we restrict to $A^2=0$, then the higher group reduces to the higher group in $T[S^1]$ \cite{Gukov:2020btk}.

Another way to see the higher-group symmetry is by studying the correlation function of the isometry defects and the generators of the symmetries for $B_3,B_2,B_1$, following the method in \cite{Barkeshli:2022edm,Hsin:2022heo}.
For instance, consider the ``dislocation" domain wall that generates the isometry $\theta^1\rightarrow \theta^1+\varphi^1$, $\theta^2\rightarrow \theta^2+\varphi^2$.
For $C$ to be invariant, $B_3$ must be shifted by
\begin{equation}
B_3\rightarrow B_3-\left(    \frac{1}{2\pi}d\varphi^i B_2^i+\frac{1}{(2\pi)^2}\left(d\varphi^1d\varphi^2 +d\theta^1 d\varphi^2+d\varphi^1 d\theta^2\right)B_1\right)~.
\end{equation}
The first and second term represent an 't Hooft anomaly, since they do not depend on dynamical fields. The third and fourth terms are ``operator-valued anomaly'' that depends on $\varphi^i$ and they represent higher group symmetry.

\subsubsection{Higher group symmetry from large diffeomorphisms}
\label{sec:Ttransform}

The above discussion can be generalized to any diffeomorphisms. Note that the full theory $T[M_d]$ often depends on the metric on $M_d$ together with some additional structures. When a diffeomorphism cannot be represented by an isometry, it is usually only a symmetry of a certain subsector of the theory, and the discussion below would apply to this subsector.

For instance, consider the element in the mapping class group $T=\left(
\begin{array}{cc}
 1    & 1 \\
  0   & 1
\end{array}
\right)$,
which can never be represented by an isometry. Nonetheless, we can still treat it in ways similar to before. In the presence of the background integer cocycle $z$ for such ``symmetry,'' the cohomology on $T^2$ becomes twisted cohomology. In particular, the cocycles $d\theta^1/2\pi,d\theta^2/2\pi$ for the two circles are replaced by the integer cochain $\omega^1$ and the integer cocycle $\omega^2$ that satisfy
\begin{equation}
    d\omega^2=0,\quad d\omega^1=z\omega^2~.
\end{equation}
To see this, we note that under $z\rightarrow z+d\phi$, $\omega^1\rightarrow \omega^1+\phi\omega^2$, for $\phi=1$ the transformation generates the action of $T$.
Then for $C=B_3+\omega^iB_2^i+\omega^1\omega^2 B_1$, 
\begin{equation}
    dC=dB_3-\omega^1 dB_2^1-\omega^2\left(z B_2^1+dB_2^2\right)+\omega^1\omega^2 dB_1~.
\end{equation}
Thus, $dC=0 \pmod{2\pi}$ translates to
\begin{equation}
    dB_3 =0,\quad dB_2^1=0,\quad dB_2^2=-z B_2^1,\quad dB_1=0\quad  \pmod{2\pi}~.
\end{equation}
This describes the background for a semi-direct product of the one-form symmetry and the 0-form $T$ ``symmetry.''

We note that the combination $\omega^i B_2^i$ is invariant under the action of $T$, and therefore it is closed $d(\omega^i B_2^i)=d\omega^i B_2^i-\omega^i dB_2^i=0$. This means that while $\omega^i,B_2^i$ are cochains that obey twisted cocycle conditions, the conditions compensate each other so that the bilinear form $\omega^i B_2^i$ is an ordinary cocycle. A similar discussion applies to the $S$-transformation, which is of finite order and can be realized as an isometry for the specific value of $\tau=i$. See \cite{Bashmakov:2022uek} for a systematic investigation into phenomena associated with special moduli of Riemann surfaces.

\subsection{Higher genus}

We have seen in the previous example of $M_2=T^2$ that one has to keep a compact scalar in order to see the full symmetry of $T[M_2]$. In general, one would get a similar compact scalar for any higher-genus $M_2$, given by the ``holonomy'' of the non-Abelian 2-form $B$-field on $M_2$. Without including it, again it would be not possible to see the $0$-form and $2$-form symmetries of $T[M_2]$ at the level of the moduli space, with the former permuting the naive SCFT ``at the origin'' with the other SCFTs, and the latter related to torsion in the orbifold fundamental group of the moduli space generated by ``large loops.''  Therefore, it would be again necessary to keep this compact scalar for the purpose of making all symmetries of $T[M_2]$ manifest. On the other hand, not keeping it might lead to enhancement of global symmetries that $T[M_2]$ itself should not have, including the well known $SO(8)$ symmetry when $G=SU(2)$ and $M_2$ being the four-holed sphere, the $E_6$ symmetry when $G=SU(3)$ and $M_2$ the three-holed sphere, and the $U(1)_r$ part of the R-symmetry. 

Beside symmetry considerations, there are more reasons to keep the compact scalar, as taking the decompactification limit is often a discontinuous process. For example, to get the correct partition function of the 6d theory from $T[M_2]$, one has to remember that there are other SCFTs for non-zero values of the compact scalar. This problem is more serious compared with the case of 6d $(2,0)$ theory $M_2=T^2$, whose BPS sector is conjectured to be exactly equivalent to 5d $\CN=2$ SYM theory on $S^1$ (see e.g.~\cite{Kim:2011mv} for a test and references therein). Also, one expects to get better behavior under cutting and gluing if one kept this scalar (cf.~Section~\ref{Sec:CutGlue}). 

To better understand the full moduli space of the theory $T[M_2]$ beyond the SCFT limit, we will mainly employ the following two tools:\begin{itemize}
    \item the free Abelian 6d (2,0) theory, whose compactification on any manifold is a free theory;
    \item  the moduli space of $T[M_2\times S^1]$, which has a fibration over the moduli space of $T[M_2]$, with the fiber parametrized by vevs of certain line operators in $T[M_2]$ wrapping the $S^1$.
\end{itemize} 

We will begin the discussion with $g=0$, which is not technically speaking ``higher genus'' but still serves as a good starting point for studying the moduli spaces in more general cases.

\subsubsection{$T[S^2]$}
When we compactify a free $(2,0)$ tensor multiplet on $S^2$ with a partial topological twist, in the massless spectrum, there will be a compact scalar from the holonomy of $B$ and three non-compact scalars. The other two among the five scalars in 6d will become 1-forms on $S^2$, and as there are no harmonic 1-forms, they will only give arise to massive modes after compactifying on $S^2$. Then the total moduli space is 
\begin{equation}
    \CM_{\text{full}}[S^2,U(1)]\simeq \R^3\times S^1.
\end{equation}
The string in 6d will not give rise to line operators in 4d when compactified on $S^2$, and therefore this is also the moduli space after further reduction on $S^1$,
\begin{equation}
    \CM_{\text{full}}[S^2,U(1)]\simeq\CM_{\text{full}}[S^2\times S^1,U(1)]\simeq\C\times \C^*.
\end{equation}
This might be confusing at first sight as the compact scalar should be part of the Coulomb branch, but Coulomb branch usually double its dimension upon compactification to 3d. Being hyper-K\"ahler and invariant after compactification, this moduli space looks like the Higgs branch. In fact, both properties are characteristics of the 4d tensor multiplet, which is exactly what we get after reducing on $S^2$. It is dual to a hypermultiplet, therefore the moduli space is unchanged after the $S^1$ reduction, but with non-standard transformation ($\bf1+\bf3$ as opposed to $\bf2+\bf2$) under $SU(2)_R$. The tensor multiplet reduced on $S^1$ will become a standard 3d vector multiplet, whose moduli space (a.k.a.~Coulomb branch) is indeed $\C\times \C^*$, where the compact direction is due to the dual photon.

In the non-Abelian case, one again has 
\begin{equation}
    \CM[S^2]=\CM[S^2\times S^1]
\end{equation}
as complex manifolds, equating the tensor branch in 4d and Coulomb branch in 3d,
and one can identify this moduli space by analyzing the 5d SYM theory on $S^2$. In general, one should view the 5d theory as a relative theory coupled to a bulk 6d theory, as we are choosing a polarization on $S^2$, not $S^1$. We will first choose the group to be $U(N)$. Then via string duality, one can turn the system into $N$ D4-branes suspended between two D6-branes, and argue that the moduli space is  
\begin{equation}
    \CM[S^2]=\CM[S^2\times S^1]\simeq\text{moduli space of $N$ SU(2) monopoles},
\end{equation}
by relating it to the NS5-D3 system of \cite{Hanany:1996ie} upon dimensional reduction. This indeed becomes $\C\times \C^*$ for $N=1$. The exact metric of the moduli spaces will depend on the size and shape of the $S^2$ (and $S^2\times S^1$). In particular, we can map the system to D3-D1 in type IIB with $\tau$ dependent on the relative size of $S^2$ and $S^1$, as well as how much $S^2$ is twisted after going around $S^1$. Then the metric on the moduli space of the monopoles will also depend on these as it depends on $\tau$. This aspect is analogous to the system studied in \cite{Seiberg:1996nz} except that we have here a different 4d lift of the 3d vector multiplet.  

Now we can get the answer for $SU(N)$ by decoupling the ``center-of-mass'' direction. This is not a unique process, and the ambiguity exactly reflects the choice of a polarization. For $N=2$, we have the $U(2)$ moduli space being 
\begin{equation}
    \CM[S^2,U(2)]\simeq \R^3\times (S^1\times \tilde\CM_{\rm AH})/\Z_2
\end{equation}
where $\tilde{\CM}_{\rm AH}$ is the double cover of the Atiyah--Hitchin manifold. $\tilde{\CM}_{\rm AH}$ has a $\Z_2$ symmetry, and decoupling the center-of-mass motion for $\CM[S^2,U(2)]$ will leave either $\tilde{\CM}_{\rm AH}$ or $\CM_{\rm AH}=\tilde{\CM}_{\rm AH}/\Z_2$.\footnote{In this case, making this choice is equivalent to asking whether we want a natural inclusion from $\CM[S^2,SU(2)]$ to $\CM[S^2,U(2)]$, or a projection from  $\CM[S^2,U(2)]$ to $\CM[S^2,SU(2)]$.} The former is the moduli space of the theory with $L^{(0)}=\Z_2$ which has a $\Z_2$ symmetry, while the latter is the moduli space of the theory with $L^{(2)}=\Z_2$ which has a $\Z_2$ 2-form symmetry. 

There are more polarizations on $S^2\times S^1$, and some of them don't come from these on $S^2$, meaning that they give rise to $T[S^2\times S^1]$ which cannot be obtained by compactifying an absolute $T[S^2]$. However, there are not more moduli spaces, as the other choices will only differ from the two mentioned above by the spectrum of line operators in 3d which is not detected by the moduli space. The two moduli spaces are also the Coulomb branches of the 3d $\CN=4$ SO(3) and SU(2) theory respectively.

Working in complex geometry, it better to view the monopole moduli space as the Kostant reduction of $T^*G_\C$ by the two $G_\C$ action on both the left and right. In this way, one can easily generalize it to other groups and polarizations. 

Although, from the point of view of the 4d theory $T[S^2]$, this moduli space is the tensor branch (and Coulomb branch from the 3d perspective), interestingly, it appears in the study of how the Higgs branch behaves under cutting-and-gluing. It was first expected that cutting and gluing will encounter some difficulties in low genera \cite{Moore:2011ee}, but it was later shown in \cite{Braverman:2017ofm} that the procedure will actually work, and one has to associate exactly the monopole moduli space to $S^2$.

\subsubsection{Higher-genus Riemann surfaces}

When $M_2=\Sigma$ is a genus-$g$ Riemann surface, the theory $T[\Sigma]$ from the Abelian 6d theory will have one tensor multiplet and $g$ vector multiplets. The moduli space is therefore
\begin{equation}
    \CM_{\text{full}}[\Sigma,U(1)]\simeq \C^g\times \R^3\times S^1\simeq \C^{g+1}\times\C^*.
\end{equation}
After compactifying on $S^1$, we will have an additional factor of $(T^2)^{g}$, which can be viewed as a $g$-dimensional Abelian variety specified by the period matrix of $\Sigma$. However, we will work with another complex structure and express the moduli space as
\begin{equation}
    \CM_{\text{full}}[\Sigma\times S^1,U(1)]=\C\times(\C^*)^{2g+1}.
\end{equation}
This now has the right dimension to be hyper-K\"ahler, and it is easy to check that for $g=0$ and $1$ it reduces to the cases discussed previously. One remark is that for $g=1$, this way of dividing up the moduli space is different from the usual gauge perspective, where one would say that the Coulomb branch is $\C^*\times T^2$ and the Higgs branch is $\C^2$. The latter perspective, although useful (and used previously to understand the moduli space for various $T[T^2]$), is not very natural from the point of view of the physical system in the sense that the $SU(2)_R$ subgroup of R-symmetry in the latter description is not geometric, and the four scalars in the hypermultiplet in fact have different origins. 

In the non-Abelian case, one would naively expect
\begin{equation}
    \CM_{\text{full}}[\Sigma,U(1)]\stackrel{?}{=}(\frak{t}_{\mathbb{C}}^{g+1}\times \mathbb{T}_{\mathbb{C}})/W,
\end{equation}
and
\begin{equation}\label{NaiveSS1}
    \CM_{\text{full}}[\Sigma\times S^1,U(1)]\stackrel{?}{=}(\frak{t}_{\mathbb{C}}\times\mathbb{T}_{\mathbb{C}}^{2g+1})/W.
\end{equation}
Although this matches the moduli spaces for both $g=0$ and 1, it is too simple a description for higher genera and can only be correct away from fixed points of $W$. One expects that the two moduli spaces are still fibrations over the base $(\frak{t}_{\mathbb{C}}\times\mathbb{T}_{\mathbb{C}})/W$, but, at the fixed points of $W$, as there will be enhancement of gauge symmetry, the fiber would respectively be larger.

For example, the moduli space of flat $G_{\mathbb{C}}$-connections (or equivalently, the moduli space of $G_{\mathbb{C}}$-Higgs bundles) on $\Sigma$ is part of $\CM_{\text{full}}[\Sigma\times S^1]$ and is expected to be the fiber over central elements in $(\frak{t}_{\mathbb{C}}\times\mathbb{T}_{\mathbb{C}})/W$. Taking this into account, we propose that the full moduli space $\CM_{\text{full}}[\Sigma\times S^1]$ is given by the total space of fibration over $(\frak{t}_{\mathbb{C}}\times\mathbb{T}_{\mathbb{C}})/W$, with the fiber over $(a,g)$ being the moduli space of $G'_{\C}$-Higgs bundles $\CM_H(\Sigma,G')$ where $G'$ is the subgroup of $G$ stabilizing both $a$ and $g$. The statement is similar for $\CM_{\text{full}}[\Sigma]$, but the fibers will not be the total moduli space of Higgs bundles, but just the base of the Hitchin fibration. In other words, the fibration of $\CM_{\text{full}}[\Sigma\times S^1]$ over $(\frak{t}_{\mathbb{C}}\times\mathbb{T}_{\mathbb{C}})/W$ factors through $\CM_{\text{full}}[\Sigma]$. The geometry of the former moduli space is arguably more interesting and will be what we focus on later.

\begin{figure}[htb!]
    \centering
\includegraphics[width=0.6\linewidth]{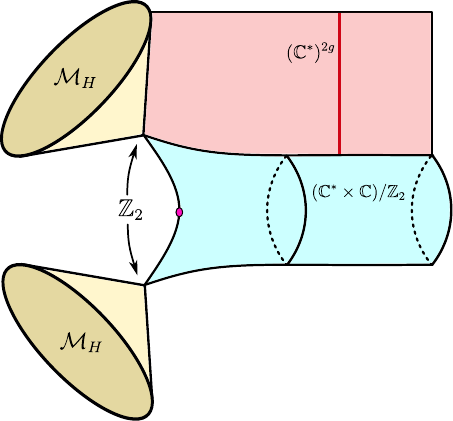}
    \caption{The full moduli space $\CM_{\rm full}$ of $T[\Sigma\times S^1,SU(2)]$ for a particular polarization. The space is a fibration over the tensor/Higgs/Kibble branch $(\C^*\times \C)/\Z_2$ (green). The generic fiber (red) is $(\C^*)^{2g}$ which can be viewed as the moduli space of Abelian Higgs bundle on $\Sigma$. Over the two singularities, the fiber becomes the moduli space $\CM_H$ of $SL(2,\C)$-Higgs bundles (yellow). This, from the point of view of the SCFT, can be identified with the Coulomb branch. From the Abelian locus on $\CM_H$, one can deform onto the mixed branch (pink), which actually contains the Higgs/Kibble branch. The exact global form of the moduli space $\CM_H$ will depend on the choice of polarization. Also for another set of polarizations, the moduli space will be a $\Z_2$ quotient, with a new singularity arising from the fixed point (purple). The moduli space for $T[\Sigma,SU(2)]$ is a similar fibration but the fibers are now only the base of the Hitchin fibration of the moduli space of (Abelian) Higgs bundles. The generic fiber thus becomes $\C^g$.}
    \label{fig:FullModuli}
\end{figure}

For the $A_1$ case, $\CM_{\text{full}}[\Sigma\times S^1]$ is illustrated in Figure~\ref{fig:FullModuli}. Over the 2-dimensional base, the generic fibers are $2g$-dimensional $(\C^*)^{2g}$, but there are two points with fiber being the $(6g-6)$-dimensional moduli space of Higgs bundles.

\begin{figure}[htb!]
    \centering
    \includegraphics[width=0.8\linewidth]{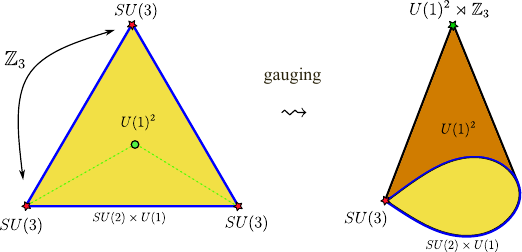}
    \caption{An illustration of the moduli space of $T[T^2]$ for the $A_2$ case. The triangle on the left represents $\mathbb{T}/W\simeq (S^1\times S^1)/S_3$ with the three vertices being the $\Z_3$ center. There, the low-energy description is given by a 4d $\CN=4$ $SU(3)$ theory. One can deform from it to get either a family of $SU(2)\times U(1)$ theories, or a two-parameter family of $U(1)^2$ Abelian theories. The $\Z_3$ symmetry of the theory acts by a $\frac{2\pi}{3}$-rotation, with a single fixed point denoted by the green dot. After gauging this $\Z_3$ symmetry, the moduli space can be identified with the fundamental domain below the green dashed lines, with the two dashed lines identified. This is illustrated on the right. The green dot becomes a theory with gauge group $U(1)^2\rtimes \Z_3$.}
    \label{fig:SU3Mod}
\end{figure}

For higher rank, the tensor branch will have higher dimensions, and it is easier to view itself as a fibration over the Weyl alcove $\mathbb{T}/W$. The partial order for stabilizers on $\mathbb{T}/W$ gives the hierarchy of singularities. The $A_2$ case is illustrated in Figure~\ref{fig:SU3Mod}. 

\subsubsection{Relations with other moduli spaces}\label{sec:OtherMod}

Inside $\CM_{\text{full}}[\Sigma\times S^1]$, there are more familiar moduli spaces. For example, as was mentioned, it contains moduli spaces of $G'_\C$-Higgs bundles for various stabilizer subgroups $G'\subset G$. Also, there is a bigger moduli space of flat $G_\C$-connections on $S^1\times\Sigma$ sitting inside. This is because, for the theory $T[\Sigma \times T^2]$, part of the solution to the BPS equation is the flatness of the connection on $\Sigma \times S^1$. As the low-energy theory of $T[\Sigma\times S^1]$ is described by a sigma model to $\CM_{\text{full}}[\Sigma\times S^1]$, this moduli space is also expected to be that of the 2d theory $T[\Sigma\times T^2]$, at least away from singularities. (This is in contrast with the 4d theory, whose low-energy description is in terms of a gauge theory, as opposed to a sigma model, and new directions of the moduli space would open up once compactified on a circle.)  This moduli space of flat connections is also a fibration, but more naturally over $\mathbb{T}_{\mathbb{C}}/W$ parametrizing the holonomy over the $S^1$, which represents a central element in $\pi_1(\Sigma\times S^1)$. The fibers are again the moduli spaces of flat $G'_\C$-connections on $\Sigma$ with $G'$ being the stabilizer of the chosen element in  $\mathbb{T}_{\mathbb{C}}/W$. The one missing direction can be obtained by looking at the full BPS equations, which involves another complex adjoint scalar. In other words, one has two equations when compactifying the 5d super--Yang--Mills theory on a three-manifold $M_3$, 
\begin{eqnarray}
    \mathcal{F}_{\mathcal{A}}=0,\nonumber\\
    d_{\mathcal A}\sigma=0.\label{FullBPS}
\end{eqnarray}
The first is the flatness for the complex $G_\C$-connection $\mathcal{A}$ \cite{Dimofte:2010tz}, while the second equation demands that the complex adjoint scalar $\sigma$ is covariantly constant. $\sigma$ comes from two of the five real-valued adjoint scalars in the 5d (and also 6d) theory that transforms trivially under the $SO(3)\subset$ $SO(5)_R$ subgroup of the R-symmetry used for the topological twist on $M_3$. In the case of $M_3=\Sigma\times S^1$ that we will focus on later, these equations can be broken down into several ones for various components of the fields (see \cite{Yonekura:2013mya,xie2014moduli} as well as \cite{Bershadsky:1995vm,kapustin2007electric,gukov2007gaugetheoryramificationgeometric,Gaiotto:2009hg} for related work).

On the other hand, there is another moduli space that doesn't miss any directions locally, but instead misses some global features. This is the moduli space for the class-$S$ theory, which one expects to obtain when zooming into the most singular points on   $(\frak{t}_{\mathbb{C}}\times\mathbb{T}_{\mathbb{C}})/W$. One such point is the ``origin'' $(0,e)$. After zooming in, the geometry will look like a fibration over $(\frak{t}_{\mathbb{C}}\times\frak{t}_{\mathbb{C}})/W$. The fibers are still ``non-linear,'' but one can also take the local geometry near the origin to get the moduli space of the ``3d class-$S$ theory'' analyzed in e.g.~\cite{Benini:2010uu}.

For the $A_1$ case, there are two singular points, with the local geometry being the same. The base is $\C^2/\Z_2$, and the fiber at the origin is the $(6g-6)$-dimensional moduli space of SL$(2,\C)$-Higgs bundles, while the fiber away from the origin will be a $2g$-dimensional $(\C^*)^{2g}$. The structure agrees with what we know about the moduli spaces of class-$S$ theories, whose Coulomb branches are the moduli spaces of SL$(2,\C)$-Higgs bundles and have a one-dimensional Higgs branch when $\Sigma$ is without punctures. What seems to be a discrepancy is that the Higgs branch of the class-$S$ theory is expected to be $\C^2/D_{g+1}$ as opposed to just $\C^2/\Z_2$ \cite{HN2010}. We now make some remarks relevant to this point.
\begin{enumerate}
    \item In the full theory $T[\Sigma]$ or $T[\Sigma\times S^1]$, it is not straightforward to define the Higgs branch in the moduli space. As $U(1)_r$ (or $SU(2)_N$ for the 3d theory) is not present in the UV and only emerges in the IR SCFT, one cannot simply define the Higgs branch as part of the moduli space where the relevant part of R-symmetry acts trivially. From the point of view of fields, it is actually more natural to refer to the base $\C^2/\Z_2$ as the tensor branch.
    \item On the other hand, for class-$S$ theories associated with closed Riemann surfaces, the Higgs branches themselves are unusual as they are contained in the mixed branch. (In contrast, the mixed branch intersects the Coulomb branch on a $2g$-dimensional sub-locus.) Physically, the gauge symmetry will not be completely broken there, and sometimes the name ``Kibble branch'' is used to refer to such a subspace of the moduli space 
    (e.g.~in \cite{HN2010}). 
    \item Combining the previous two points, it is expected that the identification of the Higgs branch in the IR as a subspace of the mixed branch, if this makes sense, is a non-trivial problem. It is, in the best case, a particular (multi-)section of the fibration away from the singularity, and one shouldn't naively expect that it is just isomorphic to the base.
    \item Another complication is that the description \eqref{NaiveSS1} fails at the singularity due to the enhancement of gauge symmetry to a non-Abelian one with new directions of the moduli space opening up, which has to be taken into account when determining the singularity on the Higgs branch. In particular, as the Coulomb branch of the 3d SCFT has an Sp$(g)$ global symmetry \cite{Benini:2010uu}, the Higgs branch should have at least $g$ deformations (from the structure of the 3d $\CN=4$ current multiplet), already illustrating that the singularity has to be worse then $\C^2/\Z_2$.
    \item In this case, the 3d SCFT moduli space is in a sense a ``counter-example'' to one of the  usual statements of symplectic duality that deformations on the Coulomb/Higgs match exactly the global symmetries on the other side, as $Sp(g)$ is of rank-$g$ while the $D_{g+1}$ singularity has $g+1$ deformations. The mismatch in this case can be understood as coming from discrete symmetries. The mirror of the 3d SCFT is given by a 3d $\CN=4$ $SO(3)$ theory with $g$ adjoint hypermultiplets,\footnote{It is usually assumed that the gauge group of the mirror theory is $SU(2)$. However, to match the index computation in \cite{HN2010}, which implies that the singularity is of type $\C^2/D_{g+1}$, the gauge group should be $SO(3)$ instead (see Lemma 6.9 of \cite{braverman2016towards} for a computation of the type of the singularity). We thank H.~Nakajima for very helpful discussion regarding this point and other ones. } This theory has a $\Z_2$ topological/magnetic symmetry acting on the Coulomb branch, disallowing a deformation that is not $\Z_2$ invariant. This can be understood very concretely in the $g=0$ case, which is already non-trivial. In this case, there is no symmetry on the Higgs branch of the pure $SO(3)$ theory, but the Coulomb branch is a $D_1$ surface (also identified with the double cover of the Atiyah--Hitchin manifold that we have encountered previously) with a one-parameter family of hyper-K\"ahlar deformations parametrized by $\lambda$ given by \cite{dancer1994family}
    \begin{equation}
    x^2-zy-1-\lambda y=0.
    \end{equation} 
    However, as $\lambda$ is odd under the $\Z_2$ symmetry, this deformation cannot be turned on if we want to preserve the $\Z_2$ symmetry of the theory. A similar scenario occurs for the other choice of the polarization where the gauge group is $SU(2)$. There, with $g$ adjoints, the singularity on the Coulomb branch is $D_{2g}$ \cite{braverman2016towards}. The Higgs branch still only has the same rank-$g$ symmetry, smaller than the naive $2g$ deformation parameter for the Coulomb branch. However, the $\Z_2$ electric 1-form symmetry of the theory tells us that part of the singularity should not be resolved. In other words, to make sure that the resolved geometry is compatible with the 1-form symmetry, it should be itself a $\Z_2$ quotient, which, in contrast to the complete resolution of the $D_{2g}$ singularity, should have a remaining singularity. This is similar to a phenomenon that we will discussed in the context of the ``4d symplectic duality'' in Section~\ref{sec:SCFTVOA}.
\end{enumerate}

If one goes beyond the $A_n$ cases, there are other locally maximal points similar to the case of $T[T^2]$ discussed previously. The SCFTs there are expected to be again of class-$S$ type but with different gauge groups. For example, in the $E_8$ case one can get a class-$S$ theory associated with Spin(16)$/\Z_2$. However, the massive spectrum of the full theory associated with $E_8$, even near the Spin(16)$/\Z_2$ point, is expected to be different from that associated with the $T[\Sigma]$ obtained from the 6d Spin(16)$/\Z_2$ theory.

The moduli space $\CM_{\text{full}}[\Sigma\times S^1]$ is also related to the moduli space of multiplicative Higgs bundles studied in \cite{Elliott:2018yqm}, which is motivated by the compactification of the 5d $\CN=2$ gauge theory on $\Sigma$ when it is Calabi--Yau ($T^2$ if compact). For more general Riemann surfaces, the two moduli spaces appear to be in general different. 

\subsubsection{Symmetry of $\CM_{\text{full}}$ and polarizations}

The symmetries of the full moduli space come from 0-form and 1-form symmetries of the 4d theory. The action of the latter is only on the Hitchin fibers and is relatively well understood (including how to gauge these to access different polarizations). Here we will focus on the former part. These will act on the base $(\frak{t}_{\mathbb{C}}\times\mathbb{T}_{\mathbb{C}})/W$ as translations of $\mathbb{T}_{\mathbb{C}}$ by central elements. This is exactly the same as the case of $g=1$, and the structures of the new singularities that emerge after quotienting by the symmetry are identical to that case analyzed previously. What is different now is that the theory at the singularities are no longer 4d $\CN=4$ theories, but instead $\CN=2$ ones. We now explain this in some examples.

\paragraph{The $A_1$ case.} The moduli space is a fibration over $(\C\times\C^*)/\Z_2$ before the quotient, which has a $\Z_2$ symmetry permuting the two SCFT points at $(0,0)$ and $(0,\pi)$ where the low-energy effective theories are both the class-$S$ theory of $A_1$ type. After dividing by the $\Z_2$, a new singularity at $(0,\pi/2)\sim (0,-\pi/2)$ emerges. This theory, again similar to the $g=1$ case, will have a disconnected gauge group which is now of rank-$g$ given by $U(1)^g\rtimes \Z_2$. This is an $\CN=2$ theory, as there is only a single (as opposed to $g$, which would be the adjoint) hypermultiplet, on which the $\Z_2$ acts by a minus sign.

\paragraph{The $A_{N-1}$ and the general case.} For $A_{N-1}$, similarly, the moduli space has $N$ SCFT points where the low-energy effective theory is the corresponding class-$S$ SCFT. These singularities are permuted by a $\Z_{N}$ symmetry and, after gauging it, a new singularity will appear where the local theory is given by a 4d $\CN=2$ $U(1)^{g(N-1)}\rtimes{\Z_{N}}$ theory with $N-1$ hypermultiplets. The hypermultiplets can be combined into a diagonal traceless $N\times N$ matrix, with $\Z_N$ acting as permutations of the diagonal entries. This is the adjoint in the $U(1)^{N-1}\rtimes{\Z_{N}}$ sense, but not for $U(1)^{g(N-1)}\rtimes{\Z_{N}}$, and hence the theory is not enhancing to $\CN=4$. Furthermore, there are in general new types of singularities, where the ``gauge group'' can be $S(U(k)^{N/k})^{g}\rtimes \Z_{N/k}$. This should be interpreted as obtained from $N/k$ copies of class-$S$ theories associated with $U(k)$ by gauging the $\Z_{N/k}$ part of the permutation symmetry and then decoupling the ``center-of-mass motion.'' For other types of Lie algebras, there will be similar singularities. For example, in the $g=1$ case, for $E_6$, we have seen new singularities with the gauge group being $(A_2)^3 \rtimes \Z_3$. It is natural to expect that the higher-genus generalization is given by gauging the $\Z_3$ part of the permutation symmetry acting on three copies of class-$S$ theories of $A_2$ type.

Before moving on to the next topic, we remark that, if one zooms in on a particular SCFT point, then indeed part of the polarization will not be relevant, and the problem reduces to choosing a maximal isotropic subgroup in $H^1(\Sigma,D)$. However, the geometry of the full moduli space depends crucially on the choice on the ``forgotten part'' of polarization concerning the degree-0 and -2 pieces of the cohomology, and there will be new SCFTs arising for certain choices. 

\subsection{Riemann surfaces with punctures}\label{Sec:CutGlue}

We now continue the investigation into the moduli space of $T[\Sigma]$ theories but now allow $n$ boundary components or punctures on $\Sigma$. 

Although we will be relatively brief, focusing on the moduli space, this setup should be the ideal playground for exploring many aspects of $T[M_d]$ when the internal manifold $M_d$ has a boundary or contains a defect, as discussed in Section~\ref{sec:PolBoundary}. 

\subsubsection{The combined moduli space}

One can again use the set of BPS equations \eqref{FullBPS} on $\Sigma\times S^1$. The ``local version'' (i.e.~taking the SCFT limit) of this has been carried out in the literature \cite{yonekura2014supersymmetric,xie2014moduli}, and our focus will be on the global aspects as well as the new phenomena that arise when $\Sigma$ is no longer closed. 

The second equation $d_{\mathcal A}\sigma=0$ demands that $\sigma$ is covariantly constant and generally has no solution when $\CA$ is irreducible. For $\CA_3$, the component of the complexified gauge field along the $S^1$ direction, there is a similar statement. When $\CA$ is Abelian, $\sigma$ and $\CA_3$ are constant. Then it is tempting to conclude that, similar to the non-punctured case, there is a fibration of the moduli space over the hyper-K\"ahler tensor branch $(\frak{t}_\C \times \mathbb{T}_{\C})/W$, with fiber being the moduli space of parabolic Higgs bundles, which generically is $\mathbb{T}_{\C}^{2g}$. However, there is now a problem about whether we should actually regard the constant $\sigma$ and $\CA_3$ as moduli. This is because the values of these fields on the boundary, which fix their values in the bulk, should be viewed as deformation parameters as opposed to moduli. 

This is similar to the more familiar story of the ``ramification parameters'' that describe the singular behavior of the gauge and Higgs fields around the punctures  \cite{gukov2007gaugetheoryramificationgeometric}. They parametrize deformations of the moduli space of Higgs bundles which can be identified with the Coulomb branch of class-$S$ theory on $S^1$ \cite{Gaiotto:2009hg}, as opposed to themselves being part of the moduli.

From the perspective of ``blowing up the punctures,'' i.e.~the interpretation of $T[M_D\backslash M_d]$ as defects in $T[M_d]$, the theory $T[\Sigma\times S^1]$ is a codimension-1 (or 2) defect in $n$ copies of $T[T^2]$ (or $T[S^1]$). In general, many deformation parameters of $T[M_D]$ can be lifted to scalar fields in the bulk theory $T[M_d\times S^{D-d-1}]$, and they are part of the moduli space of the bulk theory, not of the defect theory $T[M_D\backslash M_d]$. In the present case, the ``bulk theory'' $T[\bigcup_n T^2]=T[T^2]^{\otimes n}$ has moduli parametrized by $n$ copies of $(\frak{t}_\C\times\frak{t}_\C\times \mathbb{T}_\C)/W$. And we should interpret these as deformations of $T[\Sigma\times S^1]$. 
For each puncture, there is a $\frak{t}\times\frak{t}\times \mathbb{T}$ that can be identified with the three ramification parameters $(\alpha,\beta,\gamma)$ of the Higgs bundle.\footnote{Notice that this construction naturally leads to ``tame ramifications'' (a.k.a~regular punctures). It would be interesting to generalized it to include wild ramifications or irregular punctures.} To see this, one can first reduce the system on $S^1$, leading to a 5d $\cN=2$ gauge theory on $\Sigma$ with $n$ boundary circles. Three of the five real adjoint scalars are covariantly constant and form an $SU(2)_R$ triplet---they become $\sigma$ and the imaginary part of $\CA_3$ once compactified on another $S^1$ to connect with \eqref{FullBPS}---while the other two become the Higgs field $\phi$ which is a one-form on $\Sigma$. Once these two scalars acquire vevs in the 5d theory on a boundary $S^1$,  there is a singularity from the point of view of $\Sigma$ when we shrink the $S^1$,
\begin{equation}
    \phi\sim \beta\frac{dr}{r} -\gamma d\theta +\text{regular},
\end{equation}
with the two scalars identified with the tangent and normal components of the 1-form $\phi$. Note that, in this subsection, we assume that the sizes of the boundary $S^1$'s are small. Otherwise, there will be KK-modes of these bulk fields $\beta(\theta)$ and $\gamma(\theta)$ that are also parameters in $T[\Sigma\times S^1]$.

Similarly, the holonomy of the gauge field along a boundary component becomes a singularity for the gauge field on $\Sigma$
\begin{equation}
    A\sim \alpha d\theta+\text{regular}.
\end{equation}
This deformation parameter is indeed only present for $T[\Sigma\times S^1]$ but not for $T[\Sigma]$. To talk about $A_3$ in \eqref{FullBPS}, one needs to actually consider $T[\Sigma\times T^2]$, which requires two more compact scalars to specify the boundary condition at each $T^3$ boundary. From the point of view of the $T[T^3]$ bulk theory, one is from the holonomy of the gauge field in $T[T^2]$ on this additional circle and gives the boundary value for $A_3$. The holonomy of the dual gauge field of $T[T^2]$---or, equivalently, the vev of the ``dual photon'' in $T[T^3]$---becomes a ``quantum parameter'' $\eta\in \mathbb{T}^\vee$ \cite{gukov2007gaugetheoryramificationgeometric}, which describes the B-field in the sigma-model description of $T[\Sigma\times T^2]$.\footnote{In the gauge-theory approach for the geometric Langlands correspondence \cite{kapustin2007electric,gukov2007gaugetheoryramificationgeometric}, $\sigma$ and $\CA_3$ are often turned off both over $\Sigma$ and at the punctures, leaving only the quartet $(\alpha,\beta,\gamma,\eta)\in \mathbb{T}\times \frak{t}\times \frak{t}\times \mathbb{T}^\vee$. The discussion in the present work can be viewed as the starting point of an ``untruncated version'' of the geometric Langlands program, which would interesting to explore further.} 

Another way of understanding these new scalar parameters is that these are background 1- and 2-form fields in the 4d theory $T[\Sigma]$, which can only become scalars after compactification. From the 5d bulk $T[S^1]^{\otimes n}$  point of view, these come from the restriction of the gauge fields and its magnetic duals to the defect. Such background field leads to the interesting phenomenon that, although the moduli space of $T[\Sigma\times S^1]$ and $T[\Sigma\times T^2]$ are almost identical, due to the fact that the former is described by a sigma model at low energy, the two theories actually have different spaces of deformation parameters, as there are non-scalar background fields in $T[\Sigma\times S^1]$ at low energy.

The above analysis tells us that the moduli space of $T[\Sigma\times S^1]$ is no longer like the case with $\Sigma$ being closed. We still have a base $\left((\frak{t}_\C\times\frak{t}_\C\times \mathbb{T}_\C)/W\right)^{\times n}$, but it is now the space of parameters from the point of view of $T[\Sigma\times S^1]$. However, since it is the moduli space for $T[\partial \Sigma \times S^1]$, we will still refer to the total space of the fibration as the ``combined moduli space.'' When $n>1$, the moduli space for the defect theory is empty over a generic point of this parameter space, as there is no solution to the generic boundary values since they conflict with each other. The true base, $\CM'_{\rm base}$, is inside the fiber product 
\begin{equation}
    \CM'_{\rm base}\subset \CM_{\rm base}:=\bigtimes^n_{\frak{t}^3/W}(\frak{t}^5\times \mathbb{T})/W\subset \CM_{T[\partial \Sigma \times S^1]}.
\end{equation} 
When all of the three scalars are zero, this recovers $n$ copies of $(\frak{t}^2\times \mathbb{T})/W$, (e.g.~one triple $(\alpha,\beta,\gamma)$ for each puncture). Over  $p\in\CM_{\rm base}$, the fiber $\CM_{T[\Sigma\times S^1]}(p)$ contains the moduli space of parabolic Higgs bundles with gauge group compatible with the three scalars. The reason for the ``true base'' being generally inside $\CM_{\rm base}$ is that $\CM_{T[\Sigma\times S^1]}(p)$ can still be empty in certain cases. This can happen, for example, when either $\frak g$ is not semisimple or broken by the three scalars to a subalgebra that is not semisimple. 

The moduli space of parabolic Higgs bundles can be identified with the Coulomb branch on $\R^3\times S^1$ in the low-energy description of $T[\Sigma]$. What about the Higgs branch associated with the punctures? We believe that they are generally absent in the moduli space for the full theory but are emergent after we flow to the IR SCFT. Although our setup is different from the usual one involving codimension-2 defect in the 6d SCFT, turning on the ramification parameters seems to have the same effect in the low-energy gauge theory description, reproducing the same Coulomb branch. Therefore, it is reasonable to expect that the IR SCFT at singular points of the moduli space is the usual class-$S$ theory associated with $\Sigma$ with punctures, which would imply that there are in general emergent Higgs branches (see e.g.~\cite{Gaiotto:2009we,Chacaltana:2010ks,Benini:2010uu,Moore:2011ee,Chacaltana:2012zy} for discussion on various aspects, especially the Higgs branch, of class-$S$ theories). It would be very interesting to study this RG flow in greater detail and to better understand how new directions of the moduli space emerge in the IR. 

We now give some examples where one can be more explicit about the structure of the combined moduli space of $T[\Sigma\times S^1]$.

\subsubsection{The rank-1 Abelian case}

The adjoint action is trivial, and the values of the three globally constant scalars parametrize an $\R^3$, which combines with the $n$ triples of ramification parameters to give $\CM_{\rm base}\simeq\R^3\times (\R\times \C^*)^n\subset (\C\times \C\times \C^*)^{n}$. This is the locus where the values of the three scalars are the same across all boundary components. 

To get the true base, one needs to remove a copy of $\R\times \C^*$, as the ramification parameters at one of the punctures are determined by those at the other punctures. Then $\CM'_{\rm base}\simeq\R^3\times (\R\times \C^*)^{n-1}$ is the subspace of the moduli space of $T[T^2]^{\otimes n}$ that gives consistent boundary conditions on $\partial\Sigma$. To summarize, the nested inclusion $\CM'_{\rm base}\subset \CM_{\rm base}\subset\CM_{T[\partial \Sigma]}$ in this case is given by
\begin{equation}
    \R^3\times (\R\times \C^*)^{n-1}\subset \R^3\times (\R\times \C^*)^n\subset (\C\times \C\times \C^*)^{n}.
\end{equation}

The combined moduli space is a fibration over $\CM'_{\rm base}$, with fiber $\CM_{T[\Sigma\times S^1]}(p)\simeq (\C^*)^{2g}$ identified with the Coulomb branch of a rank-$g$ Abelian gauge theory. Therefore, one has
\begin{equation}
    \CM_{\rm combined}\simeq \R^3\times (\R\times \C^*)^{n-1} \times (\C^*)^{2g}.
\end{equation}

We see the $n$ copies of 0-form $U(1)$ symmetry in $T[T^2]$, which comes from a 1-form symmetry of $T[S^1]$, acting on $\CM_{\rm base}$, with the subgroup $S(U(1)^n)$ preserving $\CM'_{\rm base}$, as well as the $U(1)^{2g}$ 1-form symmetry of the theory $T[\Sigma]$ (which now becomes a 0-form symmetry for $T[\Sigma\times S^1]$) acting on the fiber. Although no choice of polarization is needed in the present case, these symmetries are respectively analogues of $L^\vee$, $(L/\partial (L_\delta))^\vee$, and $L_{\delta,\mathrm{ker}}^\vee$ that featured previously in Section~\ref{sec:SymmetryBoundary}.  In particular, the first two fit in a short exact sequence,
\begin{equation}
    S(U(1)^n)\to U(1)^n\to U(1),
\end{equation}
with the last $U(1)$ being a $(-1)$-form symmetry on the boundary descending from the bulk 0-form symmetry, making it a counterpart of $(\partial (L_\delta))^\vee$. 

The background gauge field for this $(-1)$-form symmetry is a $U(1)$-valued parameter in $T[\Sigma\times S^1]$ that measures the combined holonomy of the $n$ punctures. This can also be interpreted as the ``mismatch'' of the holonomies, as a flat $U(1)$ connection on $\Sigma$ can only exist if this parameter is the identity.  When it is non-trivial, the theory will have no supersymmetric ground states and will exhibit spontaneous symmetry breaking in the infrared. However, as the theory is free, the dynamics is not affected. At the level of action in the 3d $\CN=2$ superspace, this can be thought of as adding a linear superpotential for a free chiral multiplet to shift the vacuum energy, where the chiral is a combination of the Lagrange multipliers enforcing the Dirichlet boundary condition at various boundary components.

\subsubsection{The $A_1$ case}

Now we have $D=\Z_2$ and do need to choose a polarization. We will start with a ``maximal one''---in the sense that the moduli spaces are the largest so that one can construct the moduli spaces for other polarizations as quotients. 

This is a polarization with $L=H^0(\partial\Sigma=\amalg_n S^1,D=\Z_2)$ and $L^{(0)}_\delta=H^0(\Sigma,D)$. With this choice, we expect to have an $L^\vee\simeq \Z_2^n$ 1-form symmetry in the $T[\partial\Sigma]$ bulk, with a $(\partial (L_\delta))^\vee\simeq \Z_2$ quotient descending to a 0-form symmetry of the defect/boundary theory $T[\Sigma]$. 
These symmetries fit into a short exact sequence,
\begin{equation}
    \Z_2^{n-1}\to \Z_2^n\to \Z_2,
\end{equation}
with the leftmost term $(L/\partial (L_\delta))^\vee\simeq\Z_2^{n-1}$ being the symmetry shared by the bulk and boundary theory. There is an additional choice of $L^{(1)}_\delta\subset H^1(\Sigma)$ that determines the 1-form symmetry of $T[\Sigma]$, which is always abstractly $L_{\delta,\mathrm{ker}}^\vee\simeq \Z_2^g$ in the present case. The symmetries of $T[\Sigma\times S^1]$ (and of the $T[\partial\Sigma\times S^1]$ bulk) can be obtained from these by reduction, and one of our tasks is to understand their relation with the combined moduli space.

With such a polarization, the space $\CM_{\rm base}$ is almost a $\Z_2$ quotient of the rank-1 abelian case, given by a fibration with the fiber being $(S^1 \times\R^2 )^{\times n}$ over a generic point on the base $\R^3/\Z_2$. However, over the origin, the fiber is instead $\left(\frac{\R^2\times S^1}{\Z_2}\right)^{\times n}$. 

To get $\CM'_{\rm base}$, similar to the abelian case, one will need to remove a copy of $S^1 \times\R^2$ in the generic fiber. Now the combined moduli space is a fibration over $\CM'_{\rm base}$, with fiber $\CM_{T[\Sigma\times S^1]}(p)$ being generically $(\C^*)^{2g}$. Over special points with the three scalars vanishing, one obtains in the fiber the hyper-Kähler moduli space of parabolic SL$(2,\C)$-Higgs bundles on $\Sigma$, with ramification parameters given by the remaining $\left(\frac{S^1 \times\R^2}{\Z_2}\right)^{\times n}$ (i.e.~one triple for each puncture). 

For various $U(1)$ symmetries in the Abelian case, only a $\Z_2$ subgroup for each survives. Each copy of the $\Z_2^n$ 0-form symmetry acting on $\CM_{\rm base}$ has fixed loci, given by the special ramification parameter $(\frac14,0,0)$ at the corresponding puncture.\footnote{From the point of view of the moduli space of $T[T^2]$, this is exactly the $\theta=\frac{\pi}{2}$ case discussed previously except that, in the convention here, we have divided by $2\pi$ such that $\int A=2\pi\alpha=\frac{\pi}{2}$ around the puncture.}
The $\Z_2$ 0-form symmetry of $T[T^2]$ descends to a $\Z_2$ $(-1)$-form symmetry for $T[\Sigma\times S^1]$, which is basically a $\Z_2$-valued parameter labeling the two moduli spaces at $\alpha$ and $\alpha+1/2$.  

Gauging this bulk-boundary pair of $\Z_2$ results in a $\Z_2$ 2-form symmetry shared by $T[T^2]$ and $T[\Sigma\times S^1]$. The old ``$\alpha$-circle'' is a double cover of the new one, where $\alpha$ and $\alpha+1/2$ are identified. Over a point $\tilde \alpha$ on the new circle, the moduli space of $T[\Sigma\times S^1,\tilde\alpha]$ is a disjoint union of the moduli spaces of $T[\Sigma\times S^1,\alpha]$ and $T[\Sigma\times S^1,\alpha+1/2]$, and this is indeed characteristic of a 3d theory with a $\Z_2$ 2-form symmetry. The expectation value of the symmetry generator, which is now a point operator, distinguishes the two worlds. 

\begin{figure}[htb!]
    \centering
    \includegraphics[width=0.85\linewidth]{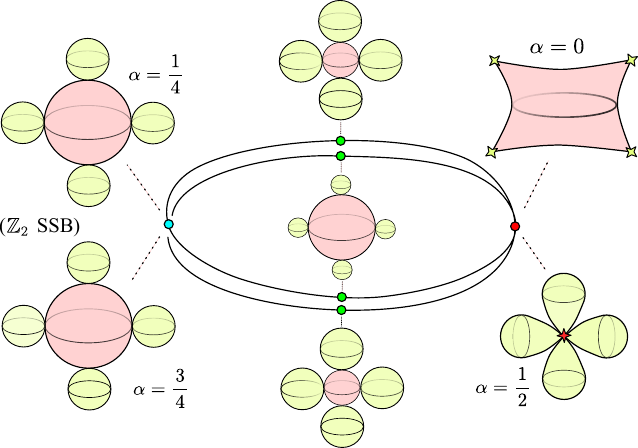}
    \caption{An illustration of the combined moduli space obtained from compactifying the 6d $(2,0)$ theory of type $A_1$ on $\Sigma\times S^1$ with $\Sigma=T^2\backslash D^2$ being a one-holed torus. Only a circle parametrized by $\alpha\in[0,1)$ is taken from the base $\CM_{T[\partial \Sigma\times S^1]}$, while only a real slice (known as the nilpotent cone) of the fiber $\CM_{T[\Sigma\times S^1]}$ is shown. The polarization is the ``maximal'' one on $\Sigma\times S^1$ (which is actually different from the reduction of the maximal one on $\Sigma$ used in the main text). The base has a $\Z_2$ symmetry, which descends to a $\Z_2$ $(-1)$-form symmetry of the fiber, meaning that they appear in pairs parametrized by $\alpha$ and $\alpha+\frac12$. After gauging the $\Z_2$, one obtains a $\Z_2$ 2-form symmetry shared by the 4d bulk and the 3d boundary theory. The new base circle, parametrized by $\tilde \alpha\in [0,\frac12)$, is double-covered by the old one, and the fiber at each $\tilde\alpha$ now has two components, characteristic of 3d theory with a 2-form symmetry. At $\tilde\alpha=0$, the two moduli spaces both has nodal (aka $A_1$ or double point) singularities marked by stars. Another special point is $\tilde\alpha=\frac14,$ for which the two components are isomorphic. The bulk theory at this point at low energy has an $O(2)$ gauge group, while the boundary theory will be in a phase with spontaneously broken $\Z_2$ symmetry. }
    \label{fig:CombinedMod}
\end{figure}

There is a special point at $\tilde\alpha=\frac14$ with all other parameters being zero. In the bulk, $T[T^2]$ at this point, as we have explained previously, is described by a 4d $\CN=4$ $O(2)$ theory at low energy. The $\Z_2$ gauge field obeys Dirichlet boundary condition at $T[\Sigma\times S^1]$, and the boundary theory should be in a phase with $\Z_2$ spontaneous symmetry breaking at low energy. Indeed, the two components of the moduli spaces are isomorphic, as $\alpha=\frac 14$ and $\frac 34$ are related by the affine Weyl group symmetry, and they are the two vacua (though each being in fact a collection of vacua) related by the $\Z_2$. This $\tilde\alpha$-family of moduli spaces is illustrated in Figure~\ref{fig:CombinedMod} for the case of $\Sigma=T^2\backslash D^2$.

In addition, there is a $\Z_2^{n-1}$ 0-form symmetry---the part of $\Z_2^n$ that doesn't descend---shared between $T[\Sigma\times S^1]$ and $T[\partial \Sigma\times S^1]$. This is a rather interesting symmetry, although it is only present when $n>1$. In the bulk, this action shifts even numbers of $\alpha_i$'s, relating points in the parameter spaces that have isomorphic fibers. Notice that it acts as symmetry of the combined moduli space and is generally not a symmetry on the individual fibers. Instead, it gives isomorphisms between different fibers, which become a genuine action at special locus with a pair of $\alpha$'s being both $\frac14$. To see this more explicitly, consider the case of $\Sigma$ being the four-holed sphere.\footnote{Another interesting case for general $\frak{g}$ is the three-holed sphere, with this symmetry being related to a ``center 1-form'' symmetry of the (equivariant) Verlinde algebra \cite{Gukov:2015sna,Gukov:2016lki,Gukov:2021swm}. For the $A_1$ case, although the symmetry is still non-trivial at the level of quantum field theories, the moduli space $\CM_{T[\Sigma\times S^1]}$ will be a point and becomes a slightly degenerate example for illustrating phenomena arising from the symmetry.} When all the other parameters except for the four $\alpha$'s are zero, $\CM_{T[\Sigma\times S^1]}(\alpha_{1,2,3,4})$ is an elliptic surface with a singular fiber of type $D_4$ (see \cite{Andersen:2016hoj,Huang:2024mtw} for more detailed descriptions of the geometry of this moduli space), with the volumes of the four exceptional divisors, assuming none of the $\alpha_i$ is ``too large,'' given by
\begin{align}
h_1&=|1-\alpha_1-\alpha_2-\alpha_3-\alpha_4|,\nonumber \\
h_2&=|\alpha_1+\alpha_2-\alpha_3-\alpha_4|,\nonumber \\
h_3&=|\alpha_1-\alpha_2+\alpha_3-\alpha_4|,\nonumber \\
h_4&=|\alpha_1-\alpha_2-\alpha_3+\alpha_4|.  
\end{align}
The shift of parameters, after the Weyl group action, is given by sending a pair of $\alpha_i$ to $\frac{1}{2}-\alpha_i$, which indeed permutes these $h_i$'s, swapping the affine node with one of the ordinary nodes. For example, the action on $\alpha_1$ and $\alpha_2$ will lead to
\begin{equation}
    h_1\leftrightarrow h_2,\quad h_3\leftrightarrow h_4,
\end{equation}
which demonstrates that the identification of the two moduli spaces with parameters related by the $\Z_2$ action is indeed non-trivial. Now assume $\alpha_1=\alpha_2=\frac 14$, and the moduli space at this point now enjoys a genuine $\Z_2$ symmetry, compatible with the equalities of the volumes, $h_1=h_2$ and $h_3=h_4$. When $\alpha_3$ is also $\frac14$, all the four exceptional divisors will have the same volume,
\begin{equation}
    h_1=h_2=h_3=h_4=\left|\frac14-\alpha_4\right|,
\end{equation}
and the moduli space $\CM_{T[\Sigma\times S^1]}$ with this parameter will have a $\Z_2\times\Z_2$ symmetry, with the action on homology compatible with the outer-automorphism of the affine $D_4$ Dynkin diagram, $\Z_2\times \Z_2\simeq\mathrm{Aut}(\tilde D_4)/\mathrm{Aut}(D_4)$.

How to think about the action of the symmetry on the moduli space? This symmetry can be viewed as creating a pair of punctures with central monodromy and merging each of them with one of the four punctures. As points in the moduli space are parametrized by the conjugacy classes of the SL$(2,\C)$ holonomies along three chosen cycles (subject to one algebraic relation), we can explicitly find how different conjugacy classes are flipped (i.e.~multiplied by the center of SL$(2,\C)$) under the action. With one choice of the cycles, an element of $\Z_2\times \Z_2$ will flip two holonomies. Very concretely, the moduli space in one of its complex structures is the following algebraic surface \cite{goldman2009trace},
\begin{equation}
    x^2+y^2+z^2-xyz=4-4\cos^22\pi\alpha_4,
\end{equation}
and the three generators of $\Z_2\times \Z_2$ flip even numbers of $x,y$ and $z$.

For the most special case with all $\alpha_i$'s being $\frac 14$, the full $\Z_2^3$ will be a symmetry of $T[\Sigma\times S^1]$. The moduli space, which is now described by the Cayley cubic, will have four singularities as $h_i=0$ for all $i$. It resembles the fiber at $\alpha=0$ in Figure~\ref{fig:CombinedMod}. The action on the moduli space permutes the four singularities through the quotient $\Z_2\times \Z_2$. One $\Z_2$ factor, which corresponds to flipping all $\alpha$'s, acts trivially on the moduli space, as it leaves all holonomies invariant. However, it still acts non-trivially on the theory. One way to confirm this is by observing that it genuinely acts on non-flat connections (e.g.~it flips the holonomy along a circle with radius $r>0$ centered at a puncture).

There is another pair of bulk-boundary symmetries---a $\Z_2$ 0-form symmetry of $T[\Sigma\times S^1]$ descending from bulk 1-form symmetries. This is a close cousin of the 0-form--$(-1)$-form pair, as both can be lifted to a 0-form symmetry of $T[\Sigma]$ descending from a 1-form symmetry in $T[S^1]$. However, This $\Z_2$ also seems to act trivially on the moduli space, as the action, even on non-flat connections, coincides with a gauge transformation, and it would interesting to understand whether the action on the physical theory $T[\Sigma\times S^1]$ is trivial or not. 

Lastly, the 4d theory $T[\Sigma]$ also has a $L_{\delta,\mathrm{ker}}^\vee\simeq\Z_2^g$ 1-form symmetry by itself. Gauging such a symmetry (possibly with a topological term) amounts to changing the $L_{\delta}^{(1)}$ part of the polarization. To give a concrete example, consider again $\Sigma=T^2\backslash\{{\rm pt}\}$, then the moduli space $\CM_{T[\Sigma\times S^1]}$ over a generic value of $\alpha$ can be the left or right side of Figure~\ref{fig:D4} for two different choices of polarization on $\Sigma\times S^1$. None of them actually comes from a polarization on $\Sigma$, which can only realize the ``intermediate'' spaces with only one $\Z_2$ quotiented out. The three choices of $\Z_2\subset \Z_2\times\Z_2$ label the three classes of polarizations on $\Sigma$.   

This concludes the discussion for case of the 6d $(2,0)$ theory of type $A_1$. This interplay between symmetries of the combined moduli space and polarizations is a general phenomenon, which we investigate next.

\subsubsection{Symmetries and polarizations}

We will use the machinery developed in Section~\ref{sec:PolBoundary}, specialized to the pair $(\Sigma,\partial\Sigma)$ and again focusing on pure polarizations. Now that $\Sigma$ has boundaries, a pure polarization would involve a choice
\begin{equation}
    L_i\subset H_*(S^1_i,D)
\end{equation}
for each boundary component, and 
\begin{equation}
    L_\delta\subset H_*(\Sigma,\partial\Sigma ;D).
\end{equation}
The degree-1 piece, $L_\delta^{(1)}\subset H_1(\Sigma,\partial\Sigma ;D)=D^{2g}$, is subject to the maximal isotropic condition identical to the case without boundaries. The effect of different choices, if the 6d theory is a $(2,0)$ SCFT, can be again detected via the moduli space of (parabolic) Higgs bundles. 

The compatibility condition requires that $L_\delta^{(0)}\subset H_2(\Sigma,\partial\Sigma;D)\simeq H^0(\Sigma,D)=D$, under the map $\partial$ to $H_1(\partial \Sigma)=D^n$, coincides with $\mathrm{im}(\partial)\cap L$. As $\partial$ is the ``diagonal map,'' to satisfy this condition, we must have $L_\delta^{(0)} \subset L_i^{(0)}$ for each $i$ and, furthermore,  $\bigcap_{i=1}^nL_i^{(0)}=L_\delta^{(0)}$. (Notice that these are well-defined relations once we identify $L_\delta^{(0)}$ and $L_i^{(0)}$'s as subgroups of $D$.)

From the viewpoint of charged operators, this equality can be understood in the following way. 
The 5d theory $T[S^1_i]$ has line operators labeled by $L^{(0)}_i$  (e.g.~can be Wilson lines in the low-energy gauge theory description) obtained from wrapping the 6d strings on $S^1$. The configuration of a string in 6d wrapping $\Sigma$ requires that a ``diagonal line'' in the bulk theory $T[S^1]^{\otimes n}$ should be able to end in $T[\Sigma]$. The diagonal line is labeled by elements in $\bigcap_{i=1}^nL_i^{(0)}$, while the possible end points in $T[\Sigma]$ are classified by $L_\delta^{(0)}$. The compatibility condition ensures that the two agrees.

The choice of $L_i$ determines the global form of the $i$-th factor $T[S^1]$ of the bulk theory. For the moduli space of $T[S^1]$ (part of the parameter space for $T[\Sigma]$), in the maximally supersymmetric case, this amounts to choosing a quotient of $(\frak{t}_\C\times\frak{t}_\C\times \mathbb{T}_\C)/W$ by the center symmetry of the $\mathbb{T}$ factor. The bulk theory $T[S^1]^{\otimes n}$ would have $\big(L^{(0)}\big)^\vee=\bigoplus_i\big(L_i^{(0)}\big)^\vee$ 1-form symmetry and $D^n/L^{(0)}= \bigoplus_iD/L_i^{(0)}$ 2-form symmetry. The former will become a 0-form symmetry of $T[T^2]^{\otimes n}$ and act on the moduli space, while the latter is related to singularities on the moduli space. 

For two points $p$ and $p'=g\cdot p$ related by $g\in \big(L^{(0)}\big)^\vee$, the fiber at $p'$ in the combined moduli space is not necessarily the same as that over $p$, and it is labeled by the image of $g$ under $\big(L^{(0)}\big)^\vee\rightarrow \big(L_\delta^{(0)}\big)^\vee$---a manifestation of the $(-1)$-form symmetry. The consistency of this picture requires the compatibility condition, with the first part (1a) ensuring different components of the moduli spaces of $T[\Sigma\times S^1]$ having a place to live in the combined moduli space, while (1b) guarantees that there are enough labels for distinct fibers in the entire $\big(L^{(0)}\big)^\vee$-orbit. If (1a) is violated, there will be more ``sectors''  in $T[\Sigma\times S^1]$---or, more precisely, versions of the theory labeled by different discrete theta angles---than what the bulk theory can accommodate. On the other hand, if (1b) is violated, there could be fibers labeled by more general elements in $\big(\im(\partial)\big)^\vee$, giving rise to more discrete theta angles of the boundary theory than it actually possesses.

Part of the short exact sequence \eqref{SESBulk}, in the present case, becomes
\begin{equation}
    \big(L^{(0)}/\partial(L_\delta)\big)^\vee\to \big(L^{(0)}\big)^\vee\rightarrow \big(L_\delta^{(0)}\big)^\vee,
\end{equation}
and the subgroup $\big(L^{(0)}/\partial(L_\delta)\big)^\vee$---given by the kernel of $\big(L_i^{(0)}\big)^\vee\rightarrow \big(L_\delta^{(0)}\big)^\vee$ dual to the inclusion $L_\delta^{(0)}\subset L^{(0)}$---also acts as 0-form symmetries on the 3d theory $T[\Sigma\times S^1]$ and its moduli space $\CM_{T[\Sigma\times S^1]}$

In fact, as this subgroup consists of symmetries shared between the bulk and boundary theory, it acts on the \textit{combined} moduli space, relating points on the base with fibers being necessarily isomorphic. For a fixed point $p$ of $g\in \big(L^{(0)}/\partial(L_\delta)\big)^\vee$ on $\CM_{\rm base}$, the fiber over it, $\CM_{T[\Sigma\times S^1]}(p)$, can have a non-trivial action under $g$.  

Understanding the symmetries allows one to get the moduli space for a given pure polarization from the ``maximal one'' with $L_i^{(0)}=L_\delta^{(0)} =D$. The latter is a geometric polarization given by $W_3$ being a handlebody and $W_2$ a collection of two-disks on its surface. For $\CM_{\rm base}$, one simply takes the quotient of the action of a subgroups of the $D^n$ 1-form symmetry, given by $L_i'^{(1)}\subset D$ for the new polarization. The fibers in the orbit of the ``traceless'' part $D^{n-1}$ are identified, but are in general classified by the image under $D^n\to D$. This leads to, after the quotient, components of the fiber labeled by $\bigvee_i L_i'^{(1)}$ (sum/coproduct of all  $L_i'^{(1)}\subset D$), whose dual, $D/\bigcap_{i} L_i'^{(0)}$, is the group of 2-form symmetries of $T[\Sigma\times S^1]$ after gauging the 0-form--$(-1)$-form pair. Notice that the compatibility condition is automatically satisfied after this procedure, as $L_\delta'^{(0)}=\bigcap_{i} L_i'^{(0)}$, and a smaller (or larger) than maximal $L_\delta$ can never be constructed in this way, as it requires gauging symmetries only on the defect $T[\Sigma]$ (or only in the bulk $T[\partial \Sigma]$) in a way incompatible with the boundary conditions for the background gauge fields. 

However, for the part of the polarization concerning $L'_{\delta,{\rm ker}}\subset H^*(\Sigma,D)$---or in $H^*(\Sigma\times S^1,D)$ as we are more than often actually using the polarization on $\Sigma\times S^1$---one can change it by gauging with no such obstructions, as it involves entirely symmetries that exist only on the boundary. The maximal choice is to take $L_{\delta,{\rm ker}}=H^1(\Sigma,D)\simeq \ker(\partial)\cap H^{1}(\Sigma\times S^1,D)$, so that $\CM_{T[\Sigma\times S^1]}$ has the most 0-form symmetry, with the action of the entire $L_{\delta,{\rm ker}}^\vee\simeq H^1(\Sigma,D)$---now viewed as a summand of $H^2(\Sigma\times S^1)$ via the Künneth decomposition. To change it to $L'_{\delta,{\rm ker}}$, one simply gauges $L_{\delta,{\rm ker}}'^{(2)}\subset H^2(\Sigma\times S^1,D)$, which is a subgroup of $L_{\delta,{\rm ker}}^\vee$ and can be thought of as the dual of $L_{\delta,{\rm ker}}/L'^{(1)}_{\delta,{\rm ker}}$, the part of $L_{\delta,{\rm ker}}$ that is not in $L'_{\delta,{\rm ker}}$.

There are numerous intriguing directions to explore concerning both the physics and geometry of the full theory $T[\Sigma]$ and their IR SCFTs, with or without punctures. However, delving into them here would take us beyond the central focus of this work---symmetry. We will discuss only one such topic next, but we hope to return to some of the other interesting questions---such as how $\CM_{\rm combined}$ behaves under cutting and gluing of $\Sigma$---in the future. 

\subsection{On the SCFT/VOA correspondence and ``4d symplectic duality''}\label{sec:SCFTVOA}

If one chooses a particular SCFT point on the full moduli space of $T[\Sigma]$, then only a part of the polarization enters the physics of the IR SCFT, which has been discussed in detail in \cite{Tachikawa:2013hya}. If one then only focuses on the spectrum of local operators, naively global aspects would not matter at all, as they only affect the spectrum of line operators in the SCFT. However, the point operators do not form an isolated part of the theory, but instead constitute an organic component that interacts with the rest of theory. Therefore, one expects to see some shadow of the global data even when examining the local operators in the SCFT. In this subsection, we discuss an instance of this related to the SCFT/VOA correspondence \cite{Beem:2013sza}. 

This correspondence states that, given a 4d $\cN=2$ SCFT $\CT$, there is a protected subsector of local operators that form a vertex operator algebra (VOA), $\chi_\CT$. As the construction is oblivious to global aspects, for different versions of $\CT$ associated with different polarizations, one always has the same $\chi_\CT$. However, it was proposed in \cite{Fredrickson:2017yka} that there is an intriguing connection between the category of $\chi_\CT$-modules with the geometry of the Coulomb branch $\CM_\CT$ of $\CT$ on $\R^3\times S^1$, the latter of which does depend on the polarization. See \cite{Fredrickson:2017jcf,Dedushenko:2018bpp} for a related observation made at around the same time, and \cite{Shan:2023xtw, Shan:2024yas} for some interesting later developments. This relation was termed ``categorical SCFT/VOA correspondence'' in \cite{Fredrickson:2017yka} as it can be viewed as a relation between the category of $\C^*$-equivariant coherent sheaves on $\CM_\CT$ and that of $\chi_\CT$-modules. Two of the authors of the present paper have been referring to it as ``4d symplectic duality'' over the years, since it is about a connection between the Coulomb branch and the Higgs branch (for which $\chi_\CT$ can be regarded as a ``chiral quantization''), analogous to the usual symplectic duality relating the Coulomb and Higgs branch of the same 3d SCFT. The phrase ``mirror symmetry'' is sometimes used in related contexts (e.g.~in \cite{Shan:2023xtw}), which is also partly justified, since the correspondence is actually related to the geometric Langlands correspondence and one should in fact use the SYZ mirror space $\tilde\CM_\CT$ as argued in \cite{Fredrickson:2017yka}.

In many known examples (e.g.~class-$S$ theories), the difference between $\tilde\CM_\CT$ and $\CM_\CT$ is exactly global and via a change of polarization. Therefore one can say that $\chi_\CT$, which is formed out of local operators, ``knows'' about the choice of polarization via its category of modules. So which moduli space is the ``right'' one? 

For class-$S$ theories, the family of Coulomb branches $\CM_{T[\Sigma\times S^1,\CP]}$ parametrized by a polarization $\CP\in\Pol(\Sigma\times S^1)$ can be constructed from the ``maximal'' one with $\CP_0$---a pure polarization that has $L^{(1)}= H^1(\Sigma\times S^1,D)$---by quotienting a subgroup of the $(L^\vee)^{(2)}=H^2(\Sigma\times S^1,D)$ 0-form symmetry acting on the Coulomb branch.\footnote{Notice that many $\CP$'s would have the same Coulomb branch as their spectra of point operators are identical. As a consequence, for the purpose of this subsection, we can restrict to just pure polarizations. We will also assume here that $\Sigma$ is closed without boundaries or punctures, but the classification of pure polarization, as well as how the moduli space behaves upon changing the polarization, is very similar to the discussion about the $L_{\delta,{\rm ker}}$ part in the previous subsection.} The theory $T[\Sigma\times S^1,\CP_0]$ has the largest 0-form symmetries and the most point operators, but no 1-form symmetries descending from the 6d 2-form symmetry and the least line operators. On the opposite end, the polarization $\tilde\CP_0$ with $L^{(1)}=0$ has the smallest moduli space, the least populated spectrum of local operators, but the most densely populated spectrum of line operators. ``Halfway'' between them, there are polarizations that give theories that are absolute already in 4d labeled by $L\subset H^*(\Sigma,D)$.

As the modules of $\chi_\CT$ can be reproduced by surface operators in $\CT$ and, for the $S^1$-compactified (and potentially also topologically twisted) theory, line operators that can end on a special boundary condition (see e.g.~\cite{Costello:2018fnz,Jeong:2019pzg,Dedushenko:2023cvd} for more details of the latter perspective), it might be tempting to think that there is a family of categories $\CC_{\CP}$ of $\chi_\CT$-modules also parametrized by $\CP$ with a partial order, with $\CC_{\CP_0}$ having the least amount of simple objects, while $\CC_{\tilde\CP_0}$ the most. However, this clashes with the observation that $\CM_{\CP_0}$ has the most components of $\C^*$-fixed loci, while  $\CM_{\tilde\CP_0}$ has the least.

In fact, one should not expect that all of the line operators can end on the special boundary condition where $\chi_\CT$ lives. Indeed, as the boundary condition is formed out of capping off the $S^1$ with a disk, we have a polarization (again assuming it is pure) whose  boundary part is given by $L_\delta\subset H^*(\Sigma\times D^2)$. Line operators that can end are labeled by $L^{(2)}$ that are in the image $\partial (L_\delta)$. However, there is a single generator for the image of $\partial$ in $H^2(\Sigma\times S^1,D)$ (or dually the image of $\partial: \,H_2(\Sigma\times D^2,\Sigma\times S^1;D)\to H_1(\Sigma\times S^1,D)$) given by the Poincaré dual of $S^1$. This describes the ``axion string'' in the 4d $T[\Sigma]$ theory (i.e.~directly coming from a string excitation in 6d) which is indeed a surface operator and can wrap $D^2$ to give rise to a line operator in the 3d theory ending on the boundary. However, this is an operator that only exist in the full theory but not in the spectrum of the IR SCFT.

In other words, none of the charged line operators that can appear in the 4d SCFT with any polarization can actually be in the image! From this observation, one might conclude that the right 4d $T[\Sigma]$ theory to use for 4d symplectic duality is the one with no charged line operators at all and no 1-form symmetry,\footnote{To be precise, here we are only talking about symmetries obtained from the reduction of the 2-form symmetry in 6d and charges with regard to such symmetries. The theory can of course has other symmetries and objects charged under those.} which unfortunately doesn't exist. With different polarizations, the 4d theory will always allow half of the charges, and the hypothetical theory with no charged line operators is not expected to be a physical theory. However, there exists a 3d theory $T[\Sigma\times S^1]$ with no charged point operator and no 0-form symmetry, and as far as the moduli space is concerned, it looks exactly like the $S^1$-compactification of the unphysical theory with no 1-form symmetry. This 3d theory is the one given by the polarization $\tilde\CP_0$. The analysis above tells us that the most natural moduli space to use for 4d symplectic duality is  $\CM_{T[\Sigma\times S^1,\tilde\CP_0]}$ associated with the polarization $\tilde\CP_0$.

Alternatively, one can state the duality in a polarization-independent way as follows. Choose a polarization for $\CT=T_{\rm SCFT}[\Sigma]$ (the IR SCFT of $T[\Sigma]$ at the ``origin'') with some 1-form symmetry, which leads to a moduli space $\CM_{T[\Sigma\times S^1,\CP']}$ with the action of a 0-form symmetry. Then what should be related to objects in $\chi_\CT$--mod are not arbitrary $\C^*$-equivariant coherent sheaves on $\CM$, but those that are invariant under the 0-form symmetry. This is similar to the subtlety that we encountered in the 3d version of symplectic duality in Section~\ref{sec:OtherMod}, where discrete symmetries are also needed to make the statement of duality precise.

\begin{figure}[htb!]
    \centering
    \includegraphics[width=0.85\linewidth]{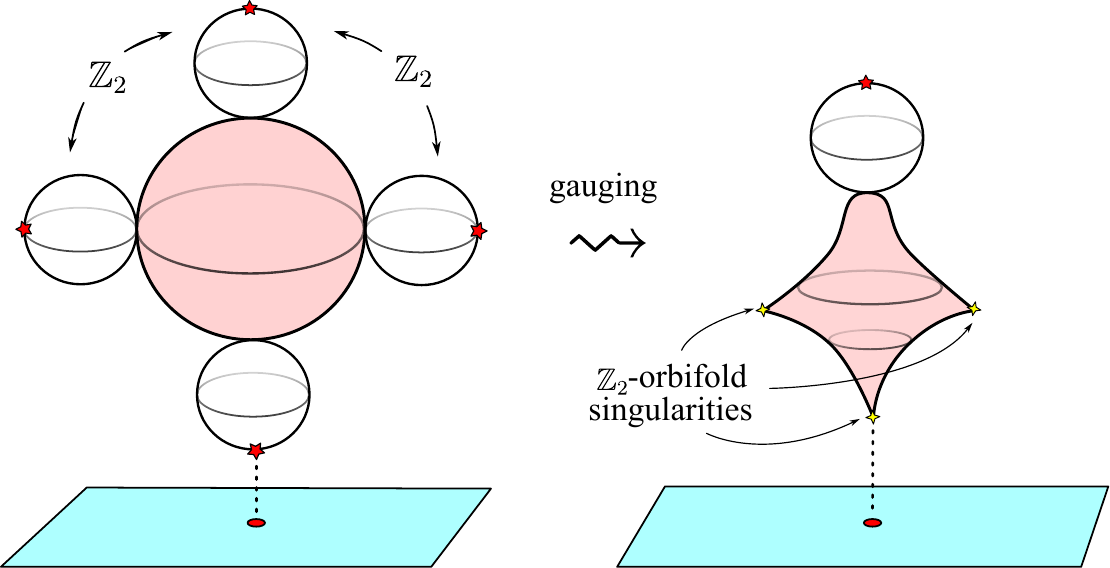}
    \caption{An illustration of the moduli spaces $\CM_{T[\Sigma\times S^1,\CP_0]}$ (left) and $\CM_{T[\Sigma\times S^1,\tilde\CP_0]}$ (right) in the $A_1$ case with $\Sigma=T^2$. A real mass $\alpha$ preserving half of the supersymmetry is turned on to resolve some singularities. This corresponds to adding a puncture on $\Sigma$ and taking the fiber at a generic value of $\alpha$ in $\CM_{\rm combined}$ (green dot in Figure~\ref{fig:CombinedMod}). The two moduli spaces are both elliptic fibrations over $\C^2$, with the only singular fiber---the nilpotent cone---at the origin. On the left, the moduli space is smooth with the singular fiber of type $D_4$ ($I_0^*$ in Kodaira's classification). The fixed loci of the $\C^*$ Hitchin action, colored by red, have five components consisting of four isolated points (red stars) and the central $\mathbb{P}^1$. The action of the $\Z_2\times \Z_2$ symmetry of the moduli space relates the four isolated fixed points. After quotienting by the $\Z_2\times\Z_2$ symmetry, one obtains the moduli space on the right, with three nodal singularities ($\C^2/\Z_2$ orbifold points) marked by yellow stars. On this moduli space, the $\C^*$-fixed loci have only two components, agreeing with the number of simple modules of the corresponding VOA. }
    \label{fig:D4}
\end{figure}

The distinguished role played by $\tilde\CP_0$ is in fact compatible with existing ``experimental evidence.'' For example, in the 4d $\CN=4$ theory associated with $\Sigma=T^2$ for $\frak g=A_1$, the VOA $\chi_\CT$ only has two simple modules, but the biggest Coulomb branch labeled by $\CP_0$, which is the SL$(2,\C)$-Higgs bundle moduli space, has five connected components of fixed loci after mass deformation. The nilpotent cone is of Kodaira type $I_0^*$ and the fixed points are illustrated in the left of Figure~\ref{fig:D4}. However, the Coulomb branch for $\tilde\CP_0$ is the PSL$(2,\C)$ moduli space, and is obtained by modding out the $\Z_2\times \Z_2$ symmetry acting on the Coulomb branch, which relates the four isolated points (see also~\cite{Gukov:2010sw} for a detailed description of the geometry of the two spaces). After that, there are only two components of fixed points, compatible with the number of simple modules of $\chi_\CT$ \cite{Beem:2017ooy,Pan:2024hcz}.

Is there then any role played by the other polarizations in 4d symplectic duality? For a line operator that cannot end, as we discussed in Section~\ref{sec:boundary}, the problem is that some flux from a would-be end point has no where to go. As a consequence, the end point is itself attached to another line operator living on the boundary. The simplest option is to choose this line operator to be the same as the bulk one. In the 3d setup for the SCFT/VOA correspondence, the bulk theory can be topologically twisted, and the bulk line operators are topological, so one gets on the boundary a point operator attached to a topological line. From the VOA point of view, this looks like a twisted module. Therefore, if one wants to realize $\CC_\CP$ for other polarizations so that $\CM_{T[\Sigma\times S^1,\CP]}$ has a counterpart in the 4d symplectic duality for any $\CP$, one possibility is to expand $\CC_{\tilde\CP_0}\simeq \chi_\CT$--mod to include certain twisted modules. It remains to be seen whether this can actually be done and lead to something non-trivial (i.e.~not a product of $\CC_{\tilde\CP_0}$ and the category of line operators).

\subsection{$T[S^2]$ revisited: Higher group symmetry}

One thing that was hard to see in the analysis of the moduli spaces above is the higher group symmetry, which we now study for $\Sigma=S^2$.

We turn on background gauge field $A$ for the $SO(3)$ isometry of $S^2$. The gauge field is characterized by the second Stiefel-Whitney class $w_2(A)$. To see how the volume form on $S^2$ is modified, we note that the sphere can be represented using the auxiliary system given by the Higgs phase of $U(1)$ gauge field $a$ coupled to two complex scalars that condense, and the volume form of $SU(2)/U(1)=S^2$ is identified with the first Chern class $\frac{da}{2\pi}$. The isometry on $S^2$ is identified with the $SO(3)$ flavor symmetry that rotates the scalar.
The symmetry of the action is $\left( U(1)_\text{gauge}\times SU(2)_\text{global}\right)/\mathbb{Z}_2$.
Thus in the presence of the $SO(3)$ background gauge field the first Chern class of the gauge bundle is modified to be a half integer
\begin{equation}
    \Omega_2(A)\equiv\frac{1}{2}w_2(A)\pmod{\mathbb{Z}}~.
\end{equation}

We decompose the 6d three-form gauge field $C$ as
\begin{equation}
C=B_3+B_1\Omega_2(A)~.
\end{equation}
Under a gauge transformation $B_1\rightarrow B_1+d\lambda$, in order for the holonomy $e^{i\oint C}$ to be gauge invariant, the background $B_3$ must transform as
\begin{equation}
    B_3\rightarrow B_3 + \lambda\frac{1}{2}dw_2(A)=B_3+\lambda W_3(A)~,
\end{equation}
where $W_3(A)$ is the third integral Stiefel-Whitney class of the $SO(3)$ bundle.
Thus the reduced $4d$ theory has 3-group symmetry that combines the two-form symmetry (with background $B_3$), 0-form symmetry (with background $B_1$) and the $SO(3)$ 0-form symmetry from the isometry of $S^2$.
The backgrounds satisfy
\begin{equation}
    dB_3=\frac{1}{2} d B_1 w_2(A)~. 
\end{equation}

The anomaly of the higher-group symmetry is given by the reduction of the $CdC$ theory in 7d.

\paragraph{The Cartan part.} One might think that $S^2$ with its $SO(3)$ isometry is too special. However, even if we keep just the Cartan part of $SO(3)$, it still participates non-trivially in the higher group symmetry. This would be useful if we want to do cutting and gluing, e.g.~viewing the $S^2$ as two disks glued together along the equator, that only preserves a $U(1)$ subgroup. If we treat $S^2$ as $S^1$ fibered over an interval $[0,1]$, the compactification on $S^1$ gives symmetry with backgrounds $B_3$ and $B_2$ that satisfy
\begin{equation}
    dB_3=B_2 \frac{dA^{U(1)}}{2\pi}~.
\end{equation}
On the other hand, compactification on $S^2$ gives symmetry with backgrounds $B_3,B_1$ satisfy
\begin{equation}
    dB_3=\frac{1}{2}dB_1 w_2(A^{SO(3)})~.
\end{equation}
If turn on background for $U(1)$ subgroup, it is
\begin{equation}
    dB_3=\frac{1}{2}dB_1 \frac{dA^{U(1)}}{2\pi}~.
\end{equation}
The symmetries are related by
\begin{equation}
    B_2=\frac{1}{2}dB_1,\quad w_2(A^{SO(3)})=\frac{dA^{U(1)}}{2\pi}\text{ mod }2~.
\end{equation}
In other words, the boundary condition breaks the one-form symmetry $B_2\rightarrow B_2+d\lambda$, which is not a background gauge transformation of $B_1$. Instead, it fixes $B_2$ to be a global two-form $B_2=dB_1/2$. If $D$ does not contain a $\mathbb{Z}_2$ subgroup, then this boundary condition for background $B_2$ is setting $B_2$ to be trivial.

\subsection{Boundary conditions of 5d bulk TQFT and 4d discrete theta angles}

One aspect that we continue to emphasize throughout this two-part series is the connection between polarizations on $M_d$ and topological boundary conditions of $\CT^{\rm bulk}[M_d]$. We now describe the latter in greater detail for $M_d=T^2$ in some simple cases. 

Consider $D=\mathbb{Z}_N$ with the action of the 7d theory being
\begin{equation}
    \frac{N}{4\pi}\int_{7d}CdC~.
\end{equation}
Upon torus compactification with $C=\alpha_1 B_2^1+\alpha_2 B_2^2+\ldots$, where $\alpha_1,\alpha_2$ generate the $H^1(T^2,\mathbb{Z}_N)=\mathbb{Z}_N\times\mathbb{Z}_N$, the action reduces to
\begin{equation}\label{eqn:5daction}
\frac{N}{2\pi}\int_{5d}B_2^1 d B_2^2~.
\end{equation}
There is another term with 3-form and 1-form proportional to $\int B_3dB_1$ from the $H^0$ and $H^2$ part, whose effects we have already studied quite extensively in the previous subsection, and we now focus on the $H^1$-part.

The topological boundary condition for the theory is also discussed in \cite{Hsin:2021unpubvn} in the context of loop Toric code lattice model. 
The gapped boundaries correspond to 
condensation of the loop excitation (the loop excitations are described by the surface operators $e^{iq_e\int b^1+iq_m\int b^2}$), which gives rise to boundary particle excitations (described by line operator on the boundary where the surface operator ends) and the refined choice of whether the boundary particle is a boson or fermion. From the point of view of polarization, this is about choosing first $L$ and then the quadratic refinement.

When $N=2$, there are six gapped boundaries: the condensed excitations are either the electric loop $(q_e=1,q_m=0)$, magnetic loop $(q_e=0,q_m=1)$ or dyonic loop $(q_e=1,q_m=1)$.
The boundary has particle excitation that come from the end point of the condensed loops. Furthermore, for each choice of condensed loop, the end point can be either boson or fermion, giving rise to $2\times 3=6$ gapped boundaries. This matches perfectly with the classification on the ``$H^1$-part'' of the polarization. Namely, there are 6 choices regarding the spectrum of line operators that can be referred to as $SU(2)$, Spin-$SU(2)$, and $SO(3)_{0,1,2,3}$ theories in the context of type $A_1$. This is exactly the same as Pol$(S^3\times S^3)$ which we analyzed in Part I, as follows from the fact that the middle cohomology groups of $T^2$ and $S^3\times S^3$ are isomorphic, with the intersection forms both being antisymmetric.

For general $N$, the gapped boundaries can be constructed from the following basic building blocks:
\begin{itemize}
    \item  
The action has boundary variation
\begin{equation}
\frac{N}{2\pi}\int_{4d}B_2^1 \Delta B_2^2~.
\end{equation}
If there is no additional boundary term, then the boundary equation of motion imposes the boundary condition $B_2^1|=0$. 

Furthermore, for even $N$ we can add the boundary term without modifying the equation of motion mod $N$
\begin{equation}
    \frac{N^2}{4\pi}\int B_2B_2~.
\end{equation}
This changes the spin of the line operator which is the bounds the surface operator $\int B_2^1$ from boson to fermion.
For odd $N$ such boundary term requires a spin structure and is not well-defined.

\item 
We can add the boundary term, 
\begin{equation}
-\frac{N}{2\pi}\int B_2^1B_2^2~,
\end{equation}
which amounts to an $S$-transformation in the bulk $(B_2^1,B_2^2)\rightarrow (B_2^2,-B_2^1)$, as the action (\ref{eqn:5daction}) changes as
\begin{equation}
    \frac{N}{2\pi}\int_{5d} \left(B_2^1 dB_2^2 -d(B_2^1 B_2^2)\right)=-\frac{N}{2\pi}\int_{5d} B_2^2 dB_2^1~.
\end{equation}
The equation of motion is $B_2^2|=0$. As before, for even $N$ we can further add $\frac{N^2}{4\pi}\int B_2^2B_2^2$ on the boundary to change the spin of the line that bounds the surface operator $\int B_2^2$ from boson to fermion.

\item We can add the boundary term
\begin{equation}
\frac{Np}{4\pi}\int_{4d}B_2^1B_2^1~.
\end{equation}
This is the discrete theta angle for $\mathbb{Z}_N$ two-form gauge field given by Pontryagin square \cite{Aharony:2013hda,Kapustin:2013qsa,Kapustin:2014gua,Hsin:2018vcg,Hsin:2020nts}.
The equation of motion implies
\begin{equation}
    B_2^1|_\partial=pB_2^2|_\partial~.
\end{equation}
As before, for even $N$ we can further add $\frac{N^2}{4\pi}\int B_2^1B_2^1$ on the boundary to change the spin of the line that bounds the surface operator $\int B_2^2$ from boson to fermion. This is equivalent to changing $p$ to $p+N$.

\end{itemize}

This again matches perfectly with the analysis of polarization in Part I. There are multiple interesting questions that one can continue to explore via the perspective of topological boundary, such as a complete classification of mixed polarizations and how they transform under mapping class group. We will not pursue these here and instead will turn our attention to the compactification on 3-manifolds.

\section{Compactification to 3d}
\label{sec:3d}

In this section, we consider $T[M_3]$ obtained by compactification on 3-manifolds. Given a $M_3$, one can again work out $\mathrm{Pol}(M_3)$ using the general recipe discussed in Part I. However, the emphasis of this section is not about classifying polarizations on general 3-manifolds. In fact, for $M_3=\Sigma\times S^1$, we have already discussed the classification, though not in great depth, in the previous section when studying the moduli space $\CM_{T[\Sigma\times S^1]}$. Instead, we will take the opportunity to discuss the following: 
\begin{itemize}
    \item Physics of the 4d TQFT $\CT^{\text{bulk}}[M_3]$ and its boundary conditions.
    \item Higher-group symmetry from isometries of $M_3$.
    \item Anomalies of $T[M_3]$ and its relation with the geometry of $M_3$.
    \item Effects of the mapping class group of $M_3$ on $T[M_3]$.
    \item The classification of charged objects in $T[M_d]$ when $H^*(M_d)$ has torsion.
    \item Applications to quantum invariants of 3-manifolds.
\end{itemize}

\subsection{Polarization and boundary condition of bulk TQFT}

An interesting feature when $d=3$ compared to the previous cases with $d<3$ is that $H^*(M_d)$ now often has torsion, providing us with an interesting playground to test our general proposal and observe new phenomena in the presence of torsion. We now assume that the 6d theory is of type $A_{n-1}$, and study polarizations on lens spaces $L(k,1)$, arguably the simplest oriented manifolds with torsion in cohomology. 

With $D=\Z_n$, the relevant homology groups with $D$ coefficients are $H^0=H^3=\Z_n$, $H^1=H^2=\Z_{\text{gcd}(n,k)}$.\footnote{One way to see this is via the universal coefficient theorem, which states that $H_1\simeq H^2$ with $\Z_n$ coefficient is isomorphic to $H_1(M_3,\Z)\otimes \Z_n\simeq \Z_k\otimes\Z_n\simeq\Z_{\text{gcd}(n,k)}$, while $H^1(M_3,\Z_n)\simeq \mathrm{Hom}(H_1(M_3,\Z),\Z_n)\simeq \Z_{\text{gcd}(n,k)}$ as well.} One choice of polarization is to take $L$ to be $H^0\oplus H^2$. This is a nice geometric polarization given by filling in $L(k,1)$ with an ALE space based on the resolved $\Z_k$ singularity, or equivalently, a 2-handlebody obtained by attaching a 2-handle with framing $k$ to a 0-handle. This would lead to a theory with $(-1)$- and 1-form symmetries. The theory with the ``opposite'' polarization will then have $0$- and $2$-form symmetries. In general, we also have polarizations ``in between.'' At low energy, the sector of the theory involving the $(-1)$- and 2-form symmetry decouples,\footnote{This sector is associated with the discrete flux of the 3-form $C$ field on $M_3$. This becomes a $D=\Z_n$-valued discrete theta angle for a $(-1)$-form symmetry of the theory $T[M_3]$. The 2-form symmetry is about the existence of sectors of the theory with difference fluxes for the $C$-field in space-time. It is a rather familiar story---which we have also seen in the $T[M_2]$ case---that the two are related by discrete gauging. For example, in one direction, the theta angle can be viewed as a way of summing over the different sectors. This phenomenon is sometimes referred to as ``decomposition'' (see e.g.~\cite{Sharpe:2014tca}). Some further details of this is given in the next subsection.} and the rest of $T[L(k,1),\CP_L]$ for the geometric polarization is then a 3d $\CN=2$ $SU(n)$ Chern--Simons theory at level $k$ with an adjoint chiral multiplet. See \cite{Gukov:2015sna,Pei:2015jsa} for some study of this theory in the context of the 6d compactification and the 3d-3d correspondence. Indeed, it has symmetries given by $(L^\vee)^{(1)}=\Z_{\text{gcd}(n,k)}$, which is actually only part of the $\Z_n$ 1-form symmetry of the $SU(n)_k$ Chern--Simons theory. However, the bigger $\Z_n$ has an 't Hooft anomaly when $n\nmid k$, and $\Z_{\text{gcd}(n,k)}$ is the anomaly-free subgroup.\footnote{Recall that the anomaly for the $L^\vee$ symmetry can be identified with the obstruction of lifting $L^\vee\subset H^*/L$ to $H^*$. As long as $(L^\vee)^{(i)}$ equals the entire $H^i$, this $(2-i)$-form symmetry should always be anomaly free as no lifting is needed and there is no obstruction.} This is exactly the kind of ``predictable accidental symmetry'' discussed in Section~\ref{sec:accidental}. In general, when part of $L^\vee$ is given by $H^i(M_d,D)$ as a sum of quotients of $D$, the true $(i-1)$-form symmetry of the theory can be larger than this, while $H^i(M_d,D)$ is only an anomaly-free subgroup.

The reduction of the 7d theory on manifolds with torsion is studied more carefully in Appendix~\ref{app:reduction}. Here we will take a shortcut and start with the 7d theory with the action
\begin{equation}
    -\frac{n}{4\pi}\int C dC,
\end{equation}
which only differs from the actual theory given by the Cartan matrix coupling by an invertible theory. As discussed in Section~\ref{sec:oneformsymmetrycompactification}, reducing the theory on $L(k,1)$ leads to the following term
\begin{equation}
    -2\pi\cdot\frac{k}{2n}\int B_2^2,
\end{equation}
where $B_2$ is the integral of $C$ on the torsion 1-cycle, normalized by a $\frac{2\pi}{n}$ factor, i.e.,~we change the normalization where $B_2$ has a holonomy taking values $0,1,\cdots,n-1 \pmod n$.

This is exactly describing the anomaly of the $\Z_n$ 1-form symmetry. In fact, starting with the 7d theory defined by the Cartan matrix and integrating out fields, one ends up with 
\begin{equation}
    2\pi\cdot\frac{k(n-1)}{2n}\int B_2^2,
\end{equation}
which only differ from the previous one by an invertible fermionic theory.

One can get other pure polarizations by gauging a subgroup of $L^\vee$. The ``opposite'' polarization (i.e.,~with smallest 1-form symmetry but largest 0-form symmetry) corresponds to gauging the entire $\Z_{\text{gcd}(n,k)}$. After that, the gauge group still has a center, and one can only have a theory with $PSU(n)$ gauge group if $n|k$.

Again, the discussion about polarization can be done in the framework of topological boundary conditions of $\CT^{\text{bulk}}[M_3]$, which we discuss below.

\subsubsection*{Bulk TQFT reduction on 3-manifolds: full theory and boundary conditions}

Consider again ${\frak g}=\frak{su}(N)$. 
In the following, we will use the continuum notation for the gauge fields, by embedding the gauge fields into $U(1)$ gauge fields.
From Appendix \ref{app:reduction}, the bulk TQFT from the reduction on 3-manifold is the sum of ``free'' part and ``discrete'' part, where the 3-form gauge field reduces as
\begin{equation}
    C_3=\left(B_3^F+\alpha_3 B_0^F+\alpha_1^i B_2^{F,i}+\alpha_2^i B_1^{F,i}\right)
    +
    \left(\tau_3^i\hat B_0^i+\tau_2^i \hat B_1^i+\hat \tau_1^i B_2^i+\hat\tau_2^i B_1^i
    \right)~,
\end{equation}
where $\int \hat\tau_1^i \wedge\tau_{2}^j=\int \hat \tau_2^i\wedge \tau_1^j=\delta^{ij}$, and $\alpha_3$ is the volume form on the 3-manifold. They satisfy the relation $d\hat \tau_1^i=n^{ij}\tau_2^j$, $d\hat \tau_2^i=m^{ij}\tau_3^j$ where $n^{ij},m^{ij}$ are integers, i.e. $\tau_2^j,\tau_3^j$ are torsion 2- and 3-cocycles, respectively. We are going to focus on 3-manifolds without torsion 2-cycles for simplicity.
In particular,
\begin{equation}
    dC_3\supset d\left(\tau_3^i\hat B_0^i+\tau_2^i \hat B_1^i+\hat \tau_1^i B_2^i+\hat\tau_2^i B_1^i
    \right)
    =
    \tau_3^i \left(d\hat B_0^i+m^{ji}B_1^j\right)+\tau_2^i\left(d\hat B_1^i+n^{ji}B_2^j\right)-\hat\tau^i_1 dB_2^i~.
\end{equation}
From the decomposition it is straightforward to compute the reduction of the action.

The free part depends on the fields $B_1^F,B_2^F,B_3^F,B_0^F$ reduced on free cycles:
\begin{equation}
-\frac{N}{2\pi}\sum_{i,j} I_{ij}^F\int B_2^{F,i}dB_1^{F,j}-\frac{N}{2\pi}\int B_0^FdB_3^F~,
\end{equation}
where $I^F_{i,j}=\int \alpha_1^i\wedge \alpha_2^j$ is the intersection form on 3-manifold between the $i,j$th basis of free 1-cycles and free 2-cycles, respectively.

The discrete part depends on the fields $B_1^i,\hat B_i^i,B_2^i,B_3^i,\hat B_0^i$,
\begin{align}
    -\int \frac{N}{4\pi}CdC\supset \frac{N}{2\pi} \int dB_2^i \hat B_1^i-\frac{N}{4\pi}n^{ji}\int B_2^i B_2^j~.
\end{align}

There are several polarizations: these polarizations correspond to different gapped boundaries of the TQFT. Since we have a field theory description, they can be obtained by adding different boundary topological terms for the fields.
The boundary terms can be parametrized as follows. Let us focus on the discrete part for now:
\begin{align}
    \frac{\alpha^{ij}}{4\pi} B_1^i dB_1^j+\frac{N\beta^{ij}}{2\pi}B_2^jB_1^j~,
\end{align}
where $\alpha,\beta$ are integer matrices.

The boundary equation of motion gives
\begin{equation}
     N\beta^{ij}B_2^j+\alpha^{ij}dB_1^j=0,
     \quad 
(\delta^{ij}+\beta^{ij})  B_1^j=0~.
\end{equation}
\begin{itemize}
    \item 
The second equation implies that the one-form gauge group is broken on the boundary, and the boundary has $\text{Im}(\mathbf{1}+\beta)$ 0-form symmetry where we reduce the integer matrix to $\mathbb{Z}_N$ coefficient matrix.

\item
The first equation tells us that some two-form gauge group is broken, and the boundary has one-form symmetry.
For example, if $\beta^{ij}=\delta^{ij}$, the first equation implies that $B^i$ has a holonomy given by an integer multiple of $\frac{2\pi}{N/\gcd(N,\alpha^{ij})}$ where the gcd is that of $N,\{\alpha^{ij}\}$ for a given fixed $i$. Thus the one-form symmetry on the boundary for each $i$ is $\mathbb{Z}_{\gcd(N,\alpha^{ij})}$.

\end{itemize}
For example, when the 3-manifold is a lens space, the symmetries and their anomalies of the theory are discussed in Section~\ref{sec:oneformsymmetrycompactification}.

\subsection{Higher-group symmetry}

We now discuss higher group symmetries for $T[M_3]$ in several examples where $M_3$ is $T^3$, $S^1\times S^2$, and the lens space $S^3/\Z_p$.

\subsubsection{Example: compactification on $T^3$}

Denote the coordinate of the circles by $\varphi^i$  and the corresponding $U(1)$ gauge fields by $A^i$ for $i=1,2,3$. Consider the decomposition
\begin{equation}
    C_3=B_3 + B_2^i\frac{d\varphi^i-A^i}{2\pi}
    +B_1^{ij}\frac{d\varphi^i-A^i}{2\pi}\frac{d\varphi^j-A^j}{2\pi}+B_0^{ijk}\frac{d\varphi^i-A^i}{2\pi}\frac{d\varphi^j-A^j}{2\pi}\frac{d\varphi^k-A^k}{2\pi}~.
\end{equation}
The fields have the following gauge transformations that leave invariant $C_3$:
\begin{align}
    &B_0^{ijk}\rightarrow B_0^{ijk}+2\pi n^{ijk}\cr
    &B_1^{ij}\rightarrow B_1^{ij}+3n^{ijk}A^k+d\lambda_0^{ij}\cr
    &B_2^{i}\rightarrow B_2^i+3n^{ijk}{A^jA^k\over 2\pi}+2{d\lambda_0^{ij}A^j\over 2\pi}+d\lambda_1^i\cr
    &B_3\rightarrow B_3+{n^{ijk}A^iA^jA^k\over (2\pi)^2}-{d\lambda_0^{ij}A^iA^j\over (2\pi)^2}+{d\lambda_1^iA^i\over 2\pi}~.
\end{align}
The background fields obey
\begin{align}\label{eqn:familyhighergroup}
    &dB_3={1\over (2\pi)^3}dB_0^{ijk}A^iA^jA^k
    -{1\over (2\pi)^2}B_1^{ij}A^iA^j
    +{1\over 2\pi}B_2^iA^i\cr
    &dB_2^i={3\over (2\pi)^2}dB_0^{ijk}A^jA^k
    +{2\over 2\pi}B_1^{ij}dA^j\cr
    &dB_1^{ij}={3\over 2\pi}dB_0^{ijk}A^k~.
\end{align}

\paragraph{Family of higher group symmetries.}

We note that the above equation (\ref{eqn:familyhighergroup}) can be interpreted as a family of 3-group symmetries parametrized by the periodic scalar $B_0^{ijk}$. In general, a family of higher groups parameterized by the space ${\cal M}$ has Postnikov classes
\begin{equation}
    \{\Theta_{n+1}\in H^{n+1}(X_{n+1},G^{(n)})\} ~,
\end{equation}
where locally $X_{n+1}\sim {\cal M}\times  \prod_{0\leq k\leq n-1} B^kG^{(k)}$ for $k$-form symmetry $G^{(k)}$. 
Examples of family of higher groups involving a parameter have been discussed in \cite{Hsin:2022iug}.

\subsubsection{Example: compactification on $S^1\times S^2$}

$S^1 \times S^2$ is an important example because it is a model example for all 0-surgeries (which, in turn, have many important applications to SPC4 and other problems in topology).

Consider the decomposition
\begin{equation}
    C_3=B_3+B_2\eta_1+B_1\eta_2(A')+B_0\eta_1\eta_2(A')~,
\end{equation}
where we turn on background $A'$ for the $SO(3)$ isometry of $S^2$. Denote the background for the $U(1)$ isometry of $S^1$ by $A$.
Consider the gauge transformation
\begin{equation}
    A\rightarrow A+d\lambda~,
\end{equation}
to compensate for the transformation,
\begin{equation}
    B_1\rightarrow B_1+B_0 \frac{d\lambda}{2\pi},\quad
    B_3\rightarrow B_3+B_2\frac{d\lambda}{2\pi}~.
\end{equation}
Similarly, a gauge transformation $B_0\rightarrow B_0+2\pi n$ induces
\begin{equation}
    B_2\rightarrow B_2+\pi n w_2(A')~.
\end{equation}
On the other hand, $B_1\rightarrow B_1+d\lambda_B$ gives
\begin{equation}
    B_3\rightarrow B_3+ \frac{1}{2}d\lambda_B w_2(A')~.
\end{equation}
Thus we find that
\begin{equation}
    dB_1=B_0\frac{dA}{2\pi},\quad 
    dB_2=\frac{1}{2}B_0 d w_2(A'),\quad 
    dB_3=\frac{1}{2} B_1 dw_2(A')+dB_2\frac{A}{2\pi}~.
\end{equation}
One can verify that the last equation is consistent with $d^2B_3=0$.
Thus in general the theory has three-group and two-group symmetries. The first equation indicates that the isometry symmetry is extended by the 0-form symmetry from the two-form symmetry of the six-dimensional theory.

Consider the anomaly of the above symmetries. To compute the anomaly, we need to further equivariantize $\eta_1$ into $\eta_1(A)$.
\begin{align}
    dC&=dB_3+dB_2\eta_1(A)-B_2\frac{dA}{2\pi}+dB_1\eta_2(A')
    \cr
    &-B_1\frac{dw_2(A')}{2}+dB_0\eta_1(A)\eta_2(A')+B_0(-\frac{dA}{2\pi})\eta_2(A')+B_0\eta_1(A)\frac{dw_2(A')}{2}~.
\end{align}

The anomaly of the symmetry is proportional to
\begin{align}
    \frac{1}{4\pi}\int C_3dC_3=
    &\frac{1}{4\pi}\int\left(B_3-B_2\frac{A}{2\pi}-B_1\frac{w_2(A')}{2}+B_0\frac{A}{2\pi}\frac{w_2(A')}{2}\right)dB_0
    \cr
    &+\frac{1}{4\pi}\int B_0\left(
    dB_3-d(B_2\frac{A}{2\pi})-dB_1\frac{w_2(A')}{2}-B_1\frac{dw_2(A')}{2} \right)\cr & +\frac{1}{4\pi}\int \left( dB_0\frac{A}{2\pi}\frac{w_2(A')}{2}+B_0\frac{dA}{2\pi}\frac{w_2(A')}{2}-B_0\frac{A}{2\pi}\frac{dw_2(A')}{2}
    \right)
    \cr
    &+\frac{1}{4\pi}\int \left(B_2-B_0 \frac{1}{2}w_2(A')\right)\left(dB_1-dB_0\frac{1}{2\pi}A-B_0\frac{-dA}{2\pi}\right)\cr
    &+\frac{1}{4\pi}\int\left(B_1-B_0\frac{A}{2\pi}\right)
    \left(dB_2-dB_0\frac{1}{2}w_2(A')-B_0\frac{dw_2(A')}{2}\right)\cr
    =&
    \frac{1}{2\pi}\int \int\left(B_3-B_2\frac{A}{2\pi}-B_1\frac{w_2(A')}{2}+B_0\frac{A}{2\pi}\frac{w_2(A')}{2}\right)dB_0
    \cr
    &\!\!\!\!\!\!\!\!\!\!+\frac{1}{2\pi}\int\left( B_2dB_1
    -B_0\frac{A}{2\pi}dB_2
    -B_0\frac{w_2(A')}{2}dB_1+B_0\frac{w_2(A')}{2}d(B_0\frac{A}{2\pi})
    \right).
\end{align}
Note that for $B_0=0$, only $B_2,B_1$ have a mixed anomaly.

\paragraph{Compactification of the Abelian theory.}

For the rank-1 Abelian theory, let us first reduce it on a circle, this gives a $U(1)$ gauge theory with gauge field $a$ and field strength $F=da$.
Then the compactification on $S^2$ gives the following.
The magnetic flux on $S^2$ labels different sectors in the 3d theory, which are summed over
\begin{equation}
    n=\oint_{S^2}\frac{F}{2\pi},\quad Z_{\rm 3d}=\sum_n Z_n~.
\end{equation}
More generally, when the gauge group in 5d is $G$, they are labeled by $\pi_1(G)$.
It is a topological operator with discrete values.

In addition, the theory has theta angle 
\begin{equation}
    B_0=\oint_{S^2} B_2~,
\end{equation}
which means that the 5d gauge bundle is twisted on $S^2$, where the structure group becomes a quotient.
The theta angle is labelled by the electric one-form symmetry $Z(G)$ in 5d.

The 6d two-form symmetry gives a one-form center symmetry and a 0-form magnetic symmetry of the resulting one-form gauge theory in 3d. In addition, there is also a two-form symmetry in 3d, where the two-form symmetry has generator $n$ given by the magnetic flux of the one-form gauge field on the internal $S^2$. 
$n$ is a topological operator, $dn=0$, and thus $n$ is locally a constant. The value of $n$ labels different superselection sector of the theory. The 2d interfaces separating different superselection sector transform as linear representations under the two-form symmetry. 

Let us consider the symmetry from the isometries.
The isometry of the circle is the 5d instanton number symmetry with current $\frac{1}{8\pi^2}\star F\wedge F$, and in the compactified 3d theory on $S^2$ it becomes the symmetry
\begin{equation}
    j=\frac{1}{2\pi} n\cdot(\star F),\quad n=\oint_{S^2} \frac{F}{2\pi}~.
\end{equation}

If we turn on background $A',B_2$, where $A'$ is the $SO(3)$ isometry on the internal $S^2$, and $B_2$ couples to the $U(1)$ one-form symmetry in the 5d $U(1)$ gauge theory, the volume form on $S^2$ is modified to be $\Omega_2(A')$, and the field strength can be decomposed as
\begin{equation}
    F_{5d}= (2\pi n+B_0) \Omega(A')+ F_{3d},\quad \oint  F_{3d}=\oint B_2\text{ mod }2\pi~,
\end{equation}
where $d\Omega(A')=\frac{dw_2(A')}{2}$ is the equivariant volume form on $S^2$ in the presence of $A'$ background, and we can treat it as $\Omega(0)+\frac{w_2(A')}{2}$. The instanton number symmetry becomes
\begin{equation}
   Q_\text{inst}=\int \star j=\frac{1}{8\pi^2}\int F_{5d}F_{5d}=\frac{1}{2\pi}\int (n+\frac{B_0}{2\pi}) F_{3d}~,
\end{equation}
where $\oint F=B_2$ mod $2\pi$. The charge is not invariant under the background gauge transformations and this indicates an anomaly: this comes from the mixed anomaly between the instanton number symmetry and one-form symmetry. Moreover, the fractional part also depends on the dynamical field $n$ due to the quantization $\int F_{3d}=\int B_2$, in addition to the backgrounds, so the symmetries combine into higher group symmetry. This reproduces $dB_3\supset dB_2\frac{A}{2\pi}$.

\subsubsection{Example: compactification on lens spaces}

The integral cohomology of the lens space $L(p,q)\simeq S^3/\mathbb{Z}_p$ is
\begin{equation}
H^*(S^3/\mathbb{Z}_p,\Z)=\mathbb{Z},0,\mathbb{Z}_p,\mathbb{Z}~.
\end{equation}

Consider first $p=1$. We can decompose
\begin{equation}
C_3=B_3+B_0\Omega_3~.
\end{equation}
Since $S^3\cong SU(2)$, $\Omega_3$ is also the Wess-Zumino term in $SU(2)$ WZW model at level one. If we turn on background for the $SO(4)=\left(SU(2)\times SU(2)\right)/\mathbb{Z}_2$, the Wess-Zumino term produces an anomaly given by the Chern--Simons term.
Thus under the transformation $B_0\rightarrow B_0+2\pi$, in order to preserve the flux of $C_3$, $B_3$ must be shifted by the $SO(4)$ Chern--Simons term at level one.
Thus the reduction of two-form symmetry produces three-group symmetry, with background satisfies
\begin{equation}\label{eqn:3groupS3/Zp}
dB_3=\frac{B_0}{8\pi^2}\text{Tr }F\wedge F~,
\end{equation}
where $F$ is the field strength of the $SO(4)$ isometry gauge field.

When $p$ is nontrivial, there is also continuous isometry as discussed in \cite{McCullough:2000}. Let us focus on the lens space $L(p,1)$ and the continuous isometry, while the discrete disconnected part will be discussed later.
\begin{itemize}
    \item For $p=2$, the continuous isometry is $SO(3)\times SO(3)=SO(4)/\mathbb{Z}_2$. The periodicity of $B_0$ is $4\pi$: if we shift $B_0$ by $2\pi$, $B_3$ will not be transformed by a well-defined amount, since $SO(3)_{1/2}\times SO(3)_{1/2}$ is not well-defined.
    
    \item For $p>2$ and even $p$, the continuous isometry is $O(2)\times SO(3)$. The periodicity of $B_0$ is $2\pi$.

    \item For $p>2$ and odd $p$, the continuous isometry is $\left(Pin^-(2)\times SU(2)\right)/\mathbb{Z}_2$. The periodicity of $B_0$ is $4\pi$: this is because the quotient Chern--Simons term $\left(Pin^+(2)_1\times SU(2)_1\right)/\mathbb{Z}_2$ is not well-defined. See e.g. \cite{Cordova:2017vab} for more properties about Chern--Simons theories with orthogonal gauge algebra.

\end{itemize}
For these cases, we have the three-group symmetry (\ref{eqn:3groupS3/Zp}), with the right hand side being the theta term of the continuous isometry groups with theta angle $B_0$. When $B_0$ is shifted by $2\pi$, $B_3$ is shifted by a Chern--Simons term.

\subsection{Mapping class group symmetry  $\text{MCG}(M_3)$}
\label{sec:MCGsymmetry}
We now discuss the action of the mapping class group of $M_3$ on $T[M_3]$. We will focus again on the Abelian case, which are already non-trivial examples to illustrate this general phenomenon.

\subsubsection{Example: $M_3=L(p,q)$}
\label{sec:lens}

Consider the example $M_3=L(p,q)$ with $\gcd(p,q)=1$ and $p>1$.\footnote{The lens space
$L(p,q)$ is defined as $S^3/\mathbb{Z}_p$ by the action $(z_1,z_2)\sim (e^{2\pi i/p}z_1,e^{2\pi i q/p}z_2)$ for $z_1,z_2\in \mathbb{C}^2$ with $|z_1|^2+|z_2|^2=1$ describing $S^3$.
} Without loss of generality we take $pq$ to be even, which can be achieved by the redefinition $q\rightarrow q+p$ which defines the same lens space $L(p,q)$.
The mapping class group for $L(p,q)$ is listed in Table~\ref{tab:Lens} (see {e.g.}~\cite{McCullough:2000}).

\begin{table}[ht]
	\begin{centering}
		\begin{tabular}{|c||c|c|}
			\hline
			~$\phantom{\int^{\int^\int}} \text{Lens space}~L(p,q) \phantom{\int_{\int}}$~ & ~$\pi_0 \text{Diff} (L(p,q))$~ & ~Orientation preserving~ \tabularnewline
			\hline
			\hline
			$\phantom{\int^{\int^\int}} p=2 \phantom{\int_{\int}}$ & $\Z_2$ & \xmark
			\tabularnewline
			\hline
			$\phantom{\int^{\int^\int}} p>2,~q\equiv\pm 1~\pmod p \phantom{\int_{\int}}$ & $\Z_2$ & \cmark
			\tabularnewline
			\hline
			$\phantom{\int^{\int^\int}} q>2,~q^2 \not\equiv \pm 1~\pmod p \phantom{\int_{\int}}$ & $\Z_2$ & \cmark
			\tabularnewline
			\hline
			$\phantom{\int^{\int^\int}} q^2\equiv +1~\pmod p,~~q\not\equiv \pm 1~\pmod p \phantom{\int_{\int}}$ & $\Z_2 \times \Z_2$ & \cmark
			\tabularnewline
			\hline
			$\phantom{\int^{\int^\int}} q^2\equiv -1~\pmod p,~~q\not\equiv \pm 1~\pmod p \phantom{\int_{\int}}$ & $\Z_4$ & \xmark~ (on order-4 el'ts)
			\tabularnewline
			\hline			
		\end{tabular}
		\par\end{centering}
	\caption{\label{tab:Lens} Mapping class groups of lens spaces.}
\end{table}

The theory $T[M_3,U(1)]$ consists of an Abelian Chern--Simons theory and a decoupled chiral multiplet (it is decoupled since the adjoint representation of $U(1)$ is trivial).
The Abelian Chern--Simons theory has the coefficient matrix given by (2.6) of \cite{Gadde:2013sca} with $[a_1,\cdots ,a_n]$ being the continued fraction expansion of $-p/q=a_1-1/(a_2-\cdots)$. 
This Chern--Simons theory describes the minimal Abelian TQFT (see, e.g.,~\cite{Hsin:2018vcg}) ${\cal A}^{p,-q}$ with $\mathbb{Z}_{p}$ fusion algebra generated by a line of spin $-\frac{q}{2p}$ mod 1, and there is also the transparent fermion line $\psi$ that satisfies $\psi^2=1$.
The line operators in the theory can be labelled by $Q\in\mathbb{Z}_p$ and $\ell\in\mathbb{Z}_2$.
The line $(Q,\ell)$ has spin
\begin{equation}
h[(Q,\ell)]=\frac{\ell}{2}-\frac{qQ^2}{2p}\text{ mod }1~.
\end{equation}

Since an Abelian TQFT is specified by the spins and the fusion algebra of the line operators, any permutation action on the set of line operators such that it preserves this data is a (0-form) global symmetry of the Abelian TQFT. If the symmetry is unitary, this requires the spin of the line operators to be preserved under the symmetry transformation.
If the symmetry is anti-unitary, since the transformation contains complex conjugation, the topological spin is complex conjugated $e^{2\pi ih}\rightarrow e^{-2\pi ih}$ and thus the spin $h$ becomes $-h$ mod 1.

Let us compare the mapping class group of $L(p,q)$ with the symmetry in $T[L(p,q),U(1)]$.
\begin{itemize}

\item $p=2$: the Chern--Simons theory is $U(1)_2$, and the $\mathbb{Z}_2$ symmetry corresponds to the time-reversal symmetry in the spin Chern--Simons theory.

The time-reversal symmetry is anomalous. The anomaly is classified by $\nu\in \mathbb{Z}_{16}$, and the theory has $\nu=\pm 2$ anomaly.

\item $p>2,q\equiv \pm1\pmod p$: the $\mathbb{Z}_2$ symmetry corresponds to the charge conjugation symmetry in the TQFT
\begin{equation}
r_1:\quad Q\rightarrow -Q~.
\end{equation}
Note since $\ell$ is defined mod 2, $\ell\rightarrow-\ell$ is a trivial permutation.

In this case, the TQFT is $U(1)_p,SU(p)_1$ or their time-reversal images.
The $\mathbb{Z}_2$ symmetry is non-anomalous, and one can gauge the symmetry following orbifold of RCFT.
(For odd $p$, $U(1)_p$ is a spin TQFT, and gauging the charge conjugation symmetry is discussed in Appendix F of \cite{Cordova:2017vab}).

\item $q^2\equiv +1\pmod p$, $q\not\equiv \pm 1\pmod p$: the theory has $\mathbb{Z}_2\times\mathbb{Z}_2$ symmetry generated by $r_1,r_2$ with
\begin{equation}
r_2:\quad (Q,\ell)\rightarrow (qQ,\ell+\alpha Q)\quad\text{where }\alpha=q(q^2-1)/p\text{ mod }2~.
\end{equation}
Note $\alpha p\equiv 0\pmod 2$ since $pq\equiv0\pmod 2$, and thus it is well-defined for $Q\in\mathbb{Z}_p$.

\item $q^2\equiv-1\pmod p$, $q\not\equiv  \pm 1 \pmod p$: the theory has $\mathbb{Z}_4$ symmetry generated by
\begin{equation}
r_2':\quad (Q,\ell)\rightarrow (qQ,\ell+\alpha' Q)\quad\text{where }\alpha=q(q^2+1)/p\text{ mod }2~.
\end{equation}
This is a time-reversal symmetry since it flips the sign of the spin
\begin{equation}
h[(Q,\ell)]=-h[r_2'(Q,\ell)]\text{ mod }1~.
\end{equation}
This agrees with the mapping class group element that reverses the orientation of $M_3$.\footnote{
Note reversing the orientation on $M_3$ corresponds to performing a orientation-preserving 6d Lorentz symmetry to also reverse the orientation on the 3d spacetime.
}
Note $(r_2')^2=r_1$. Thus the symmetry squares to the charge conjugation symmetry $Q\rightarrow -Q$.

For odd $p$, examples of the TQFT are discussed in \cite{Barkeshli:2017rzd,Benini:2018reh}, and it was shown in \cite{Barkeshli:2017rzd} the $\mathbb{Z}_4$ symmetry does not have anomaly by constructing boundary of the TQFT preserving the symmetry.

\end{itemize}

\subsubsection{Example: $M_3=S^3/G$ of $T[M_3,U(1)]$}

Next, we consider the family of manifolds $S^3/G$ where we used $S^3\cong SU(2)$ and the ADE classification for finite subgroups $G\subset SU(2)$. 
The ADE classification and the mapping class group for $S^3/G$ (see e.g.~table 2 of \cite{McCullough:2000}) are given as follows 
\begin{equation}
\begin{array}{|c|c|c|}
\hline
G & \text{ADE} & \text{MCG}(S^3/G)\\ \hline
2D_4=Q_8 & D_4 & S_3\\
2D_{2n>8} & D_{n+2>4} & \mathbb{Z}_2\\
2T_{12} & E_6 & \mathbb{Z}_2\\
2O_{24} & E_7 & \text{trivial}\\
2I_{60} & E_8 & \text{trivial} \\\hline
\end{array}~
\end{equation}
In the above table, $G=2D_m$ is the double cover of the the dihedral group of order $m$ (since $SU(2)\cong S^3$ is the double cover of $SO(3)$), and similarly $G=2T_{12},2O_{24},2I_{60}$ are the double cover of tetrahedral group, the double cover of octahedral group, and the double cover of icosahedral group respectively.
The cases classified by the A-series are not listed in the above table, since they correspond to lens space and it is already discussed in Section~\ref{sec:lens}.

The Abelian Chern--Simons theory in $T[M_3,U(1)]$ for each $G$ in the table has coefficient matrix given by the Cartan matrix of the corresponding ADE Lie algebra. Such TQFT is known to be equivalent to the Chern--Simons theory with the ADE (universal covering) gauge group at level one.

Let us identify the mapping class group with the symmetry in $T[M_3,U(1)]$. The non-trivial cases are the following
\begin{itemize}
\item $G=Q_8$: the $S_3$ mapping class group is identified with the $S_3$ triality symmetry that permutes the three fermion lines in Spin(8)$_1$.

\item $G=2D_{2n>8}$: the $\mathbb{Z}_2$ mapping class group is identified with the $\mathbb{Z}_2$ charge conjugation symmetry in Spin$(2n+4)_1$ that exchanges the two spinor nodes\footnote{The theory 
Spin$(2n+4)_1$ has three non-trivial lines: one in the vector representation, and the other two in the spinor representations.
The two spinor lines are exchanged by this $\mathbb{Z}_2$ charge conjugation symmetry.
}.

\item $G=2T_{12}$: the $\mathbb{Z}_2$ mapping class group is identified with the $\mathbb{Z}_2$ Dynkin diagram automorphism that acts as $\mathbb{Z}_2$ symmetry in $(E_6)_1$.\footnote{
$(E_6)_1\leftrightarrow SU(3)_{-1}$ has two non-trivial lines, and they are exchanged by this $\mathbb{Z}_2$ symmetry.
}

\end{itemize}

\subsection{Anomaly in $T[M_3]$}

Much of the anomaly of symmetries of the $T[M_3]$ theory in (2+1)d descends from the anomaly of the two-form symmetry in the 6d theory, via a process which, of course, also depends on the choice of $M_3$ and the polarization.

This kind of anomaly is rather straightforward to study via the bulk action. Namely, one takes the 7d bulk to be $M_3\times Y_4$ with $\partial Y_4=Y_3$ the space-time of the $T[M_3]$ theory. The bulk action for the anomaly of the $T[M_3]$ theory is then obtained by reducing the 7d $CdC$ theory on $M_3$. 

In fact, we have already discussed the anomaly for the $\mathbb{Z}_n$ one-form symmetry of $T[L(k,1)]$ from this perspective in Section~\ref{sec:oneformsymmetrycompactification}. We now consider some anomalies that are not captured in this way.

\subsubsection{Mixed anomaly between mapping class group and one-form symmetry}

Consider the example $G_{6d}=U(1)$ and $M_3=L(p,1)$ with $p=4m+2>2$.
As discussed in \ref{sec:lens}, the mapping class group is $\mathbb{Z}_2$ and it acts on $T[M_3]$ by charge conjugation symmetry of $U(1)_{-p}$. On the other hand, there is a mixed anomaly between the charge conjugation symmetry and the $\mathbb{Z}_2$ subgroup one-form symmetry in $U(1)_{-p}=SO(2)_{-p}$ for $p=2$ mod $4$ given by the SPT phase \cite{Cordova:2017vab}\footnote{
One way to see this anomaly is that gauging the charge conjugation symmetry extends the $\mathbb{Z}_2$ subgroup one-form symmetry to be $\mathbb{Z}_4$. This extension is already observed in the $\mathbb{Z}_2$ orbifold of the chiral algebra of $U(1)_{p}$ in \cite{Dijkgraaf:1989hb}.
}
\begin{equation}
\pi\int B_2\text{Bock}(X)~,
\end{equation}
where $B_2$ is the background the $\mathbb{Z}_2\subset\mathbb{Z}_p$ subgroup one-form symmetry and $X$ is the background for the $\mathbb{Z}_2$ charge-conjugation symmetry (which is identified with the mapping class group 0-form symmetry), and Bock is the Bockstein homomorphism for the short exact sequence $1\rightarrow\mathbb{Z}_2\rightarrow\mathbb{Z}_4\rightarrow\mathbb{Z}_2\rightarrow 1$.

\subsubsection{Gravitational anomalies and the Rokhlin invariant}

Theories in (2+1)d can have a gravitational Chern--Simons term, which can be expressed as a gravitational or thermal response.\footnote{One alternative point of view is that one can cancel the gravitational response of the theory with a background gravitational Chern--Simons term. However, when the level being used is fractional, this can lead to a framing anomaly, similar to the discussion in \cite{Witten:1988hf}. The invariant being discussed here can be understood as such a framing anomaly given by the fractional part of the gravitational Chern--Simons level.} One can attempt to compute such a gravitational background term by reducing the part of the 7d TQFT action that involves the spin and R-symmetry connections. However, the integral over the internal $M_3$ is expected to depend on its metric, and it is not clear from this point of view how to get any metric-independent information.

One common approach for defining invariants of 3-manifolds is to start with one for 4-manifolds and then consider a relative version of it, which can lead to a ``secondary invariant'' for 3-manifolds. Here, we will follow a similar procedure, starting with invariants of 4-manifolds given by the gravitational anomaly of the $T[M_4]$ theory. 

For 6d $(1,0)$ theories, there is a unique topological twist that works for general $M_4$, and the gravitational anomaly $d\in \Z$ for the free multiplets are (see e.g.~\cite{Gukov:2018iiq} for further discussions):
\begin{align*}
  \text{tensor}: & \;  \frac{\chi+5\sigma}{2},\\
    \text{hyper}: &  \; -\frac{\sigma}{4},\\
      \text{vector}: & \;  -\frac{\chi+\sigma}{2}.
\end{align*}
  To give an interacting example, the gravitation anomaly for the compactification of the E-string theory is $d=-\frac{11\chi+31\sigma}{2}$.

For 6d $(2,0)$ theories, there are three different twists. For the Vafa--Witten twist, one has $d=\frac{\chi+3\sigma}{2}$ for both the free tensor multiplet and the $A_1$ theory. This is in line with the intuition that the 2d theory can be thought of as a sigma model to the ``moduli space of the 6d theory on $M_4$,'' and the moduli spaces associated to $U(1)$ and $SU(2)$ often have the same dimensions, with the latter sometimes related to the form via a discrete quotient. 

One can try to find the secondary invariant for a particular 6d theory, but as the gravitational anomaly is just given as a combination of $\chi$ and $\sigma$, we cannot expect to have any new invariant by considering more exotic theories. (On the other hand, one does expect to get interesting secondary invariants starting with more interesting quantities of $T[M_4]$ such as the partition function.) 

As for $\chi$, it actually cannot be used to define a non-trivial secondary invariant as any 3-manifold can bound a 4-manifold with arbitrary Euler characteristic. Naively, the situation for $\sigma$ is similar, but one can get something non-trivial by working in the spin case. There, a famous theorem of Rokhlin states that the signature of a spin 4-manifold is divisible by 16. This allows to define a $\mathbb{Z}_{16}$-valued secondary invariant $\mu(M_3,\frak{s})$ of 3-manifolds equipped with spin structure $\frak{s}$ as the signature of a spin 4-manifold bounded by $(M_3, \frak{s})$ mod 16. 

From the point of view of compactifications of 6d theories, this procedure amounts to characterizing the gravitational Chern--Simons term on $T[M_3]$ via the gravitational anomaly of the boundary theory $T[M_4]$. Naively, this gives a $\Z$-valued gravitational Chern--Simons level, but as $T[M_3]$ is itself a boundary of the 4d $\CT^{\rm bulk}[M_3]$ bulk theory and $T[M_4]$ an interface on the boundary with the other side being $\CT^{\rm bulk}[M_4]$ (see Figure~\ref{fig:TMopen}), such a level is in general ambiguous. However, for the 6d hypermultiplet, which is sensitive to the spin structure, the gravitational anomaly is $\frac14\Z$-valued, giving rise to a mod-4 invariant for $T[M_3]$, which is exactly four times the Rokhlin invariant. 

The fact that we are only getting $4\mu$ but not $\mu$ is related to the hypermultiplet being ``quaternionic'' with four fermions always appearing together. In general, we expect that a 6d theory gives $c\cdot\mu$ with $c\in\Z_{16}$ being actually always divisible by $4$. This gives a $\Z_4$-grading for 6d theories, with non-spin theories living in the zeroth degree.\footnote{Note that the property of being non-spin and the value of the $\Z_4$ can also depend on the choice of backgrounds of R-symmetry and other global symmetries. With the twisted R-symmetry background,  the $(1,0)$ vector and tensor are non-spin.} 

It is straightforward to translate the behavior of $\mu$ to the QFT side. For example, if $M_4$ is a cobordism from $M_3^+$ to $M_3^-$, then $T[M_4]$ is a two-dimensional wall that separates two three-dimensional theories $T[M_3^+]$ and $T[M_3^-]$. We have
\be
\mu (M_3^+) - \mu (M_3^-) \; = \; \sigma (M_4) \mod 16,
\ee
and, if multiplied by $c$, it describes the canceling of gravitational anomaly on the interface via the anomaly inflow from the bulk. Similarly, if $M_4$ can be represented as a union of two 4-manifolds $M_4^{\pm}$ glued along a common boundary $M_3$, then $T[M_4]$ can be described as the effective 2d theory of a three-dimensional slab $T[M_3]$ with boundary conditions $T[M_4^+]$ and $T[M_4^-]$. Then the fractional part of the gravitational anomalies on the boundaries and of the gravitational Chern--Simons level of $T[M_3]$ have to match. 

Now, getting back to the original question about obtaining this background term from the 7d TQFT by reducing on $M_3$, a relevant result that can aid with this computation is the following. The Rokhlin invariant can be written as a linear combination of $\eta$-invariants without local correction terms \cite{MILLER1987301},
\begin{equation}
\mu (M_3,\frak{s}) \equiv - \eta_{\text{sign}} - 8 \eta_D \quad \text{mod}~16.
\end{equation}
This gives a conceptual explanation of why reducing the 7d topological theory on $M_3$ can lead to the Rokhlin invariant of $M_3$ multiplying the gravitational Chern--Simons term in the $T[M_3]$ theory.

There are many other interesting statements about the Rokhlin invariants that should have nice physics interpretations from the $T[M_d]$ point of view. For example, when $M_4$ is not spin, one can still compute $\mu$ via \cite{Melvin1994}
\begin{equation}
    \mu(M_3,\frak{s})\equiv \sigma(M_4) - C\cdot C + 8\mathrm{Arf}(C)\pmod{16}
\end{equation}
where $C$ is the characteristic surface (``$w_2$-defect'') giving the obstruction for extending $\frak{s}$ into $M_4$. When $M_3$ is a lens space $L(p,q)$, its Rokhlin invariant is understood very thoroughly. For example, when $p$ is odd (i.e.~there is a unique spin structure), $\mu$ is proportional to the Dedekind symbol, which also appears in the study of $T[M_3]$ theories and the 3d-3d correspondence for lens spaces  (see \cite{Gukov:2015sna,Pei:2015jsa} and references therein).

\subsection{The Gluck twist}

In this subsection, we will focus on a particular element in the mapping class group of  $M_3=S^2\times S^1$, known as the Gluck twist.

The mapping class group for $S^2\times S^1$ is $\mathbb{Z}_2\times\mathbb{Z}_2\times\mathbb{Z}_2$, where the first two $\mathbb{Z}_2$'s act on $S^2$ and $S^1$ respectively in an orientation-reversing way, while the last $\mathbb{Z}_2$ is the Gluck twist \cite{gluck1961},
\begin{equation}
\varphi: \qquad (x,\theta) \mapsto (\text{rot}_{\theta} (x),\theta) \,, \quad \text{where}~x \in S^2 \,, \quad \theta \in S^1,
\end{equation}
where $\text{rot}_{\theta} (x)$ is a rotation of $S^2 \subset \mathbb{R}^3$ by angle $\theta$ about the $z$-axis. Note that this $\varphi$ is in the kernel of the $\text{MCG} (S^1 \times S^2)$ action on $H^* (S^1 \times S^2)$. On the torus given by the product of the equator of $S^2$ and $S^1$, the Gluck twist generates the Dehn twist transformation.

The Gluck twist is often used to produce candidates for exotic 4-spheres, {i.e.}~counterexamples to the smooth Poincar\'e conjecture in four dimensions (SPC4).
For example, an infinite family of Cappell--Shaneson homotopy 4-spheres \cite{Cappell1976} is obtained with the help of Gluck twist as follows.
First, for each choice of $m \in \mathbb{Z}$, they construct a mapping torus of $T^3$,
\begin{equation}
M_4 \; = \; T^3 \times [0,1] / (x,0) \sim (Ux,1)
\end{equation}
with the monodromy matrix (sometimes called the Cappell--Shaneson matrix)
\begin{equation}
U=\left(\begin{array}{ccc}
0 & 1 & 0 \\
0 & 1 & 1 \\
1 & 0 & m+1
\end{array}
\right)~.
\end{equation}
Note, $\det (U - {\bf 1}) = 1$, so that $H_1 (M_4,\mathbb{Z}) = \mathbb{Z}$ is generated by $S^1 = \{ 0 \} \times S^1 \subset M_4$. Its complement in $M_4$ is $M_4 \setminus (\{ 0 \} \times S^1)$, with boundary $S^2 \times S^1$. The candidate 4-sphere is then obtained by performing a surgery
\begin{equation}
\Sigma_m \; := \;
\left( M_4 \setminus (\{ 0 \} \times S^1) \right)
\, \bigcup_{\varphi} \,\left( S^2 \times D^2 \right).
\end{equation}
This family is labeled by $m \in \mathbb{Z}$ and a choice of $\varphi$, which one usually takes to be either ${\bf 1} \in \text{MCG} (S^1 \times S^2)$ (called the ``easy'' choice of framing) or the Gluck twist (called the ``hard'' choice of framing).

In 1984, Aitchison and Rubinstein \cite{Aitchison1984} showed that all $\Sigma_m$ are standard with easy framing, and in 2009 Akbulut showed the same is true for hard choice of framing \cite{akbulut2009cappellshanesonhomotopyspheresstandard}. This family of homotopy 4-spheres admits generalization where $U \in SL(3,\mathbb{Z})$ is a more general matrix with the property $\det (U - {\bf 1}) = \pm 1$. Many of the resulting Cappell--Shaneson homotopy 4-spheres were also shown to be standard by Gompf using the ``fishtail surgery trick'' \cite{gompf2010more}, but many candidates still remain open.

Let us consider compactification of the three-form $C_3$ on $S^2\times S^1$ with background $A$ for the Gluck twist (embedded in a $U(1)$ gauge field). We can decompose
\begin{equation}
    C_3=B_3+\left(\omega_1-\frac{A}{2\pi}\right)B_2+  \left(\omega_2- \frac{1}{2}\frac{dA}{2\pi}\right) B_1+\left(\omega_1-\frac{A}{2\pi}\right)\left(\omega_2- \frac{1}{2}\frac{dA}{2\pi}\right)B_0~,
\end{equation}
where $\omega_1,\omega_2$ are the volume forms on $S^1,S^2$, respectively.
For $C_3$ to be closed, the backgrounds satisfy $dB_2=0,dB_1=0$ and
\begin{equation}
    dB_3=\frac{dA}{2\pi}B_2~.
\end{equation}

 \paragraph{Comparison with the $T$ transform on torus.}

 If we think about $S^1 \times S^2 \to [0,1]$ as a $T^2$ fibration over the interval, then it is natural to compare the mapping class group of the central fiber $\text{MCG} (T^2) = SL(2,\mathbb{Z})$ to that of the ambient space $\text{MCG} (S^1 \times S^2) = \mathbb{Z}_2 \oplus \mathbb{Z}_2 \oplus \mathbb{Z}_2$. In particular, $\mathbb{Z} \subset SL(2,\mathbb{Z})$ generated by $T$-transformation via a mod-2 reduction gives $\mathbb{Z}_2$ generated by the Gluck twist. It would be interesting to relate the higher group symmetries in the two cases.

\subsubsection{The Gluck twist as a domain wall}

The element of the mapping class group of $S^1\times S^2$ corresponding to the Gluck twist can be viewed as a domain wall in the theory $T[S^1 \times S^2]$. One interesting question is how to characterize this wall in the quantum field theory language, which would also give us a way of distinguishing it from the trivial domain wall. Recall that for the analogous wall corresponding to the $T$ element of MCG$(T^2)=$
\,SL$(2,\Z)$, the domain wall will carry a Chern--Simons term \cite{Acharya:2001dz,Witten:2003ya}. Similarly, one can ask whether there is any topological term on the ``Gluck wall.'' We will not complete this task in this paper, but instead focus on clarifying one question: is this wall a $\Z_2$ wall (i.e.~two such wall stacked on top of each other would cancel and become the identify wall) given that the Gluck twist is an order-two action? Readers who followed the journey all the way to this point would most likely anticipate that there is some subtlety involved. 

Indeed, as the theory $T[S^1\times S^2]$ (or $T[M]$ in general) depends on the metric of $S^1\times S^2$, and the Gluck twist is only order-2 topologically, there is a potential problem. Namely, if we start with the natural metric compatible with the Cartesian product, after the Gluck twist, the pullback metric is no longer that of a Cartesian product. Instead, a term $d\varphi dx_6$ can be generated in the metric where $x_6$ is the coordinate on $S^1$ while $\varphi$ is the azimuthal angle for $S^2$. Unlike the $T^2$ case where a change of coordinate/basis can cancel this term, for $S^1\times S^2$, this is expected to be a genuinely different metric. So if we define the Gluck twist in this naive way, it won't be a symmetry of $T[S^1\times S^2]$ but will instead change the theory. To make it a symmetry, one should try to either find a way of deforming the metric back to the standard one, or define the Gluck twist differently. Indeed, as the Gluck twist is referring to a mapping class, one needs to fix a representative in order to talk about its action on the space of metrics. But in this kind of situations, as the space of metrics before modding out by diffeomorphisms is contractible, it is more convenient to view the Gluck twist as a loop in the space of metrics quotient by diffeomorphisms. One way to construct such a loop is to form a path between two metrics $g_1$ and $g_2$ which are related by a Gluck twist. This can be done by introducing a $d\varphi dx_6$ term in the metric and tuning its coefficient such that it can be canceled after pullback along the Gluck twist.

Another way, which is expect to be equivalent via a change of coordinates, is to realize the Gluck twist as the following loop on the space of metrics on $S^1\times S^2$ characterized by a twist angle $\delta$ which one uses to identify the two $S^2$ on the two ends of $S^2\times [0,1]$ via a $\delta$ rotation. Reaching $\delta=2\pi$ will give back the metric of the Cartesian product. Defined in this way, can the Gluck twist be a $\Z_2$ action? It turns out that this is still a $\Z$ action, which we will illustrate with the simplest example of free scalars, meaning that we are only looking at fields of the 6d theory that become scalars on $S^1
\times S^2$ after topological twist. 

The KK modes of a complex scalar $\phi$ are characterized by three integral quantum numbers $(l,m,k)$, where $l\ge0$ and $m=-l,\ldots,l$ specify a spherical harmonic function and $k$ is the momentum along $S^1$. So we have
\begin{equation}
    \phi (x_{1,2,3},\theta,\varphi,x_6) =\sum\phi_{lmk}(x_{1,2,3})Y_l^m(\theta,\varphi)e^{2\pi i k x_6/r}.
\end{equation}
Going from $\delta=0$  to $\delta=2\pi$  transforms the modes via
\begin{equation}\label{eqn:ModeShift}
    \phi_{lmk}\rightarrow \phi_{l,m,k-m}.
\end{equation}
Now it is obvious that this is indeed a $\Z$-action. 

But how is this compatible with the Gluck twist being of order two? 
In fact, the $\Z$-action on the level of quantum field theory is not automatically implying that the loop itself can't be of finite order. This is because if a loop (e.g.~obtained by going from $\delta=0$ to $4\pi$) in the space of metrics is trivial, the null homotopy in general can involve metrics that are not twisted products and quantum numbers $(l,m,k)$ stop being relevant in characterizing eigenstates of the Laplacian.  Therefore the square of action \eqref{eqn:ModeShift} 
\begin{equation}
    \phi_{lmk}\rightarrow \phi_{l,m,k-2m}.
\end{equation}
being non-trivial on the modes is not in fact an obstruction for the loop to be of order 2. 

This might be slightly counter-intuitive, as the spectrum gets permuted, with some energy levels crossing each other. Shouldn't such crossing be protected under homotopy? The point is exactly that they are not protected, as one is not able to distinguish real crossing vs ``touching then separating'' using any quantum numbers. As the latter is homotopic to the identity, having such a crossing in the spectrum is not an invariant notion under deformation. In other words, although after going from $\delta=0$ to $4\pi$ (or even $2\pi$), the spectrum appears to have been shifted with various crossings, one could in principle deform this loop in the space of general metrics such that there is in the end no crossings and it becomes just a trivial loop.

After dealing with this subtlety, one can then ask how to detect the Gluck twist from the domain wall. One idea is to look for background terms involving the $SO(3)$ symmetry. It is natural to expect that on the domain wall in certain $T[S^2\times S^1]$, a topological term such as $w_2(SO(3))$, which is indeed of order two, can be generated after the Gluck twist. It would be interesting to understand this more precisely.

Why do we expect that there are any non-trivial effects? Can it be that this loop given by the Gluck twist is just trivial for any $T[S^1\times S^2]$? This is because it can be easily non-trivial in the ``dual'' perspective, when considering the action of the Gluck twist on the $S^1\times S^2$ Hilbert space of a 4d theory.

For example, one can consider the 6d theory on the geometry of $M_2\times S^1\times S^2\times \R$, and the domain wall for the Gluck twist cannot trivial if it acts non-trivially on the Hilbert space of the 6d theory on $M_2\times S^1\times S^2$. But this action can also be detected in the 4d theory $T[M_2]$ as part of the MCG action on the $S^1\times S^2$ Hilbert space.

Indeed, for a state in this Hilbert space generated by a line operator inserted along the core of $S^1\times D^3$ (of which $S^1\times S^2$ is a boundary), the Gluck twist is a change of framing labeled by $\pi_1(SO(3))$ that detect whether this operator is bosonic and fermionic (cf.~Section~\ref{sec:framing}).
Whenever the 4d theory has a fermionic line---which is extremely common and we will give an example next---via changing the order of compactification, this would imply that the Gluck twist gives rise to a non-trivial domain wall in the $T[S^1\times S^3]$ .  

\subsubsection{Example for detecting Gluck twist}

To give one of the simplest example for a 4d theory where the Gluck twist can be detected, we can use (3+1)d $\mathbb{Z}_2$ gauge theory with a fermionic Wilson line on the space of topology $S^1\times S^2$. Since there is only one nontrivial 1-cycle, there are two ground states, or a single logical qubit. The Gluck twist acts on the ground states by some single-qubit gate $U$.

If there is a $\mathbb{Z}_2$ Wilson line on $S^1$, the Gluck twist mapping class group action rotates the framing of the fermion particle by $2\pi$ and thus produces a minus sign. 
Let us choose a basis where the Wilson line is the Pauli $Z$ gate. Then $U$ satisfies
\begin{equation}
    UZ|\Psi\rangle=-ZU|\Psi\rangle~,
\end{equation}
which means that before or after applying the Wilson line operator on a ground state $|\Psi\rangle$ the action of the Gluck twist differs by a sign.
On the other hand, if there is only magnetic flux on $S^2$, and we perform the Gluck twist, nothing happens. The operator that creates magnetic flux is Pauli $X$, and thus $U$ commutes with Pauli $X$. We conclude that the Gluck twist acts on the ground states simply by the Pauli $X$ gate up to a constant phase.

Although this example might seem too simple, it can be a sector of a bigger and more ``realistic'' theory which would exhibit a similar behavior. For example, one can take the 6d $A_1$ $(2,0)$ theory and consider the Spin-$SU(2)$ version of the $T[T^2]$ theory, which similarly have a fermionic Wilson line for the low-energy $SU(2)$ gauge group. In fact, almost all non-trivial interacting theories obtained from 6d known to us have fermionic line operators.

\subsection{Spectrum of operators from torsion reduction}\label{sec:TorsionSpectrum}

So far, it might appear that the discussion about polarizations and symmetries in the presence of torsions is almost completely in parallel with that of the torsion-free case. However, we will encounter one important subtlety caused by torsion as we study the spectrum of charged operators.

\subsubsection{Diagnosing the problem}

The charges of strings in the relative 6d theory are classified by a lattice $\Lambda$ with a Dirac pairing,
\begin{equation}
    \Lambda \times \Lambda \rightarrow \mathbb{Q}/\Z,
\end{equation}
which then leads to the classification by $D=\Lambda/\Lambda^\vee$. One would naively expect that the spectrum for charged operators after compactification is given again by $H_*(M_d,\Lambda)$, and choosing a polarization just amounts to changing the coefficient of some of the homology groups from $\Lambda$ to a sublattice $\Lambda'$ between $\Lambda$ and $\Lambda^\vee$. However, this cannot be correct as this procedure would not change the order of the homology group. For instance, assume that there is a torsion subgroup $\Z_k\subset H_*(M_d,\Z)$, then we have the above procedure generating $\Lambda'\otimes\Z_k$, which is isomorphic to either $\Lambda\otimes\Z_k$ or $\Lambda^\vee\otimes\Z_k$, since they are all free $\Z$-modules of the same rank. This would contradict the expectation that one should get more operators starting with strings charged in $\Lambda$ compared with only those $D$-neutral ones in $\Lambda^\vee$, at least for some polarizations. Notice that this problem only arises for torsion subgroups, since, for the free part, it is perfectly fine to have a sublattice abstractly isomorphic to the lattice containing it.

One can make this point more precise by identifying some operators that are missing from the reduction of the symmetry generators. Namely, from
\begin{equation}\label{DShortExact}
    \Lambda^\vee\rightarrow \Lambda\rightarrow D,
\end{equation}
one has a long exact sequence,
\begin{equation}\label{DLongExact}
    \ldots\rightarrow H_{i+1}(M,D)\rightarrow H_{i}(M,\Lambda^\vee)\rightarrow H_i(M,\Lambda)\rightarrow \ldots
\end{equation}
This being exact prevents one from including the reduction of a three-dimensional symmetry operator on a $(i+1)$-cycle with $D$ coefficient (which is also $(2-i)$-dimensional) as a part of $H_i(M,\Lambda)$. Therefore, the naive procedure is missing something. Notice that this is again a problem that only arises with torsion. 
Operators labeled by elements in $H_{i+1}(M,D)$ that come from the free part of $H_i(M,\Z)$ can end, which is described by the map $H_{i+1}(M,\Lambda)\rightarrow H_{i+1}(M,D)$. For elements associated with torsion, this map is zero, and the next map to $H_{i}(M,\Lambda^\vee)$ enables us to view it as a $(2-i)$-dimensional operator with a charge in $\Lambda^\vee$.

There are several other ways to see that the naive procedure is wrong. One is to realize that the 6d charged objects are not really charged under the lattice $\Lambda$, but a set $\Lambda/W$ formed from a quotient, and ``$H_*(M_d,\Lambda/W)$'' doesn't really make sense unless the $W$ action is compatible with the reduction to $k$-torsion, in which case it becomes $H_*(M_d,\Lambda)/W^N$ for some integer $N$. For 6d $(2,0)$ theories labeled by $\frak{g}$, $\Lambda\simeq\Lambda_{\frak{g}}$ is the weight lattice of $\frak{g}$ and $W$ is the Weyl group. When $\frak{g}=A_{n-1}$, this can also be understood as decoupling the center-of-mass degree of freedom does not always commute with a mod-$k$ reduction. 

Notice that for a $d$-manifold, $H_d$ and $H_{d-1}$ are both free, so this subtlety starts to manifest itself for line operators in $T[M_3]$, but also affects line and point operators in $T[M_4]$, as well as point operators (and, to some extent, the space-filling line operators) in $T[M_5]$. We will try to keep $d$ general first and specify to $d=3$ later.

\subsubsection{Reading between torsion lines}

To understand the spectrum for all polarizations, it is enough to understand it for one, and obtain the others via gauging (which is of course itself a non-trivial procedure and quite case-specific). Without loss of generality, one can consider one $\Z_k$ factor in $H_i(M_d,\Z)$ at a time, which can contribute to both $H^i(M_d,D)$ and $H^{i+1}(M_d,D)$. We will choose $L$ for our pure polarization to stay clear of the $H^{d-i-1}(M_d,D)$ part but to contain the $H^{d-i}(M_d,D)$ part. This will maximize the spectrum for the relevant $(2-i)$-dimensional operators.

The right procedure, for each $\Z_k$ factor, we claim, is not to tensor every term in \eqref{DShortExact} with $\Z_k$ and then take quotient, but instead replace it with
\begin{equation}
    \Lambda_k^\vee\rightarrow \Lambda_k\rightarrow D,
\end{equation}
where $\Lambda_k$ is a truncation of a $W$-quotient of  $\Lambda$ and $\Lambda_k^\vee$ is a subset. For a $(2,0)$ theory, $\Lambda_k$ turns out to be the set of integrable highest weights at level $k$ and $\Lambda_k^\vee$ is the subset that also belongs to the root lattice. 

The set $\Lambda_k^\vee$ classifies charges of operators that come from strings not charged under $D$, and one expects that
\begin{equation}
    \Lambda_k^\vee\simeq (\Lambda^\vee\otimes\Z_k)/W
\end{equation}
agrees with the naive reduction. This is because those strings are not attached to a three-dimensional topological operator and turn out to be unaffected by the subtlety, the reason for which will become gradually clear. For a 6d $(2,0)$ theory labeled by an ADE-type Lie algebra, this indeed gives the elements of the root lattice that lives in the Weyl alcove.\footnote{This follows from the fact that the mod-$k$ reduction with only long roots can be combined with the $W$ action to get the action of the affine Weyl group, $W_{\rm aff}=W\rtimes k\Lambda_{\rm coroot}=W\rtimes k\Lambda^\vee$, for which the level-$k$ Weyl alcove is a fundamental domain. When $\frak{g}$ is not of type ADE, which would be relevant for us later when we discuss the $\hat Z$-invariant, this won't be the case, as the root lattice no longer coincides with the coroot lattice. Also, it would be interesting to study whether $\Lambda_k$ and $\Lambda^\vee_k$ have nice interpretations for various class of 6d $(1,0)$ theories. \label{foot:nonADE}}  On the other hand, 
\begin{equation}
    \Lambda_k=(\Lambda/k\Lambda^\vee)/W
\end{equation} 
is formed by first modding out by $k\Lambda^\vee$ and then quotienting the action of $W$.

How does this solve the various problems we remarked about with the naive reduction? First, this is indeed larger than $\Lambda^\vee\otimes \Z_k$ by a factor of $|D|$, compatible with the expectation from reduction of string operators in 6d. More importantly, it now includes the missing operators, as \eqref{DLongExact} is now replaced, for this $\Z_k$ factor, with a non-exact sequence of maps
\begin{equation}
D\otimes\Z_k\rightarrow\Lambda^\vee_k\rightarrow \Lambda_k.
\end{equation}
As both maps are inclusions, we no longer get zero after composing them. Instead, the reductions of the symmetry operators on torsion cycles are now correctly included in the spectrum. This reduction procedure is illustrated in Figure~\ref{fig:TorsionReduction} and will be revisited in the context of quantum invariants of three-manifold later in Section~\ref{sec:ZhatGeneral}.

\subsubsection{Spectrum of line operators in $T[M_3]$}

A good class of examples to better illustrate this construction is $T[M_3]$.
Some of the simplest examples for an oriented manifold with torsion are lens spaces $M_3=L(k,1)$ with $H_1(M_3,\Z)=\Z_k$. As the description above works for each individual factor in homology, the restriction to $L(k,1)$ is without loss of generality.\footnote{Of course there is additional information carried by the homology, such as the linking pairing, that gives rise to additional structures beyond just the spectrum, such as the braiding between the line operators. We will refrain from discussing these in detail here, although later we will make use of a small part of the braiding information.} We will first consider the 6d $(2,0)$ theory with $\frak{g}=\frak{su}(2)$.  Then the weight lattice is $\Lambda=\Z$, the root lattice $\Lambda^\vee=2\Z$, and the defect group $D=\Z_2$ is also identified with the center of $SU(2)$. We then have 
\begin{equation}
    \Lambda^\vee_k=\Z_k/\Z_2.
\end{equation}
This is a set of $\lfloor\frac{k}{2}\rfloor+1$ elements which can be labeled as $\left\{0,2,\ldots,2\lfloor\frac{k}{2}\rfloor\right\}$. On the other hand,  
\begin{equation}
    \Lambda_k=\Z_{2k}/\Z_2\simeq \{0,1,\ldots,k\}
\end{equation}
is a set of $k+1$ elements. This exactly coincides with the set of independent Wilson/vortex lines in the $T[L(k,1)]$ theory, which, at low energy, is a 3d $\CN=2$ $SU(2)_k$ Chern--Simons theory with adjoint matter. The subset $\Lambda^\vee_k$ are Wilson lines with trivial $\Z_2$ 1-form charge. When $k$ is even, one can gauge a $\Z_2$ 1-form symmetry generated by the $k$-th line to arrive at the $SO(3)_{k/2}$ theory. However, notice that the spectrum of the new theory is not $\Lambda^\vee_k$, as one also needs to consider the fusion with the 1-form symmetry generator \cite{MOORE1989422}. It is better to view $\Lambda^\vee_k$ as a intermediate step from where the spectrum for any given polarization can be more easily constructed, without itself being associated with any particular polarization. 

\begin{figure}[htb!]
    \centering
\includegraphics[width=0.8\linewidth]{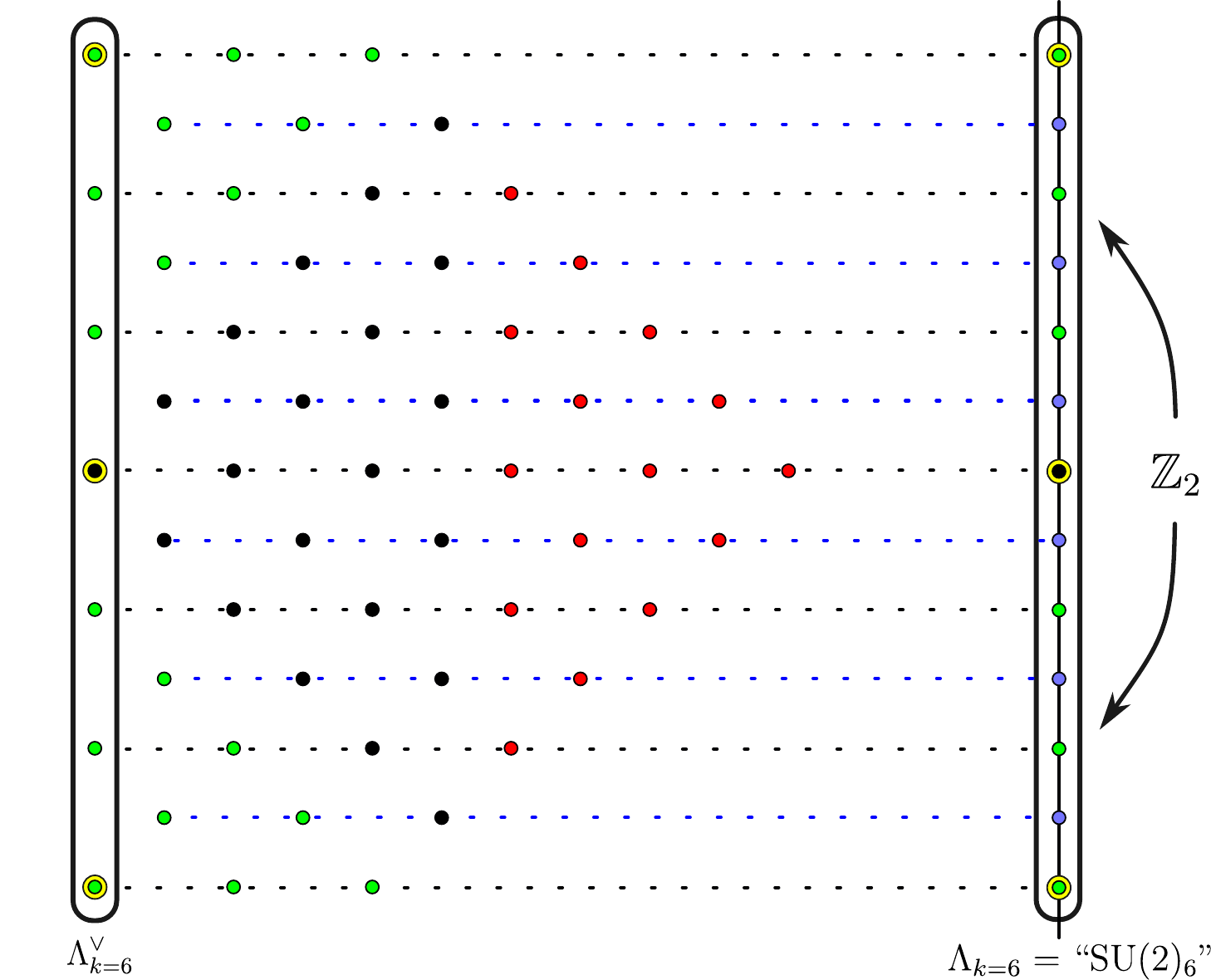}
    \caption{An illustration of $\Lambda^\vee_k$ and $\Lambda_k$ for the case of $k=6$ and $\Lambda$ being the weight lattice of SU$(2)$. The black and red dots form a $\Z_k\times \Z_k$, which is a fundamental domain of the two shift-by-$k$ actions on $\Z\times\Z$. For either the embedding of $\Lambda^\vee_k$ or the projection to $\Lambda_k$, it is more natural to choose a different fundamental domain given by the black and green dots. The projection to the line on the right then gives the set of charges for the $(2-i)$-dimensional operators coming from a $\Z_k$ factor in $H_i(M_d,\Z)$ in the ``SU(2) version'' of the theory. It contains charges (green dots plus one black dot for the zero charge) shared with the ``SO(3) version'' of the theory, and some additional ones (blue dots). A double circle with a yellow ring denotes charges that can be lifted to $H_{i+1}(M_d,D)$, which contains topological operators. The $\Z_2$ Weyl group acts by flipping this picture horizontally.  }
    \label{fig:TorsionSpectrum}
\end{figure}

As mentioned before, for the $A_{n-1}$ series, one can arrive at the same conclusion by ``decoupling the center.'' This is because, for the $U(n)$ theory, the Weyl group is actually compatible with the mod-$k$ reduction, as it acts only by permutation. For $n=2$, from a $k$-torsion cycle, we first get $\Z_k\oplus \Z_k$. Then there are two ways of removing a $\Z_k$ factor. One is by taking the ``off-diagonal'' $\Z_k$ generated by $(1,-1)$. This is in fact the mod-$k$ reduction of the embedding of the character lattice for $SO(3)$ into that of $U(2)$, and it is perfectly compatible with the $\Z_2$ Weyl group symmetry. This is illustrated for $k=6$ in Figure~\ref{fig:TorsionSpectrum}. On the contrary, for the ``$SU(2)$ version,'' there is not such an embedding, but instead a projection from the character lattice of $U(2)$ to that of $SU(2)$,
\begin{equation}
    \Z\oplus \Z\rightarrow \Z
\end{equation}
given by $(a,b)\mapsto a-b.$ Then the question is how to perform a mod-$k$ reduction of it. The naive way corresponds to tensoring $\Z_k$ to both sides to arrive at 
\begin{equation}
    \Z_k\oplus \Z_k\rightarrow \Z_k.
\end{equation}
As explained before, this has multiple problems, including having the wrong cardinality and missing some operators. The latter issue can now be understood from the geometry of M2-branes ending on two M5-branes. 

To obtain line operators in $T[M_3]$, the boundaries of M2-branes will be 1-dimensional along $M_3$. If the boundary wraps a $k$-torsion cycle $\gamma$ $a$ times along the first M5 and $b$ times along the second, we say that it has charge $(a,b)$. The set $\Lambda^\vee_k$ is then obtained by forcing $a=-b$. As the root of $
\frak{su}(2)$ is given by $(1,-1)$, we see that $\Lambda^\vee_k$ is indeed the level-$k$ truncation of the root lattice. For the ``$SU(2)$ version'' of decoupling the center, one allows $(a,b)$ to be arbitrary, but only keeps track of the difference $a-b$. Now we consider an M2-brane of charge $(k,0)$. It is tempting to say that, as $k\gamma$ is null homologous,  one can just shrink it and detach the M2-brane. However, the ``movie'' for shrinking $k\gamma$ traces out a 2-cycle with $\Z_k$ coefficient, and after that, the M2-brane will wrap this 2-cycle.\footnote{Notice that, although this torsion 2-cycle is represented by an open manifold with boundary $k\gamma$ in $M_3$, the M2-brane we started with also has a bulk part ending on $k\gamma$. Hence the brane now wraps a closed 2-cycle in the full geometry.} This is a non-trivial topological line operator that can have braiding phases with other operators, which cannot be eliminated, even if one pushes the M2-brane away from the M5-branes, as it originates from the topological term $\int C\wedge G\wedge G$ in the M-theory bulk. There is, however, no problem identifying it with $(-k,0)$, by first shifting $(-k,0)$ to $(0,k)$ via a center-of-mass degree of freedom, and then using the Weyl symmetry swapping $a$ and $b$. The latter can be realized as a geometric process by letting the M2-brane touch the other M5-brane, first along a small bounding curve, but then follow the ``reversed movie'' to let the boundary wrap $\gamma$ $k$ times along the other M5-brane.
 This tells us that we have equivalences of the kind $(k,0)\sim (-k,0)$ but not with $(0,0)$, which, in a sense, doubles the periodicity to $2k$. When $k$ is even, one can also label the charge of the topological operator as $(\frac{k}2,-\frac k2)$ up to a center-of-mass shift, and it is now also in the root lattice.  

Then we see that tensoring with $\Z_k$ indeed kills more charges by identifying physically distinct ones. It would identify $(0,0)$ with $(0,k)$ and $(k,0)$, and, by linearity, $(a,b)$ with $(k+a,b)\sim(a,k+b)$, reducing the total number of distinct charges by a factor of 2. There are several equivalent prescriptions to fix this problem. One is by demanding that the representative $(m,n)$ in $\Z\times \Z$ for any element of $\Z_k\times \Z_k$ is chosen to satisfy $m+n\in [0,k)$. Then $m-n$ is well defined mod $2k$, as one can no longer simply shift $m$ by $k$, but have to compensate with a $-k$ shift for $n$. This gives the desired map
\begin{equation}\label{SU2Map}
    \Z_k\oplus \Z_k\rightarrow \Z_{2k}.
\end{equation}
This is not a group homomorphism but a map between sets, which should not worry us for at least two reasons. First of all, we will quotient both sides with the $\Z_2$ action to arrive at a map between sets anyway,
\begin{equation}
    (\Z_k\oplus \Z_k)/\Z_2\rightarrow \Lambda_k\simeq \Z_{2k}/\Z_2.
\end{equation}
Secondly, the charges of line operators generally won't form a group due to the fusion rules being non-Abelian. The projection \eqref{SU2Map} amounts to choosing a different fundamental domain for the $\Z\times\Z$ under the action of shifting by $k$. This is illustrated in Figure~\ref{fig:TorsionSpectrum} for $k=6$. For other even values of $k$, the picture would be completely analogous. For an odd $k$, $\Lambda^\vee_k$ will not contain the symmetry generator. This is when the $\Z_2$ 1-form symmetry is anomalous and can't be gauged.\footnote{Notice that for $T[L(k,1)\times S^1]$, the anomalous 1-form symmetry in $T[L(k,1)]$ will become a one-form and a zero-form symmetry which are themselves non-anomalous but have mixed 't Hooft anomaly. Therefore, one can gauge one of them and realize the spectrum of the ``$SO(3)$ version'' in 2d. In particular, if one just gauge the 1-form symmetry, the spectrum of point operator is given by $\Lambda^\vee_k$. However, from the point of view of the 6d reduction, both symmetries are accidental for odd $k$, and their gauging cannot be done via changing the polarization. This is a good example in which not all global forms of $T[M_d]$ come from choices of polarization. See \cite{Gukov:2015sna,Gukov:2021swm,Gu:2025gtb} for more detailed discussions about symmetries, anomalies, and the spectrum of operators in these theories.}

For general $\frak{g}$, the story is very similar. (See \cite{Gukov:2016lki} for related discussions about some aspects of the $\frak{g}=A_{n-1}$ case.) On a three-manifold $M_3$, one can put together all the subgroups of $H_1(M_3,\Z)$ to arrive at $\Lambda_{M_3}$---the set of charges for line operators in $T[M_3]$. It fits in a sequence of maps of sets,
\begin{equation}\label{LambdaM3}
    H_2(M_3,D)\rightarrow \Lambda^\vee_{M_3}\simeq H_1(M_3,\Lambda^\vee)/W^N \rightarrow \Lambda_{M_3}.
\end{equation}
For the middle term, the quotient is by the action of copies of $W$, one for each $\Z_k$ or $\Z$ subgroup of $H_1(M_3,\Z)$.

A choice of polarization would determine a subgroup $L^{(2)}$ of $H^2(M_3,D)$ and $L^{(1)}$ of $H^1(M_3,D)$, with the condition that they pair trivially with each other and are maximal subgroups with this property. The image of $L^{(1)}$ under \eqref{LambdaM3} to $\Lambda_{M_3}$ are charges for topological line operators that generate a $(L^\vee)^{(1)}=H^1(M_3,D)/L^{(1)}$ 1-form symmetry. One should gauge this symmetry by condensing these line operators. The remaining line operators will be coming from those that braid trivially with the symmetry generators. Their charges give a subset $\Lambda_{M_3,L}^\vee$ between $\Lambda_{M_3}^\vee$ and $\Lambda_{M_3}$. This is also the pre-image of $\Lambda_{M_3}\rightarrow L^{(2)}\subset H^2(M_3,D)\simeq H_1(M_3,D)$.  Another way to think about this set is by looking at each $\Z_k$ factor, 
\begin{equation}
    \Lambda_k^\vee\rightarrow \Lambda_k\rightarrow D.
\end{equation}
Then $D\otimes\Z_k$, viewed as a subgroup of $H_1(M_3,D)$, can pair with $L^{(1)}$. The kernel for this pairing is lifted to a subset $\Lambda_{k,L}^\vee\subset\Lambda_k$. 
To obtain $\Lambda_{M_3,L}$ which describe the spectrum of line operators in $T[M_3,\CP_L]$, one also needs to take into account of the fusion with the 1-form generators, identifying lines related by fusion, but also introduce twisted lines when a line is fixed under fusion. Patching these together for all subgroups of $H_1(M_3,\Z)$ gives $\Lambda_{M_3,L}$ for the polarization $L$ on the three manifold $M_3$.

The reason for having this two-step process, from the point of view of polarization, is that one cannot change just the $L^{(1)}$ or $L^{(2)}$ part by themselves, but has to modify them together to make sure that $L$ is ``maximal isotropic.'' Also, this is compatible with the prescription for gauging 1-form symmetry in 3d theories \cite{MOORE1989422}. In contrast, if we had $d=4$, then $L^{(2)}$ and $L^{(3)}$ could be chosen independently, and both $\Lambda_{k,L}^\vee$ and $\Lambda_{k,L}$ (and one more by only gauging the $(L^\vee)^{(2)}$ 0-form symmetry in $T[M_4]$) could be realized by physical theories.

For mixed polarizations or different quadratic refinements associated with the same choice of $L$, additional topological terms need to be included when gauging. We will not attempt to classify these here. Given a specific term, it is usually straightforward to determine the spectrum of line operators $\Lambda_{M_3,\CP}$ after gauging.

\subsection{Symmetries and the $\hat Z$ invariants} \label{sec:ZhatSymmetry}

In this subsection, we study aspects of quantum invariants of three-manifolds, as an application (and illustration) of how polarizations, symmetries, and the spectrum of operators in our previous discussion manifest themselves in quantum topology. We focus on the $\hat Z$ (a.k.a.~GPPV) invariants of 3-manifolds \cite{Gukov:2016gkn,Gukov:2017kmk}, but the reader should keep in mind that many statements, stemming from symmetries of $T[M_3]$, also apply to other quantum invariants whose definition can be expressed in terms of the $T[M_3]$ theory.

\subsubsection{$\hat{Z}$ and non-invertible symmetries}

There are some long-standing questions about $\hat{Z}_a$ since its invention: How can one understand or characterize the label $a$ in the infrared description of the 3d $\CN=2$ theory $T[M_3]$? Can it always be realized as a boundary condition of $T[M_3]$ on $D^2\times S^1$? And for $M_3$ such that all $a$ can be realized as physical boundary conditions, how can one canonically find these boundary conditions? 

We will reflect on these questions from the perspective of symmetry, which helps to relate the UV or M-theory definition with the description in the low-energy effective theory.

We first recall the meaning of $a$ in the M-theory definition of $\hat{Z}_a$. The invariant $\hat{Z}$ counts certain M2-branes ending on $N$ M5-branes supported on $\R^2\times S^1 \times M_3$. To contribute to a $\hat{Z}$, an M2-brane will wrap $S^1 \times \gamma$, where $\gamma\subset M_3$ is a curve in the 3-manifold $M_3$. Then naturally one gets a label $a$ recording the homology class of $\gamma$ in $H_1(M_3,\Z^N)$. If one decouples the center-of-mass motion of the $N$ M5-branes, $a$ is then valued in $H_1(M_3,\Lambda^\vee)$, with $\Lambda^\vee$ the root lattice of $SU(N)$ \cite{Gukov:2017kmk}.\footnote{Ideally, one should divide by copies of the Weyl group, one for each ``block'' of the linking form. However, it is often convenient not to take the quotient and let them act as automorphisms of the Abelian group $H_1(M_3,\Lambda^\vee)$. Notice that in the convention of \cite{Gukov:2017kmk}, only one ``global copy'' is modded out.}

From the point of view of polarization, we know that there are in fact different ways of decoupling the center-of-mass motion, corresponding to replacing $\Lambda^\vee$ with another $\Lambda'\subset\Lambda$. In fact, the minimal and most universal (e.g.~works for any choices of $M_3$) is to take $\Lambda$ itself. Then the set of charges is $\Lambda_{M_3}$ that includes $\Lambda_{M_3}^\vee$ as a subset (cf.~equation \eqref{LambdaM3}).\footnote{Notice that the notation $\Lambda_{M_3}$ was also used in \cite{Gukov:2017kmk}, which, in our notation here, is actually the smaller set $\Lambda_{M_3}^\vee$ (up to action of copies of the Weyl group).} However, as operators in the sector $a\in \Lambda_{M_3}\backslash\Lambda_{M_3}^\vee$ carry charges under the symmetry $H^1(M,D)$, $\hat Z_a$ vanishes due to the selection rule associated with this symmetry. This justifies considering only these $a\in \Lambda_{M_3}^\vee$. In fact, to have a BPS state of finite energy, one should additionally require $a$ to be a torsion element. The vanishing of $\hat Z_a$ for everything outside of Tor$\, \Lambda_{M_3}^\vee$ imposes a collection of constraints on the $T[M_3]$ theory.

Now that we identify $a$ as the label for charge sectors for line operators in $T[M_3]$, we want to understand how it can be realized as a boundary condition. A simpler question is the dimensional reduction of this one, namely how to find a vector $v_a$ in the Hilbert space $\CH_{T[M_3]}(T^2)$ on $T^2$ viewed as the boundary of the spacetime $D^2\times S^1$. Notice that if the theory were topological, this would be a much simpler task, as the states in the $T^2$ Hilbert space can then be labeled by line operators and hence $a$.\footnote{One doesn't have to assume that there is a unique or distinguished line for each element in $\Lambda_{M_3}$. When there are multiple line operators, one can simply sum over them. }

If, after deforming to a gapped phase, a set of line operators $\Lambda_{M_3}$ survives and becomes topological in the infrared, then $\Lambda_{M_3}$ is realized as a non-invertible symmetry in the IR. From this perspective, one can view the difficulty of defining the analogue of $v_a$ in the theory $T[M_3]$ as being related to the breaking of this non-invertible symmetry. Such a breaking can be understood as a consequence of having modes from the motion of M5-branes mixing with modes from M2-branes, and in general there is no way of effectively telling them apart. In other words, if one can freeze all the M5 modes, the remaining theory will in general have a larger non-invertible symmetry generated by topological lines labeled by (the torsion part of) $\Lambda_{M_3}$. However, turning on the M5-brane modes generally breaks this symmetry, making it difficult to define $v_a$ in the $T[M_3]$ theory. 

This gives a conceptual explanation for the following observation. For $M_3$ which we know about the set of boundary conditions $\{B_a\}$ of $T[M_3]$ (e.g.~$L(k,1)$ and a degree-$k$ circle bundle over a Riemann surface discussed in \cite{Gukov:2017kmk}), the theory always admits a deformation (e.g.~via turning on mass parameters associated with global symmetries) to a gapped phase. This perspective also instructs us that, to find the set of boundary conditions $\{B_a\}$, one should look for deformations that kill the M5 degrees of freedom. In fact, as we only need $v_a$ to compute $\hat Z_a$, a similar deformation for $T[M_3\times S^1]$ or $T[M_3\times T^2]$ would work equally well.

\subsubsection*{Example with $T[L(k,1)]$ }

We now illustrate the discussion above in a concrete example where $M_3=L(k,1)$ is a lens space. 

In this case $T[M_3]$ is an $SU(N)_k$ Chern--Simons theory with adjoint matter. The adjoint matter comes from the motion of M5-branes, and if we freeze it (e.g.~by turning on a mass associated with a $U(1)$ global symmetry), we indeed get a topological theory. The Wilson lines of this topological theory, labeled by level-$k$ representations of $SU(N)$, generate a non-invertible symmetry. 

However, if we unfreeze the adjoint chiral multiplet, it will break this non-invertible symmetry down to a ``center symmetry'' $\Z_N\otimes\Z_k$, as vortex lines valued in the center of the group cannot be screened by the adjoint matter.

This is the $H^1(M_3,D)$ symmetry that is generally present in $T[M_3]$ with a suitable polarization, as we have discussed in the previous subsection. Next, we will turn to this remaining symmetry and explore its role in the study of $\hat{Z}$.

\subsubsection{$\hat{Z}$ and generalized symmetries}\label{sec:ZhatGeneral}

We start with the polarization such that $T[M_3]$ has as much 1-form symmetry as possible. This can be achieved with the geometric polarization given by a 2-handlebody $W_4$, which picks an $L\subset H^*(M_3,D)$ that completely spares the $H^1$ part. Then the theory has an anomaly-free 1-form symmetry given by $H^1(M_3,D)$. This, in general, looks like a quotient of copies of $D$. The $a$-label for $\hat Z_a$ then takes values in $\mathrm{Tor}H_1(M_3,\Lambda^\vee)$ with $\Lambda^\vee$ the root lattice of the simply-connected gauge group $G$.\footnote{From the perspective of this paper, the torsion subgroup is not very special and one can certainly talk about operators in $T[M_3]$ charged under the non-torsion part. This is also an interesting subject in the study of $\hat Z$. See e.g.~\cite{Costantino:2021yfd} for discussion about the physics of $\hat Z_a$ with $a$ being non-torsion. Still, the situation with $a$ taking values in the torsion part is better understood and what we will focus on in this subsection.}  Pick a $\gamma\in H^1(M_3,D)$. This is a generator of 1-form symmetry that acts non-trivially on $\Lambda_{M_3}$ but trivially on $\Lambda_{M_3}^\vee$. However, there is another (``S-dual'') action by fusion, as the generator is also a line operator. This gives rise to the shift symmetry,
\begin{equation}\label{ShiftSym}
    \hat Z_a=\hat Z_{\text{``$a+\gamma$''}},
\end{equation}
except that one needs to be more careful about the meaning of $\gamma+a$. 

As before, using the connecting / Bockstein homomorphism associated with $\Lambda^\vee\rightarrow\Lambda\rightarrow D$, we have
\begin{equation}\label{gammaAction}
    H^1(M_3,D)\rightarrow H^2(M_3,\Lambda^\vee)\simeq H_1(M_3,\Lambda^\vee).
\end{equation}
As $D$ is finite, the image is in the torsion part, enabling us to interpret $\gamma$ as an element in Tor$H_1(M_3,\Lambda^\vee)$. This is also injective from the ``torsion part'' of $H^1(M_3,D)$ given by the co-kernel of $H^1(M_3,\Lambda)\rightarrow H^1(M_3,D)$. As a consequence, the action of $\gamma$ is free as long as it is non-trivial in the co-kernel. Composing with the map $\Lambda^\vee_{M_3}\rightarrow\Lambda_{M_3}$, one also gets the action of $\gamma$ on general line operators.  

The only piece remaining needed to make sense of \eqref{gammaAction} is to verify that it is compatible with the $W$ action. Namely, for two different representatives $b,w\cdot b\in\Lambda_{M_3}$ related by a Weyl group action, $b+\gamma$ and $w\cdot b+\gamma$ are also related by a Weyl group action. This can be check for each $\Z_k$ factor, which follows from a fact,
\begin{equation}
    w\cdot \gamma-\gamma \equiv 0 \in \Lambda_k.
\end{equation}
This relation, via the exponential map, can be interpreted as the conjugation action of the Weyl group becoming trivial on the center of the group.

How does our theoretical prediction \eqref{ShiftSym} compare with ``experiments''? Indeed, this symmetry can be verified in various cases where the computation for $\hat Z$ is available. For example, for a class of Seifert manifolds, Chung observed a ``center symmetry'' for $\hat Z$ \cite{Chung:2019jgw}, which exactly comes from this shift symmetry. 

This discussion is quite general and applies not only to the 3d theory $T[M_3]$ obtained from the reduction of a 6d theory, but also to the 2d theory from a reduction of a 5d theory, which is enough to define $\hat Z$ as the geometry with the 6d theory contains a circle anyway. If the 5d theory is the $\CN=2$ super--Yang--Mills theory with a---not necessarily simply-laced---gauge group $G$, one can just replace $\Lambda^\vee$ by the root lattice and $\Lambda$ by the weight lattice. The 1-form symmetry generator will be 0-dimensional, which acts on other point operators. However, one does need to deal with a subtlety mentioned in Footnote~\ref{foot:nonADE}.

The symmetry generator labeled by $\gamma$ can be interpreted as an 't Hooft flux (e.g.~vortex line with flux valued in the center of the group). To make this more explicit, again consider part of the long exact sequence in homology around the connecting map $\delta$,
\begin{equation}
\ldots\rightarrow H_2(M_3,\Lambda)\rightarrow H_2(M_3,D)\stackrel{\delta}\rightarrow H_1(M_3,\Lambda^\vee)\rightarrow\ldots,
\end{equation}
whose physical interpretation we now elaborate. 

\begin{figure}[htb!]
    \centering
\includegraphics[width=\linewidth]{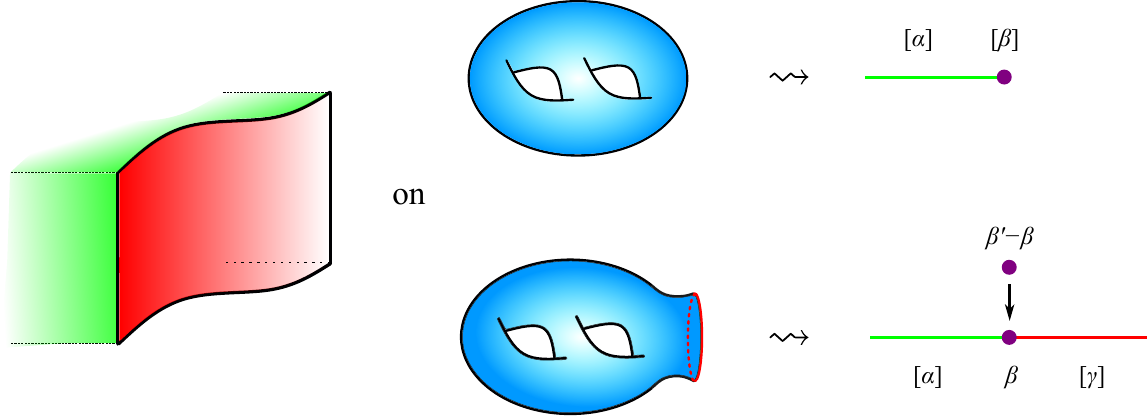}
    \caption{The reduction of string operators on a 2-cycle $[\alpha]\in H_2(M_d,D)$. On the left is an illustration of a string in the 6d theory living on the boundary of a three-dimensional topological operator. In the middle, two types of 2-cycles with $D$-coefficients are illustrated. The compactification on a geometrically closed one (upper) leads to a point operator attached to a topological line, while the compactification on a geometrically open cycle (lower) leads to a junction between a topological line operator and a ``physical'' line operator obtained from wrapping the string on the boundary cycle (red). In both scenarios, the point operator depends on a lift of $[\alpha]$, to either $[\beta]\in H_2(M_d,\Lambda)$ or $\beta\in C_2(M_d,\Lambda)$, but $[\gamma]\in H_1(M_d,\Lambda)$ won't depend on this choice. The difference between two lifts gives a genuine point operator, which can be used to modify the junction without changing the two line operators.}
    \label{fig:TorsionReduction}
\end{figure}

A $D$-valued 2-cycle in the middle term $H_2(M_3,D)$ can be geometrically represented by a 2-chain $\alpha$ with a boundary $\partial \alpha$ that is zero with $D$-coefficient. When it is actually closed, compactifying a string attached to a three-dimensional topological operator on $\alpha$ leads to a point operator labeled by a lift $[\beta]$ of $[\alpha]$ in $H_2(M_3,\Lambda)$ (modulo the $W$-action) attached to a topological line labeled by $[\alpha]\in H_2(M_3,D)$. However, when $\alpha$ is not closed geometrically, the boundary can be lifted to become a 1-chain $\gamma$ with coefficient in $\Lambda^\vee$. This is via the usual procedure (e.g.~in the context of the snake lemma): One first lifts $\alpha$ to a 2-chain with $\Lambda$-coefficient $\beta\in C_2(M_3,\Lambda)$, then the boundary $\partial \beta\in C_1(M_3,\Lambda)$ is a closed 1-chain with $\Lambda$-coefficient; since it becomes zero when reduced to $D$-coefficient, it must lift to a 1-chain $\gamma$ with $\Lambda^\vee$-coefficient. This turns out to be well defined once we pass on to homology. Wrapping a string along the boundary $\partial \alpha$ (or more precisely the lift $\partial \beta$) of this open 2-chain then leads to a line operator, labeled by $[\gamma]\in H_1(M_3,\Lambda^\vee)$. As it is with $\Lambda^\vee$-coefficient, this line operator is not on the boundary of a non-trivial topological surface operator, and since the cycle $\partial \beta$ is a boundary, this line operator can end, with the end point labeled by the (non-unique) lift $\beta\in C_2(M_3,\Lambda)$ of $\alpha$. But this is actually a junction because the string is attached to a three-dimensional symmetry generator, which, after compactifying on $\alpha$, becomes a topological line labeled by the element $[\alpha]\in H_2(M_3,D)$. This is illustrated in Figure~\ref{fig:TorsionReduction}. The existence of such a junction tells us that the topological line operator $\alpha$ actually lives in the charge sector $[\gamma]=\delta([\alpha])$, and this is the physical interpretation of the connecting morphism $\delta$. The uniqueness of $[\gamma]$ can be understood physically as follows. The only non-unique choice made in the construction is the lift $\beta$ of $\alpha$. For a different choice $\beta'$, we then have $\beta-\beta'\in C_2(M_3,\Lambda^\vee)$. This is a point operator with no topological lines attached, and stacking it to the junction won't change the line operators on either side of the junction.

The topological line operator $[\gamma]$ has the same source and similar behavior as the 't Hooft flux (more precisely its Poincaré dual)---the generator for the 1-form electric symmetry---in the 5d/4d theory $T[S^1]$ and $T[T^2]$, as they are all obtained from the compactification of the 3d topological operator in the 6d theory to a codimension-2 operator. 

\subsubsection{Polarization and $\hat Z$} 

How does changing the polarization affect $\hat Z$? If we focus again on the $H^{1}$ and $H^2$ part and only on pure polarizations, at the level of the theory $T[M_3]$, this amounts to gauging part of the anomaly-free 1-form $H^1(M_3,D)$ symmetry. This changes the spectrum of line operators via the following process \cite{MOORE1989422}. One first eliminates all lines charged under this 1-form symmetry. These are lines that braid non-trivially with the generators, and are captured by the map
\begin{equation}
    \Lambda_{M_3}\rightarrow H_{1}(M_3,D).
\end{equation}
Physically, $H_1(M_3,D)$ describes the topological surface operators that are attached to the line operators in the relative theory. They are condensed for the polarization with $L^{(2)}\simeq H^2(M_3,D)\simeq H_1(M_3,D)$, but once we choose a smaller subgroup $L'^{(2)}\subset H^2(M_3,D)$, only the pre-image of $L'^{(2)}$ are charges of genuine line operators. For a line operator outside the pre-image, it only lives on the boundary of a topological surface operator. Another equivalent way to think about this is that this map can be used to remember the braiding between lines in $\Lambda_{M_3}$ with the central lines, which are in the image of 
\begin{equation}
     H_{2}(M_3,D)\rightarrow\Lambda_{M_3},
\end{equation}
via the ``intersection pairing''
\begin{equation}
    H_1(M_3,D)\times H_{2}(M_3,D)\rightarrow \mathbb{Q}/\Z.
\end{equation}
The relevant part of this actually originates from the linking pairing on $H_1(M_3,\Z)$.

After taking these neutral lines, one also needs to identify line operators related by the fusion action of the 1-form generators. This amounts to choosing an $L^{(1)}\subset H^1(M_3,D)\simeq H_2(M_2,D)$ and quotienting $\Lambda_{M_3}$ by its image. However, as $\Lambda_{M_3}$ is not a group but just a set, the action can have fixed points,
and the third step is to add the ``twisted lines,'' according to the fixed points. One such line $(\alpha,\gamma)$ is needed when there is a $\gamma \in H_2(M_2,D)$ that fixes an $\alpha \in \Lambda_{M_3}$ via $\gamma + \alpha =w\cdot \alpha$ for some $w\in W^N$. These are line operators that will be charged under the $(L^\vee)^{(2)}\simeq H^2(M_3,D)/L^{(2)}$ 0-form symmetry.

In the end, one arrives at a set $\Lambda_{M_3,L}$ that describes the spectrum of line operators in the theory with the polarization given by $L$. We now examine what happens to $\hat Z$ in this procedure.

First of all, the label $a$ will take values in a new set, namely $\Lambda_{M_3,L}$, that is, apriori, quite different from $\Lambda_{M_3}$ after the these steps. However, if we examine the effect of each step, we find that it almost has no effect. The first step is to take a subset of $\Lambda_{M_3}$, but this subset still contains $\Lambda_{M_3}^\vee$---a quotient of $H_1(M_3,\Lambda^\vee)$---which are these values of $a$ such that $\hat Z$ is non-zero. If we restrict to this set, the first step won't affect us. The second step would lead to identifying different labels $\alpha\sim\alpha+\gamma$ via the shift action. However, as $\hat Z_{a}=\hat Z_{a+\gamma}$, it is rather straightforward to go between the original set $H_1(M_3,\Lambda^\vee)$ and the new one given by $H_1(M_3,\Lambda^\vee)/H_2(M_3,D)$, and is in a sense just a choice of convention, similar to whether one wants to quotient by (copies of) $W$. For the third step, the twisted lines are related by symmetry, and different labels should give the same value $\hat Z_{(\alpha,\gamma)}=\hat Z_{(\alpha,\gamma')}$, and it is not necessary to keep $\gamma$ as part of the label. Therefore, we see that even after changing the polarization, one can keep using the same set of labels. 

Then one can ask about the actual values of $\hat Z_a$. As gauging the 1-form symmetry amount to summing over different values of the background gauge field, in this geometry, one only need to consider different fluxes over the $D^2$. This is the same as inserting the topological line defect along $S^1$. But this is the same as the shift symmetry acting on the label $a$. 

Therefore, we conclude that changing between pure polarizations at most changes $\hat Z$ by a normalization. This agrees with observations in the literature \cite{Chauhan:2022cni,Chauhan:2023uac}.

There are still some interesting questions that one can explore. One is to include insertion of line operators as in \cite{Gukov:2017kmk} (see also \cite{Pei:2026lnl}). Given that the spectra in different theories are different, it is not obvious that they contain the same information. In particular, the insertion of a ``twisted lines'' might be interesting to consider. This, in the original theory, corresponds to having non-trivial background $\gamma$ for the 1-form symmetry along the boundary $T^2$, which is emitted from a line operator with charge $\alpha$ in the bulk with $\gamma+\alpha=w\cdot \alpha$.

We now continue to study some related aspects of $\hat Z$.

\subsection{$\hat Z$ and the volume conjecture}\label{sec:volume}

The half-index of $T[M_3]$ with different 2d $(0,2)$ boundary conditions leads to different topological invariants of $M_3$. Here we consider a class of boundary conditions that are naturally associated to roots of unity and lead to general Reshetikhin--Turaev invariants of $M_3$, beyond the most familiar ones at $q = e^{2 \pi i /k}$ related to Chern--Simons theory with compact gauge group at level $k$.

The calculations below suggest that, for $r$ odd,
\begin{equation}
\hat Z_b (q^2) \Big|_{q \to e^{2\pi i /r}}
\end{equation}
exhibits exponential growth, controlled by the hyperbolic volume. This is rather peculiar, considering that at primitive roots of unity $\hat Z$-invariants often enjoy nice finite limits.\footnote{It should be noted that since the early days of $\hat Z$-invariants it was observed \cite{Cheng:2018vpl} that not only individual $\hat Z_a (M_3,q)$ may have singular radial limits, but also their linear combinations $\sum_a c_a^{\text{WRT}} \hat Z_a (M_3, q)$ that at some roots of unity give correct values of WRT invariants, at other roots of unity behave as $\exp \left( - \frac{4 \pi^2}{\hbar} \big( \text{CS} (\alpha_*) - m_* \big) \right)$. This behavior can be easily understood with the tools of resurgent analysis and leads to a relation
\begin{equation}
\text{WRT} (M_3, k) \; = \; \left( \sum_a c_a^{\text{WRT}} \hat Z_a (M_3, q) - P_k (q) \right) \Big|_{q \to e^{2 \pi i / k}}
\label{WRTcorrections}
\end{equation}
where the ``correction terms'' $P_k (q)$ are determined by complex Chern-Simons values and the (K-theoretic) Stokes coefficients \cite{Gukov:2023cog,Costin:2023kla}. A simple example is a Seifert manifold $M_3 = - M(-2; \tfrac{1}{2}, \tfrac{1}{3}, \tfrac{1}{2})$, for which one finds vanishing correction terms at all roots of unity of odd order \cite{Cheng:2018vpl}, while for roots of unity of event order eq.~(\ref{WRTcorrections}) holds with the correction terms $P_k (q) = (-1)^{k+1} \frac{\eta^3 (q)}{\eta^2 (q^2)}$.}

In order to apply the Gauss sum reciprocity, it is important to carefully fix the conventions and normalizations.
The conventions used in \cite{Gukov:2017kmk,Gukov:2020frk} are such that the $n$-colored Jones polynomial of the unknot with framing $p$ is
\begin{equation}
J_n (\text{unknot}) \; =\; q^{\frac{p(n^2-1)}{4}}\,\frac{q^\frac{n}{2}-q^{-\frac{n}{2}}}{q^\frac{1}{2}-q^{-\frac{1}{2}}}
\end{equation}
where
\begin{equation}
q = e^{2\pi i / k}.
\end{equation}
More generally, for a link $\CL$ constructed from a plumbing graph $\Gamma$, we have
\begin{multline}
J_{\vec n} (\CL(\Gamma)) \; = \; \frac{1}{q^{1/2}-q^{-1/2}}\prod_{I\;\in\; \text{Vert}}
q^{\frac{a_I(n_I^2-1)}{4}}
\,\left(\frac{1}{q^{n_I/2}-q^{-n_I/2}}\right)^{\text{deg}(I)-1}
\times \\
\prod_{(I,J)\;\in\;\text{Edges}}
(q^{n_{I}n_{J}/2}-q^{-n_{I}n_{J}/2}).
\qquad\qquad
\label{Jones-tree}
\end{multline}
Here, Vert and Edges are the set of vertices and edges of the graph $\Gamma$, deg$(I)$ denotes the valency of the vertex $I$, and $a_I,n_I\in\Z$ are respectively the framing and coloring of the component of the link associated with the vertex $I$.

In these conventions,
\begin{equation}
J_n ({\bf 3_1^{\ell}}, q) = J_n (\text{unknot}) \cdot q^{n-1} \sum_{m=0}^{n-1} q^{mn} (q^{n-m})_m,
\end{equation}
\begin{equation}
J_n ({\bf 4_1}, q) = J_n (\text{unknot})
\sum_{i=0}^{n-1} \prod_{j=1}^i
\left( q^{(n+j)/2} - q^{-(n+j)/2} \right) \left( q^{(n-j)/2} - q^{-(n-j)/2} \right).
\label{Jfig8}
\end{equation}
These are also the conventions used in \cite{Gukov:2003na,Chen:2015wfa}, except that the factor $J_n (\text{unknot})$ is removed. This gives the normalized colored Jones polynomial.

It is also important to note that, for $SU(2)$ partition function, the range of summation (again, in these conventions) runs over $n = 1, \ldots, k-1$.
(For $SO(3)$, the sum runs over half of this range.)
In particular, various versions of the WRT invariants are suitably normalized sums over the colors \cite{Gukov:2017kmk,Gukov:2020frk}:
\begin{equation}
F^{SU(2)}[\CL]:= \sum_{n_I \in \{ 1, \ldots, k-1 \}} 
J_{\vec n} (\CL,q)
\prod_{I} \frac{q^{n_I/2}-q^{-n_I/2}}{q^{1/2}-q^{-1/2}},
\label{WRT-sum}
\end{equation}
\begin{equation}
F^{(\vec c)}[\CL]:= \sideset{}{'}\sum_{\scriptsize\begin{array}{c} 0 \leq n_I\leq 2k-1,\\ n_I=c_I\mod 2 \end{array}} 
J_{\vec n} (\CL,q)
\prod_{I} \frac{q^{n_I/2}-q^{-n_I/2}}{q^{1/2}-q^{-1/2}},\quad \vec c=\{c_I\in \{0,1\}\}
\label{RT-sum}
\end{equation}
\begin{equation}
F^\text{tot} [\CL]:= \sideset{}{'}\sum_{ 0 \leq n_I\leq 2k-1}
J_{\vec n} (\CL,q)
\prod_{I} \frac{q^{n_I/2}-q^{-n_I/2}}{q^{1/2}-q^{-1/2}}.
\end{equation}
Note, that $F^{SU(2)}[\CL] \sim F^\text{tot} [\CL]$ and $F^{SO(3)}[\CL] := F^{(1,1,\ldots,1)} [\CL]$. In these conventions, the $SO(3)$ case corresponds to odd values of $k$, namely $k\equiv1$ or 3 $\pmod 4$, whereas spin-refined $SU(2)$ TQFT has $k\equiv0$ or 2 $\pmod 4$.

Now we are ready to approach the volume conjecture for closed 3-manifolds (see Appendix~\ref{sec:FKvolume} for discussion of knot complements). The ``new'' RT invariants used by \cite{Chen:2015wfa} in these conventions can be written as
\begin{equation}
\text{RT}_r (S^3_p (K)) = - \frac{1}{2r} e^{\pi i (\frac{3}{r} + \frac{r+1}{4}) \cdot \text{sign} (p)}
\sum_{n=1}^{k-1} (q^{n} - q^{-n})^2
(-q^{\frac{1}{2}})^{p(n^2 - 1)} J_n (K,q^2)
\label{RTnew}
\end{equation}
where $J_n (K,t)$ is the normalized colored Jones (we have also assumed $\CL=K$ is now a knot) and $q = e^{2\pi i / r}$ or, equivalently, $k=r=\text{odd}$. In other words, all powers of $q$ in this sum are doubled compared to the ``ordinary'' WRT invariants.

$$
\begin{array}{c||c|c|c|c|c|c}
	p & \cdots & -2 & -1 & 1 & 2 & \cdots \\
	\hline
	\text{RT}_{51} (S^3_p ({\bf 3_1^{\ell}}))~ & ~\cdots~ & ~0.893\, -0.230 i~ & ~4.644\, -0.038 i~ & ~4.447\, +7.673 i~ & ~-0.377+8.142 i~ & ~\cdots 
\end{array}
$$

In order to express \eqref{RTnew} in terms of the $q$-series invariants $\hat Z_b (q)$, we double the range of summation
$$
\text{RT}_r (S^3_p (K)) = - \frac{1}{4r}
(-q^{\frac{1}{2}})^{-p}
e^{\pi i (\frac{3}{r} + \frac{r+1}{4}) \cdot \text{sign} (p)}
\sum_{n \in \Z / 2k \Z}
(-q^{\frac{1}{2}})^{p n^2}
(q^{n} - q^{-n})^2 J_n (K,q^2)
$$
and formally write $x = q^n$,
$$
\big( q^{\frac{n}{2}} - q^{-\frac{n}{2}} \big)^2 J_n (K,q) \; = \; \big( x^{\frac{1}{2}} - x^{-\frac{1}{2}} \big) F_K (x,q)
\; = \; \sum_{\ell} q^{n \ell} F_{\ell} (q)
$$
$$
\big( q^{n} - q^{-n} \big)^2 J_n (K,q^2) \; = \; \big( x - x^{-1} \big) F_K (x^2,q^2)
\; = \; \sum_{\ell} q^{2 n \ell} F_{\ell} (q^2)
$$
where we introduced $F_K (x,q)$ as in \cite{Gukov:2019mnk}; see Appendix~\ref{sec:FKvolume} for $F_K$ volume conjecture. In other words, compared to the familiar surgery formulae, both $x$ and $q$ have doubled exponents.

Next we need to apply the Gauss sum to a term with a particular value of $\ell$. Using $(-1)^{pn^2} = (-1)^{pn}$, we can write such a generic term as
\begin{equation}
\sum_{n \in \Z / 2k \Z}
\exp \left( \frac{\pi i}{r} p n^2 + \frac{\pi i}{r} (4 \ell + pr) n \right)
F_{\ell} (q^2).
\end{equation}
Using the Gauss sum\footnote{Recall, we keep using $k = r$.}
\begin{equation}
\sum_{n \in \Z / 2k \Z} \exp \left( \frac{\pi i}{2k} M n^2 + \frac{\pi i}{k} n \tilde \ell \right)
\; = \;
\frac{e^{\frac{\pi i}{4} \text{sign} (M)} \sqrt{2k}}{\sqrt{M}}
\sum_{a \in \Z / M \Z}
\exp \left( - \frac{2\pi i k}{M} \big( a + \frac{\tilde \ell}{2k} \big)^2 \right)
\end{equation}
with $M=2p$ and $\tilde \ell = 4 \ell + pr$, we get
\begin{multline}
\frac{e^{\frac{\pi i}{4} \text{sign} (p)} \sqrt{r}}{\sqrt{p}}
\sum_{a \in \Z / 2p \Z}
\exp \left( - \frac{\pi i r}{p} \big( a + \frac{4 \ell + pr}{2r} \big)^2 \right) =
\\
= \frac{e^{\frac{\pi i}{4} \text{sign} (p)} e^{- \frac{\pi i}{4} pr} \sqrt{r}}{\sqrt{p}}
\sum_{a \in \Z / 2p \Z}
(q^2)^{- \frac{\ell^2}{p}} \exp \left(
-\frac{i \pi  a^2 r}{p} - \frac{4 \pi i}{p} ab - \pi i r a \right)
\end{multline}
where
\begin{equation}
b : = \ell \bmod p.
\end{equation}

Therefore, with the $\hat Z_b (q)$ defined as
\be
\hat Z_b (S^3_p (K)) \; = \; \sum_{\ell \in p \Z + b}
q^{- \frac{\ell^2}{p}} F_{\ell} (q),
\ee
we can write \eqref{RTnew} in terms of $\hat Z$-invariants as
\begin{multline}
\text{RT}_r (S^3_p (K)) = -
(-q^{\frac{1}{2}})^{-p} \;
\frac{e^{\pi i (\frac{3}{r} + \frac{r+2}{4}) \cdot \text{sign} (p)}
e^{- \frac{\pi i}{4} pr}}{4 \sqrt{rp}} \times
\\
\sum_{a \in \Z / 2p \Z} \; \sum_{b \in \Z_p} e^{-\frac{i \pi  a^2 r}{p} - \frac{4 \pi i}{p} ab - \pi i r a }
\, \hat Z_b (q^2) \Big|_{q \to e^{2\pi i /r}}.
\end{multline}

Note, for $p=-1$ (and, similarly, for $p=1$) this expression simplifies to
\be
\text{RT}_r (S^3_{-1} (K)) = 
- \frac{e^{- 2\pi i /r}}{2 \sqrt{r}} 
\hat Z_0 (q^2) \Big|_{q \to e^{2\pi i /r}}.
\ee
In particular, $S^3_{-1} ({\bf 3_1}^{\ell}) = \Sigma (2,3,5)$ and $S^3_{-1} ({\bf 4_1}) = - \Sigma (2,3,7)$. Also, $S^3_{-p/r} (\text{unknot}) = - L(p,r)$ and for $S^3_{-p} (\text{unknot}) = - L(p,1)$:
\be
\hat Z_0 (q) \; = \; - 2q^{\frac{p-3}{4}}
\,, \qquad
\hat Z_1 (q) \; = \; 2q^{\frac{p^2-3p+4}{4p}}.
\label{ZLens}
\ee
where $p$ different Spin$^c$ structures are labeled by $b=0,\ldots, p-1$.
By directly evaluating \eqref{RTnew} for the $(-1)$-surgery on the unknot, we find
\be
\text{RT}_r (S^3_{-1} (\text{unknot})) =
\frac{1}{i \sqrt{r}} (e^{2\pi i /r} - e^{-2\pi i /r}),
\ee
which means that a factor of $(q - q^{-1})$ is missing in the normalization.

Actually, the two terms in $(q - q^{-1})$ correspond to the contributions of $\hat Z_0$ and $\hat Z_1$, respectively. So, if we keep both $b=0$ and $b=1$ contributions even for $|p|=1$, we get the correct expression without changing the normalization much. Indeed, at $p=1$ we have $\hat Z_0 (q^2) + \hat Z_1 (q^2) \sim (q-q^{-1})$.

Using the above analysis, let us define
\be
c_{r,p,b} = (-1)^{p+1}
e^{-\pi i \frac{p}{r}}
\frac{e^{\pi i (\frac{3}{r} + \frac{r+2}{4}) \cdot \text{sign} (p)}
	e^{- \frac{\pi i}{4} pr}}{4 \sqrt{r|p|}}
e^{\frac{\pi i}{4} (1-\text{sign} (p))}
\sum_{a \in \Z / 2p \Z}
e^{-\frac{i \pi  a^2 r}{p} - \frac{4 \pi i}{p} ab - \pi i r a }.
\label{crpb}
\ee
Then,
$$
p>0: \qquad
\text{RT}_r (S^3_{p} (\text{unknot})) =
e^{\pi i (p-3)/r} \left( - 2 e^{- \pi i (p-3)/r} c_{r,p,0} + 2 e^{- i \pi \frac{p^2 - 3p + 4}{rp}} c_{r,p,1} \right) ,
$$
$$
p<0: \qquad
\text{RT}_r (S^3_{p} (\text{unknot})) =
e^{-\frac{\pi i}{2}+\pi i (p-3)/r} \left( - 2 e^{- \pi i (p-3)/r} c_{r,p,0} + 2 e^{- i \pi \frac{p^2 - 3p + 4}{rp}} c_{r,p,1} \right).
$$

Next, let us consider
\be
\Sigma (2,3,7)
= S^3_{-1} ({\bf 3_1^r})
= S^3_{+1} ({\bf 4_1})
= - S^3_{+1} ({\bf 3_1^{\ell}}).
\ee
In these and many other similar examples ({e.g.}~below), the $\hat Z$-invariants are linear combinations of false theta-functions:
\begin{gather}
\tilde \Psi^{(a)}_p (q) \;  :=  \; \sum_{n=0}^\infty \psi^{(a)}_{2p}(n) q^{\frac{n^2}{4p}} \qquad \in q^\frac{a^2}{4p}\,\Z[[q]],
\label{falsetheta} \\
\psi^{(a)}_{2p}(n)  =  \left\{
\begin{array}{cl}
\pm 1, & n\equiv \pm a~\pmod{2p}\,, \\
0, & \text{otherwise}.
\end{array}\right. \nonumber
\end{gather}
Their limiting values at general roots of unity can be easily found using modular properties and has been discussed in a closely related context {e.g.}~in \cite{MR1701924,MR2191375,Kucharski:2019fgh,Chung:2019jgw}:
\be
\tilde \Psi^{(a)}_p (e^{2\pi i m/k})
\; = \;
\sum_{n=0}^{pk}
\left( 1 - \frac{n}{pk} \right)
\psi_{2p}^{(a)} (n) e^{\pi i \frac{m n^2}{2pk}}.
\ee
In particular, for $m=2$ we get
\be
\tilde \Psi^{(a)}_p (e^{4\pi i/k})
\; = \;
\sum_{n=0}^{pk}
\left( 1 - \frac{n}{pk} \right)
\psi_{2p}^{(a)} (n) e^{\pi i \frac{n^2}{pk}}.
\ee
Using this, we get
$$
\text{RT}_r (S^3_{+1} ({\bf 4_1})) =
- c_{r,+1,0} e^{-\frac{2\pi i}{r}} e^{\frac{4\pi i}{r} \frac{83}{168}} \left( \tilde \Psi^{(1)}_{42} (q^2) - \tilde \Psi^{(13)}_{42} (q^2) - \tilde \Psi^{(29)}_{42} (q^2) + \tilde \Psi^{(41)}_{42} (q^2) \right)
\Big|_{q \to e^{2\pi i /r}}
$$

$$
\begin{array}{c||c|c|c|c|c|c}
r & 1 & 3 & 5 & 7 & 9 & \cdots \\
\hline
~\text{RT}_{r} (S^3_{+1} ({\bf 4_1}))~~ & ~~~0~~~ & ~~~1~~~ & ~1.214\, -0.5 i~ & ~0.091\, -0.399 i~ & ~-0.945-0.637 i~ & ~\cdots 
\end{array}
$$

A more interesting example of a surgery on the trefoil knot which is {\it not} a homology sphere is
\be
M_3
\; = \; S^3_{-3} ({\bf 3_1^r})
\; = \; - S^3_{+3} ({\bf 3_1^{\ell}}).
\label{int31surgery}
\ee
The corresponding $\hat Z$-invariants again can be expressed in term of the false theta-functions \cite{Cheng:2018vpl,Gukov:2019mnk}:
\begin{subequations}
	\label{Zhatint31}
	\be
	\hat Z_0 (q) = q^{\frac{71}{72}}
	\left( \tilde \Psi_{18}^{(1)} + \tilde \Psi_{18}^{(17)}  \right)
	= q + q^5 - q^6 - q^{18} + q^{20} + \ldots,
	\ee
	\be
	\hat Z_1 (q) = - q^{\frac{71}{72}}
	\left( \tilde \Psi_{18}^{(5)} + \tilde \Psi_{18}^{(13)}  \right)
	= - q^{4/3} \left( 1 + q^2 - q^7 - q^{13} + q^{23} + \ldots \right).
	\ee
\end{subequations}
We find
\be
\text{RT}_r (S^3_{-3} ({\bf 3_1^r})) =
e^{- \pi i /2}
\left( c_{r,-3,0} \hat Z_0 (q^2) + c_{r,-3,1} \hat Z_1 (q^2) \right)
\Big|_{q \to e^{2\pi i /r}}
\ee
where the parity reversal relation $\text{RT}_{r} (S^3_{-3} ({\bf 3_1^r})) = \overline{\text{RT}_{r} (S^3_{+3} ({\bf 3_1^{\ell}}))}$ was used.

$$
\begin{array}{c||c|c|c|c|c|c}
r & 1 & 3 & 5 & 7 & 9 & \cdots \\
\hline
~\text{RT}_{r} (S^3_{-3} ({\bf 3_1^r}))~~ & ~~~0~~~ & ~~~1~~~ & ~0.425\, +0.309 i~ & ~0.336\, -0.833 i~ & ~0.154\, +0.266 i~ & ~\cdots 
\end{array}
$$

Note, the extra factor of $e^{- \frac{\pi i}{2}}$ is the same as we found for surgeries on the unknot at negative $p$. Therefore, from all these examples we conclude that for $p>0$ the coefficients $c_{r,p,b}$ work on the nose, and for $p<0$ they need to be accompanied by a factor of $e^{- \frac{\pi i}{2}}$:
\bea
p>0: & \qquad & c_{r,p,b}, \\
p<0: & \qquad & e^{- \frac{\pi i}{2}} c_{r,p,b}.
\eea
This basically means that the phase factor $e^{\frac{\pi i}{4} (1-\text{sign} (p))}$ that was introduced in \eqref{crpb} to account for the square root of the denominator needs to be removed.

To summarize, we find
\be
\text{RT}_r (S^3_{p} (K)) \; = \; \sum_{b \in \Z_p}
c_{r,p,b} \;
\hat Z_b (q^2) \Big|_{q \to e^{2\pi i /r}}
\ee
with
\be
c_{r,p,b} = (-1)^{p+1}
e^{-\pi i \frac{p}{r}}
\frac{e^{\pi i (\frac{3}{r} + \frac{r+2}{4}) \cdot \text{sign} (p)}
	e^{- \frac{\pi i}{4} pr}}{4 \sqrt{r|p|}}
\sum_{a \in \Z / 2p \Z}
e^{-\frac{i \pi  a^2 r}{p} - \frac{4 \pi i}{p} ab - \pi i r a }.
\ee

Based on the analysis here and in \cite{Kucharski:2019fgh,Chung:2019jgw} it is natural to expect that all other Reshetikhin-Turaev invariants at more general roots of unity are similarly expressed as linear combinations of $\hat Z_b (q^m)$. It would be interesting verify this explicitly and work out the coefficients of such a linear relation.

\section{Compactification on $M_4$}\label{sec:2d}

When compactifying on $M_4$, one encounters for the first time situations in which no polarizations exist, and one must work with relative $T[M_4]$ theories. One way to see this is that $M_4$ can belong to a non-trivial cobordism class in $\Omega_4$, making it generally impossible to obtain a geometric polarization by finding a $W_5$ whose boundary is $M_4$. (In contrast, this is always possible for $M_d$ for $d=1,2,3$, as the cobordism groups are trivial in these dimensions). Given a particular $M_4$, one can in principle classify $\Pol(M_4)$ by working out all the boundary conditions of $\CT^{\text{bulk}}[M_4]$, similar to what we did in Part I (e.g.~with $\Pol(\mathrm{pt})$). However, since such a classification depends on both the choice of the 6d theory and the manifold $M_4$, and the goal of this paper is to discuss general aspects of compactification, we will not attempt to classify all polarizations (see, e.g.,~\cite{Chen:2023qnv} for a recent study). Instead, we work with the relative theory $T[M_4]$, focusing on the case where the 6d theory has $(2,0)$ supersymmetry and the twist on $M_4$ is the 6d lift of the Vafa--Witten twist. Then the theory $T[M_4]$ will have a protected subsector VOA$[M_4]$ \cite{Gadde:2013sca}. We use the bulk TQFT to make predictions about the structure of its modules.

\subsection{Modules of VOA$[M_4]$}
VOA$[M_4]$ is describing a chiral theory obtained from $T[M_4]$ by a holomorphic twist. Such a procedure will not change the bulk theory, and the chiral theory will have central charge\footnote{The full $T[M_4]$ is not a conformal theory, and $c_L-c_R$ here should be understood as $-\frac{d}{2}$ where $d$ is its gravitational anomaly.} $c=c_L-c_R$ and couple to the same 3d TQFT $\CT^{\text{bulk}}[M_4]$. 

The line operators of the 3d TQFT are labeled by $H_2(M_4,D)$, and originates from wrapping the three-dimensional operators in the 7d theory on two-cycles on $M_4$. These lines will in general have non-trivial braiding and spins controlled by the intersection form $(\cdot,\cdot)$ and its quadratic refinement $q$ on $H_2(M_4,D)$. 

For the boundary condition given by the relative theory $T[M_4]$, all such lines can end, and their boundaries are local operators that have non-trivial spins and braidings. After the holomorphic twist, all the lines should still be able to end, and one has a collection of modules for VOA$[M_4]$ labeled by $H_2(M_4,D)$, with again the spins and mutual braidings controlled by the line operators of the 3d TQFT. This is a property that one can use to check whether a proposed VOA associated with a given 4-manifold has the chance of being correct. In other words, one should be doubtful if the candidate VOA doesn't have a category of modules with the expected spins and braidings. 

Before making more general statements, we will first illustrate this in an example.

\subsubsection{Examples: VOAs for $\mathbb{CP}^2$ and $\overline{\mathbb{CP}^2}$}
We will consider the case of $\frak{g}=A_1$ and  $M_4=\overline{\mathbb{CP}^2}$ to compare the predictions from the TQFT with VOA. In fact, there are two closely related VOAs relevant for this case. One is $\hat{\frak{su}}(2)_1$ while the other is known as the Urod algbra that emerges from the study of Nakajima--Yoshioka blowup equations \cite{Bershtein:2013oka}. They in fact have the same underlying vertex algebra but different stress-energy tensor. This motivate the question: why is the Urod algebra better than the more familiar $\hat{\frak{su}}(2)_1$ whose stress-energy tensor is just the one given by the Sugawara construction?

If fact, at the level of characters, they are related by \begin{equation}
    \chi(U_0) = q^{-1/4}
\chi(V_1), \quad\text{and}\quad \chi(U_1) = q^{
-1/4}\chi(V_0),
\end{equation}
where $V_0$ and $V_1$ are the two irreducible integrable modules for $\hat{\frak{su}}(2)_1$ while $U_0$ and $U_1$ are the two corresponding modules for the Urod algebra. The pre-factors with $q^{\pm\frac{1}{4}}$ might look innocuous, but they do change the spin of the modules, from $(0,\frac{1}{4})$ to $(-\frac{1}{4},0)$.

The 7d TQFT can be taken to be 
\begin{equation}
    \frac{1}{2\pi}\int CdC
\end{equation}
using a duality between $SU(2)_1$ and $U(1)_2$. If we were compactifying on $\mathbb{CP}^2$, the part that couples to the 0-form symmetry on the boundary would be 
\begin{equation}
    \frac{1}{2\pi}\int AdA,
\end{equation}
which describes a $U(1)_2$ Chern--Simons theory. The only non-trivial line should have spin $\frac{1}{4}$, which matches that of the $\hat{\frak{su}}(2)_1$. However, for $\overline{\mathbb{CP}^2}$, we get $SU(2)_{-1}=U(1)_{-2}$, where the spin of the non-trivial line is reversed. This exactly agrees with the Urod VOA. So we arrive at the conclusion that the Urod algebra is indeed compatible with the TQFT, while the more familiar $\hat{\frak{su}}(2)_1$ would not.

Strictly speaking, the Urod algebra should be associated to $\overline{\mathbb{CP}^2}$ minus a point, as the blow-up operation is topologically taking connected sum with $\overline{\mathbb{CP}^2}$. However, to get the TQFT in 3d, we have only kept the ``most interesting'' part that is sensitive to $H_2(M_4)$, and subtracting a point is not affecting the analysis.

Notice that the two theories are almost identical, with $U(1)_{-2}$ being dual to $U(1)_2$ with the difference being copies of the invertible fermionic theory $U(1)_1$ \cite{Seiberg:2016rsg}. This is reminiscent of how the Urod algebra is constructed out of $\hat{\frak{su}}(2)_1$ by modifying the stress energy tensor.

For $\mathbb{CP}^2$, a VOA has been proposed in \cite{Feigin:2018bkf}, and it is indeed expected to have two simple modules with the only non-trivial irreducible module having spin $\frac{1}{4}$. It would be interesting to check the prediction from the bulk TQFT on the representation theory of VOA$[M_4]$ in more examples and apply it to help identify the correct VOA when there are multiple plausible candidates.

\subsubsection{Extensions from topological interfaces}

One can then ask about what happens when $T[M_4]$ can be made absolute, or, in other words, when $\CT^{\text{bulk}}[M_4]$ has topological boundary conditions.

When this is the case, there will be a collection of the line operators with trivial spin and trivial mutual braiding that can end on the topological boundary. These lines will define a maximal isotropic subgroup of $H_2(M_4)$. Then, at the level of VOAs, one expects that the process of colliding the $T[M_4]$ boundary condition with the topological one to form an absolute 2d theory can be interpreted as an extension by this collection of mutually local modules. One effect of the extension is to kill the rest of the modules that have non-trivial braidings with some modules that we are extending by, and, as a result, one ends up with a holomorphic VOA.

One can also have something in between the two extreme cases. Instead of having a topological boundary condition that kills half of the dimensions of the charge lattice, one can consider a topological interface that only condenses a collection $C\subset H_2(M_4,D)$ given by an isotropic but non-maximal subgroup. This will be an interface with a ``smaller'' 3d TQFT with fewer lines, which describes the modules of the new VOA, denoted as VOA$[M_4,C]$, obtained after extending VOA$[M_4]$ by modules in $C$.

To be democratic to all choices of $C$, one should consider all theories $T[M_4,C]$ and VOA$[M_4,C]$ on the equal footing. And even if one is just interested in an extreme case either when $C=0$ or when the VOA is holomorphic, using fully the relations between the web of theories and VOAs should allow one to gain insights into the system.

\subsection{$\Z_2^{U}$ in $T[M_4]$ from 6d $(1,0)$ theories}

Now we also give some remarks for the $(1,0)$ case. As we have explained, 6d $(1,0)$ theories have a universal $\Z_2^U$ symmetry, which will lead to $|H_1(M_4,\Z_2)|$ different versions of 2d theories, each with a $\Z_2$ symmetry. Alternatively, one can view this as a single theory with a discrete theta angle valued in $H_1(M_4,\Z_2)$. The $\Z_2$ symmetry in 2d can be interpreted as the composition of two R-parities. One is $(-1)^F$, while the other comes from the center of $SU(2)_R$ of the 6d theory. They become the same when acting on the supersymmetry algebra, but in general are different when acting on the full theory $T[M_4]$.  Existence of such a family of theories and such a symmetry in 2d are constraint for $T[M_4]$ obtained from 6d $(1,0)$ theories. 

One simple example of this is when we take the 6d theory to be the free hypermultiplet. Then the 2d theory consists of $(0,1)$ Fermi and chiral multiplets coming from left- and right-moving spinors on $M_4$ (or, equivalantly, sections of $S^\pm$ respectively). Then the choice of the $\Z_2^U$ holonomy can be identified with a choice of a spin structure. The spectrum of the Dirac operator will depend on such a choice, and one indeed gets different $T[M_4]$ theories in general. Each theory will have different mass spectrum, and even the number of massless modes, given by left- and right-moving harmonic spinors can be different, only with the difference $h^+-h^-=\sigma/4$ fixed.

\section{New 4-manifold invariants: a version of Vafa--Witten theory with two \texorpdfstring{$q$}{q}'s}
\label{sec:2dqq}

Similar to Part I, we conclude with a ``case study'' that combines results obtained and techniques developed in earlier sections. The goal of this section is to argue for, and present evidence supporting, new families of 4-manifold invariants that depend on two ``$q$-parameters'' $q_j = e^{2\pi i \tau_j}$, $j=1,2$.\footnote{Not to be confused with qq-charactes \cite{Nekrasov:2015wsu} which also involve two $q$-variables, but of a different nature, in a context rather different from general 4-manifolds.} It is based on a crucial claim that 6d $(2,0)$ theory on a 2-torus $T^2$ is {\it not} the maximally supersymmetric $\CN=4$ Yang--Mills theory or, equivalently, that the elliptic genus of 2d theory $T [M_4]$ is {\it not} equal to the Vafa--Witten partition function $Z_{\text{VW}} (M_4;q)$.
There have been several clues in the literature pointing to this crucial fact, including the cutting-and-gluing (surgery) operations \cite{Gadde:2013sca,Feigin:2018bkf}, modular properties, and of course the symmetries and anomalies that are in the center of our attention here.

$$
\begin{tikzpicture}
\node at (0,0) {6d $(2,0)$ theory on $T^2 \times M_4$};
\node at (4,-2.5) {2d $(0,2)$ theory $T[M_4]$ on $T^2$};
\node at (-4,-2.5) {4d Vafa--Witten theory on $M_4$};
\node at(0,-2.5) {$\stackrel{\textcolor{red}{?}}{\cong}$};
\draw[->] (-2,-.5) to node [left=0.25] {\textcolor{red}{?}~~} (-4,-2);
\draw[->] (2,-.5) to node [right=0.25] {~~~} (4,-2);
\end{tikzpicture}
$$

Let us start with the gravitational anomaly that controls the modular properties of the resulting partition function. In \cite{Vafa:1994tf}, Vafa and Witten consider two versions of the partition function on $M_4$, denoted $Z_{M_4}$ and $\widehat Z_{M_4}$. (The dependence on a single $q$-variable is not manifest in those notations.) Here, we denote these two versions $\ZVW (M_4)$ and $\ZVW' (M_4)$, respectively, in part to avoid confusion with 3-manifold invariants introduced in \cite{Gukov:2016gkn,Gukov:2017kmk} and discussed in an earlier section. Then,
\be
\ZVW' (M_4) \; = \; \eta^{-w} \ZVW (M_4)
\ee
where $\frac{w}{2}$ is the modular weight of $\ZVW (M_4)$, and
\be
\ZVW (M_4) = q^{-s} \sum_{n \in \Z_{\ge 0}} q^n \chi \left( \CM_n \right),
\ee
with $\chi(\CM_n)$ being the Euler characteristics of the instanton moduli space $\CM_n$ of instanton number $n$.\footnote{Note that there is another refinement obtained by replacing the Euler characteristics by the $\chi_{y^2}$-genus \cite{Alexandrov:2019rth}, which is different from the present one, and, therefore, it would be interesting to see whether they can be combined.}
This means that the shift from integer powers of $q$ in $\ZVW' (M_4)$ is equal to $-s - \frac{w}{24}$. In other words, the transformation of the Vafa--Witten invariant under $S$ and $T$ generators of the modular group are controlled by two numbers, $s$ and $w$, which depend on $M_4$ and the gauge group. The parameter $s$ also depends on fluxes, i.e.~on $c_1$ of the gauge bundle, but that dependence comes in the form $s + \frac{N-1}{2N} v \cdot v$, and following \cite{Vafa:1994tf} here we focus on the constant offset $s$.

Based on the analysis of several examples, it was suggested in \cite{Vafa:1994tf} (and further verified in \cite{Labastida:1999ij}) that, for $\mathfrak{g} = \mathfrak{su} (2)$,
\bea
w & = & - \chi, \\
s & = & - \frac{\chi}{12}~. \notag
\eea
For example, when $M_4$ is a K3 surface, $w = -24$ and $s=-2$. Note that both $s$ and $w$ depend only on the Euler characteristic of $M_4$, and not on the signature $\sigma := \sigma (M_4)$. Therefore, even if we work with $\ZVW' (M_4)$, for which the gravitational anomaly is moved into transformation under $T$, it is still proportional to the Euler characteristic, $-s - \frac{w}{24} = \frac{1}{8} \chi$.

On the other hand, the central charges of the 2d $(0,2)$ theory $T[M_4; \mathfrak{su} (2)]$ can be obtained directly by integrating the anomaly polynomial of the 6d $(2,0)$ theory of type $A_1$ and verified in a number of ways, see {e.g.}~\cite{Gukov:2016gkn,Dedushenko:2017tdw}:
\be
c_L = 13 \chi + 18 \sigma
\quad,\qquad
c_R = \frac{1}{2} (27 \chi + 39 \sigma).
\label{cLcR}
\ee
The elliptic genus of $T[M_4]$, computed as a trace over the Ramond sector, is not a modular form, but rather a modular function transforming under $SL(2,\Z)$ with a certain multiplier system that depends on the gravitational anomaly $c_R - c_L = d/2$. Multiplying it by $\eta^d$ makes it into a true modular form of weight
\be
\bar c := c_R - c_L = \frac{1}{2} \chi + \frac{3}{2} \sigma.
\ee
The term proportional to the Euler characteristic compares well to $- \frac{w}{2}$ in the Vafa--Witten partition function. As we explain below, the `anomalous' term $\frac{3}{2} \sigma$ comes from the Kaluza--Klein modes on $T^2$, which are the gist of the distinction between the ordinary Vafa--Witten theory on $M_4$ and its close cousin with two $q$'s. The elliptic genus of $T[M_4]$ and $\ZVW (M_4)$ arise as two different limits of this more general invariant of $M_4$.

The existence of the two parameters can be understood as two independent $(-1)$-form symmetries in four dimensions when we reduce the 6d theory on $T^2$, with one being emergent, giving a one-parameter family of deformations for $T[T^2]$ (see also \cite{Najjar:2024vmm} for related observations). Both parameters originate from symmetries of the 5d theory $T[S^1]$, where one can define two closely related $U(1)$ symmetries. One is $U(1)_{S^1}$, which can be identified geometrically as the rotation of the $S^1$. The other is the instanton symmetry $U(1)_{\text{inst}}$ associated with the fact that the 5d theory has a gauge theory description. For the $(2,0)$ theory, there is evidence that, in the BPS sector, the two symmetries coincide \cite{Kim:2011mv}.\footnote{For general 6d $(1,0)$ theories with 5d gauge theory descriptions for their KK reduction,  they can be different and there is possibility of having one more parameter in the game. For example, as we have seen previously, the $U(1)_{\text{inst}}$ in 5d gauge theory description of E-string theory should not be able to capture all KK-modes from the Higgs branch.} After the compactification to 4d to get $T[T^2]$, $U(1)_{S^1}$ leads to the natural parameter $\tau_{\text{geom}}$ that can be identified with the complex structure of $T^2$. In the effective description where $T[T^2]$ is viewed as a gauge theory coupled to matter, one also has the gauge theory parameter $\tau_{\text{gauge}}$ that keeps track of the instanton number. If one does not keep the KK modes, it is expected that the two are identical, leading to the familiar statement that $T[T^2_\tau]$ in certain limit becomes 4d super--Yang--Mills with coupling constant $\tau$. However, once KK modes are included, their masses also depends on $\tau_{\text{geom}}$, and one can make the two parameters independent by adjusting the masses and gauge coupling separately. Another way of saying this is that the ``current'' $n_{\text{geom}}$ and $n_{\text{inst}}$ are related by
\begin{equation}
    n_{\text{geom}} = n_{\text{inst}} +n_{\text{KK}},
\end{equation}
where $n_{\text{inst}}$ is the instanton number from the gauge field, while $n_{\text{KK}}$ counts the KK-momentum for other matter content of the theory.

There are several benefits of having one additional parameter. First of all, one can hope that the invariant of 3- and 4-manifolds obtained will be stronger. Secondly, a one-parameter family can relate different theories and their corresponding invariants as different limits, providing new insights into them. There are in fact two natural limits, one is $q_{\text{inst}}=q_{\text{KK}}$, where one gets undeformed $T[T^2]$, while the other is when $q_{\text{KK}}=0$ where the KK tower is killed. As we emphasized in previous sections, in this limit, the theory is still not just 4d $\CN=4$ super--Yang--Mills theory as the moduli space is different.\footnote{Recall that this is due to a group-valued scalar. As we have seen in previous sections and well known in the context of 3d-3d correspondence, the group-valued (rather than Lie-algebra-valued) scalar fields generally play an important role in $T[M_d]$ for $d>1$. They are also crucial for proper understanding of the 2d $(0,2)$ theory $T[M_4]$ of our interest here. In particular, $G$-valued scalars, one for each generator of $H^2 (M_4)$, are the ``main carriers''---via either the center or $\pi_1$ of $G$---of the 0-form symmetries of $T[M_4]$. These winding-momentum pairs come with 't Hooft anomalies, given by the intersection form on $H^2 (M_4)$. For a discussion about symmetry in general sigma model, see {e.g.}~\cite{Hsin:2022heo}.} But at the level of the partition function, under the Vafa--Witten twist, it is likely that it decomposes into a sum of Vafa--Witten partition functions of the super--Yang--Mills theory. For the $A_n$ series, the moduli space only has maximal singularities of $A_n$ type, and one expects that the partition function will be a multiple of the usual Vafa--Witten partition function for $A_n$ super--Yang--Mills theory.

To demystify the additional parameter in a way that is as explicit as possible, here, we study the deformed partition function in the Abelian case, where one can integrate out the KK modes to compute the gravitational background couplings of the effective 4d theory.\footnote{Notice that the partition function of the 6d theory on $M_4\times T^2$ will depend on the gravitational and R-symmetry backgrounds. But in the first step, when we are reducing the theory on $T^2$, we are assuming that the background fields are independent of $T^2$ and there are no R-symmetry holonomies along it. Therefore the effective coupling will only depend on $q_{S^1}$. In the second step when we compute the partition function, we will assume that the backgrounds are these of the Vafa--Witten twist. The reader should keep in mind that the existence of the two $q$'s is not unique to the Vafa--Witten backgrounds but in fact present for any generic backgrounds.} The fermions in the Abelian 6d $(2,0)$ theory transform as $(4,4)$ under Spin$(5,1)\times$Sp$(2)_R$, where the first 4 is the Weyl spinor of Spin$(5,1)$. When reduced on $T^2$, they give rise to a tower of KK modes labeled by a pair of integers $(n_1,n_2)$, with the mass $m$ of $\psi^{(n_1,n_2)}$ dependent on $n_1$,$n_2$ and $\tau=\tau_{\text{KK}}$. As the mass is in general complex, integrating out $\psi^{(m,n)}$ will generate an effective coupling by evaluating $2\pi\int\text{Tr}_r {\hat A}(R)e^{F_R/2\pi}$ for a representation $r$ of the massive fermion $\psi$.
A single Dirac fermion in the fundamental $\mathbf{4}$ of Sp$(2)_R=$Spin(5)$_R$ with mass $m=|m|e^{i\varphi}$ generates the topological theta angle $\theta_R=\varphi$ for Spin$(5)_R$ and gravitational theta term $\frac{\varphi\sigma}{2}$, with $\sigma =\frac{1}{24\pi^2}\int\text{Tr}\left(R\wedge R\right)$.
Thus the tower of massive fermions produces the term
\be
\frac{1}{\pi} \text{Arg}(m(n_1,n_2;\tau))\left(\frac{1}{8\pi}\text{Tr}\left( F_R\^F_R\right)+\frac{\pi}{4}\sigma\right)~,
\ee
where we used the fact that the fundamental representation has Dynkin index $1/2$.

When we integrate over all the KK modes, the total phase equals to that of 
\be
\prod_{(n_1,n_2)\neq (0,0)}(n_2-n_1\tau)=C\cdot\prod_{n_1\in \Z_+}\sin (\pi n_1\tau )=C'\cdot\prod_{n_1\in \Z_+}e^{-\pi i n_1\tau}\left(1-e^{2\pi i n_1\tau}\right)=C''\eta(\tau)
\ee
where $C$, $C'$ and $C''$ are $\tau$-independent constants and we have used the zeta regularization for $\sum_{n\in\Z_{+}}n=-1/12$.

If the partition function in the end is  analytic in $\tau$ (which would be the case for topologically twisted theories), one then expects a factor  
\be
\eta(q_{KK})^{\#\int_{M_4}\left(\frac{1}{8\pi^2}\Tr F_R\^F_R+ \frac{\sigma}{4}\right)}~.\label{eqn:FF+RR}
\ee

The Vafa--Witten twist identifies the $SU(2)_+$ subgroup in the Lorentz group Spin$(4)=SU(2)_+\times SU(2)_-$ with the diagonal Spin$(3)\subset $ Spin(4) subgroup inside the Spin(5) $R$-symmetry. This corresponds to the following background of $R$-symmetry, 
\begin{equation}
 F_R=R+\*R   ~.
\end{equation}
More explicitly, 
\begin{equation}
    (F_R)_{ab}^I=\frac{1}{2}J^I_{cd}R_{ab}^{\;\;cd}~,
\end{equation}
where $J^I_{ab}=\eta^I_{cd}e^c_ae^d_b$ are self-dual two-forms with $\eta^I,e^a$ being the 't Hooft symbol and the vielbein, respectively. Explicitly, $J^1=e^2e^3+e^1e^4$, $J^2=e^3e^1+e^2e^4$, $J^3=e^1e^2+e^3e^4$. (See { e.g.}~\cite{BenettiGenolini:2017zmu} for a review.)
These coefficients satisfy the identity 
\begin{equation}
    \sum_I J^I_{ab}J^I_{cd}=g_{ac}g_{bd}-g_{ad}g_{bc}+\epsilon_{abcd}~.
\end{equation}

We note that such a choice of background is not invariant under the parity transformation.

For such a background, the theta term of the field strength for the $R$-symmetry becomes\footnote{We note that the $SU(2)$ theta angle is normalized differently compared to Spin(3), since the vector representation is the adjoint representation of $SU(2)$, $\theta_{\rm Spin(3)}=2\theta_{SU(2)}$.}
\begin{align}
 &\frac{1}{8\pi^2}\int \text{Tr }F_R^{\rm Spin(5)}\wedge F_R^{\rm Spin(5)}=2\cdot \frac{1}{8\pi^2}\int \text{Tr }F_R\wedge F_R\cr
 &=\frac{1}{8\pi^2}\int \text{vol}_4\sum_I \frac{1}{4!} \epsilon_{abcd}(F_R)^I_{ab}(F_R)^I_{cd}=\frac{1}{768\pi^2}\int \text{vol}_4
 \epsilon_{abcd}\sum_I J^I_{ef}J^I_{gh} R_{abef}R_{cdgh}\cr 
 & =\frac{1}{768\pi^2}\int \text{vol}_4
 \epsilon_{abcd}\left(R_{abef}R_{cdef}-R_{abef}R_{cdfe}+\epsilon_{efgh}R_{abef}R_{cdgh}\right)=\frac{1}{12}\left(2\chi+3\sigma\right)\!,
\end{align}
where $\text{vol}_4$ is the volume form of the four manifold $M_4$, and
we have used
\begin{align}
  &\frac{1}{4(2\pi)^2} \int_{M_4}\text{Tr } {R\^*R}=\frac{1}{128\pi^2}\epsilon^{abcd}\epsilon^{efgh}\int \text{vol}_4 R_{abef}R_{cdgh}=\chi~,\cr
 & \frac{1}{3!(2\pi)^2} \int_{M_4}\text{Tr } {R\^R}=\frac{1}{96\pi^2}\epsilon^{cdef}\int \text{vol}_4 R_{abcd}R_{abef}=\sigma~.
\end{align}

Therefore the factor is given by 
\be
\eta(q_{KK})^{\#\left(\chi(M_4)+3\sigma(M_4)\right)}~,
\ee
where $\#$ is an overall coefficient.

We remark that the Euler characteristic $\chi$, which is even under parity, can arise from the parity-odd expression (\ref{eqn:FF+RR}) (for the usual parity transformation on the theta terms) due to the property that the
twisting condition $F_R=R+\star R$ does not respect parity.

A similar computation can be done for each of the hyper, tensor and vector multiplets of 6d $(1,0)$ theory with the Donaldson--Witten tiwst, and the corresponding factors are given by
$\eta^{\sigma/8}$, $\eta^{(\chi+5\sigma)/4}$, and $\eta^{-(\chi+\sigma)/4}$, agreeing with the analysis using the gravitational anomaly.

In the Abelian theory, for a generic metric, there are no Abelian instantons, and therefore there won't be a dependence on $q_{\text{gauge}}$. However, when the metric is not generic, one gets in addition the theta function $\theta_{\Lambda}(q_{\text{gauge}})$ for the lattice $\Lambda=H^{2+}(M_4,\R)\cup H^2(M_4,\Z)$ in the variable $q_{\text{gauge}}$ and therefore the partition function will depend on both $q_{\text{gauge}}$ and $q_{KK}$. 

The fact that the partition function factorizes is expected to be a special phenomenon for the non-interacting Abelian theory where the KK modes are uncharged under the gauge group. However, in the non-Abelian case, this would not be the case and the two $q$'s are expected to be intermingled in a non-trivial way. One should be able to see this also from the 2d point of view and verify explicitly in simple examples, which we hope to investigate in future work. 

Another very interesting problem is to see the two $q$-parameters in the Seiberg--Witten geometry of $T[T^2]$: does changing $q_{\rm KK}$ only change the effective gravitational couplings or can there be a more substantial effect? One can ask this question at the level of the Donaldson--Witten partition function. Namely, when the R-symmetry background is that of the Donaldson--Witten twist, the partition function of the 6d theory is expected to be given by a sum over the Seiberg--Witten invariants
\begin{equation}
    Z_{\rm DW}[M_4\times T^2] = \sum_{\lambda\in \mathrm{spin}^c} C_\lambda\cdot \mathrm{SW}(\lambda)
\end{equation}
when $b_2^+(M_4)>1$ (see \cite{EString} for a recent study of this partition function with more details on this decomposition). With the deformation turned on, $C_\lambda$ is expected to depend on both $q_{\rm gauge}$ and $q_{\rm KK}$. It is reasonable to expect that this deformation allows one to better distinguish between contributions from different spin$^c$ structures. In particular, if a combination of the two $q$'s can detect the dimension of the moduli space of Seiberg--Wittten equations, $n(\lambda)=\lambda^2-\frac{1}{4}(2\chi+3\sigma)$, then the deformed partition function can be used to gain insights into the simply-type conjecture \cite{Witten:1994cg,Kronheimer_Mrowka_2007}, which states that $\mathrm{SW}(\lambda)=0$ unless $n(\lambda)=0$ for simply-connected 4-manifolds with $b_2^+>1$. 

We conclude this section by pointing out an intriguing connection to the recent work \cite{Bringmann:2019vyd} on modular completions of false theta functions. In physics, this work can be interpreted as a study of (non-)modular properties of the $\hat Z$-invariants discussed in Section~\ref{sec:ZhatSymmetry} and \ref{sec:volume} by introducing the second modular parameter. The two modular parameters, denoted by $\tau$ and $w$ in \cite{Bringmann:2019vyd}, play similar roles as $\tau_{\text{gauge}}$ and $\tau_{\text{KK}}$ in our discussion here. One can ask, under the modular group, whether both $(\tau,w)$ and $(\tau_{\text{gauge}},\tau_{\text{KK}})$ indeed transform in the same way: 
\begin{equation}
\tau_{\text{gauge}} \stackrel{?}\to \frac{a \, \tau_{\text{gauge}} + b}{c \, \tau_{\text{gauge}} + d} \,, \qquad
\tau_{\text{KK}} \stackrel{?}\to \frac{a \, \tau_{\text{KK}} + b}{c \, \tau_{\text{KK}} + d}.
\end{equation}
It would be interesting to explore this potential connection further, in particular to see if the error function $\text{erf} (z)$, that plays an important role in \cite{Bringmann:2019vyd} (see also \cite{Alexandrov:2016enp,Pioline:2025xgf}), is also a natural object in the present context of 4-manifold invariants.

\section*{Acknowledgement}

We thank Cyril Closset, Thomas Dumitrescu, Anton Kapustin, Hiraku Nakajima, Pavel Putrov, Nathan Seiberg, Dan Xie, Tian Yang, and Bingyu Zhang for discussions. We thank the audience of our talks based on this work given on various occasions throughout the past five years for their feedback and encouragements. SG was supported by the Simons Collaboration Grant on ``New Structures in Low-Dimensional Topology,'' by the NSF grant DMS-2245099, and by the U.S. Department of Energy, Office of Science, Office of High Energy Physics, under Award No. DE-SC0011632. 
The work of P.-S.\ H.\ was supported by the U.S.~Department of Energy, Office of Science, Office of High Energy Physics, under Award Number DE-SC0011632, and by the Simons Foundation through the Simons Investigator Award, by the Simons Collaboration of Global Categorical Symmetry, and also by Department of Mathematics
King’s College London. The work of D.P.~is partly supported by research grant 42125 from Villum Fonden, ERC-SyG project No.~810573 ``Recursive and Exact New Quantum Theory,'' and Simons Collaboration on ``New Structures in Low-Dimensional Topology.'' D.P.~also want to thank the Yau Center for Mathematical Sciences at Tsinghua University for hospitality during his visits.

\appendix

\section{Frequently used notations}
\label{sec:notations}

In this appendix, we list some notations used throughout the paper and its prequel for quick reference. 

\begin{itemize}
\item[$\CT^{\text{bulk}}$:] A seven-dimensional 3-form Abelian Chern--Simons theory.
    \item[$D$:] Defect group of the 7d TQFT that classifies  3-dimensional operators in the theory.
    \item[$M_d$:] A  connected $d$-dimensional (smooth) manifold.
    \item[{$\CT^{\text{bulk}}[M_d]$}:] A $(7-d)$-dimensional theory obtained by reducing the 7d TQFT on $M_d$.
    \item[$T$:] A six-dimensional quantum field theory living on the boundary of $\CT^{\text{bulk}}$. It has 2-form $D$ symmetry whose anomaly is described by $\CT^{\text{bulk}}$.
    \item[{$T[M_d]$}:] A $(6-d)$-dimensional theory obtained by reducing the 6d theory on $M_d$, which might be a relative theory living on the boundary of $\CT^{\text{bulk}}[M_d]$.
    \item[$H^i(M_d,D)$:] The $i$-th cohomology of $M_d$ with $D$ coefficients. It classifies $(3-i)$-dimensional topological operators in $\CT^{\text{bulk}}[M_d]$.
    
    \item[$\CH(M_6)$:] The Hilbert space of $\CT^{\text{bulk}}$ on $M_6$ or, alternatively, the Hilbert space of the 1d TQFT $\CT^{\text{bulk}}[M_6]$ on a single point.

    \item[$\langle\cdot,\cdot\rangle$:] An anti-symmetric bilinear form on $H^3(M_6,D)$ (with $M_6$ implicit from the context). It measures non-commutativity of operators (labeled by elements in $H^3(M_6,D)$) in the 1d TQFT $\CT^{\text{bulk}}[M_6]$ acting on $\CH(M_6)$.
    \item[$\L$:] A maximal isotropic subgroup of $H^3(M_6,D)$ with respect to $\langle\cdot,\cdot\rangle$, often referred to as a ``polarization.'' It is a set of maximal commuting operators in $\CT^{\text{bulk}}[M_6]$. This was used primarily in Part I and only appears a couple of times in the present paper. Notice that in Section~\ref{sec:TorsionSpectrum}, $\Lambda$ is used instead to denote the lattice of string charges in the 6d theory.

    \item[$q$:] A quadratic function on $\L$ that refines certain (possibly degenerate) symmetric bilinear form on $\L$. Together with $\L$, it leads to a well-defined partition function of the 6d theory $T$ on $M_6$.
     
    \item[$\Pol(M_6)$] The set of polarizations on $M_6$. 
    \item[$\tilde \Pol(M_6)$] The set of refined polarizations $(\L,q)$ on $M_6$. It also classifies topological boundary conditions of $\CT^{\text{bulk}}[M_6]$.
    
    \item[{$T[M_6,(\L,q)]$}:] An absolute 0-dimensional theory constructed from $T[M_6]$ with the refined polarization $(\L,q)$.
    
    \item[$\L^\vee$:] The Pontryagin dual of $\L$. It is the group of $(-1)$-form symmetries of $T[M_6,(\L,q)]$. It is isomorphic to $H^3(M_6,D)/\L$.
    \item[$\bar\L$:] A lift of $\L^\vee$ to $H^3(M_6,D)$, which then can be decomposed into $\L\oplus\bar\L$. A choice of $\bar\L$ leads to an explicit set of basis for the partition vector of $T$  on $M_6$. 
    
    \item[$\tilde \Pol(M_d)$:] The set of refined polarizations on $M_d$. It also classifies topological boundary conditions of $\CT^{\text{bulk}}[M_d]$.
    
    \item[$\CP$:] A refined polarization on $M_d$ (with the manifold understood from the context).
    
    \item[{$T[M_d,\CP]$}:] An absolute $(6-d)$-dimensional theory constructed from $T[M_6]$ with refined polarization $\CP$. We also sometimes use this notation without fully specifying the refinement when it can be ignored for the topic being discussed. 
    
    \item[$\CS(\CP)$:] A subgroup of $H^*(M_d,D)$ classifying charged objects in $T[M_d,\CP]$.
    
    \item[$\CS(\CP)_{\text{ind}}$:] A subgroup of $\CS(\CP)$ classifying charged objects that are independent, {e.g.}~those which
    exist without the need to be attached to higher-dimensional objects.
    
    \item[$L$:] A maximal isotropic subgroup of $H^{d-3\le*\le 3}(M_d,D)$. It is a sum of graded pieces $L^{(i)}$.  Alternatively, one can regard $L$ as a subgroup of the Poincar\'e dual $H_{d-3\le*\le 3}(M_d,D)$. Then $L^{(i)}$ will be a subgroup of $H_{d-i}(M_d,D)$.
    
    \item[$\CP_L$:] A ``pure polarization'' labeled by $L$. It satisfies $\CS(\CP_L)=\CS(\CP)_{\text{ind}}=L$. The theory $T[M_d,\CP_L]$ has $(2-i)$-dimensional charged objects classified by $L^{(d-i)}\subset H_{i}(M_d,D)$. 
    
    \item[$L^\vee$:] The Pontryagin dual of $L$, which is isomorphic to $H^{d-3\le*\le 3}(M_d,D)/L$. It describes the symmetries of the theory $T[M_d,\CP_L]$. More precisely, the theory has a $(L^\vee)^{(i)}$ $(2-i)$-form symmetry. The $U(1)$-valued pairing between $L^{(d-i)}\subset H_{i}(M_d,D)$ and $(L^\vee)^{(i)}$ describes the action of the symmetry generator on the charged objects.
    
    \item[$\bar L$:] A lift of $L^\vee$ to $H^{d-3\le*\le 3}(M_d,D)/L$. Existence of such a lift is equivalent to the $L^\vee$ symmetry of $T[M_d,\CP_L]$ being anomaly-free.
    
\end{itemize}

\section{Reduction of 7d three-form Chern--Simons theory}
\label{app:reduction}

In this appendix we study the compactifications of the 7d three-form Abelian Chern--Simons theory with action
\begin{equation}
\sum_{I,J}\frac{K_{IJ}}{4\pi}\int C^IdC^J~,
\end{equation}
where $C^I$ are three-form $U(1)$ gauge fields, 
$K_{IJ}=K_{JI}$ are integer symmetric matrix, and we compactify the theory on manifolds of the form $S^1\times Y_6$, $M_2\times Y^5$, $M_3\times Y_4$, $M_4\times Y_3$, $M_5\times Y_2$, $M_6\times Y_1$ with the subscripts labeling their dimension. We decompose the three-form gauge field $C$ as
\begin{equation}
C^I=\sum_{i,J}\left(\alpha^{IJ}_i B^{F,J}_{3-i} + \tau^{IJ}_i {\hat B}^J_{3-i}+{\hat \tau}^{IJ}_i B^J_{3-i}\right)~,
\end{equation}
where $d\hat \tau_i= n^{(i+1)}\tau_{i+1}$. $B_i$ is an off-shell $\mathbb{Z}/n^{(4-i)}\mathbb{Z}$ gauge field, while other fields are off-shell $U(1)$ higher-form gauge fields. 

Consider the manifold to be $M_6\times Y_1$. The reduction of $\frac{K_{IJ}}{4\pi}\int C^IdC^J$ gives
\begin{equation}
\frac{K_{IJ}}{2\pi} \int  (B_0^{I})^T\left(n^{(3)}B_1^{J}+d\hat B_0^{J}\right)
+\frac{K_{IJ}}{4\pi}\int (B_0^{F,I})^TdB_0^{F,J}~.
\end{equation}

Consider the manifold to be  $M_5\times Y_2$. 
On 5-manifold we have the isomorphism
\begin{equation}
\text{Tor }H_1(M_5)\cong \text{Tor }H^2(M_5)\cong 
\text{Tor }H^4(M_5)\cong \text{Tor }H_3(M_5),\quad
\text{Tor }H_2(M_5)\cong \text{Tor }H^3(M_5)~.    
\end{equation}
Thus $n^{(2)}=n^{(4)}$, and $\int \hat \tau_1\wedge \tau_4=\int \tau_2\wedge \hat\tau_3$. 
The reduction of $\frac{K_{IJ}}{4\pi}\int C^IdC^J$ gives
\begin{align}
&\frac{K_{IJ}}{2\pi}\int B_1^{F,I}dB_0^{F, J}+\frac{K_{IJ}}{2\pi}
\int \left(n^{(2)}B_2^I-d\hat B_1^I\right)B_0^J\cr
 &-\frac{K_{IJ}}{4\pi}\int\left( \left(B_1^{I}+(n^{(3)})^{-1}d\hat B_0^I\right)^{T}n^{(3)}\left(B_1^{J}+(n^{(3)})^{-1}d\hat B_0^J\right)-(d\hat B_0^I)^T(n^{(3)})^{-1}d\hat B_0^J\right)
~.\cr
\end{align}

Consider the manifold to be $M_4\times Y_3$.
On $4$-manifold we have $\text{Tor }H_0=\text{Tor }H^1=\text{Tor }H^4=\text{Tor }H_3$, and $\text{Tor }H_1=\text{Tor }H^2=\text{Tor }H^3=\text{Tor }H_2$.
Thus $n^{(1)}=n^{(4)}$, $n^{(2)}=n^{(3)}$. Denote the intersection pairing by $Q,\tilde Q$.
The reduction of $\frac{K_{IJ}}{4\pi}\int C^IdC^J$ gives
\begin{align}
&\frac{K_{IJ}}{2\pi}
\left(
\int \alpha_0^id\alpha_3^j\int B_0^{Ii}B_3^{Jj}
-\int \alpha^i_1 \alpha_3^j\int B_2^{Ii}dB_0^{Jj}\right)
+\cr
&\frac{K_{IJ}}{4\pi}\left(\int \alpha_2^i\alpha_2^j\int B_1^{F,Ii}dB_1^{F,Jj}-2\int \alpha^i_1 \alpha_3^j\int B_2^{F,Ii}dB_0^{F,Jj} \right)~.
\end{align}
The last term is an anomaly for the 0-form symmetry, given by the intersection form for two-cycles on $M_4$. 
For ${\frak g}={\frak u}(1)$ the 0-form symmetry in $T[M_4]$ is one $U(1)$ for each element in $H^2(M_4)$, and this reproduces the anomaly (2.6) and (2.7) of \cite{Dedushenko:2017tdw}.
The other terms are Berry phases in the bulk \cite{Hsin:2020cgg,Hsin:2022iug}, which can be interpreted as an anomaly in the space of coupling \cite{Cordova:2019jnf} $B_0^{Ii}$ . For connected $M_4$ we can drop the first term.

Consider the manifold to be  $M_3\times Y_4$. On 3-manifold we have $\text{Tor } H_0(M_3)\cong \text{Tor } H^1(M_3)\cong \text{Tor } H_2(M_3)\cong \text{Tor } H^3(M_3)$, and thus $n^{(3)}=n^{(1)}$.
The reduction of $\frac{K_{IJ}}{4\pi}\int C^IdC^J$ gives
\begin{align}
&\frac{K_{IJ}}{4\pi}\left(\int\alpha^i_1\alpha^j_2\int B_2^{F,Ii}dB_1^{F,Jj}
+\int \alpha_3^i\alpha_0^j\int B_0^{F,Ii}dB_3^{F,Jj}\right)\cr
&+\frac{K_{IJ}}{2\pi} \int\left(\left(-B_1^{I}n^{(1)}+d\hat B_0^I\right)^T Q B_3^{J}+B_1^{I}QdB_2^{J}\right)\cr
&+\frac{K_{IJ}}{4\pi}\int\left(
\left(B_2^{I}+(n^{(2)})^{-1}d\hat B_1^I\right)^T(\tilde Q\otimes n^{(2)})
\left(B_2^{J}+(n^{(2)})^{-1}d\hat B_1^J\right)
-d(\hat B_1^I)^T(\tilde Q\otimes (n^{(2)})^{-1})d\hat B_1^J
\right)
~.\cr
\end{align}
Thus we recover the statement that the linking form on $M_3$ gives the 't Hooft anomaly of the one-form symmetry in $T[M_3]$ (weighted by the coefficient $K$ in the 7d three-form Chern--Simons theory) \cite{Eckhard:2019jgg}. The last term is a mixed anomaly between the one-form and ordinary symmetry from the reduction of the 6d two-form symmetry, and the anomaly coefficient is given by the intersection form between one and two cycles on $M_3$ (the free parts in the homology contributes). The third term is a bulk Berry phase that presents an anomaly in the space of coupling.

Consider the manifold to be  $M_2\times Y_5$.
The reduction of $\frac{K_{IJ}}{4\pi}\int C^IdC^J$ gives
\begin{equation}
\frac{K_{IJ}}{2\pi}\left(\int \alpha^i_0d\alpha^j_1 \int B_3^{I,i}B_2^{J,j}+\int \alpha_0^i\alpha_2^j\int B_3^{Ii}dB_1^{Jj}\right)
-\frac{K_{IJ}}{4\pi}\int\alpha^i_1\alpha^j_1\int B_2^{I,i}dB_2^{J,j}~.
\end{equation}
The last term represents anomaly of one-form symmetry, given by the intersection form between one-cycles on $M_2$.
For instance, for $M_2=T^2$ and the polarization that the $C$ field has free boundary component along both $a,b$ cycles on the torus, this is a mixed anomaly between the electric and the magnetic one-form symmetries that generalizes the anomaly in the $U(1)$ gauge theory discussed in \cite{Gaiotto:2014kfa}.
For connected $M_2$ the first term is trivial and can be dropped, while for $M_2$ with several components the theory $T[M_2]$ can have two-form symmetry depending on the boundary condition for the $C$ field.

Consider the manifold to be  $M_1\times Y_6$.
The reduction of $\frac{K_{IJ}}{4\pi}\int C^IdC^J$ gives
\begin{equation}
\frac{K_{IJ}}{2\pi}
\int \alpha_0^i\alpha_1^j\int B_3^{Ii}dB_2^{Jj}
-
\frac{K_{IJ}}{4\pi}\int \alpha_0^id\alpha_0^j\int B_3^{Ii}B_3^{Jj}=
\frac{K_{IJ}}{2\pi}
\int \alpha_0^i\alpha_1^j\int B_3^{Ii}dB_2^{Jj}~.
\end{equation}
The first term describes a mixed anomaly between the two-form and one-form symmetries, given by the intersection form between 0-and 1-cycles on $M_1$.
For instance, if $M_1=S^1$ and the boundary condition for the $C$-field has free components along $S^1$ and transverse to $S^1$, then the first term describes the mixed anomaly between the electric one-form symmetry and the magnetic two-form symmetry in the 5d gauge theory from compactification on $M_1$.
The last term is an anomaly for the two-form symmetry, given by the linking form on $M_1$ between torsion 0-cycles. There are no torsion 0-cycles for one-manifold $M_1$ and thus such anomaly is not present in the 5d theory.

\section{Completeness of constraints for polarizations on manifolds with boundary}
\label{sec:completeness}

In this Appendix, we will show that the three constraints on the pair $(L_\delta,L)$ of polarization data on manifolds with a boundary in Section~\ref{sec:3Constraints} is in fact sufficient. In other words, any pair $(L,L_\delta)$ satisfying the three constraints gives a pure polarization on the manifold $M_{d}$ with boundary $M_{d-1}$ in the sense that the further reduction on $(N_{7-d},N_{6-d})$ gives a polarization on the 6-manifold 
\begin{equation}
Y_6=(M_{d-1}\times N_{7-d}) \bigcup_{M_{d-1}\times N_{6-d}} (M_{d}\times N_{6-d}).
\end{equation}

 Let $i^*$ denote the map $H^*(M_{d-1}\times N_{7-d})\oplus H^*(M_{d}\times N_{6-d})\rightarrow H^*(M_{d-1}\times N_{6-d}) $ in the Mayer--Vietoris sequence \eqref{MVOpen}. (Here and below, all omitted coefficients are in $D$.) To fix a Lagrangian subgroup $\Lambda$ of $H^3(Y_6)$, one needs to first make \eqref{MVOpen} a short exact sequence by quotienting out the image of $i^*$ and taking the kernel of $i^*$,
 \begin{equation}
     0\rightarrow H^2(M_{d-1}\times N_{6-d})/\mathrm{im}(i^2) \rightarrow H^3(Y_6)\rightarrow \mathrm{ker}(i^3) \rightarrow0.
 \end{equation}
 The image of $i^*$ are cocycles in $M_{d-1}\times N_{7-d}$ that come from restrictions of cocycles on either sides, while the kernel is given by a pair of cocycles from the two sides that can be glued along $M_{d-1}\times N_{7-d}$. 
 
 More precisely, the map $i^*=j^*-j'^*$ is the difference of the restriction map
 \begin{equation}
     j^*:\quad H^*(N_{7-d}\times M_{d-1})\rightarrow H^*(N_{6-d}\times M_{d-1})
 \end{equation}
 and the map
 \begin{equation}
     j'^*:\quad H^*(N_{6-d}\times M_{d})\rightarrow H^*(N_{6-d}\times M_{d-1}).
 \end{equation} Therefore, its kernel is the extension of the intersection of the images of the two maps above by the sum of the two kernels,
 \begin{equation}
     0\rightarrow \mathrm{ker}(j^*)\oplus  \mathrm{ker}(j'^*)\rightarrow\mathrm{ker}(i^*)\rightarrow \mathrm{im}(j^*)\cap  \mathrm{im}(j'^*)\rightarrow 0.
 \end{equation}
 For the purpose of discussion about pairings and isotropic subgroups, one can regard all short exact sequences of being split.\footnote{In fact, it is easy to see that if a non-degenerate bilinear  form on an Abelian group $G$ remains non-degenerate on a subgroup $H\subset G$, then $G=H\oplus G/H$ as one has a map $G\simeq G^\vee \rightarrow H^\vee \simeq H$ for which the inclusion $H\rightarrow G$ is a section.}  
 
 The pairing on $H^3(Y_6)$ is induced from that on ker$(j^3)$, ker$(j'^3)$ and the one between  $\mathrm{im}(j^3)\cap  \mathrm{im}(j'^3)$ and $H^2(M_{d-1}\times N_{6-d},D)/\mathrm{im}(i^2)$. As elements in ker$(j^3)$ (or ker$(j'^3)$) are relative 3-cocycles (i.e.~vanishing on boundary), there is a well defined non-degenerate intersection pairing. The last pairing is induced from that on $H^*(N_{6-d}\times M_{d-1})$ between degree 2 and 3. When im$(i^2)$ is modded out, the pairing remains perfect if one restricts to a subgroup of $H^3(N_{6-d}\times M_{d-1})$ that pairs trivially with im$(i^2)$, and it is not hard to see that this subgroup is exactly $\mathrm{im}(j^3)\cap  \mathrm{im}(j'^3)$ .
 
 Then a polarization on $(M_d,M_{d-1})$ is a family of choices, functorial with respect to $(N_{7-d},N_{6-d})$, of three maximal isotropic subgroups for the three pairings.
 
 When considering pure polarization, it is more convenient to express all relevant groups with coefficients in $H^*(M_d)$ and $H^*(M_{d-1})$.  
 We have 
 \begin{equation}
     H^n(M_{d-1}\times N_{7-d})\oplus H^n(M_{d}\times N_{6-d})\simeq H^*(N_{7-d},H^*(M_{d-1}))\{n\}\oplus H^*(N_{6-d},H^*(M_{d}))\{n\}
 \end{equation}
 and then map $i^*=j^*-j'^*$ is the difference between
 \begin{equation}
     j^*:\quad H^*(N_{7-d},H^*(M_{d-1}))\rightarrow H^*(N_{6-d},H^*(M_{d-1}))
 \end{equation}
 and the map
 \begin{equation}
     j'^*:\quad H^*(N_{6-d},H^*(M_{d}))\rightarrow H^*(N_{6-d},H^*(M_{d-1})) 
 \end{equation}
 induced from $H^*(M_{d})\rightarrow H^*(M_{d-1})$. 
 
 Just as in the case of closed manifolds, a choice of $L\subset H^*(M_{d-1})$ determines a maximal isotropic subgroup for the kernel of $j^3$ in a functorial way. On the other hand, a choice of $L_\delta$ gives a subgroup in the kernel of $j'^3$ and a subgroup of $\mathrm{im}(j^3)\cap  \mathrm{im}(j'^3)$, by decomposing the image of the map
 \begin{equation}
     H^*(N_{6-d},L_\delta)\rightarrow H^*(N_{6-d},H^*(M_{d})).
 \end{equation}
 The condition $\partial L_\delta \subset L$ guarantees that the image of the above map, after applying $j'^3$, is also in the image of $j^3$, and the maximal isotropy condition on $L_\delta$ ensures that the subgroup in ker$(j'^3)$ given by $L_\delta$ is maximal isotropic. The last thing to check is that the subgroup in $H^2(M_{d-1}\times N_{6-d},D)/\mathrm{im}(i^2)$ given by $L$ pairs trivially with the subgroup in $\mathrm{im}(j^3)\cap  \mathrm{im}(j'^3)$ given by $L_\delta$ and is maximal. This is guaranteed by the condition $\partial L_\delta = L\cap \im(\partial)$ in \eqref{LLdelta}. After this, one has specified all data needed to define a maximal isotropic subgroup $\L\subset H^3(Y_6)$, and such choice is obviously functorial as no data associated with $N_{7-d}$ and $N_{6-d}$ were used.

\section{5d discrete theta angle and Witten's $Sp(n)$ anomaly}

In this appendix we discuss the 5d discrete theta angle for $Sp(n)$ gauge field that describes Witten's anomaly in 4d (if the gauge field is promoted to be dynamical), by embedding the discrete theta angle into 5d $SU(2n+1)$ Chern--Simons term.
We will focus on the case $n=1$, first treating the gauge field as background fields.

Let us start by discussing the anomaly of $SU(2)$ symmetry in 3+1 dimensions.
Consider three Weyl fermions in the fundamental of $SU(3)$.
In the instanton background, there is a fermion zero mode
\begin{equation}
    n_L-n_R=1~.
\end{equation}
Let us consider an $SU(2)$ subgroup inside $SU(3)$. Then in the $SU(2)$ instanton background, which is also $SU(3)$ instanton background, there is the same zero mode. But then the zero modes in the context of $SU(2)$ is the Witten anomaly: the path integral is not invariant under fermion parity, which is an element in the $SU(2)$ gauge group \cite{Witten:1982fp,Wang:2018qoy}.
The zero mode is also associated with the chiral anomaly for $SU(3)$ symmetry. Thus we conclude that the $SU(2)$ Witten anomaly can also be interpreted as $SU(3)$ chiral anomaly.

\subsubsection*{5d perspective: SPT phase} The Witten anomaly is described by an SPT phase of $SU(2)$ symmetry in 5d, given by the discrete theta angle, by the anomaly inflow mechanism or bulk-boundary correspondence.
On the other hand, the chiral anomaly of $SU(3)$ symmetry is described by the 5d Chern--Simons term for $SU(3)$ symmetry at level one. Thus we conclude that the Witten anomaly can be embedded in the $SU(3)$ Chern--Simons term under the inclusion $SU(2)\subset SU(3)$.

The relation can also be understood as follows.
The Witten anomaly depends on the spin structure as
\begin{equation}
    \int_{5d} {d\text{CS}_3^{SU(2)}\over 2\pi}\eta~,
\end{equation}
where $\eta$ is the $\mathbb{Z}_2$ one-cochain describes the spin structure. The anomaly inflow on the boundary from this action implies that in the background with an odd $SU(2)$ instanton number, the theory is not invariant under fermion parity symmetry, which is Witten's anomaly \cite{Witten:1982fp}.
If we turn on a background spin$^c$ connection $A$, which satisfies $\oint dA\equiv\pi\oint w_2 \pmod {2\pi}$, then after integration by parts, the action can be described as
\begin{equation}\label{eqn:5ddiscrete}
\int_{5d}     \text{CS}_3^{SU(2)}  {dA\over 2\pi}~.
\end{equation}

On the other hand, the $SU(3)$ Chern--Simons term depends on the spin structure as follows. If we turn on a background spin$^c$ connection $A$, then the Chern--Simons term is \cite{Gukov:2020btk}
\begin{equation}
\int_{5d}    \text{CS}_5^{SU(3)}+\text{CS}_3^{SU(3)}\frac{dA}{2\pi}~.
\end{equation}
If we substitute $SU(2)\subset SU(3)$ gauge field, the first term vanishes, and we recover the discrete theta angle (\ref{eqn:5ddiscrete}). 

The discussion can be generalized to $SU(N)$. For even $N$, the fermion parity can be identified as $\mathbb{Z}_2$ subgroup in the center of $SU(2)$ and $SU(N)$. Thus we can consider ${SU(N)\times \text{Lorentz}\over \mathbb{Z}_2}$ gauge field, instead of $SU(N)$ gauge field. The argument about the anomaly remains the same.

\section{Topological term from KK modes}

Let us start with a fermions coupled to a gauge field $A$ in $(d+1)$ spacetime dimensions,
\begin{equation}
    i\bar\psi \gamma^\mu (\partial_\mu-iA_\mu)\psi~,
\end{equation}
and reduce it on a circle or torus. We will investigate how the low-energy theory depends on the holonomy of the gauge field.

\subsubsection*{$S^1$ reduction}

Let us denote the coordinate for the $(d+1)$st dimension by $z\sim z+2\pi R$. We decompose the fields as
\begin{equation}
    \psi=\sum \psi_n e^{in z/R},\quad A^\mu=\sum_m A^\mu_m e^{im z/R}~,
\end{equation}
where $n$ is integer or half integer depending on the spin structure along the circle. Let us take $\psi_n$ to be an eigenvector of $\gamma^{d+1}$ (here we take the spacetime index to be $\mu=1,\cdots,d+1$):
\begin{equation}
    \gamma^{d+1}\psi_n=\psi_n~,
\end{equation}
and similar for another KK fermion with a minus sign on the right hand side.
Then the KK mode $\psi_n$ has action
\begin{equation}
    \sum_n i\bar\psi_{-n}\left(\gamma^{\mu'}\partial_{\mu'}+i{n\over R}\right)\psi_n
    +\sum_{n,m}\left( i\bar\psi_{-n-m}\left(-iA^{d+1}_{m}\right)\psi_{n}
    +i\bar\psi_{-n-m}\left(-i\gamma^{\mu'}A^{\mu'}_{m}\right)\psi_{n}\right)~,
\end{equation}
where $\mu'=1,\cdots,d$.

We note that $\psi_n$ for non-zero $n$ are massive fermions: the mass that couples $\psi_{-m},\psi_n$ is
\begin{equation}
    M_{m,n}=\delta_{m,n}\frac{n}{R}-A^{d+1}_{m-n}~.
\end{equation}

In particular, the scalar mode $A^{d+1}_0$ contributes to the Yukawa coupling
\begin{equation}
    M_{n,n}=\frac{n}{R}-A^{d+1}_0~.
\end{equation}
Thus when the value of $A^{d+1}_0$ passes through integers $\frac{n}{R}$, the fermion mass changes sign and there is additional topological term generated.
Denote
\begin{equation}
    f(x)=\sum_n h(x-{n\over R})~.
\end{equation}
We note that $f'(x)$ is the density for an energy spectrum with equal spacing $1/R$.
The
 low-energy action contains
\begin{equation}
\pi f(A^{d+1}_0)\text{Tr}\left(\hat A(R)e^{F'/(2\pi)}\right)
~.
\end{equation}

\subsubsection*{$T^2$ reduction}

Let us compactify the theory on a torus with complex parameter $\tau=\tau_1+i\tau_2$. Let us denote its angle by $\varphi$.
Denote the circle coordinates by $u,v$, they are related to the Cartesian coordinates $x,y$ by
\begin{equation}
    x=\cos\varphi u - \sin\varphi v,\quad 
    y=\sin\varphi u+\cos\varphi v~.
\end{equation}
The derivatives are
\begin{equation}
    \partial_x=\cos\varphi \partial_u -\sin\varphi \partial_v,\quad
    \partial_y=\sin\varphi \partial_u +\cos\varphi \partial_v~.
\end{equation}

We decompose the fields as
\begin{equation}
    \psi=\sum_{n,m}\psi_{n,m}e^{in u+im v},\quad 
    A^\mu =\sum_{n,m}A^\mu_{m,n}e^{inu+imv}~,
\end{equation}
where in the decomposition of $\psi$, $n,m$ are integers or half-integers depending on the spin structure on the circles.
We take the KK mode $\psi_{m,n}$ to be eigenvectors of $\gamma^{d+1}$. Since $\gamma^{d+1}$ anticommute with $\gamma^{d}$, they cannot be both diagonalized, $\psi_{m,n}$ in general will have two components in the eigenbasis of $\gamma^d$. Then the kinetic term written in terms of $u,v$
is
\begin{align}
  &  i\bar\psi^+_{-m,-n}\left(\gamma^{\mu'}\partial_{\mu'}+
    i\frac{n\cos\varphi-m
    sin\varphi}{R}+i\frac{n\cos\varphi+m\sin\varphi}{R'} \right)\psi^+_{m,n}=0,\quad \cr
 &   i\bar\psi^-_{-m,-n}\left(\gamma^{\mu'}\partial_{\mu'}
    +i\frac{n\cos\varphi-m
    sin\varphi}{R}-i\frac{n\cos\varphi+m\sin\varphi}{R'} 
    \right)\psi^-_{m,n}=0~,
\end{align}
where $\psi_{m,n}^\pm$ are the two components that correspond to two eigenvalues of $\gamma^d$.

The KK tower of the fermion generates the topological term proportional to (denote $\tau=R/R'$)
\begin{equation}
    \sum_{n,m} \text{sign}\left(\frac{n/R+m/R'}{n/R-m/R'}\right)=\sum_{n,m}\text{sign}\left(\frac{n+m\tau}{n-m\tau}\right)~.
\end{equation}

\section{$\hat Z$ for knot complements and the volume conjecture (by Sunghyuk Park)}\label{sec:FKvolume}

This appendix complements section~\ref{sec:volume} with the analysis of volume conjecture for $\hat Z$ invariants of knot complements.

Let $J_{K,n}(q)$ be the $n$-colored Jones polynomial, normalized so that $J_{K,1}(q) = 1$ and $J_{\text{unknot},n}(q) = 1$ for all $n$.
Define 
\begin{equation}
J_K(y,q) := \sum_{n \geq 1} J_{K,n}(q)y^{-n}
\end{equation}
to be their generating series. Let 
\begin{equation}
F_K(x,q) = \sum_{n\geq 0}F_{K,n}(q)x^n
\end{equation}
be the $\hat Z$ invariant for knot complements \cite{Gukov:2019mnk}, normalized so that $F_{\text{unknot}}(x,q) = 1$. This series is well-defined for closures of homogeneous braids \cite{Park}, in which case the coefficients $F_{K,n}(q)$ are Laurent polynomials; in the analysis below, we will assume that $K$ is a homogeneous braid knot.  

The sequence of polynomials $J_{K,n}(q)$ and $F_{K,n}(q)$ are in the kernel of the same $q$-difference operator $\hat{A}(\hat{x},\hat{y})$, and as a result, they have many similar features. 
However, while the colored Jones polynomials have been studied for several decades, the polynomials $F_{K,n}(q)$ are relatively new and haven't been studied as much. 
For instance, it is well-known that one can obtain the hyperbolic volume of a knot complement from a certain asymptotics of $J_{K,n}(q)$, while the analogous statement for $F_{K,n}(q)$ hasn't appeared in the literature to the best of our knowledge.  
One of the motivations of this appendix is to fill in this gap by studying asymptotic series associated to the polynomials $F_{K,n}(q)$ analogous to that of $J_{K,n}(q)$. 

From the sequence of polynomials $J_{K,n}(q)$, we can obtain various perturbative expansions associated to branches of the A-polynomial curve $A_K(x,y) = 0$. 
This is a well-studied subject, and we summarize some of the relevant facts below.
For a hyperbolic knot $K$, by taking the large $n$ asymptotics of $J_{K,n}(e^{\frac{2\pi i}{n}})$, we get \cite{Gukov:2003na}:
\begin{equation}\label{eq:volumeconj}
J_{K,n}(e^{\frac{2\pi i}{n}}) \underset{n \rightarrow \infty}{\sim} e^{\frac{V_K}{2\pi}n}n^{\frac{3}{2}} Z_{\text{pert}}^{\alpha_1} \qty(K, \frac{2\pi i}{n}),
\end{equation}
where $V_K$ is the complexified volume of the knot complement, and $Z_{\text{pert}}^{\alpha_1}(K, h)$ is the formal power series
\begin{equation}
Z_{\text{pert}}^{\alpha_1}(K, h) \in \overline{\mathbb{Q}}[[h]]
\end{equation}
associated to the geometric branch $y^{\alpha_1}(x)$ of the A-polynomial curve at $x = 1$. 
For instance, 
\begin{equation}
Z_{\text{pert}}^{\alpha_1}(\mathbf{4}_1, h) = \frac{1}{\sqrt[4]{3}}\qty(1 + \frac{11}{72\sqrt{-3}}h + \frac{697}{2(72\sqrt{-3})^2}h^2 + \frac{724351}{30(72\sqrt{-3})^3}h^3 + \cdots) \in \frac{1}{\sqrt[4]{3}} \mathbb{Q}(\sqrt{-3})[[h]],
\end{equation}
where $\mathbb{Q}(\sqrt{-3})$ is the trace field of $\mathbf{4}_1$.

When $K$ is not hyperbolic, the right-hand side of \eqref{eq:volumeconj} is in general a combination of several perturbative trans-series contributions associated to flat connections whose real part of classical action is $0$.
For instance, for the left-handed trefoil $\mathbf{3}_1$, 
\begin{equation}
J_{\mathbf{3}_1,n}(e^{\frac{2\pi i}{n}}) \underset{n\rightarrow \infty}{\sim}
e^{\frac{2\pi i}{24}n}n^{\frac{3}{2}}
Z_{\text{pert}}^{\alpha_1} \qty(\mathbf{3}_1, \frac{2\pi i}{n})
+ Z_{\text{pert}}^{\alpha_0} \qty(\mathbf{3}_1, \frac{2\pi i}{n}),
\end{equation}
where
\begin{align*}
Z_{\text{pert}}^{\alpha_1} (\mathbf{3}_1,h) &= e^{-\frac{2\pi i}{8}}\qty(1 -\frac{23}{2^3 \cdot 3}h + \frac{529}{(2^3 \cdot 3)^2} \frac{h^2}{2!} -\frac{12167}{(2^3 \cdot 3)^3}\frac{h^3}{3!} +\frac{279841}{(2^3 \cdot 3)^4}\frac{h^4}{4!} +\cdots ) = e^{-\frac{2\pi i}{8}} q^{-\frac{23}{24}},\\
Z_{\text{pert}}^{\alpha_0} (\mathbf{3}_1, h) &= 1 + 0h + 2\;\frac{h^2}{2!} + 12\;\frac{h^3}{3!} + 146\; \frac{h^4}{4!} + 2580\; \frac{h^5}{5!} + 63722\; \frac{h^6}{6!} +\cdots.
\end{align*}
The Melvin-Morton-Rozansky expansion gives
\begin{equation}\label{eq:MMR}
	J_{K,n}(e^h) \underset{\substack{n \rightarrow \infty \\ x = e^{n h} \text{ fixed}}}{\sim} \frac{1}{\Delta_K(x)} + \frac{P_1(x)}{\Delta_K(x)^3}\frac{h}{1!} + \frac{P_2(x)}{\Delta_K(x)^5}\frac{h^2}{2!} + \cdots,
\end{equation}
where $P_n(x)$ are some Laurent polynomials in $x$. This is the perturbative series associated to the trivial branch $y^{\alpha_0}(x) = 1$ of the A-polynomial curve. 
Since in some sense $F_K(x,q)$ is a non-perturbative completion of the Melvin-Morton-Rozansky expansion, we will often write the right-hand side of \eqref{eq:MMR} as $F_K(x,e^h)$.
Note, $Z_{\text{pert}}^{\alpha_0} (\mathbf{3}_1, h)$ in the previous example is exactly $F_{\mathbf{3}_1}(1,e^h)$.

Taking the perturbative expansion of the generating series, we get
\begin{equation}\label{eq:yMMR}
	J_K(y,e^h) = \frac{1}{y-1} + \frac{Q_1(y)}{(y-1)^2}\frac{h}{1!} + \frac{Q_2(y)}{(y-1)^3}\frac{h^2}{2!} + \cdots,
\end{equation}
where $Q_n(y)$ are some Laurent polynomials in $y$.
There are a few ways to further specialize this series. One is to set $y=-1$: 
\begin{equation}
		J_K(-1,e^h) = -\frac{1}{2}F_K(1,e^h). 
\end{equation}
Another is to take the residue of $\frac{1}{y}J_K(y,e^h)$ either at $y=0$ or at $y=1$: 
\begin{equation}
		\mathrm{Res}_{y=1}\qty(\frac{1}{y}J_K(y,e^h)) = -\mathrm{Res}_{y=0}\qty(\frac{1}{y}J_K(y,e^h)) = F_K(1,e^h). 
\end{equation}

Now, we can do the same for the sequence of polynomials $F_{K,n}(q)$, completely in parallel. 
By taking the large $n$ asymptotics of $F_{K,n}(e^{\frac{2\pi i}{n}})$, in case $K$ is hyperbolic, we conjecture the following

{\bf Conjecture} (Volume conjecture for $F_K$):
\begin{equation}\label{eq:FKvolconj}
		F_{K,n}(e^{\frac{2\pi i}{n}}) \underset{n \rightarrow \infty}{\sim} e^{\frac{V_K}{2\pi}n}n^{\frac{1}{2}} F_{\text{pert}}^{\alpha_1} \qty(K, \frac{2\pi i}{n}),
\end{equation}
\emph{where $V_K$ is the complexified volume of the knot complement, and $F_{\text{pert}}^{\alpha_1} (K, h)$ is the formal power series}
\begin{equation}
		F_{\text{pert}}^{\alpha_1} (K, h) \in \overline{\mathbb{Q}}[[h]]
\end{equation}
\emph{associated to the geometric branch $x^{\alpha_1}(y)$ of the A-polynomial curve at $y = -1$.}

{\bf Remarks:}
In comparison with the volume conjecture for colored Jones polynomials,
\begin{itemize}
\item The exponential term is still $e^{\frac{V_{K}}{2\pi}n}$. 
\item The power of $n$ after the exponential factor is $\frac{1}{2}$ here, which is different from $\frac{3}{2}$ of the asymptotic expansion of colored Jones polynomials.
\item In the examples we have considered, the perturbative part $F_{\text{pert}}^{\alpha_1} (K, h)$ is the same as $Z_{\text{pert}}^{\alpha_1} (K, h)$ up to sign. 
\end{itemize}

For example, in case of the $\mathbf{4}_1$ knot, 
\begin{align*}
F_{\mathbf{4}_1}(x,q) 
&= -x -3x^2 -(q^{-1}+6+q)x^3 -(2q^{-2}+3q^{-1}+11+3q+2q^2)x^4 -\cdots,
\end{align*}
and more explicitly, 
\begin{equation}
		F_{\mathbf{4}_1,n}(q) = -\sum_{0\leq i\leq j \leq n-1}\qbin{j+i}{2i}.
\end{equation}
By studying the asymptotics of $F_{\mathbf{4}_1,n}(e^{\frac{2\pi i}{n}})$, \eqref{eq:FKvolconj} can be numerically verified, with
\begin{equation}
F_{\text{pert}}^{\alpha_1} (\mathbf{4}_1, h) = -\frac{1}{\sqrt[4]{3}}\qty(1 + \frac{11}{72\sqrt{-3}}h + \frac{697}{2(72\sqrt{-3})^2}h^2 + \frac{724351}{30(72\sqrt{-3})^3}h^3 +\cdots).
\end{equation}
Note, this is exactly $-Z_{\text{pert}}^{\alpha_1} (\mathbf{4}_1, h)$.
In fact, in this case, the exponential part of the volume conjecture can be proved analytically:

{\bf Theorem:}
\emph{We have}
\begin{equation}
\lim_{n \rightarrow \infty} \frac{\log F_{\mathbf{4}_1, n}(e^{\frac{2\pi i}{n}})}{n}
=
\frac{V_{\mathbf{4}_1}}{2\pi}.
\end{equation}

\emph{Proof:}
Setting $k=j-i$ and simplifying the expression, we need to show that
\[
\lim_{n\rightarrow \infty} \frac{1}{n} \log 
\left(
\sum_{\substack{0\leq i\leq \lfloor \frac{n-1}{2} \rfloor \\ 0 \leq k\leq n-1-2i}} \exp \sum_{l=1}^{2i} 
\left(
\log \sin \left(\pi \frac{k+l}{n}\right) - \log \sin \left( \pi \frac{l}{n}\right) 
\right)
\right)
= \frac{1}{\pi} D(e^{\frac{2\pi i}{6}}),
\]
where $D$ is the Bloch-Wigner function; $D(e^{\frac{2\pi i}{6}})$ is the hyperbolic volume of the regular ideal tetrahedron. 
Since the outside summation is a summation of positive numbers over $\sim \frac{n^2}{4}$ pairs $(i,k)$, the left-hand side is equal to 
\begin{align*}
&\lim_{n\rightarrow \infty} \frac{1}{n} 
\log 
\left(
\max_{\substack{0\leq i\leq \lfloor \frac{n-1}{2} \rfloor \\ 0 \leq k\leq n-1-2i}} \exp 
\sum_{l=1}^{2i} 
\left(
\log \sin \left(\pi \frac{k+l}{n}\right) - \log \sin \left( \pi \frac{l}{n}\right) 
\right)
\right)\\
&= 
\frac{1}{\pi}
\lim_{n\rightarrow \infty}  
\max_{\substack{0\leq i\leq \lfloor \frac{n-1}{2} \rfloor \\ 0 \leq k\leq n-1-2i}}
\frac{\pi}{n}
\sum_{l=1}^{2i} 
\left(
\log \sin \left(\pi \frac{k+l}{n}\right) - \log \sin \left( \pi \frac{l}{n}\right) 
\right).
\end{align*}
Using Euler-Maclaurin formula, we can replace the summation into an integral, and setting $\theta = \pi \frac{2 i}{n}$ and $\theta' = \pi \frac{k}{n}$, the above expression becomes
\begin{align*}
&\frac{1}{\pi} \max_{\substack{0 \leq \theta, \theta', \\ \theta + \theta' \leq \pi}}
\left(
- (\Lambda(\theta+\theta') - \Lambda(\theta')) + \Lambda(\theta)
\right)\\
&= 
\frac{1}{\pi} \max_{\substack{0 \leq \theta, \theta', \\ \theta + \theta' \leq \pi}}
\left(
\Lambda(\theta) + \Lambda(\theta') + \Lambda(\pi - \theta - \theta')
\right),
\end{align*}
where $\Lambda(\theta) := -\int_{0}^{\theta} \log |2\sin(x)|dx$ is the Lobachevsky function. 
Since $\Lambda(\theta) + \Lambda(\theta') + \Lambda(\pi - \theta - \theta')$ is exactly the hyperbolic volume of the ideal tetrahedron with dihedral angles $\theta, \theta', \pi - \theta-\theta'$, which is maximized exactly for the regular ideal tetrahedron, we conclude that this is equal to $\frac{1}{\pi} D(e^{\frac{2\pi i}{6}})$. ~~~$\Box$

When $K$ is not hyperbolic, the right-hand side of \eqref{eq:FKvolconj} is in general a combination of transseries associated to contributions of flat connections whose real part of classical action is $0$. 
For instance, 
\begin{align*}
	F_{\mathbf{3}_1,n}(e^{\frac{2\pi i}{n}}) \underset{n\rightarrow \infty}{\sim} 
	& e^{\frac{2\pi i}{24}n}n^{\frac{1}{2}}
    F_{\text{pert}}^{\alpha_1} \qty(\mathbf{3}_1, \frac{2\pi i}{n}) 
	+ F_{\text{pert}}^{\alpha_a} \qty(\mathbf{3}_1, \frac{2\pi i}{n}) 
	+ e^{-\frac{2\pi i}{3}n}F_{\text{pert}}^{\alpha_b} \qty(\mathbf{3}_1, \frac{2\pi i}{n})
\end{align*}
where
\begin{align*}
	F_{\text{pert}}^{\alpha_1} (\mathbf{3}_1, h) &= e^{-\frac{2\pi i}{8}}\qty(1 -\frac{23}{2^3 \cdot 3}h + \frac{529}{(2^3 \cdot 3)^2} \frac{h^2}{2!} -\frac{12167}{(2^3 \cdot 3)^3}\frac{h^3}{3!} +\frac{279841}{(2^3 \cdot 3)^4}\frac{h^4}{4!} +\cdots ) = e^{-\frac{2\pi i}{8}} q^{-\frac{23}{24}},\\
	F_{\text{pert}}^{\alpha_a} (\mathbf{3}_1, h) &= \frac{i}{\sqrt{3}}\qty(1 -\frac{2}{3}h + \frac{8}{3^2} \frac{h^2}{2!} +\frac{22}{3^3} \frac{h^3}{3!} +\frac{1136}{3^4}\frac{h^4}{4!}  + \cdots),\\
	F_{\text{pert}}^{\alpha_b} (\mathbf{3}_1, h) &= \frac{i}{2\sqrt{3}}\qty(1 -\frac{11}{2^2 \cdot 3}h + \frac{122}{(2^2 \cdot 3)^2} \frac{h^2}{2!} -\frac{1358}{(2^2 \cdot 3)^3}\frac{h^3}{3!} + \frac{15176}{(2^2 \cdot 3)^4}\frac{h^4}{4!} + \cdots).
\end{align*}
Note, $F_{\text{pert}}^{\alpha_1} (\mathbf{3}_1, h) = Z_{\text{pert}}^{\alpha_1} (\mathbf{3}_1, h)$. 

It is natural to ask whether $F_{K,n}(q)$ have Melvin-Morton-Rozansky-like expansion. Unlike $J_{K,n=\frac{u}{h}}(e^h)$, whose coefficients (as a power series in $h$) are polynomials in $u$, the coefficients of $F_{K,n=\frac{v}{h}}(e^{h})$ (as a power series in $h$) grow exponentially in $v$. 
Therefore, there is no naive analogue of Melvin-Morton-Rozansky expansion for $F_{K,n}(q)$. 
Still, it is an interesting question whether we can obtain the expansion \eqref{eq:yMMR} as some asymptotic expansion of $F_{K,n}(q)$. 

{\bf Question:} Does a relation $F_{K,n}(e^{h}) \overset{?}{\sim} J_K(e^{n h},e^{h})$ hold in some asymptotic expansion?

The perturbative expansion of $F_K(x,e^{h})$ is the same as the Melvin-Morton-Rozansky expansion \eqref{eq:MMR} of the colored Jones polynomials
\begin{equation}
	F_K(x,e^{h}) = \frac{1}{\Delta_K(x)} + \frac{P_1(x)}{\Delta_K(x)^3}\frac{h}{1!} + \frac{P_2(x)}{\Delta_K(x)^5}\frac{h^2}{2!} + \cdots.
\end{equation}
There are a few ways to further specialize this series. One is to set $x=1$:
\begin{equation}
		F_K(1,e^h) = -2 J_K(-1,e^h).
\end{equation}
Another is to take residues of $\frac{1}{x}F_K(x,e^{h})$ at some root of $\Delta_K(x) = 0$. In this way, we obtain some perturbative series that appear in the asymptotics of $F_{K,n}(q)$.
For instance, the series $F_{\text{pert}}^{\alpha_a} (\mathbf{3}_1, h)$ that we saw earlier can be obtained as a residue:
\begin{equation}
		F_{\text{pert}}^{\alpha_a} (\mathbf{3}_1, h) = -\mathrm{Res}_{x=e^{\frac{2\pi i}{6}}}\qty(\frac{1}{x} F_{\mathbf{3}_1}(x,e^h) ) = \mathrm{Res}_{x=e^{-\frac{2\pi i}{6}}}\qty(\frac{1}{x} F_{\mathbf{3}_1}(x,e^h)).
\end{equation}

\bibliographystyle{utphys}
\bibliography{biblio}{}

\end{document}